\documentclass[
  amssymb, amsmath,
  floatfix,
 reprint,%
 aip,jcp,%
]{revtex4-1}
\usepackage{xcolor}

\usepackage{graphicx}
\usepackage{scrextend}
\usepackage{simplewick}

\usepackage{bm}
\usepackage[colorlinks=true,linkcolor=blue]{hyperref}
\expandafter\ifx\csname package@font\endcsname\relax\else
 \expandafter\expandafter
 \expandafter\usepackage
 \expandafter\expandafter
 \expandafter{\csname package@font\endcsname}
\fi
\begin{document}

\title{Equation-of-Motion Coupled-Cluster Theory based on the 4-component Dirac--Coulomb(--Gaunt) Hamiltonian. Energies for single electron detachment, attachment and electronically excited states}

\author{Avijit Shee}
\email{ashee@umich.edu}
\affiliation{Department of Chemistry, University of Michigan, 930 N.\ University, Ann Arbor, MI 48109-1055, USA} 
\affiliation{Universit\'e de Lille, CNRS, UMR 8523 -- PhLAM -- Physique des Lasers, Atomes et Mol\'ecules, F-59000 Lille, France Fax: +33-3-2033-7020; Tel: +33-3-2043-4163}

\author {Trond Saue}
\email{trond.saue@irsamc.ups-tlse.fr}
\affiliation{Laboratoire de Chimie et Physique Quantiques, UMR 5626 CNRS --- Universit\'e Toulouse III--Paul Sabatier, 118 route de Narbonne, F-31062 Toulouse, France Fax: +33-5-6155-6065; Tel: +33-5-6155-6031}

\author {Lucas Visscher}
\email{l.visscher@vu.nl}
\affiliation{Division of Theoretical Chemistry, Faculty of Sciences, Vrije Universiteit Amsterdam, De Boelelaan 1083, 1081 HV Amsterdam, The Netherlands, Tel: +31-20-598-7624}

\author{Andr\'e Severo Pereira Gomes}
\email{andre.gomes@univ-lille.fr (corresponding author)}
\affiliation{Universit\'e de Lille, CNRS, UMR 8523 -- PhLAM -- Physique des Lasers, Atomes et Mol\'ecules, F-59000 Lille, France Fax: +33-3-2033-7020; Tel: +33-3-2043-4163}

\date{\today}
\revised{\today}

\begin{abstract}
We report in this paper an implementation of 4-component relativistic Hamiltonian based Equation-of-Motion Coupled-Cluster with singles and doubles (EOM-CCSD) theory for the calculation of ionization potential (IP), electron affinity (EA) and excitation energy (EE). In this work we utilize previously developed double group symmetry-based generalized tensor contraction scheme, and also extend it in order to carry out tensor contractions involving non-totally symmetric and odd-ranked tensors. Several approximated spin-free and two-component Hamiltonians can also be accessed in this implementation. We have applied this method to the halogen monoxide (XO, X= Cl, Br, I, At, Ts) species, in order to assess the quality of a few other recent EOM-CCSD implementations, where spin-orbit coupling contribution has been approximated in different degree. Besides, we also have studied various excited states of CH$_2$IBr, CH$_2$I$_2$ and I$_3^-$ (as well as single electron attachment and detachment electronic states of the same species) where comparison has been made with a closely related multi-reference coupled-cluster method, namely Intermediate Hamiltonian Fock Space Coupled-Cluster singles and doubles (IHFS-CCSD) theory.          
\end{abstract}

\maketitle

\section{Introduction}

Theoretical approaches based on molecular quantum mechanics\cite{grimme_calculation_2004,Helgaker2012,correlation-szalay-cr2011-112-108,Lyakh:2012cn,Casida:2012gy} have grown into increasingly important tools to help experimentalists understand species in their electronically excited states in the gas-phase\cite{GIBSON:2006hd} or in complex environments\cite{senn_qm/mm_2009,env-gomes-arpcspc2012-108-222,Gordon:2012dk}, and with that address speciation (oxidation states of specific centers, structures)\cite{core-bagus-ssr2013-68-273,Geckeis:2013jb,Wilson:2016hb,Seidel:2016iga,Purgel:2011ds,Sures:2017hd} as well as the underlying factors driving photochemical processes (reactivity, photodissociation etc.)\cite{Liu:2009hf,SaizLopez:2012is,Xing:2014gp,Starik:2015gz,Hochlaf:2017cw}. 

Computational models are particularly important for species containing (a) first-row transition metals, since then one often encounters dense electronic spectra due to many low-lying quasi-degenerate states arising from the partially-filled $d$ shells which requires the treatment of both dynamical and non-dynamical electron correlation\cite{Shavitt2009}; and (b) heavy elements (e.g.\ those with atomic number Z = 31 or higher) for which the manifestations of relativistic effects\cite{Pyykko1988,saue:hamprimer,pyykko:2012bt,pyykko:2012kh,Autschbach2012,dyall-faegri-book-2007,reiher_book} significantly alter the species' electronic structure and, by extension, their excited states and other molecular properties\cite{saue:molprop,Saue2005,nrsbook}. However, as electron correlation and relativistic effects are non-additive, in order to achieve a balanced description of the electronic states of heavy element species both have to be treated on the same footing. 

Among the available relativistic (multireference) methods\cite{Timo_overview}, those based on the coupled-cluster ansatz\cite{bartlett:reviewcc,Lyakh:2012cn} and in particular the Fock-space coupled-cluster (FS-CC)\cite{correlation-Visscher-JCP2001-115-9720} method and its intermediate Hamiltonian (IHFS-CC) formulations\cite{correlation-Landau-JCP2000-113-9905,correlation-Landau-JCP2001-115-6862,Landau:2004dd,Eliav:2012gp}, have shown to be among the most reliable and cost-effective ones, and allowed the treatment of systems with complex, open-shell ground-states~\cite{actinide-infante-jcp2006-125-074301,actinide-Infante-JCP2007-127-124308,actinide-Gomes-PCCP2008-10-5353,ruiprez_ab_2009,ATO-Rota-JCP2011-135-114106,PereiraGomes:2014iy,Denis:2015ez,Parmar:2014kn,Figgen:2008hd,Weigand:2009dq,Nikoobakht:2015ko} as well as simpler, closed-shell ones\cite{real09,gomes10,pawel1,Tecmer:2012fc,actinide-Gomes-PCCP2013-15-15153,Tecmer:2014fs}.
The appeal of (IH)FS-CC resides in the fact that they show a similar scaling to the of single-reference approaches (e.g.\ O(N$^6$) for (IH)FS-CC with coupled cluster singles doubles (CCSD)-based wavefunctions, can be implemented in a straightforward manner starting from a single-reference coupled-cluster code\cite{correlation-Visscher-JCP2001-115-9720} and yield several excited states simultaneously. Furthermore, different oxidation states belonging to different sectors of Fock space\cite{Haque:1984dk} can be obtained from a single calculation, a useful feature for actinide chemistry, where often spectra of the same species in different oxidation states are to be analyzed.
There are, however, some important drawbacks in the requirement to define suitable model spaces\cite{Lindgren:1974kc,Lindgren:1987in} and the problem of intruder states\cite{Kaldor:1988wp,Kaldor:1991cd} which may prevent convergence for one of more sectors. The latter issue can be alleviated but not completely avoided by the use of the intermediate Hamiltonian formalism\cite{Malrieu:1985ce}, having as consequence that the IHFS-CC variants are the only ones that have been widely used in molecular applications employing relativistic Hamiltonians.

For the transition energies of closed-shell ground states and for determining the wavefunctions and transition energies of open-shell cases (e.g.\ doublets) one often prefers a simpler ``black-box'' type method in which the definition of model spaces  is not necessary and in which convergence problems due to intruder states are avoided. In those situations, the equation-of-motion coupled-cluster (EOM-CC) approach\cite{Bartlett:2011ho} is an excellent alternative to IHFS-CC: the FS-CC and EOM-CC methods are formally equivalent for single electron attachment and detachment states starting from a closed-shell reference\cite{bartlett:reviewcc,Meissner:2010fq}), though EOM-CC is a computationally more robust approach as it replaces the iterative solution of the coupled cluster equations by a diagonalization. In what follows we shall consider CCSD wavefunctions exclusively, and thus employ the shorthand CC instead of CCSD for brevity.

This robusteness is also found for the (1 hole, 1 particle) sector that corresponds to singly excited states; there, however, differences in energies arising from the use of different parametrizations for the wavefunctions---linear for EOM-CC and nonlinear for IHFS-CC, with the latter having the virtue of also ensuring valence extensivity (i.e.\ with respect to active holes and particles used to define a model space)~\cite{Mukhopadhyay:1991fp,Lindgren:1987in}---become apparent and have been discussed for systems containing light\cite{Musial:2008da,fscc-vs-lrcc:musial:1,fscc-vs-lrcc:musial:2,fscc-vs-lrcc:musial:3} as well as heavy\cite{real09} elements.  

The appealing features of EOM-CC have made it an extremely popular method for light element systems, and its popularity is growing for heavier species as attested by the number of recent reports in the literature of EOM-CC implementations that take into account relativistic effects. Though some of the latter are based on solving the four-component (4C) Dirac equation for atomic and molecular systems\cite{Pathak:2016ck,Pathak:2015he,Pathak:2014gv,Pathak:2014dm,Pathak:2016dk} and therefore account rigorously for scalar relativistic (SR) effects and spin-orbit coupling (SOC), for reasons of computational efficiency most of them \cite{Klein:2008hx,Yang:2012gd,Wang:2015jw,Epifanovsky:2015hsa,Cao:2016fx,Cao:2017id,Zhang:2017jj,Akinaga:2017hv,Wang:2016hx} have been devised in a more approximate framework where SOC is treated to within different degrees of approximation e.g.\ starting from the spin-free exact two-component (sfX2C) Hamiltonian and including SOC via atomic mean-field (AMF) integrals\cite{so-Hes-CPL1996-251-365-371} or perturbatively. 

While in the aforementioned works and elsewhere in the literature\cite{gomes10,real09,ATO-Rota-JCP2011-135-114106,pawel1} one can see that approximate treatments of SOC such as in the AMF approximation can yield rather accurate excitation energies even for heavier species up to and including iodine, the situation is not as clear cut for heavier elements\cite{Zhang:2017jj}, and therefore for $5d, 5f, 6p$ elements and beyond it may be preferable to rely on approaches based on 4C Hamiltonians (that is, the Dirac--Coulomb (DC), Dirac--Coulomb--Gaunt (DCG) or Dirac--Coulomb--Breit (DCB) Hamiltonians), or on X2C approaches but using a molecular mean-field\cite{Sikkema2009} (MMF) approach, which have been shown to yield results largely indistinguishable from their 4C counterparts\cite{Tecmer:2014fs}.

Our primary goal in this work is to present the implementation in the \textsc{Dirac} code\cite{DIRAC17} of the EOM-CC approach, for obtaining the energies for electron attachment (EOM-IP), detachment (EOM-EA) and singly excited states (EOM-EE) based on 4C and accurate 2C Hamiltonians (X2C-MMF), though we note the implementation can be used with single reference wavefunctions obtained with any other Hamiltonian available in \textsc{Dirac}.  Here we shall place particular emphasis on the discussion of the exploitation of double point group symmetry, in contrast to other implementations which do not exploit or report the use of symmetry. Furthermore, we shall be able to perform for the first time a thorough comparison of the performance of the three EOM-CC approaches in comparison to IHFS-CC ones in the appropriate sectors and with equivalent relativistic Hamiltonians and basis sets. We recall that similar comparisons have been made by Musial and Bartlett for light-element systems using non-relativistic Hamiltonians~\cite{Musial:2008da,fscc-vs-lrcc:musial:1,fscc-vs-lrcc:musial:2,fscc-vs-lrcc:musial:3}. The calculation of transition moments and of excited state expectation values for the three EOM-CC variants considered here will be addressed in a subsequent publication. 

We demonstrate the use of our implementation in the study of different halogenated species: the halogen monoxides (XO, X = Cl, Br, I, At, Ts), the triiodide species (I$_3^-$) and the diiodo- (CH$_2$I$_2$) and iodobromo-methane (CH$_2$IBr) species. Our focus on halogenated species stems from the fact that these (and in particular iodine-containing ones) are of great importance to photochemical processes in the atmosphere such as ozone depletion and aerosol formation in coastal areas\cite{SaizLopez:2012is,SaizLopez:2016fx,Burkholder:2015js,SaizLopez:2014fr,GomezMartin:2013fa} -- keeping in mind aerosols are an important vector of dispersion of radioactive species in case of nuclear accidents\cite{Mehboob:2016kc,Funke:2012cc} -- and have been extensively studied theoretically and experimentally in the gas-phase. Going beyond iodine we note that species containing astatine have received considerable attention in recent years both theoretically and experimentally\cite{ato-ayed-jpc2013-117-10589,ato-ayed-jpc2013-117-5206,ato-champion-ica2009-362-2654,ato-champion-jpc2010-114-576,ato-champion-jpc2013-117-1983,PereiraGomes:2014iy,Sergentu:2016hn} due to their potential as radiotherapeutic agents so that a black-box approach such as EOM-CC may become a valuable tool to further elucidate their chemistry. Finally, by completing the monoxide series with also its heaviest member, TsO, one can investigate the growing importance of SOC on the description of the ground and excited-state wavefunctions as the charge on the halogen nucleus increases down the series.

The paper is organized as follows: in the next section we briefly review the theoretical underpinnings of EOM-CC theory and discuss implementation details. This is followed by sections outlining the computational details of the EOM-CC and IHFS-CC calculations, the presentation and discussion of our results, conclusions and perspectives. In Appendix \ref{Symmetry} we provide further information on the use of double group symmetry in tensor contractions. Finally, working equations for the determination of right and left eigenvalues and eigenvectors are given in Appendix \ref{work_eqs}. Results not shown in the manuscript are available as Supplementary Information at the publisher's website and via the zenodo repository~\cite{paper:dataset}.

\section{EOM-CC Theory: basic formulation} \label{sec:theory}
We start from the coupled-cluster ansatz
\begin{equation}
  |CC\rangle=\exp(\hat{T})|\Phi_0\rangle;\quad\hat{T}=\sum_{l=1}t_{l}\hat{\tau}_{l}, 
\end{equation}
where $\Phi_0$ is the reference (Hartree--Fock) determinant and the operator $\hat{T}$ in the present work is restricted to single and double
excitations
\begin{equation}
\hat{T}=\hat{T}_1+\hat{T}_2;\quad\hat{T}_1=\sum_{ia}t_i^aa_a^{\dagger}a_i;\quad
\hat{T}_2=\frac{1}{4}\sum_{ijab}t_{ij}^{ab}a_a^{\dagger}a_b^{\dagger}a_ja_i,
\end{equation}
thus defining the coupled-cluster singles-and-doubles (CCSD) model. Here and in the following indices $\left\{i,j,\ldots,n,o\right\}$,
$\left\{a,b,\ldots,f,g\right\}$ and $\left\{p,q,r,s\right\}$ refer to occupied (hole),
virtual (particle) and general orbitals, respectively. The energy and the cluster amplitudes are found from the equations
\begin{align}
\langle\Phi_0|\hat{\bar{H}}|\Phi_0\rangle=&\ E\\
\langle\Phi_l|\hat{\bar{H}}|\Phi_0\rangle=&\ 0;\quad |\Phi_{l}\rangle=\hat{\tau}_{l}|\Phi_{0}\rangle,\label{eq:ampeq}
\end{align}
conveniently given in terms of the similarity-transformed Hamiltonian
\begin{equation}
  \hat{\bar{H}}=\exp(-\hat{T}) \hat{H}\exp(\hat{T}).
\end{equation}

The equation-of-motion couped-cluster (EOM-CC) method is a robust theory for the calculation of multiple excited states on an equal footing, obtained by diagonalization of
the similarity-transformed Hamiltonian $\hat{\bar{H}}$ within a selected excitation manifold. The similarity-transformed Hamiltonian is non-Hermitian, so
right-handed ($R$) and left-handed ($L$) eigenvectors, determined by the solution of 
\begin{equation} \label{Eq:right}
\hat{\bar{H}} |R_{\mu}\rangle = E_{\mu} |R_{\mu} \rangle
\end{equation}
\begin{equation}
\langle L_{\mu}| \hat{\bar{H}} =  E_{\mu} \langle L_{\mu}|, \label{Eq:lambda} 
\end{equation}
for a given excited state ${\mu}$ with energy $E_{\mu}$, are therefore not simple adjoints of each other but obey the biorthogonality condition
\begin{equation}
\langle L_{\mu}|R_{\nu}\rangle = \delta_{\mu\nu}. \label{Eq:biorthogonality} \end{equation}
One should note that the resolution of equations \ref{Eq:right} and \ref{Eq:lambda} closely resembles a CI-type diagonalization, where the matrix representation of 
$\hat{H}$ has been replaced by that of $\hat{\bar{H}}$, and the right- and left-hand wavefunctions are parametrized respectively as
\begin{equation}
|\Psi_{\mu}\rangle = \exp(\hat{T})\hat{R}_{\mu}|\Phi_0 \rangle
\end{equation} 
and
\begin{equation}
  \langle \bar {\Psi}_{\mu} | = \langle \Phi_0 | \hat{L}_{\mu}\exp(-\hat{T}),
\end{equation}
via $\hat{R}$ or $\hat{L}$ operators (see below), thus defining the excited states on the basis of the coupled-cluster wave function $|CC\rangle$ for the reference state. 
To simplify notation we will in the following no longer explicitly mention the excited state  label ${\mu}$, but one should keep in mind that $\hat{R}$ or $\hat{L}$ may 
target one or more excited states.

Different choices for the $\hat{R}$ and $\hat{L}$ operators define different EOM-CC models. In the present work we have used three most popular choices of 
$\hat{R}$ (of excitation kind) and $\hat{L}$ (of de-excitation kind), defining the three lowest Fock space sectors:
\begin{itemize}
\item excited states (1h-1p) :
\begin{equation}
\hat{R}^{EE} = r_0 + \sum_{ia} r^a_i \{ a^{\dag}_a a_i^{\phantom{\dag}} \} + \sum_{i>j,a>b} r^{ab}_{ij} \{ a^{\dag}_a a^{\dag}_b a_j^{\phantom{\dag}} a_i^{\phantom{\dag}} \} 
\end{equation}
\begin{equation}
\hat{L}^{EE} = l_0 + \sum_{ia} l_a^i \{ a^{\dag}_i a_a^{\phantom{\dag}} \} + \sum_{i>j,a>b} l_{ab}^{ij} \{ a^{\dag}_i a^{\dag}_j a_b^{\phantom{\dag}} a_a^{\phantom{\dag}} \}
\end{equation}
\item ionized states (1h) : 
\begin{equation}
\hat{R}^{IP} =  \sum_i r_i \{a_i^{\phantom{\dag}}\} + \sum_{i>j,a} r_{ij}^{a} \{a^{\dag}_a a_j^{\phantom{\dag}} a_i^{\phantom{\dag}}\} 
\end{equation}
\begin{equation}
\hat{L}^{IP} =  \sum_i l^i \{a^{\dag}_i\} + \sum_{i>j,a} l^{ij}_{a} \{a^{\dag}_j a^{\dag}_i a_a\} 
\end{equation}
\item electron-attached states (1p):
\begin{equation}
\hat{R}^{EA} = \sum_a r^a \{a^{\dag}_a \} + \sum_{a>b,i} r^{ab}_{i} \{a^{\dag}_a a^{\dag}_b a_i\} 
\end{equation}
\begin{equation}
\hat{L}^{EA} = \sum_a l_a \{a_a \} + \sum_{a>b,i} l_{ab}^{i} \{a^{\dag}_i a_b a_a\} ,
\end{equation}
\end{itemize}
where curly brackets refer to normal ordering with respect to the Fermi vacuum defined by the reference $\Phi_0$, and the sets $\{\mathbf{r}\}, \{\mathbf{l}\}$   
to the amplitudes of the corresponding operators.

We have here truncated our $\hat{R}$ and $\hat{L}$ operators at the singles-doubles level since the same truncation is used for the $\hat{T}$ operators.
The equations \ref{Eq:right} and \ref{Eq:lambda}, together with the equation \ref{Eq:biorthogonality} expressing the biorthogonality of the left- and right-hand eigenvectors 
define the EOM-CC models under consideration.

\section{Implementation details} \label{sec:implementation}

The present implementation has been carried out within a development version of the DIRAC quantum chemistry package\cite{DIRAC17}. As we in this paper focus only on the determination of transition energies,
we can summarize the calculation in three steps:
\begin{enumerate}
\item Solve closed-shell ground state CCSD equations to obtain T$_1$ and T$_2$ amplitudes.
\item Construct the one and two-body intermediates based on the T$_1$ and T$_2$ amplitudes necessary for the construction of $\hat{\bar{H}}$. 
\item Diagonalization of $\hat{\bar{H}}$ in the full singles-doubles excitation space to obtain excitation energies and eigenvectors. An iterative matrix-free method is employed to avoid the explicit construction of $\bar{H}$, due to its generally very large size.
\end{enumerate}

The first step is carried out within a Kramers-unrestricted formalism\cite{CCSD-Visscher-JCP1996-105-8769-8776}, and the parallelization of the code is such that the most numerous integrals involving three or four virtual indexes are distributed over different compute nodes\cite{Pernpointner:2003fm}. As we shall see below, this scheme can be generalized to the parallelization of the EOM-EA and EOM-EE models. 

The intermediates in the second step are those originally defined by Bartlett, Gauss, Stanton and coworkers\cite{Gauss1991,Gauss1991b,Stanton:1993be,Gauss1995,Gwaltney:1996jw} in a spin-orbital basis, and for which the construction in a four-component formalism has been discussed in detail in our previous work\cite{Shee_JCP2016} on the calculation of ground-state properties. We have extensively made use of the previously developed tensor contraction routines compatible with double group symmetry. 

What follows discusses the work necessary for the third step, which required two main components: the first is a significant extension in scope of the aforementioned contraction routines, so that they were also able to carry out tensor contractions with non-totally symmetric and odd-ranked tensors while still exploiting double point group symmetry. These extensions, which are not trivial due to the handling of complex quantities, will be discussed in detail in section \autoref{subsec:extension_tc} below.  The second component is a matrix diagonalizer that can handle both real and complex general matrices. It will be discussed in more detail in section \ref{sec:davidson}.

\subsection{Extension to odd-ranked and non-totally symmetric tensor contractions} \label{subsec:extension_tc}
In our previous work\cite{Shee_JCP2016}, we have developed general-purpose tensor contraction routines to handle tensor contractions related to relativistic coupled-cluster theory. These routines handle relativistic symmetry as expressed by double point groups, where the boson irreps of single point groups are complemented by fermion irreps, spanned by functions with half-integer spin. The routines, restricted to the highest Abelian subgroup of the full symmetry group, exploit optimal blocking and sparsity of a tensor for each contractions. They assume that the product of all tensor indices belongs to boson irreps, which is true for even-ranked tensors, but excludes odd-ranked tensors in which this product belongs to a fermion irrep. 

In the present work, (L)R-vectors for IP and EA are odd-ranked and hence span fermion irreps. In order to accommodate these tensors in our tensor contraction scheme we formally increase the rank by one by introducing, as a bookkeeping device, a label for each fermion irrep representing each a continuum orbital, so that electrons are considered to be ionized(attached) to(from) this orbital.  This follows in spirit the well-known approach of adding a very diffuse gaussian basis function to an EOM-EE code to simulate the ionization to the continuum~\cite{stanton:1999continuumorbital}. The fundamental difference is that all actions are done at the contraction level only (similar to the EA-EOM-CCSD implementation of Nooijen and Bartlett~\cite{Nooijen_JCP95}) and do not require the definition of a basis function. Since the EOM-EE machinery (N$^6$ scaling) is not used the proper N$^5$ scaling of EOM-IP is obtained.

The continuum orbital will always belong to the same fermion irrep as the orbital from where/to which ionization/electron attachment occurs. In this way, and due to our restriction to Abelian double groups at the correlated level, the (L)R-vectors become totally symmetric, and we block them according to the same scheme as used for even-ranked tensors. Since we only have one continuum orbital per irrep, the size of arrays is not increased relative to the original odd-ranked arrays. In the contraction step we ensure that continuum orbitals can only be contracted with themselves.  As an illustration, we consider the following contribution to the R-sigma vector equation of EOM-IP (cf. Eq. (B2) of the Appendix)
\begin{equation} \label{eq:IP-example}
(\sigma)_{ij}^{ a} \leftarrow  - \sum_m W^{ma}_{ij} * r_m.
\end{equation}
By introducing the continuum orbital this term is rewritten as
\begin{equation}
  (\sigma)_{o1,o2}^{c1,p2} \leftarrow  - \sum_{o3} W^{o3,p2}_{o1,o2} * r_{o3}^{c1},
\end{equation}
where $c1$ represents the continuum orbital, and indices $o$ and $p$ refer to holes and particles, respectively.

The corresponding subroutine call is:  
\begin {verbatim}
 call contraction_424((/"o3","p2","o1","o2"/),&
  & (/"c1","o3"/), (/"c1","p2","o1","o2"/),&
  & sigma2,-1.0d0,1.0d0,nrep, &
  & LeftTensor=B
\end{verbatim}              
where \texttt{r1} contains the trial vector coefficients.  As explained above, \texttt{c1} and \texttt{o3} will belong to the same fermion irrep, thereby \texttt{r1} is blocked with respect to the symmetry of \texttt{o3}. The $\sigma$-vector is blocked according to the symmetry of \texttt{c1} as  well. In this manner the operation count of IP- and EA- type contractions is reduced significantly, especially for linear molecules and other molecules with high symmetry. 

The second generalization of the original implementation of tensor contractions is to allow contractions for which the product of tensor indices is not totally symmetric. Such contractions occur for EOM-EE target states that belong to non-totally symmetric irreps. We have illustrated this extension schematically in Appendix \ref{Symmetry}. 

\subsection{Davidson diagonalization for non-Hermitian matrices} \label{sec:davidson}

Diagonalization of $\hat{\bar{H}}$ in the full singles-doubles excitation space to obtain excitation energies and eigenvectors is an expensive task, since the matrix dimension is in principle huge, and we thus  employ an iterative procedure of the Davidson type\cite{Davidson:1975db}. Since $\hat{\bar{H}}$ is non-Hermitian we have in this work implemented a generalized eigensolver following the algorithm of Hirao and Nakatsuji\cite{Hirao:1982fc}, which is capable of obtaining multiple roots at a time as well as handling $\bar{H}$ which can be either real or complex depending on the double point group in use -- though operating with double precision variables instead of complex ones\cite{Morgan:1992iu}. 

We solve the left and right eigenproblems separately, using a modified Gram-Schmidt procedure~\cite{matrix-computations:book} for orthonormalizing the trial vectors during the iterative procedure (cf. Refs.~\citenum{DAVIDSHERRILL1999143,vallet:2000epciso}). This approach is often more cost-effective for an EOM-CC implementation as excitation energies are usually the only quantity sought, requiring only the solution for one side. If both left and right calculated, the left eigenvectors are rescaled to satisfy the bi-orthogonality condition (\autoref{Eq:biorthogonality}). 
             

The first and costlier part of this  algorithm is generally the formation of the left ($\boldsymbol{\sigma}^L)$ or right $(\boldsymbol{\sigma}^R)$ sigma vectors
\begin{eqnarray}
\boldsymbol{\sigma}^R &=& \bar{H}\mathbf{b} \label{right-sigma-vector} \\
\boldsymbol{\sigma}^L &=& \mathbf{b}^\dagger\bar{H} \label{left-sigma-vector}
\end{eqnarray}
where $\mathbf{b}(\mathbf{b}^\dagger)$ is the (complex conjugate) matrix of trial vectors, as it involves the contractions outlined in Appendix \ref{work_eqs}. Here we carefully avoid the possibility of generating three-body intermediates by suitably rearranging the order of contractions, which will again be reflected in the sigma vector expressions in Appendix \ref{work_eqs}. 

Another particularity of our implementation has to do with the parallelization of the most expensive and memory intensive step of the sigma vector construction. The terms which arise from the integrals involving four virtual orbitals are constructed with distributed-memory Message Passing Interface (MPI)  parallelization. As seen in Appendix \ref{work_eqs} such terms appear in the EOM-EA and EOM-EE sigma vector equations in addition to the ground-state CCSD amplitude equations. We then synchronize final sigma vectors to the master and proceed in serial mode for the rest of the Davidson iteration steps.      

The choice of the initial trial vectors is of paramount importance for the final convergence of the method. We have adopted the following routes to choose our guess vectors:

(a) We fully diagonalize the singles-singles block of the transformed Hamiltonian. Eigenvectors of that diagonalization are considered as guess vectors.

(b) We approximate the singles-singles block of the transformed Hamiltonian by its diagonal elements, that is, ($\overline{F}_a^a $ - $\overline{F}_i^i$ -  $W^{ia}_{ai}$), -$\overline{F}_i^i$ and  $\overline{F}_a^a$ for EOM-EE, EOM-IP and EOM-EA, respectively, with intermediates $\overline{F} $ and $W$ defined in Appendix \ref{work_eqs}. The corresponding unit vectors are considered as guess vectors. Since they are selected according to energy and the number of roots requested,  we refer to these as pivoted unit vectors.

Further computational savings for IP and EA calculations can be achieved by using the fact that for real and complex double groups the states are doubly degenerate due to time-reversal symmetry and each span a different irrep. This means we only need consider one of the two degenerate Kramers pair as our guess vectors for each irrep, and thus may calculate only half of the total number of $\sigma$-vectors (for excitation energy calculations similar considerations are ungainly, since symmetry-adaptation requires constructing multideterminant reference states). However, for the quaternion double groups this scheme cannot be employed in a straightforward manner, and we must request twice as many roots as we want states, irrespective of the nature of the  calculations.

Finally, the implementation allows for the use of root following using the overlap between initial and generated trial vectors\cite{Butscher:1976ku,Zuev:2014fc} during the procedure, in the case one wishes to target states with dominant (1h1p), (1h0p) or (1p0h) character, which may turn out to be higher in energy than states with (2h2p), (2h1p) or (2p1h) character.

\section{Computational Details}

All coupled-cluster calculations were carried out with a developmental version of the \textsc{Dirac} electronic structure code\cite{DIRAC17} (revisions \texttt{e25ea49} and \texttt{7c8174a}), employing Dyall's basis sets\cite{basis-Dyall-TCA2002-108-335,basis-Dyall-TCA2003-109-284,basis-Dyall-TCA2012-131-3962,dyallbasis} of triple-zeta quality (dyall.av3z) for the halogens, and Dunning's aug-cc-pVTZ sets\cite{basis-Kendall-JCP1992-96-6796-6806} for oxygen, all of which are left uncontracted. In these calculations we employed the molecular mean-field\cite{Sikkema2009} approximations to the Dirac--Coulomb ($^2$DC$^M$) and Dirac--Coulomb--Gaunt ($^2$DCG$^M$) Hamiltonians -- where in the latter the Gaunt-type integrals are explicitly taken into account only during the SCF step -- along with the usual approximation of the energy contribution from $\left(SS|SS \right )$-type two-electron integrals by a point-charge model~\cite{relat-Visscher-TCA1997-98-68}. Apart from the EOM-CC method, we have employed the intermediate Hamiltonian Fock-Space (IHFS-CC) method\cite{correlation-Visscher-JCP2001-115-9720,Landau:2004dd}. Details of the main ($P_m$) and intermediate ($P_i$) model  and complement ($Q$) spaces used will be given below for each system. 

To further simplify the notation, in what follows we abbreviate EOM-CC and IHFS-CC to EOM and IHFS respectively, adding whenever appropriate the qualifiers EE/IP/EA for the first and (1h1p)/(1h0p)/(0h1p) for the second to denote the Fock-space sector under consideration. 
We also note that in the cases of known doubly degenerate electronic states (e.g.\ the $\Omega=1/2_{(g/u)},\ldots$ for electron attachement/detachment or the $\Omega= \pm1_{(g/u)},\ldots$ for excitation energies in linear symmetry), in the EOM calculations only one has been explicitly calculated. Furthermore, in EOM calculations, unless otherwise noted, we have used pivoted unit vectors as initial trial vectors and new solution vectors were generated : a) using the root following procedure for EOM-IP; b) not using the root following procedure for EOM-EE/EA.

\subsection{Halogen monoxides radicals (XO, X = Cl -- Ts)}

The electronic states of halogen monoxide radicals have been obtained starting from the anions (XO$^-$) in order to provide a closed-shell reference determinant for electron detachment calculations for both IHFS(1h0p) and EOM-IP calculations. In all calculations $C_{{\infty}v}$ symmetry was used.
All spinors with energies between -10.0 and 100.0$\,E_h$ have been correlated, which corresponds to considering, respectively: (a) 20 electrons and 206 virtuals for the systems containing Cl; (b) 32 electrons and 246 virtuals for the systems containing Br; (c) 32 electrons and 248 virtuals for the systems containing I; (d) 46 electrons and 340 virtual spinors for the systems containing At; and (e) 46 electrons and 306 virtuals for the systems containing Ts. In terms of the nature of the occupied atomic spinors correlated, the spaces above correspond to including the $2s2p$ oxygen and the $(n)sp (n-1)spd (n-2)f$ halogen atomic shells ($n$ denoting the valence shell; $f$ shells are obviously available only for At and Ts).

In the IHFS(1h0p) case, the $P_m$ space for all species but TsO comprises the five highest occupied molecular spinors of the anion which arise from the valence ($p$-$p$) manifold ($\pi^{(2)}_{1/2} \pi_{3/2}^{(2)} \sigma_{1/2}^{(2)}  \pi_{1/2}^{*(2)} \pi_{3/2}^{*(2)}$), thus placing the $\sigma_{1/2}^{*(0)}$ and all remaining virtuals in the $Q$ space, whereas the $P_i$ space included all other occupied spinors. For TsO, we encountered convergence problems due to intruder states with the aforementioned  $P_m$ space, and had to move the lowest-lying $\pi_{1/2}^{(2)}$ into $P_i$.  In the EOM-IP case, the number of roots requested was 3 and 2 for $\Omega=1/2,\ 3/2$, respectively, which allows us to obtain the ground and low-lying states, which correspond to those obtained with IHFS(1,0) for the model space above.  Spectroscopic constants were obtained by constructing potential energy curves for each species and performing polynomial fits to energies calculated for X-O internuclear distances ranging from (a) 1.46~\AA~to 1.98~\AA~for ClO; (b) 1.58~\AA~to 2.14~\AA~for BrO; (c) 1.66~\AA~and 2.16~\AA, for IO; (d) 1.84 and 2.40~\AA, for AtO and (e) 1.86 and 2.44~\AA, for TsO, respectively, with spacings no smaller than 0.01~\AA~between points. For most of the calculated points, the Hartree-Fock self-consistent field (SCF) procedure converged to the correct state with the default start potential. For TsO with an elongated bond length (beyond 2.32~\AA) this procedure lead to a wrong SCF solution, and we needed to adjust the start potential to ensure occupation of the correct orbitals in the early stage of the SCF iterations.

\subsection{Triiodide}

We investigated the excitation energies, electron attachment and electron detachment for I$_3^-$ with EOM and IHFS starting from the closed-shell ground state of  I$_3^-$ in all cases. The geometry used is $R = 2.93$~\AA, which was used previously to compare different electronic structure methods for excitation energies~\cite{gomes10}. 
All spinors with energies between -3.0 and 12.0$\,E_h$ have been correlated, which corresponds to considering 52 electrons and 332 virtual spinors. This choice is slightly different from that of Ref.~\citenum{gomes10}, since there we employed the augmented core-valence triple-zeta basis and with that included additional virtual spinors in the complement space $Q$. In all calculations the $D_{{\infty}h}$ point group was used.

In the case of IHFS calculations we considered the same active spaces used in Ref.~\citenum{gomes10}: for the IHFS(1h0p) calculations, the $P_m$ space contained the 16 highest-lying occupied spinors ($4, 2, 6, 4$ for $\omega={1/2}_g, {3/2}_g, {1/2}_u, {3/2}_u$, respectively), with the remaining 6 occupied spinors ($4, 0, 2, 0$ for $\omega={1/2}_g, {3/2}_g, {1/2}_u, {3/2}_u$, respectively) being included in $P_i$.  For the IHFS(0h1p) calculations, the $P_m$ space contained the 20 lowest-lying virtual spinors ($6, 2, 8, 4$ for $\omega={1/2}_g, {3/2}_g, {1/2}_u, {3/2}_u$, respectively), with the subsequent 24 virtual spinors ($6, 6, 4, 4, 2, 2$ for $\omega={1/2}_g, {3/2}_g, {5/2}_g, {1/2}_u, {3/2}_u, {5/2}_u$, respectively) making up the $P_i$ space. For the IHFS(1h1p) calculations the $P_m$ and $P_i$ spaces for both calculations are constructed as the direct product of the respective spaces from the  IHFS(0h1p) and IHFS(1h0p).  The number of roots requested in the EOM calculations was: $3, 1, 3, 2$ in $\Omega={1/2}_g, {3/2}_g, {1/2}_u, {3/2}_u$ symmetries for EOM-IP; $6, 4, 2, 6, 3, 1$ in $\Omega={1/2}_g, {3/2}_g, {5/2}_g, {1/2}_u, {3/2}_u, {5/2}_u$ symmetries for EOM-EA; and $10$ in each of the $\Omega={0}_g, 1_g, 2_g, 1_u, 2_u$ symmetries for EOM-EE. In the EOM-IP calculations the root following procedure was not used.

\subsection{Dihalomethanes}

We investigated the excitation energies, electron attachment and electron detachment for the CH$_2$IBr  and CH$_2$I$_2$ systems with EOM-EE and IHFS, starting from the closed-shell ground state in both cases. All calculations were performed on a single structure obtained by a geometry optimization performed with the ADF code\cite{FonsecaGuerra1998,ADF2001,ADF2017authors}, using TZ2P basis sets and the scalar relativistic Zeroth-Order Regular Approximation (ZORA) Hamiltonian. The corresponding Cartesian coordinates and structural parameters can be found in the Supplementary Information. All calculations were performed in $C_{2v}$ (CH$_2$I$_2$) and  $C_{s}$ (CH$_2$IBr) symmetries. As C$_{2v}$ is not an Abelian double group\cite{luuk:sym}, the C$_2$ subgroup was employed in the coupled cluster calculation and defines the symmetry labels for the states that were calculated.

All spinors with energies between -3.0 and 6.0$\,E_h$ for CH$_2$I$_2$ and -4.0 and 6.0$\,E_h$ CH$_2$IBr have been correlated, which corresponds to considering, respectively: (a) 40 electrons and 364 virtual spinors for CH$_2$IBr; and (b) 40 electrons and 374 virtual spinors for CH$_2$I$_2$. In terms of the nature of the occupied atomic spinors correlated, the spaces above correspond to including the $1s$ of hydrogen, the $2s2p$ of carbon, and the $(n)sp (n-1)spd$ halogen atomic shells ($n$ denoting the valence shell). The number of roots requested for the EOM calculations is as follows: 7 and 6 of $^1E$ symmetry for EOM-IP and EOM-EA, respectively, for each of the species; and for EOM-EE, 12 of $A, B$ symmetries, respectively, for each of the species. 

For the IHFS(1h0p) calculations, the $P_m$ spaces for both species contained the 12 highest-lying spinors (6 in each of the $^1e$, $^2e$ representation), with the remaining 28 spinors (14 in each of the $^1e$, $^2e$ representations) being included in $P_i$.  For the IHFS(0h1p) calculations, the $P_m$ space for CH$_2$I$_2$ contained the 26 (13 in each of the $^1e$, $^2e$ representations) lowest-lying spinors, with 30 additional spinors (15 in each of the $^1e$, $^2e$ representations) making up the $P_i$ space, whereas for CH$_2$IBr 20 and 30 spinors make up the $P_m$ and $P_i$ spaces, respectively, and, as was the case for the (1h0p) sector, these are evenly divided between the $^1e$ and  $^2e$ representations. For the IHFS(1h1p) calculations on the CH$_2$I$_2$ and CH$_2$IBr, the $P_m$ and $P_i$ spaces for both calculations are constructed as the direct product of the respective spaces from the  IHFS(0h1p) and IHFS(1h0p). Unfortunately, IHFS(1h1p) calculations with the corresponding model space did not converge, and attempts with larger model spaces were not practically feasible due to technical constraints (the MPI implementation did not fully support the 64-bit integers needed to address a larger memory space) which occur due to the increase in storage requirements caused by the use of complex algebra in C$_s$ symmetry.

\section{Results} \label{sec:results}

\subsection{Halogen monoxides}

We begin the discussion by analyzing our results for the halogen monoxide radicals. In Table~\ref{tab:so-monoxide-radicals}, we present the spin-orbit splitting of the  $^2\Pi$ ground states, 
\begin{equation}
T_\text{so} = E(X ^2\Pi_{1/2})  - E(X ^2\Pi_{3/2}),
\end{equation}
calculated at the ground-state ($X ^2\Pi_{3/2}$) EOM-IP equilibrium structure. 

Comparing first the EOM-IP and IHFS(1h,0p) $T_\text{so}$ values in Table~\ref{tab:so-monoxide-radicals}, we see for both $^2$DCG$^M$ and $^2$DC$^M$ the expected close agreement between the two methods along the series: up to IO differences are of the order of 0.001 eV or better for both Hamiltonians, for AtO differences are slightly larger (about 0.01 eV and  0.005 eV for $^2$DC$^M$ and $^2$DCG$^M$, respectively), while for TsO discrepancies of around 0.023 eV and 0.025 eV for the $^2$DC$^M$ and $^2$DCG$^M$ Hamiltonians, respectively, are found. 
These differences between the two methods are due to use of the Intermediate Hamiltonian formalism: whereas EOM-IP and FS(1h0p) should yield exactly the same results for singly ionized states this does not hold for EOM-IP and IHFS(1h0p).
The pronounced differences between EOM-IP and IHFS(1h0p) for TsO can either be due to missing contributions from higher sectors (in particular 2h1p) or the division into $P_m$ and $P_i$ spaces in IHFS calculations. The latter is a sensitive point for our IHFS calculations since states belonging to the $P_i$ space are not dressed and therefore treated in a CI-like way~\cite{Landau:2004dd}. 

To better understand the observed trends in spin-orbit splittings, it is illustrative to look at the composition of the electronic states of XO in terms of the molecular spinors of XO$^-$ at the EOM-IP equilibrium structures (listed in the Supplementary Information). In all cases, the $X ^2\Pi_{3/2}$ and $X ^2\Pi_{1/2}$ states are dominated by ionizations from the $\pi^*_{3/2}$ and $\pi^*_{1/2}$ orbitals of the anions. For the ClO molecule these $\pi^*$ orbitals has most weight on the less electronegative oxygen atom and their energies are only slightly split by spin--orbit coupling. As the electronegativity of the halogen atom decreases along the series, the bonding $\pi$ orbital overall becomes centered on the oxygen whereas the antibonding $\pi^*$ orbital moves to the halogen. For IO the spin--orbit splitting is considerable but one may still interpret the highest $\omega=1/2$ and $\omega=3/2$ orbitals as two $\pi^*$  orbitals, now with the dominant weight on iodine. This simple picture starts to break down for AtO in which the $\omega=1/2$ orbital also contain a significant $\sigma$ contribution and has an increased oxygen participation. For the TsO molecule, spin-orbit coupling is so strong that the notion of $\sigma$ and $\pi$ orbitals is better avoided. The lowest orbital in the p-orbital valence space is the Ts $7p_{1/2}$ orbital which is relatively compact and hardly participates in chemical bonding (cf. Ref.\citenum{Faegri_JCP2001}). The $\omega=3/2$ orbitals are also virtually non-bonding and centered either on the O or the Ts. The bonding is provided by the $\omega=1/2$ component of the Ts $7p_{3/2}$ orbital which combines with the O $2p_\sigma$. These changes induced by the increasing importance of spin-orbit coupling on the bonding orbitals can be visualized by means of plotting the spinor magnetization densities (Supplementary Information). For TsO, the qualitatively different orbital structure is also reflected in the composition of the two lowest states. While the $X ^2\Pi_{3/2}$ still shows contributions from configurations where also the lower energy $\omega=3/2$ is unoccupied, the $X ^2\Pi_{1/2}$ state is nearly completely dominated by a configuration in which the $\pi^*_{1/2}$ is singly occupied. For the EOM-IP calculations we see also a small contribution from a configuration in the 2h1p sector in which this ionization is accompanied by an excitation from the $\sigma_{1/2}$ to a high-lying $\sigma^*_{1/2}$ spinor.

For the higher lying $\Omega = 1/2$ and $\Omega = 3/2$ states that arise by excitation from the bonding $\pi$ orbitals, we largely observe the same patterns with respect to their composition (see Supplementary Information) as discussed for the spin-orbit split ground state. Again there is one dominant singly ionized configuration for both EOM-IP and IHFS(1h0p) and a small 2h1p contribution for the former. The exception is again TsO, where for the third $\Omega = 1/2$ state of TsO we find the largest discrepancy in energy (1.63 eV) between EOM-IP and IHFS(1h0p) among all states considered. This is due to the fact that the lowest $\pi_{1/2}$ orbital had to be put in the $P_i$ space, which leads to a poor description of states that are dominated by ionization from this orbital. 

We note that while EOM-IP and IHFS(1h0p) perform in very nearly the same way for states dominated by singly ionizations (and belonging to $P_m$) irrespective of the Hamiltonian used, the choice of Hamiltonian does have important implications for the value of $T_\text{so}$: the $^2$DC$^M$ are larger than the $^2$DCG$^M$ ones, and this difference grows slowly along the series (0.003 eV for ClO, 0.0035 eV for BrO, 0.0041 eV for IO, 0.0047 eV for AtO), culminating in the largest difference (0.0216 eV) for TsO. However, since $T_\text{so}$ itself increases still faster as the halogen becomes heavier, in absolute terms the importance of the Gaunt interaction diminishes along the series.  The effect of the Gaunt term corresponds to about 7\% of $T_\text{so}$ for ClO (and therefore must be taken into account), whereas for AtO it corresponds to less than half percent (and therefore can be safely ignored), only to become important again for TsO, for which its contribution being just short of 2\%.

We see a rather good agreement between our results and experimental ones~\cite{IO-Gilles-JCP1992-96-8012} based on photodetachment measurements on the XO$^-$ species (and therefore similar to our computational approach), with the $^2$DCG$^M$ results being in general closer to experiment. Since our calculations do not take into account corrections for zero-point vibrations, basis set incompleteness and higher excitations, we are not in a position to make definitive quantitative statements on the 
performance of the different methods. 

The only theoretical study that has accounted for vibrational corrections is that of Peterson and coworkers~\cite{IO-Peterson-JPC2006-110-13877-13883}, though using single-reference CCSD(T) calculations. These, however, yield $T_\text{so}$ values which underestimate the experimental values, in particular for IO.  Other theoretical calculations have corrected for the single-reference approach by using multireference CI~\cite{Zhou:2017ky} or EOM-IP but using different approximate Hamiltonians, that differ both in the treatment of scalar relativistic effects and SOC: Akinaga and Nakajima~\cite{Akinaga:2017hv} used a two-component method combining the third-order Douglas--Kroll--Hess Hamiltonian (DKH3) for scalar relativistic effects and screened nuclear SOC integrals, Epifanovsky and coworkers~\cite{Epifanovsky:2015hsa} used the one and two-electronic Breit--Pauli (BP) Hamiltonian with atomic mean-field SOC integrals, and no scalar relativistic effects, whereas Cheng and coworkers~\cite{Cheng:2018ji} have used spin-free exact two-component Hamiltonian (SFX2C-1e) for scalar relativity in conjunction with BP, atomic and molecular mean-field SOC integrals (the Hamiltonian used is the SO part of modified Dirac Hamiltonian by Dyall\cite{Dyall1994}) for the SO part. Even though in all of the above mentioned methods the EOM-CC framework have been used, they differ in terms of their treatment of SOC - Akinaga and Nakajima~\cite{Akinaga:2017hv} considers SOC from the start with their approximated Hamiltonian, Epifanovsky and coworkers~\cite{Epifanovsky:2015hsa} and  Cheng and coworkers~\cite{Cheng:2018ji} both treat SOC as a perturbation to their choice of spin-free/scalar relativistic Hamiltonians. Afterwards an effective Hamiltonian is constructed with the SOC integrals at the EOM level. However, the latter two approaches differ from one another because Cheng and coworkers~\cite{Cheng:2018ji} treat amplitude relaxation due to SOC perturbation while Epifanovsky and coworkers~\cite{Epifanovsky:2015hsa} do not.   

As not all calculations use strictly the same basis sets (some use triple-zeta bases with core-valence correlating functions or quadruple-zeta bases, but in both without adding diffuse functions as done here), the differences we observe between the different EOM-IP results are not exclusively due to the different Hamiltonians used. With some precaution we, however, believe we can affirm that, first, all EOM-IP approaches show more or less the same performance for ClO and get quite close to experiment. Second, for BrO, apart from the X2C-based approach of Cheng and coworkers~\cite{Cheng:2018ji}, which shows results comparable to our $^2$DCG$^M$ ones, all others seem to underestimate the experimental $T_\text{so}$. Finally, for IO, at a first glance the BP-based approach of Cheng and coworkers~\cite{Cheng:2018ji} yields the closest results to experiment, but the good agreement between our results and their X2C-based ones indicates that this may be somewhat fortuitous and due to cancellation of errors in the Hamiltonian and the EOM-CC method. It is quite likely, though, that approximate treatments of spin-orbit interaction such as proposed by Cheng and coworkers~\cite{Cheng:2018ji} are sufficiently accurate to treat molecules containing up to iodine. As Cheng and coworkers~\cite{Cheng:2018ji} have not explicitly explored their approach for heavier species,  we can only speculate as to their general applicability for species containing $6p$ elements and beyond, where strong second-order SOC effects are expected. 

We end the comparison of theoretical results for $T_\text{so}$ by noting that our results for IO and AtO show $^2$DC$^M$ faithfully reproduce
the IHFS(1h0p) results of Gomes and coworkers~\cite{PereiraGomes:2014iy} using the 4-component DC Hamiltonian, in line with the findings of Tecmer and coworkers~\cite{Tecmer:2014fs} for small actinide species. We decided to extend this comparison here and show in Table~\ref{tab:so-monoxide-radicals} $T_\text{so}$ values for the DC Hamiltonian, along with a more detailed comparison in the Supplementary Information. We see differences in energy between the DC and $^2$DC$^M$ Hamiltonians of less than a tenth of a milli-electron volt. A closer inspection of the results in the Supplementary Information reveals that, in absolute terms, the differences between Hamiltonians are very systematic and of the order of milli-electron volts. Thus, the differences between Hamiltonians for relative energies (whether between the anion and the radical, or between the states of the radical) are in effect about two order of magnitude lower than the absolute ones, so that the molecular mean-field approach shows errors of less than a wavenumber for all excitation energies considered, as well as ionization energies.

\begin{table}[htb]
\caption{Spin-orbit splittings ($\mathrm{T_{so}}$, in eV) between the 
$\Omega=3/2$ and the $\Omega=1/2$ components of the $^2\Pi$ ground state 
for the halogen monoxides for the EOM-IP and IHFS(1h0p)
approaches employing augmented triple zeta bases and the molecular mean-field Hamiltonians with ($^2$DCG$^M$) 
and without ($^2$DC$^M$) the Gaunt interaction, 
calculated at the EOM-IP equilibrium structures for the $\Omega=3/2$ ground-state. 
For comparison we present EOM-IP results with the DC Hamiltonian, experimental values~\cite{IO-Gilles-JCP1992-96-8012}
and prior theoretical results: the SO-icMRCI of Zhou {\em et al.}~\cite{Zhou:2017ky}, 
the CCSD(T) with SOC corrections of Peterson {\em et al.}~\cite{IO-Peterson-JPC2006-110-13877-13883},
and the EOM-IP results of : 
(a) Akinaga {\em et al.}~\cite{Akinaga:2017hv}, using the DKH3 Hamiltonian and screened nucleus SOC;
(b) Epifanovsky {\em et al.}~\cite{Epifanovsky:2015hsa}, using the Breit-Pauli Hamiltonian and AMF-SOC integrals;
(c) Cheng {\em et al.}\cite{Cheng:2018ji}, using the Breit-Pauli Hamiltonian with T-relaxed SOC; and
(d) Cheng {\em et al.}\cite{Cheng:2018ji}, using the one-component X2C Hamiltonian with T-relaxed SOC.}
\begin{center}
\begin{tabular}{l c c c c c}
\hline
\hline
                   & ClO  & BrO   & IO    &  AtO    &  TsO    \\
\hline
EOM  DC              &    0.0422	&      0.1294	&	0.2776	&	0.7232	&	1.1953	\\
EOM  $^2$DC$^M$      &    0.0423	&      0.1295	&	0.2777	&	0.7233	&	1.1954	\\
IHFS $^2$DC$^M$      &    0.0423	&      0.1294  &	0.2777	&	0.7238	&	1.2194	\\
EOM $^2$DCG$^M$     &    0.0396	&      0.1260	&	0.2737	&	0.7189	&	1.1915	\\
IHFS $^2$DCG$^M$     &    0.0396	&      0.1260  &	0.2737	&	0.7194	&	1.2165  \\
                           &		&      	&		&	       &	       \\
IHFS DC~\cite{PereiraGomes:2014iy}  
                           &	    	&          	&	0.28	&	0.72	&	    	\\
MRCI     ~\cite{Zhou:2017ky}
                           &	      	&      0.1189	&	    	&	        &	        \\
EOM(a)~\cite{Akinaga:2017hv}  
                        &	0.0422	&      0.1151	&	    	&	        &	        \\
EOM(b)~\cite{Epifanovsky:2015hsa}  
                        &	0.0382	&      0.1123	&	    	&	        &	        \\
EOM(c)~\cite{Cheng:2018ji}
                       &	0.0394	&      0.1196  &	0.2533  &	        &	        \\
EOM(d)~\cite{Cheng:2018ji}
                          &	0.0395	&      0.1220  &	0.2658 	&	        &	        \\
CCSD(T)~\cite{IO-Peterson-JPC2006-110-13877-13883}
                           &	0.0388	&      0.1061	&	0.2201	&	        &	        \\
                           &		&		&		&	        &	        \\
	Exp.~\cite{IO-Gilles-JCP1992-96-8012}        
                           &	0.0397	&      0.1270	&	0.2593	&	        &	        \\
\hline
\hline
\end{tabular}
\end{center}
\label{tab:so-monoxide-radicals}
\end{table}%

\begin{figure}[h]
    \centering
\begin{minipage}[b]{0.57\linewidth}
\centering
\includegraphics[width=\textwidth]{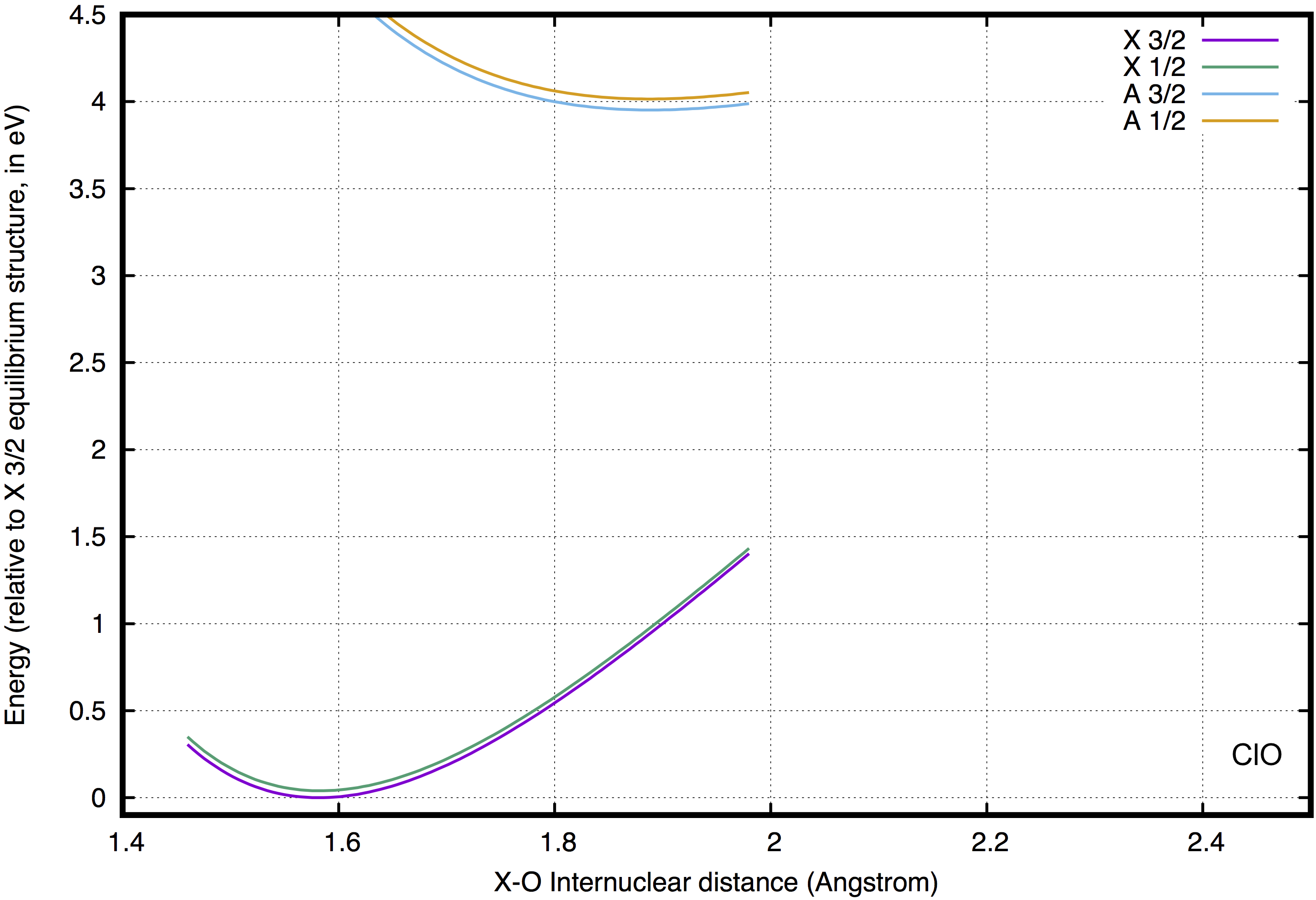}
\label{fig:potential-energy-curves-clo}
\end{minipage}
\begin{minipage}[b]{0.57\linewidth}
\centering
\includegraphics[width=\textwidth]{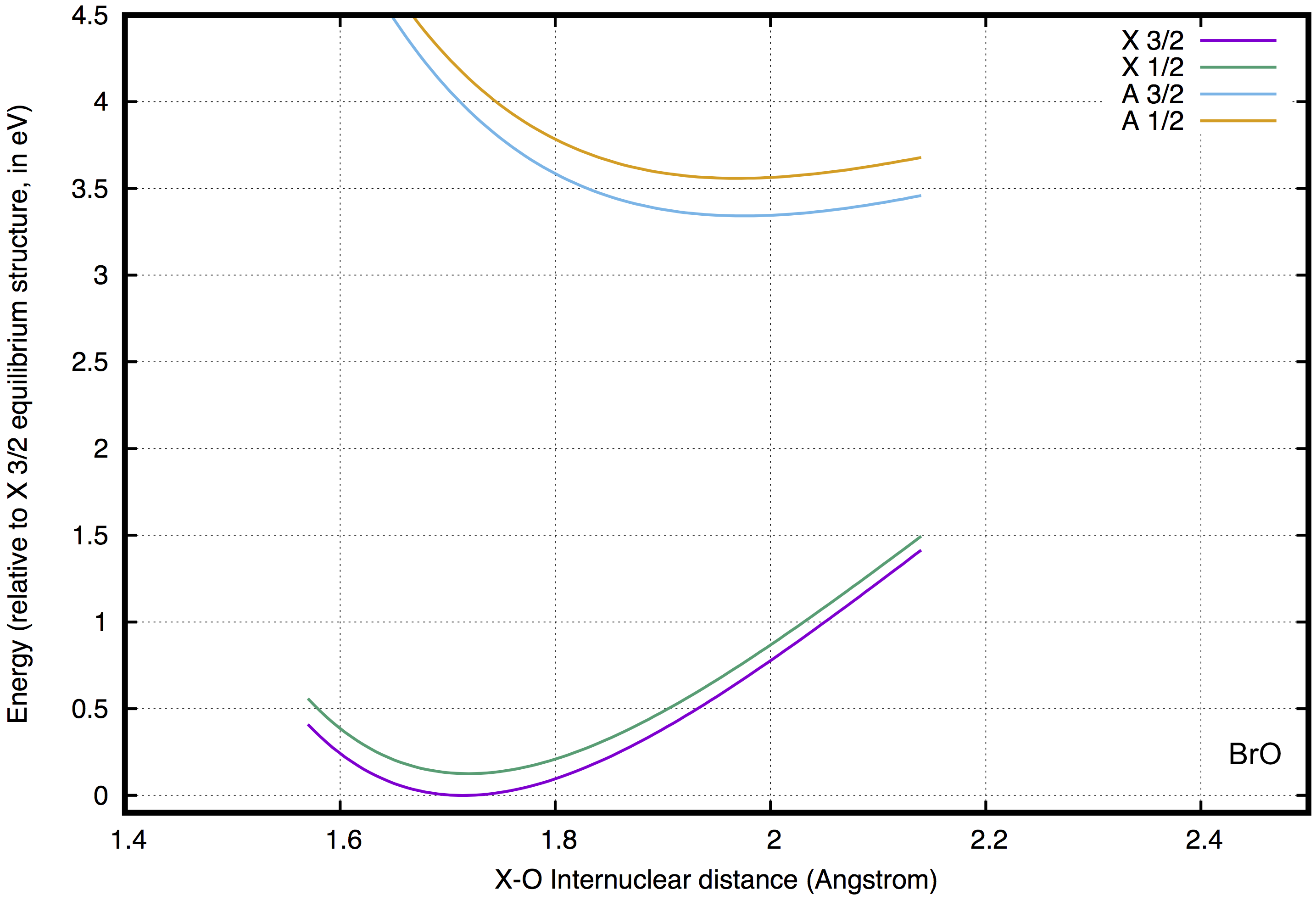}
\label{fig:potential-energy-curves-bro}
\end{minipage}
\begin{minipage}[b]{0.57\linewidth}
\centering
\includegraphics[width=\textwidth]{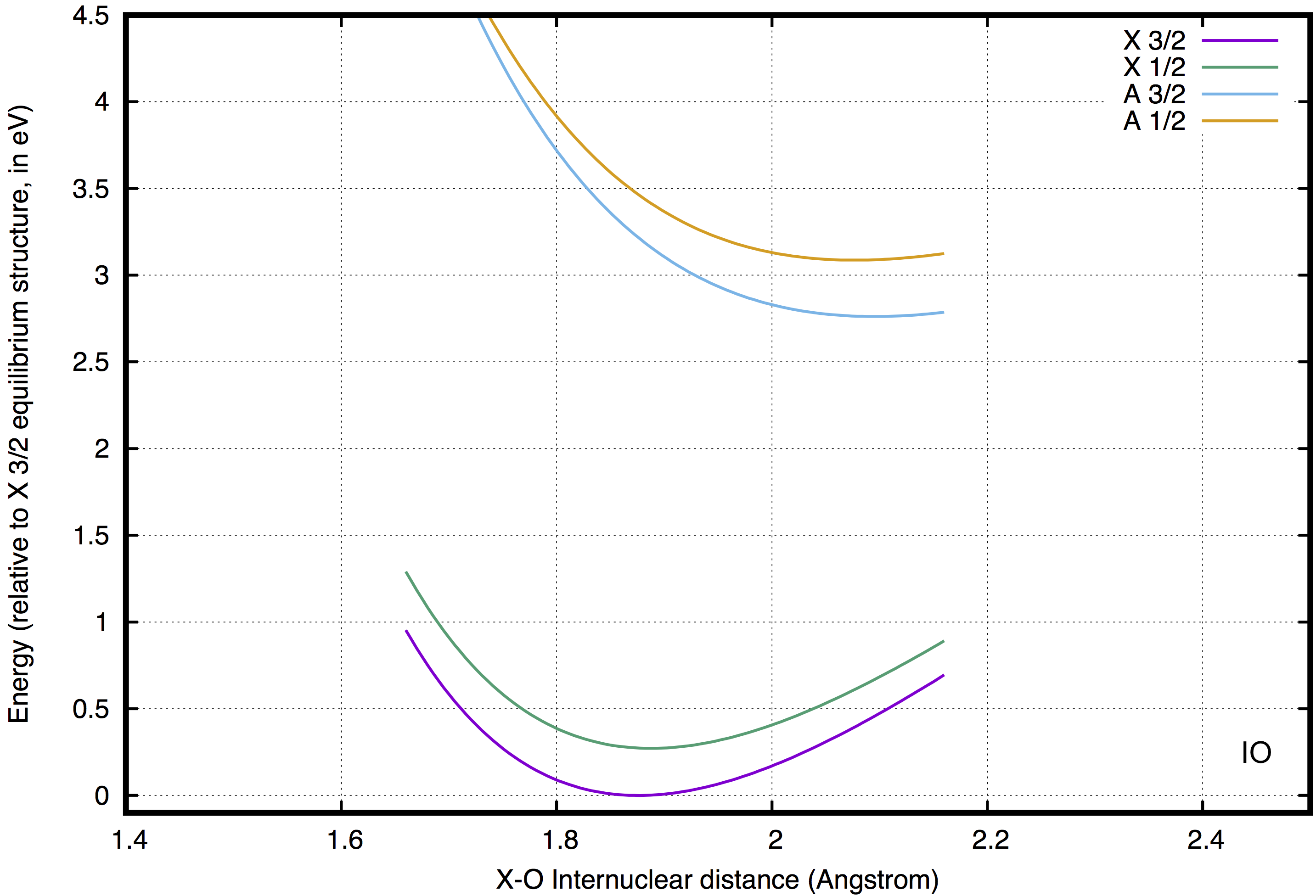}
\label{fig:potential-energy-curves-io}
\end{minipage}
\begin{minipage}[b]{0.57\linewidth}
\centering
\includegraphics[width=\textwidth]{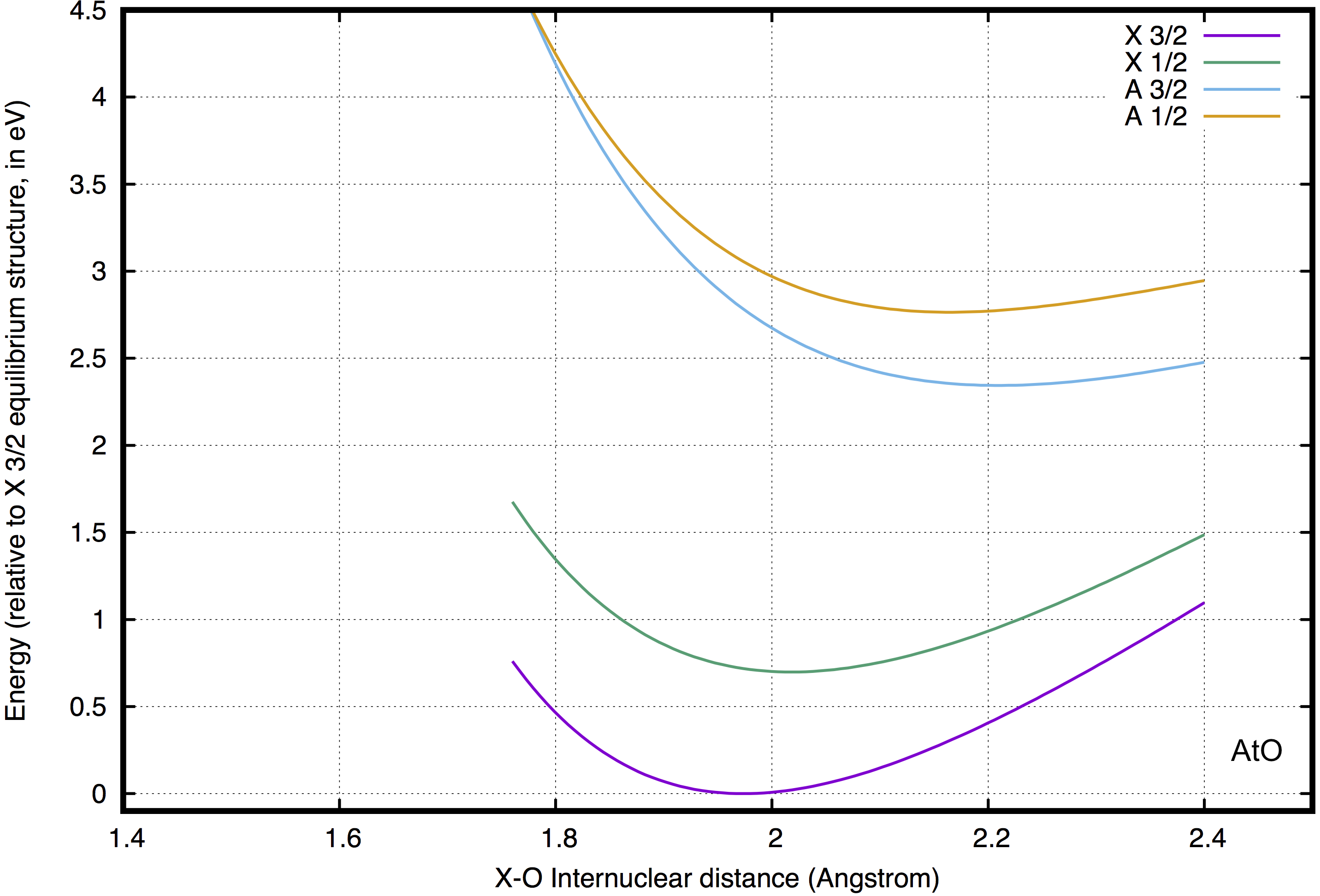}
\label{fig:potential-energy-curves-ato}
\end{minipage}
\centering
\begin{minipage}[b]{0.57\linewidth}
\includegraphics[width=\textwidth]{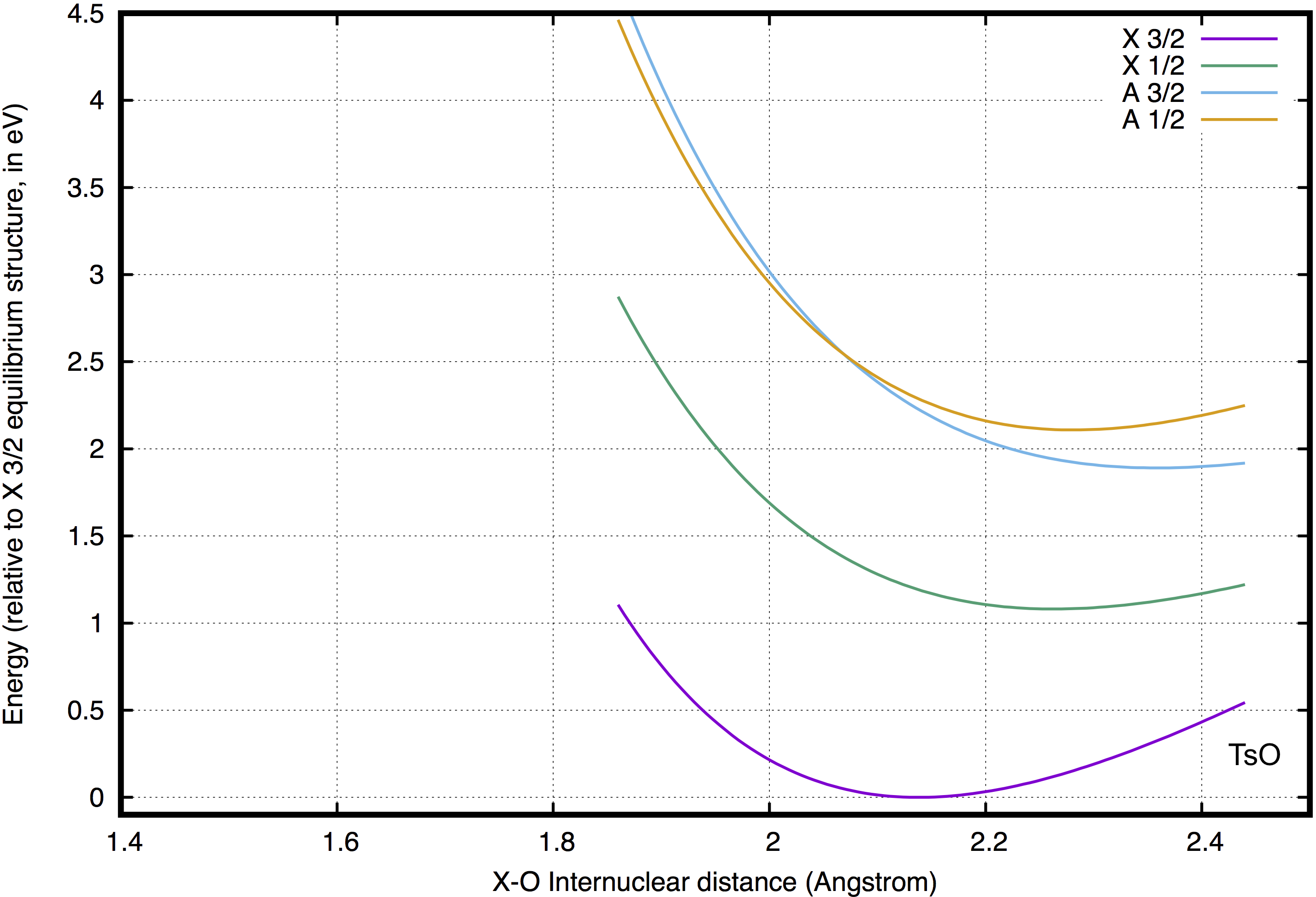}
\label{fig:potential-energy-curves-tso}
\end{minipage}

    \caption{Potential energy curves of the spin-orbit split $X ^2\Pi$ and $A ^2\Pi$ states of the XO molecules, obtained with EOM-IP and the $^2$DCG$^M$ Hamiltonian.~\cite{paper:figures}}
    \label{potential-energy-curves-xo-all}
\end{figure}

We will now consider the characteristics of the potential energy curves of the lowest electronic states of the halogen monoxides presented in Figure~\ref{potential-energy-curves-xo-all}. Due to the similarities in results between the $^2$DCG$^M$ and $^2$DC$^M$ Hamiltonians and between EOM-IP and IHFS(1h0p), we have opted to only present $^2$DCG$^M$ EOM-IP results since they seem to better represent the spin-orbit splitting. Additional data for these states (equilibrium structures, harmonic vibrational frequencies and vertical as well as adiabatic excitation energies) can be found in the Supplementary Information.

In Figure~\ref{potential-energy-curves-xo-all} the changing nature of the $\pi$ and $\pi^*$ orbitals that we discussed before is clearly visible in the curves. For the lightest two halides (ClO, BrO) the splitting of the $A ^2\Pi$ is more pronounced than that of the $X ^2\Pi$ curves as it is the bonding orbital that contains the largest halogen fraction. For iodine and for heavier halides the splitting is more pronounced for the $X ^2\Pi$  than for the $A ^2\Pi$ because for these molecules the antibonding orbital contains the largest halogen fraction. The SOC does not have a large influence on the difference in equilibrium bond lengths between the $X ^2\Pi$ and $A ^2\Pi$ states, which is relatively stable  at about 0.15~\AA. 

Another trend we observe from Figure~\ref{potential-energy-curves-xo-all} that is further evinced by the data in the Supplementary Information is the decrease of the ground-state harmonic frequencies along the series for $X ^2\Pi$, from about 900 cm$^{-1}$ for ClO to 573 and 457 cm$^{-1}$ for the $X_{3/2} {^2\Pi}$ and $X_{1/2} {^2\Pi}$ of TsO. While this is partly due to the increasing reduced mass, also the decrease in force constants indicates a reduction of the bond strength by almost a factor of two going from ClO to TsO. For the excited $A_{3/2} {^2\Pi}$ and $A_{1/2} {^2\Pi}$ states, the trend is less clear. While having significantly smaller force constants overall due to double occupancy of the $\pi^*$ instead of $\pi$ orbitals, the force constants of the $A$ states in IO are larger than these of the lighter ClO and BrO, as well as these of the heavier AtO. TsO is again special, with the two A states crossing each other and having rather different force constants. This shows that for such heavy atoms, the treatment of SOC as a minor perturbation can not be justified.

\subsection{Triiodide}

We discuss now our results for the triiodide molecule, previously investigated by some of us~\cite{gomes10} with a large array of relativistic correlated electronic structure methods (CASPT2, TD-DFT, MRCI and IHFS-CC),
and more recently by Wang and coworkers~\cite{Wang:2014es} with EOM approaches which take SOC into account in an approximate manner. As the experimental interest on this species has to do with its complex dissociation behavior
upon electronic excitations, our main interest here is in comparing our EOM-EE approach to IHFS(1h1p), which we know accurately describes the two absorbing  $0_u^+$ states in the ranges of 3.43--3.45 eV and 4.25--4.28 eV, respectively
(these excitation energies having been determined by  photofragment yield spectra~\cite{Choi:2000jv,Zhu:2001fy}), as well as how EOM-EE with molecular mean-field Hamiltonians compares to the
approximate schemes proposed by Wang and coworkers~\cite{Wang:2014es}.

\begin{table}[htb]
\caption{Vertical excitation energies (T$_v$, in eV) for I$_3^-$ 
with the EOM-EE and IHFS(1h1p) approaches employing the molecular mean-field Hamiltonians 
with ($^2$DCG$^M$) and without ($^2$DC$^M$) the Gaunt interaction at R = 2.93 \AA. 
We also present SO-CASPT2 results by Gomes and coworkers\cite{gomes10}, approximate EOM-EE 
schemes by Wang and coworkers~\cite{Wang:2014es}((a) EOM-SOC-CCSD; (b) SOC-EOM-CCSD; (c) cSOC-EOM-CCSD)
and experimental results~\cite{Choi:2000jv,Zhu:2001fy}, along with the mean absolute 
differences (MAD) and its standard deviation ($\sigma$) between EOM-EE and IHFS(1h1p) calculations and 
and those between IHFS(1h1p)-$^2$DCG$^M$ and the SO-CASPT2 and approximate EOMCC methods.} 
\begin{center}
\begin{tabular}{c c cc p{0.1cm} cc p{0.1cm} ccc p{0.1cm} c}
\hline
\hline
	&		&    \multicolumn{2}{c}{$^2$DC$^M$ } 	         	&&	\multicolumn{2}{c}{$^2$DCG$^M$ } &&      \multicolumn{3}{c}{EOM~\cite{Wang:2014es}} && CAS\\
\cline{3-4}
\cline{6-7}
\cline{9-11}
State	& $\Omega$	&	EOM 	&IHFS     &&	EOM	&IHFS	&&   (a)  &  (b)  &  (c)  && PT2\cite{gomes10}\\
\hline
1	&	2$_g$	&    2.24       &	2.07    &&      2.25    &	2.08 	&&	2.22  &	   2.36   &	2.16   &&	2.24	\\
2	&	1$_g$	&    2.37       &	2.20    &&      2.38    &	2.21 	&&	2.35  &    2.50   &	2.29   &&	2.32	\\
3	&	0$_u^-$	&    2.37       &	2.22    &&      2.38    &	2.23 	&&	2.34  &    2.49   &	2.29   &&	2.47	\\
4	&	1$_u$	&    2.38       &	2.23    &&      2.38    &	2.24 	&&	2.34  &    2.47   &	2.30   &&	2.47	\\
5	&	0$_g^-$	&    2.84       &	2.66    &&      2.84    &	2.66 	&&	2.81  &    2.96   &	2.72   &&	2.76	\\
6	&	0$_g^+$	&    2.89       &	2.71    &&      2.89    &	2.71 	&&	2.86  &    2.99   &	2.75   &&	2.82	\\
7	&	1$_g$	&    3.07       &	2.88    &&      3.07    &	2.89 	&&	3.04  &    3.20   &	2.96   &&	2.85	\\
8	&	2$_u$	&    3.32       &	3.19    &&      3.33    &	3.20 	&&	3.30  &    3.47   &	3.25   &&	3.10	\\
9	&	1$_u$	&    3.41       &	3.27    &&      3.42    &	3.27 	&&	3.39  &    3.55   &	3.34   &&	3.11	\\
10	&	0$_u^+$	&    3.66       &	3.51    &&      3.67    &	3.52 	&&	3.65  &    3.79   &     3.56   &&	3.52	\\
11	&	2$_g$	&    4.09       &	3.92    &&      4.10    &	3.93 	&&	4.04  &    4.19   &	3.98   &&	3.98	\\
12	&	0$_u^-$	&    4.08       &	3.93    &&      4.08    &	3.93 	&&	4.05  &    4.18   &	3.91   &&	3.79	\\
13	&	1$_u$	&    4.18       &	4.02    &&      4.18    &	4.02 	&&	4.15  &    4.29   &	4.01   &&	3.80	\\
14	&	1$_g$	&    4.21       &	4.03    &&      4.22    &	4.04 	&&	4.17  &    4.32   &	4.10   &&	4.06	\\
15	&	0$_u^+$	&    4.49       &	4.33    &&      4.49    &	4.33 	&&	4.50  &    4.67   &     4.42   &&	4.51	\\
16	&	0$_g^-$	&    4.69       &	4.51    &&      4.69    &	4.51 	&&	4.65  &    4.76   &	4.51   &&	4.51	\\
17	&	0$_g^+$	&    4.70       &	4.51    &&      4.70    &	4.51 	&&	4.65  &    4.82   &	4.51   &&	4.53	\\
18	&	1$_g$	&    4.90       &	4.71    &&      4.90    &	4.71 	&&	4.86  &    4.99   &     4.73   &&	4.60	\\
MAD     &               &    0.17       &               &&      0.17    &               &&      0.13  &    0.28   &     0.05   &&       0.11   \\
$\sigma$&               &    0.02       &               &&      0.02    &               &&      0.02  &    0.03   &     0.03   &&       0.08   \\
\hline
\hline
\end{tabular}
\end{center}
\flushleft
Exp.: Photofragment yield spectra~\cite{Choi:2000jv,Zhu:2001fy}:  3.43--3.45, 4.25--4.28.
\label{tab:triiodide-ee}
\end{table}%

From our results, presented in Table~\ref{tab:triiodide-ee}, we observe that EOM-EE excitation energies systematically overestimate the IHFS(1h1p) ones; we find a mean absolute deviation (MAD) of 0.17 eV for both the $^2 $DC$^M$ and $^2 $DCG$^M$ Hamiltonians, with small standard deviations ($\sigma$) of 0.02 eV in both cases. On other words, spectra obtained with EOM will show a shift in the origin with respect to IHFS, but will otherwise look the same. This behavior has been discussed previously by Musial and Bartlett for light-element systems~\cite{fscc-vs-lrcc:musial:1,fscc-vs-lrcc:musial:2,fscc-vs-lrcc:musial:3}, and has to do with the differences in parametrization for the excited-state wavefunctions in the two approaches (linear for EOM-EE and non-linear for IHFS(1h1p)). In the IHFS(1h1p) sector, the nonlinear parametrization of the wave operator, apart from the single excitation operators contains the product of the electron attachment and detachment operators whose presence assures valence extensivity~\cite{Mukhopadhyay:1991fp}, in contrast to the linear operator in EOM-EE which does not. 

Some of us had already observed the same systematic behavior of EOM in comparison to IHFS for 
another heavy element species (UO$_2^{2+}$), but since in that study a two-step SO-LRCC (thus analogous to EOM-EE) calculation~\cite{real09} was used, there could still be doubts as to whether the observed differences arose solely
from the difference in parametrization. Here, as we have used exactly the same Hamiltonians for both methods, we can affirm that the differences indeed come from the parametrization. The SO-CASPT2
results of Ref.~\citenum{gomes10}, reproduced here for the convenience of the readers, show by comparison a slightly smaller MAD but at the same time much less systematic behavior than EOM, with some states being very close to the IHFS ones and others quite far apart, as reflected by the larger $\sigma$ value of 0.08 eV.  Our view is that this underscores the lesser reliability of CASPT2 with respect to coupled-cluster approaches since it can result, for instance, in spurious inversions between states: for EOM we observe one such inversion for states 11 ($2_g$) and 12 ($0_u^{-}$), which are very close in energy in both EOM and IHFS calculations, whereas for CASPT2 we see two such inversions between states 11, 12 and 13 ($1_u$); Furthermore, CASPT2 places states such as 8 ($2_u$) and 9 ($1_u$), or 15 ($0_u^{+}$) and 16 ($0_g^{-}$) much close together than both coupled-cluster approaches.

In line with what has been established in our prior investigation~\cite{gomes10}, the EOM-EE and IHFS(1h1p) wavefunctions for the absorbing $0_u^+$ states are predominantly made up of transitions from the $\sigma_{1/2g}$ to $\sigma^*_{1/2u}$ and $\pi_{1/2g}$ to $\sigma^*_{1/2u}$ spinors. For IHFS(1h1p) the excited determinants making up the $0_u^+$ wavefunctions correspond to excitations within the main ($P_m$) model space (essentially from $\sigma_{1/2g}$ and $\pi_{1/2g}$ to the LUMO $\sigma^*_{1/2u}$), with small contributions from excitations falling within the intermediate ($P_i$) model space. For EOM the picture is very much the same as that of IHFS(1h1p), though we note there are also contributions from doubly excited (2h2p) determinants, as well as singly excited (1h1p) determinants containing high-lying virtuals, both of which fall outside of what is the IHFS(1h1p) $P$ space; individually these are all negligible contributions, but taken as a whole they represent a minor but non-negligible ($\simeq$ 4\%) contribution.

The excitation energies of triiodide have also been investigated with the three approximate EOM schemes introduced by Wang and coworkers~\cite{Wang:2014es}, which are based on the inclusion of SOC at the post-SCF step using a one-electron SOC operator ($\hat{h}^{SO}$) originating from the relativistic effective core potential (RECP) operator. In the first scheme (EOM-SOC-CCSD), $\hat{h}^{SO}$ is introduced for the solution of the ground-state coupled-cluster equations (so SOC is directly included in the $F_{ae}, F_{mi}, F_{me}, W_{mbij}$, and $W_{abej}$ intermediates and indirectly in the other intermediates through the cluster amplitudes), and the excited states are obtained by diagonalizing the corresponding similarity-transformed Hamiltonian (including $\hat{h}^{SO}$) in the space of singly and doubly excited determinants. In the second scheme (SOC-EOM-CCSD), $h^{SO}$ is included in the Hamiltonian for the EOM step only, and thus neither the Hartree-Fock not the ground-state coupled-cluster wavefunctions incorporate any SOC effects. The SOC-EOM-CCSD scheme is computationally less expensive since only the $F_{ae}, F_{mi}, F_{me}, W_{mbij}$, and $W_{abej}$ intermediates can be complex, but at the cost of having to determine the EOM excitation energies in the space of the ground-state, singly and doubly excited determinants while at the same time introducing unlinked terms involving matrix elements of the SOC Hamiltonian which make the excited state energies
not size-intensive and the ground-state energy not size-extensive. Finally, the third scheme (cSOC-EOM-CCSD) approximates SOC-EOM-CCSD by neglecting the unlinked terms in SOC-EOM-CCSD. SOC effects on the ground state are not taken into account with this approach, and interactions between double-excitation determinants and single-excitation determinants through SOC are not fully considered because of the neglect of the term where the SOC operator coupled singly and doubly excited determinants in the EOM equations. We note these schemes exploit time-reversal symmetry, and that due to the use of orbitals not including SOC, it was possible to use single-point group symmetry.

From Table~\ref{tab:triiodide-ee} we see that EOM-SOC-CCSD performs rather consistently with the four-component-based approaches presented here, with small deviations (from 0.02 to 0.05 eV) from our results which are likely due to difference in basis sets and the truncation of the correlating space in our case. As a consequence, EOM-SOC-CCSD are similarly very systematic in their deviation from IHFS(1h1p), showing a slightly better MAD value for EOM-SOC-CCSD than ours that is likely to be a fortuitous result, as the corresponding $\sigma$ value is the same as ours. A comparison to SOC-EOM-CCSD, on the other hand, shows that the latter is a rather poor approximate scheme, as not only individual excitation energies are quite different from ours (generally shifted upwards by over 0.1 eV, and with a MAD value nearly twice as large as ours for EOM-SOC-CCSD), though the systematic nature of the difference between EOM and IHFS is still roughly intact, as there is only a small increase of the $\sigma$ value to 0.03 eV. Finally, the correction to SOC-EOM-CCSD (cSOC-EOM-CCSD) does reduce the MAD value and is therefore on average close to IHFS than the more rigorous EOM schemes, but as was the case for SOC-EOM-CCSD, this obscures the fact that the errors are less uniformly distributed than for EOM-SOC-CCSD or the four-component-based EOM since one also has a $\sigma$ value of 0.03 eV. This makes the method less reliable in practice. 

In addition to the excitation energies, we present in Table~\ref{tab:triiodide-ip-ea} some of the lowest ionization energies and electron affinities of the triiodide (thus yielding the I$_3$ and I$_3^{2-}$ radicals). Our 
ionization energies show a rather good agreement with experiment for the states under consideration -- and particularly for the first -- which is not surprising, 
in the light of the good performance shown by EOM-IP for the XO species, and taking into account that the experimental results are for vertical electron detachment and the chosen bond lenght is quite close to the equilibrium structure of I$_3^{-}$. As for IO, we see that the Gaunt interaction plays a negligible role in
the ionization energies, and also that the IHFS model space is sufficiently flexible that the EOM and IHFS results are essentially identical.

\begin{table}[htb]
\caption{Vertical electron detachment ($\mathrm{IP^{\Omega}_{n}}$, in eV) and attachment 
($\mathrm{EA^{\Omega}_{n}}$, in eV) energies for the I$_3^{-}$ species, 
calculated at R = 2.93 {\AA}  with the EOM-IP/EA and IHFS(1h0p)/(0h1p) approaches, respectively, 
employing the molecular mean-field Hamiltonians with ($^2$DCG$^M$) and without ($^2$DC$^M$) 
the Gaunt interaction, as well as experimental values for the ionization energies~\cite{Choi:2000eo}}.
\begin{center}
\begin{tabular}{l cccc}
\hline
\hline
Method                        &	IP$^{3/2u}_1$	&	IP$^{1/2g}_2$	&	IP$^{1/2u}_3$	&	IP$^{3/2g}_4$ \\
\hline
EOM $^2$DC$^M$ 	      &    4.28   	&        4.47          &        4.92          &        4.99          \\ 
IHFS   $^2$DC$^M$       &    4.28   	&        4.47          &        4.92          &        4.99          \\ 
EOM $^2$DCG$^M$ 	      &    4.28   	&        4.47          &        4.91          &        5.00          \\ 
IHFS   $^2$DCG$^M$      &    4.28   	&        4.47          &        4.91          &        5.00          \\ 
Exp.~\cite{Choi:2000eo}       &    4.25         &        4.53          &        4.87          &        4.93          \\ 
\hline
                              &	EA$^{1/2u}_1$	&	EA$^{1/2u}_2$	&	EA$^{1/2g}_3$	&	EA$^{1/2u}_4$ \\ 
\hline
EOM $^2$DC$^M$ 	      &    2.51   	&         3.64          &         3.88         &        4.39            \\ 
IHFS   $^2$DC$^M$       &    2.51    	&         3.66          &         3.89         &        4.39            \\ 
EOM $^2$DCG$^M$ 	      &    2.51    	&         3.65          &         3.88         &        4.39            \\ 
IHFS  $^2$DCG$^M$      &    2.52         &         3.67          &         3.89         &        4.39            \\ 
\hline
\hline
\end{tabular}
\end{center}
\label{tab:triiodide-ip-ea}
\end{table}%

For the electron affinities the same trends as for ionization energies are observed with respect to the importance of the Gaunt interactions and the similarity of EOM and IHFS results. Unfortunately, we are unable to compare the
calculated values to experiment since, to the best of our knowledge, such results are not available in the literature.

\subsection{Dihalomethanes}

We now turn our attention to the diiodo- (CH$_2$I$_2$) and iodobromo-methane (CH$_2$IBr) species which, apart from their experimental interest, are examples of species with lower symmetry than those discussed before and therefore more costly to treat from a computational standpoint (CH$_2$IBr requiring the use of complex algebra).  

We begin with their ionisation energies shown in Table~\ref{tab:ip-halomethanes}.
We see there is hardly any difference between $^2$DCG$^M$ and $^2$DC$^M$ results (at most differences of 0.01 eV) for both CH$_2$I$_2$ and CH$_2$IBr. In all cases, the ionizations are determined to be single-particle processes, with the absence of important (2h1p) amplitudes in the EOM-IP case.
The ionization energies are in good agreement with experiment~\cite{Potts:79nRiJUm,Cartoni:2015ex,Satta:2015gp,Tsal:1975er,Lago:2005kr,Zhao:2014cu,Sandor:2016eq,Xing:2014gp,Lee:2005jt}, with typical differences being of the order of 0.1 eV. Such differences are quite far from what are the best experimental error bars avaliable~\cite{Lago:2005kr}, which are well under 0.05 eV for both species, but given that our calculations have not been performed at the experimental structures\cite{Kudchadker:1975cp} but rather on PBE-optimized ones, and still lack corrections due to basis set completeness--and probably more importantly, of higher-order electron correlation effects--we consider this accuracy to be sufficient for the purposes of this paper.

\begin{table}[htb]
\caption{Electron detachment energies ($\mathrm{IP_{n}}$, in eV) for
the CH$_2$I$_2$ and CH$_2$IBr species with the EOM-IP and IHFS(1h0p) 
approaches employing the molecular mean-field Hamiltonians with ($^2$DCG$^M$) 
and without ($^2$DC$^M$) the Gaunt interaction and augmented triple zeta bases
at the ZORA-SO/PBE/TZ2P geometry.  Experimental results and prior theoretical 
calculations are shown when available.}
\begin{center}
\begin{tabular}{l cccccc}
\hline
\hline
Method                        &	IP$_1$	&	IP$_2$	&	IP$_3$	&	IP$_4$	&	IP$_5$	 &    IP$_6$     \\ 
\hline
                  \multicolumn{7}{c}{CH$_2$I$_2$}\\
\hline
EOM $^2$DC$^M$ 	      &       9.37    &	      9.69    & 	10.12 	&	10.46 	&	12.75 	 &    13.52       \\ 
IHFS   $^2$DC$^M$       &       9.37    &	      9.69    & 	10.12 	&	10.46 	&	12.75 	 &    13.53       \\ 
EOM $^2$DCG$^M$ 	      &       9.36    &	      9.68    & 	10.12 	&	10.45   &	12.74  	 &    13.52       \\ 
IHFS   $^2$DCG$^M$      &       9.36    &       9.68    & 	10.12   &	10.45 	&	12.74 	 &    13.52       \\ 
                              &		      & 	      &	                &	        &	         &                \\
TD-B3LYP~\cite{Satta:2015gp}  &       9.46    &       9.57    &          9.74   &       10.29   &       12.65    &    13.32       \\ 
Exp.~\cite{Potts:79nRiJUm}    &       9.46    &       9.76    &         10.21   &       10.56   &       12.75    &    13.67       \\
Exp.~\cite{vonNiessen:1982fz} &       9.46    &       9.76    &         10.2    &       10.6    &       12.8     &    13.7        \\
Exp.~\cite{Lago:2005kr}       &       9.42    &        &                 &               &                &               \\
\hline
                  \multicolumn{7}{c}{CH$_2$IBr}\\
\hline
EOM           $^2$DC$^M$   &         9.65  &      10.17    &       10.79   &       10.98   &       13.33    &    14.24      \\	
IHFS       $^2$DC$^M$   &	        9.65  &      10.16    &       10.76   &       10.95   &       13.32    &    14.25      \\
EOM           $^2$DCG$^M$  &	        9.65  &      10.16    &       10.79   &       10.98   &       13.33    &    14.24      \\
IHFS       $^2$DCG$^M$   &	        9.65  &      10.16    &       10.76   &       10.95   &       13.32    &    14.25      \\
                              &		      &		      &      	      &		      &	  	       &               \\
SO-MRCI~\cite{Sandor:2014cu}       &       9.69    &      10.26    &      10.91    &      11.12    &      13.62     &               \\
Exp.~\cite{Lago:2005kr}       &     9.69      &          &               &               &                &               \\
\hline
\hline
\end{tabular}
\end{center}
\label{tab:ip-halomethanes}
\end{table}%

There have not been many other theoretical studies of ionization energies in the literature: we are aware of the TD-B3LYP~\cite{Satta:2015gp} study of Satta and coworkers for CH$_2$I$_2$ and the SO-MRCI studies of Weinacht and coworkers~\cite{GonzalezVazquez:2010ki,Geissler:2011dy,Sandor:2014cu} for CH$_2$IBr. With respect to TD-B3LYP, we observe our results are of similar quality for the first ionization energy, with B3LYP overestimating the experimental value by slightly less (0.04 eV) than we underestimate it (0.06 eV). However, for the higher ionizations we see a consistent underestimation of the experimental results by TD-B3LYP, while EOM-IP results are often 0.1 eV closer to experiment.

In the case of CH$_2$IBr, EOM-IP again underestimates the experimental result (by 0.04 eV) whereas the SO-MRCI results more closely match experiment, something that may reflect additional orbital relaxation in the SO-MRCI calculations since state-averaged (spin-free) orbitals were used. For the other ionizations the two methods yield results which are apart by more than 0.1 eV, with a notable difference of 0.29 eV for IP$_5$. Given that for CH$_2$I$_2$ the higher ionizations by EOM-IP have followed the experimental ones rather well, we wonder the extent to which the SO-MRCI results are biased towards the description of the low-lying states. Unfortunately, to our knowledge there are no experimental results for these ionizations to shed further light on the performance of the methods.

\begin{table}[htb]
\caption{Electron attachment energies ($\mathrm{EA_{n}}$, in eV) for
the CH$_2$I$_2$ and CH$_2$IBr species with the EOM-EA and IHFS(0h1p) 
approaches employing the molecular mean-field Hamiltonians with ($^2$DCG$^M$) 
and without ($^2$DC$^M$) the Gaunt interaction and augmented triple zeta bases
at the ZORA-SO/PBE/TZ2P geometry.  Experimental EAs~\cite{Modelli:1992jp} obtained by electron transmission
spectroscopy (ET) or dissociative attachment spectra (DA) are available for CH$_2$I$_2$ only.}
\begin{center}
\begin{tabular}{l l cccccc}
\hline
\hline
Method             &    EA$_1$ &   EA$_2$   & EA$_3$  & EA$_4$  & EA$_5$  & EA$_6$ \\
\hline
                  \multicolumn{7}{c}{CH$_2$I$_2$}\\
\hline
 EOM $^2$DC$^M$	&   -0.32   &    0.50    &  0.76   & 1.01   & 1.51   & 1.71 \\
 IHFS $^2$DC$^M$	&   -0.32   &    0.52    &  0.77   & 1.03   & 1.54   & 1.74 \\
 EOM  $^2$DCG$^M$	&   -0.32   &    0.50    &  0.76   & 1.01   & 1.51   & 1.71 \\
 IHFS $^2$DCG$^M$&   -0.32   &    0.52    &  0.77   & 1.03   & 1.52   & 1.74 \\
 	                &           &                &         &         &         & \\
Exp., ET~\cite{Modelli:1992jp} &  $<0$  &    0.68 &           &            &              & \\
Exp., DA~\cite{Modelli:1992jp} &  $<0$  &    0.46 &           &            &              & \\
\hline
                  \multicolumn{7}{c}{CH$_2$IBr}\\
\hline
EOM $^2$DC$^M$	&   -0.02   & 0.54 & 0.99 & 1.02 & 1.64 & 1.93 \\
IHFS $^2$DC$^M$  &    0.01 & 0.59 & 1.04 & 1.11 & 1.66 & 2.02 \\
EOM $^2$DCG$^M$	&   -0.02   & 0.54 & 0.99 & 1.02 & 1.64 & 1.93 \\
IHFS $^2$DCG$^M$  &    0.01 & 0.59 & 1.04 & 1.12 & 1.66 & 2.02 \\
\hline
\hline
\end{tabular}
\end{center}
\label{tab:ea-halomethanes}
\end{table}%

Our results for electron affinities are summarized in Table~\ref{tab:ea-halomethanes}. From these we see that for CH$_2$I$_2$ we have both EOM-EA and IHFS(0h1p) predicting a bound electron attachment state (corresponding to the CH$_2$I$_2^-$ species), in line with the experimental results for Modelli and coworkers~\cite{Modelli:1992jp}, who measured a bound first electron attachment state, and a second attachment energy. In passing we note that Modelli and coworkers found bound states for CHI$_3^-$ and CI$_4^-$ as well. Our calculations have placed the first electron attachment state (EA$_1$) at 0.32 eV below the ground state for the neutral species, and the second state (EA$_2$) at 0.50 eV, which agrees well with the value obtained via the dissociative attachment spectra (0.46 eV). 
We agree less in the interpretation of the process: with the help of MS-X$\alpha$ calculations, Modelli and coworkers have modelled it as the addition of an electron to a single virtual orbital. In our calculations, all electronic states but EA$_4$ correspond to multideterminantal wavefunctions--for the first two states, which are the most relevant ones for the comparison to the experimental results, we have that a determinant with an electron attached to the LUMO and another with an electron attached to the LUMO+1 contribute to the wavefunctions of EA$_1$ to about  56\% and 30\%, respectively; in the case of EA$_2$, these contribute by about 37\% and 59\%, respectively. 

The EOM-EA calculations for CH$_2$IBr show a similar trend, but with the first electron attachment state (EA$_1$) being only slightly bound, at 0.02 eV below the CH$_2$IBr ground state energy and with a second attachment state (EA$_2$) at around 0.54 eV above the CH$_2$IBr ground state energy. We also observe that wavefunctions for the electron attachment states are made up of more than a single determinant.  The IHFS(0h1p) calculations yield results not far from EOM-EA for all the electron attachment states considered, but that show instead a weakly unbound (0.01 eV) EA$_1$. We believe this is an artifact of the calculations, due to the impossibility of using a larger model space.  
There are unfortunately no experimental results to which compare our calculations for CH$_2$IBr, but we note that in the work of Guerra and coworkers~\cite{Guerra:1991jp}, for the series of chloromethanes (from CH$_3$Cl to CCl$_4$) only CCl$_4^-$ shows a bound state; for the series of bromomethanes (from CH$_3$Br to CBr$_4$) the CH$_2$Br$_2^-$ species is not bound (though it shows a state slightly above zero energy~\cite{Modelli:1992jp}), but further substituting hydrogens by bromines yields stable anions. These findings, taken together with those for the iodine-substituted species, indicate that the heavier halogens help stabilize the first electron attachment state for the same degree of substitution. This makes it plausible that CH$_2$IBr$^-$, by the substitution of bromine by iodine, would have its first electron attachment stable stabilized with respect to CH$_2$Br$_2^-$ and have it become (weakly) bound.

Finally, we present results for excitation energies in Tables~\ref{tab:excitations-ch2i2} and~\ref{tab:excitations-ch2ibr} for CH$_2$I$_2$ and CH$_2$IBr, respectively. We are only aware of the works of Liu and coworkers~\cite{Liu:2007cg,Liu:2006cp}, who considered SOC for these systems with the SO-CASPT2 approach.
The electronic spectra of CH$_2$I$_2$ has been well-studied experimentally, in gas-phase~\cite{Ito:1961eh,Kawasaki:1975gb} and in organic solvents~\cite{Ito:1961eh,Gedanken:1979fa}. In the gas-phase, two main features at 4.29 eV and 4.98 eV have been first identified~\cite{Ito:1961eh}, with later photodissociation 
studies~\cite{Kawasaki:1975gb} revealing additional transitions. For CH$_2$IBr we are aware of studies in the gas phase~\cite{Butler:1987il}, which yield valence transitions at 4.58 eV and 5.79 eV, which are slightly changed in the presence of a organic solvent~\cite{Lee:1982ko}. 

As for I$_3^-$, we observe in Table~\ref{tab:excitations-ch2i2} a tendency of EOM-EE results to systematically overestimate the IHFS(1h1p) ones, with a mean deviation of 0.18 eV for both Hamiltonians considered (with differences between $^2$DCG$^M$ and $^2$DC$^M$ of the order of 0.01 eV or less) and similarly low $\sigma$ values (0.03 eV), which is quite close to what is obtained for I$_3^-$ (MAD of 0.17 eV and $\sigma$ of 0.02 eV). The states considered have been found to be of a singly excited nature and the EOM and IHFS wavefunctions are dominated by the same excited determinants. We observe once more the tendency of SO-CASPT2 to show somewhat higher deviations to IHFS(1h,1p) than EOM-EE, with much more uneven errors for the different excitations than EOM-EE (MAD of 0.43 eV and $\sigma$ of 0.09 eV). Without transition moments it is difficult to comment on the accuracy with respect to the experimental values, since our results show a number of close-lying states with energies close to the experimental peak values, but at the same time this give us confidence that we shall be able to reproduce the peak positions to a tenth of an eV or less.

\begin{table}[htb]
\caption{Vertical excitation energies (T$_v$, in eV) for CH$_2$I$_2$ 
with the EOM-EE and IHFS(1h1p) approaches employing the molecular mean-field Hamiltonians 
with ($^2$DCG$^M$) and without ($^2$DC$^M$) the Gaunt interaction. We also present
the SO-CASPT2 results with ANO-RCC bases of Liu and coworkers\cite{Liu:2007cg} 
(at CASPT2 equilibrium geometries in C$_1$ symmetry) and experimental results, 
along with the mean absolute difference (MAD) and its standard deviation ($\sigma$) between the EOM-EE 
and IHFS(1h1p) approaches and between IHFS(1h1p)-$^2$DCG$^M$ and SO-CASPT2.}
\begin{center}
\begin{tabular}{c c cc p{0.2cm} cc p{0.2cm} c}
\hline
\hline
	&		&    \multicolumn{2}{c}{$^2$DC$^M$ } 	         	&&	\multicolumn{2}{c}{$^2$DCG$^M$ } &&		\\
\cline{3-4}
\cline{6-7}
State	& Symmetry	&	EOM 	&IHFS      	&&	EOM	&IHFS      	&&	CASPT2\cite{Liu:2007cg}\\
\hline
1	&	a	&	3.60 	&	3.44	&&	3.60 	&	3.44	&&	3.76	\\
2	&	b	&	3.62 	&	3.46	&&	3.62 	&	3.45	&&	3.78	\\
3	&	a	&	3.63 	&	3.47	&&	3.63 	&	3.46	&&	3.78	\\
4	&	b	&	3.85 	&	3.68	&&	3.85 	&	3.68	&&	4.03	\\
5	&	a	&	3.87 	&	3.70	&&	3.87 	&	3.70	&&	4.27	\\
6	&	b	&	3.94 	&	3.79	&&	3.94 	&	3.79	&&	4.27	\\
7	&	a	&	3.99 	&	3.83	&&	3.99 	&	3.83	&&	4.31	\\
8	&	b	&	4.06 	&	3.90	&&	4.06 	&	3.90	&&	4.38	\\
9	&	b	&	4.22   	&	4.04	&&	4.22 	&	4.03	&&	4.50	\\
10	&	a	&	4.32   	&	4.15	&&	4.32 	&	4.14	&&	4.60	\\
11	&	b	&	4.35   	&	4.17	&&	4.35 	&	4.16	&&	4.62	\\
12	&	a	&	4.49   	&	4.32	&&	4.49 	&	4.31	&&		\\
13	&	b	&	4.64   	&	4.48	&&	4.63 	&	4.47	&&		\\
14	&	a	&	4.68   	&	4.52	&&	4.68 	&	4.52	&&		\\
15	&	b	&	4.75   	&	4.58	&&	4.74 	&	4.58	&&		\\
16	&	a	&	4.91   	&	4.75	&&	4.91 	&	4.74	&&		\\
17	&	a	&	5.59   	&	5.33	&&	5.59 	&	5.33	&&		\\
18	&	b	&	5.61   	&	5.35	&&	5.61 	&	5.35	&&		\\
MAD     &               &       0.18    &               &&      0.18    &               &&      0.43    \\
$\sigma$&               &       0.03    &               &&      0.03    &               &&      0.09    \\
\hline
\hline
\end{tabular}
\end{center}
\flushleft
Exp., photodissociation of molecular beams, model\cite{Kawasaki:1975gb}: 3.98, 4.34, 4.42, 4.89, 4.97, 5.79. \\ 
Exp., absorption in vapour, maxima\cite{Ito:1961eh}: 4.29, 4.98. \\
Exp., absorption in iso-octane, maxima\cite{Ito:1961eh}: 4.25, 4.94, 5.85. \\
Exp., absorption in iso-octane, resolved\cite{Ito:1961eh}: 4.02, 4.39, 4.91, 5.88. \\ 
Exp., MCD in cyclohexane, resolved\cite{Gedanken:1979fa}: 3.97, 4.19, 4.49, 4.91, 5.48, 5.88. \\ 
\label{tab:excitations-ch2i2}
\end{table}%

For the excitation energies of CH$_2$IBr, the general observations with respect to a comparison to experiment made for CH$_2$I$_2$ apply, as we see from Table~\ref{tab:excitations-ch2ibr} that we obtain EOM-EE excitation energies that match well the energies of the experimental peak maxima, and that there's little differences between Hamiltonians (in general, differences are smaller than 0.01 eV). We note from the EOM-EE results that once more there's little difference between Hamiltonians. Unfortunately, we were unable to perform a detailed comparison to IHFS(1h1p) due to the impossibility of converging the latter calculations, and therefore only make a comparison to SO-CASPT2~\cite{Liu:2006cp}. As was the case for CH$_2$I$_2$, there is a general shift to higher energies in the SO-CASPT2 results compared to the EOM-EE $^2$DCG$^M$ ones, but which is not very systematic: shifts for the low-end of the spectrum (up to 4.8 eV) are of $\simeq$0.15-0.21 eV range, with the high-end (from about 5.8 eV onwards) showing large variations (from 0.11 to 0.42 eV) while the variations in the mid-range of the spectrum are of about 0.1 eV, which make the $\sigma$ value go up to 0.17 eV.

\begin{table}[htb]
\caption{Vertical excitation energies (T$_v$, in eV) for CH$_2$IBr
with the EOM-EE approach employing the molecular mean-field Hamiltonians 
with ($^2$DCG$^M$) and without ($^2$DC$^M$) the Gaunt interaction, using the 
correlating space which does not include the Br $3d$ spinors. We also present 
the SO-CASPT2 results with ANO-RCC bases of Liu and coworkers\cite{Liu:2006cp} 
(obtained at CASPT2 equilibrium geometries, and for C$_1$ symmetry) and experimental 
results, along with the mean absolute difference (MAD) and its standard deviation ($\sigma$) 
between EOM-EE-$^2$DCG$^M$ and SO-CASPT2.}
\begin{center}
\begin{tabular}{c c c p{0.2cm} c p{0.2cm} c}
\hline
\hline
State	& Symmetry	&   $^2$DC$^M$ 	&&    $^2$DCG$^M$	&&	CASPT2\cite{Liu:2007cg}\\
\hline
1	&	b	&	3.86	&&	3.86	&&	4.07	\\
2	&	a	&	3.86	&&	3.86	&&	4.07	\\
3	&	b	&	3.98	&&	3.98	&&	4.16	\\
4	&	a	&	4.04	&&	4.04	&&	4.25	\\
5	&	b	&	4.34	&&	4.34	&&	4.49	\\
6	&	a	&	4.45	&&	4.44	&&	4.61	\\
7	&	b	&	4.66	&&	4.66	&&	4.75	\\
8	&	a	&	4.67	&&	4.67	&&	4.77	\\
9	&	b	&	5.08	&&	5.07	&&	5.17	\\
10	&	a	&	5.13	&&	5.13	&&	5.21	\\
11	&	b	&	5.16	&&	5.16	&&	5.25	\\
12	&	a	&	5.30	&&	5.30	&&	5.39	\\
13	&	b	&	5.37	&&	5.37	&&	5.44	\\
14	&	a	&	5.39	&&	5.38	&&	5.81	\\
15	&	b	&	5.80	&&	5.80	&&	5.91	\\
16	&	a	&	5.82	&&	5.82	&&	6.24	\\
17	&	b	&	6.09	&&	6.09	&&	6.24	\\
18	&	b	&	6.10	&&	6.10	&&	6.28	\\
MAD	&		&		&&		&&	0.17	\\
$\sigma$&		&		&&		&&	0.17	\\
\hline
\hline
\end{tabular}
\end{center}
\flushleft
Exp., absorption on hexane\cite{Lee:1982ko}: 4.63, 5.82 \\
Exp., absorption in gas phase\cite{Butler:1987il}: 4.58, 5.79; 6.52 (Rydberg)
\label{tab:excitations-ch2ibr}
\end{table}%

\section{Conclusions}

In this work we have described the formulation and implementation in the \textsc{Dirac} code of the EOM-CCSD 
method for electron attachment (EOM-EA-CCSD), electron detachment (EOM-IP-CCSD) and excitation energies (EOM-EE-CCSD) 
based on four-component Hamiltonians. This implementation, which can be used with any of the Hamiltonians available in 
\textsc{Dirac}, exploits double point group symmetry for all of the above EOM variants, and yields both left and 
right eigenvectors. 

We have proceeded in validating the implementation by a careful comparison to intermediate Hamiltonian Fock-space (IHFS-CCSD) calculations on 
different classes of systems: the series of halogen monoxide radicals (XO, X = Cl--Ts); the triiodide (I$_3^-$) species and 
some of its electron detached (the I$_3$ radical) and attached (the I$_3^{2-}$ species) states; as well as the neutral, 
cationic and anionic forms of the diiodo- and iodobromethane molecules.

In all of the electron attachment and detachment cases considered (XO, I$_3$/I$_3^{2-}$, CH$_2$I$_2^{-/+}$, CH$_2$IBr$^{-/+}$) we have found 
the EOM-CCSD and IHFS-CCSD methods to differ only slightly, as expected from formal considerations. Whenever more significant discrepancies 
were found, we believe to have shown these are due to the shortcomings on the main model spaces employed in the IHFS-CCSD case, which was not 
sufficiently large. In the cases where experimental data are available, our calculations are in very good agreement with them.

For the excited states, in the cases where we compared singly excited states (I$^-_3$ and CH$_2$I$_2$) and had a ground state reference
described by a single determinant, the agreement between EOM-EE-CCSD and IHFS-CCSD is quite good, though EOM-EE-CCSD shows a tendency to systematically
overestimate the IHFS-CCSD values. From formal considerations the two methods are in general not expected to yield the same results due to the lack of
valence extensivity for EOM-EE-CCSD. The differences found here are in line with those found in other comparisons of EOM-EE-CCSD and IHFS-CCSD
for other small systems (containing or not heavy centers). 

We believe the tradeoff in EOM-EE-CCSD between losing some accuracy (in the absolute sense) by forsaking valence extensivity and the gain in robusteness 
in the calculations (due to the diagonalization-based approach used, as opposed to the iterative one used in IHFS-CCSD) is well worth taking in practical 
applications.  This is exemplified in the case of CH$_2$IBr, for which we have not been able to  converge the IHFS-CCSD(1h1p) calculations. Though
in this case we have been unable to compare the two coupled cluster approaches, we note the  performance of EOM-EE-CCSD relative to the 
SO-CASPT2 results is similar to that for CH$_2$I$_2$ and I$^-_3$.  We believe a more stringent comparison to SO-CASPT2 and experiments requires
the availability of transition moments for the EOM-EE-CCSD case, which we shall describe in a subsequent publication. 

Another important aspect not addressed in this work is that of going beyond the CCSD model for transition energies, since it 
is known for instance that molecular properties and energetics can be quite sensitive to details of the electron correlation. 
In subsequent publications we envisage to explore the inclusion of triple excitations, notably via a perturbative approach, on the different EOM 
variants~\cite{Watts:1996bi,Stanton:1996hw,Manohar:2009kza,Matthews:2016bi}.

Finally, as far as Hamiltonians are concerned we have shown first that, based on single-point comparisons of our EOM-CC results for the XO 
systems and triiodide calculations performed with the Dirac--Coulomb Hamiltonian, that the corresponding two-component molecular mean-field approach ($^2$DC$^M$) does indeed 
yield results  which are nearly indistinguishable from the former, and for this reason we strongly recommend the use of the molecular mean-field approach. 
Concerning the inclusion of the Gaunt interaction, we have determined it to have in general a small effect on the 
energetics of the species under consideration (and on the spectroscopic constants for the case of halogen monoxides), though
for ClO, BrO and TsO it is necessary to include it as it corresponds to roughly between 7\% and 2\% of that of the spin-orbit
splitting of the $^2\Pi$ ground-state.

\section{Supplementary Material}

See supplementary information for results discussed but not shown in the manuscript: (a) spectroscopic constants, vertical and adiabatic excitation energies, 
projection and wavefunction analysis for halogen monoxide radicals, along with spinor magnetization plots for the halogen monoxide anions; 
(b) DC and ${^2}$DC$^M$ energies for halogen monoxides (anions and radicals), triiodide radical and anions; (c) optimized structures for the halomethanes. 
These resources are also available via the zenodo repository~\cite{paper:dataset}, along with the outputs for the respective calculations.

\section{Acknowledgements}

The members of the PhLAM laboratory acknowledge support from the CaPPA project (Chemical and Physical Properties of the Atmosphere), 
funded by the French National Research Agency (ANR) through the PIA (Programme d'Investissement d'Avenir) under contract 
``ANR-11-LABX-0005-01'' as well as by the Ministry of Higher Education and Research, Hauts de France council and European Regional 
Development Fund (ERDF) through the Contrat de Projets Etat-Region (CPER) CLIMBIO (Changement climatique, dynamique de 
l'atmosph\`ere, impacts sur la biodiversit\'e et la sant\'e humaine).
Furthermore, ASPG acknowledges funding from the CNRS Institute of Physics (INP) via the PICS program (grant 6386), and 
computational time provided by the French national supercomputing facilities (grants DARI x2016081859, A0010801859, A0030801859), and 
both AS and ASPG acknowledge many illuminating discussions on the implementation of the matrix-free diagonalization method 
with Dr.\ Jean-Pierre Flament (PhLAM).

\appendix

\section{Use of double group symmetry in tensor contraction involving non-totally symmetric tensors}\label{Symmetry}
Let us consider a generalized tensor contraction involving tensors belonging to non-totally symmetric irreps of the following form

\begin{equation}
A_{ij..}^{k'l'..} * B_{kl..}^{i'j'..} = C_{ij\ldots kl..}^{k'l'\ldots i'j'..} \label{Eq:tensor_prod}
\end{equation}

In \autoref{Eq:tensor_prod}, when upper (primed) and lower (unprimed) indices are the same, they define a contraction.  

We define the product irreps
$\Gamma_{K_f}$ and $\Gamma_{K_c}$ for the products of all free (f) or contracted (c) ket indices of tensor $B$. The latter product is necessarily equal
to the product of contracted indices appearing in tensor $A$, but since these indices then refer to bra functions the result will be the complex conjugate 
irrep $\Gamma^*_{K_c}$. The same is true for the irrep product of the contracted bra functions of $B$: $\Gamma^*_{L_c}$, which is equal to the product
$\Gamma_{L_c}$ appearing in the ket of tensor $A$. The free indices of both tensors are different and lead to four possible product irreps: $\Gamma^*_{L'_f}$,
$\Gamma^*_{K'_f}$, $\Gamma_{L_f}$, and $\Gamma_{K_f}$.  

Using these definitions, the tensor contraction can be expressed with its explicit symmetry content as 

{\scriptsize
\begin{align}
A ((\Gamma^*_{k'} \otimes  \Gamma^*_{l'} \ldots)  \otimes (\Gamma_{i} \otimes \Gamma_{j}\ldots) \equiv 
(\Gamma^*_{K'_f} \otimes  \Gamma^*_{K_c})  \otimes (\Gamma_{L_f} \otimes  \Gamma_{L_c}) \equiv \Gamma_A )
\nonumber \\ 
\quad B ((\Gamma^*_{i'} \otimes  \Gamma^*_{j'} \ldots) \otimes (\Gamma_{k} \otimes \Gamma_{l}\ldots) \equiv
(\Gamma^*_{L'_f} \otimes \Gamma^*_{L_c}) \otimes (\Gamma_{K_f} \otimes \Gamma_{K_c}) \equiv \Gamma_B) 
\nonumber \\   
\qquad = C ((\Gamma^*_{K'_f} \otimes  \Gamma^*_{L'_f}) \otimes (\Gamma_{L_f} \otimes \Gamma_{K_f}) \equiv \Gamma_C) \nonumber.     
\end{align}.  
}

In order to use the efficient BLAS matrix multiplication routines, we first sort the tensors in a block sparse manner  

{\scriptsize
\begin{align}
A  \rightarrow A'  :  (\Gamma^*_{K'_f} \otimes  \Gamma_{L_f}) \otimes  (\Gamma^*_{K_c} \otimes  \Gamma_{L_c}) =  \Gamma_{f_A}\otimes\Gamma_{c_A} {} =& \Gamma_A \label{Eq:blas_A}\\
B \rightarrow B' :   (\Gamma^*_{L_c} \otimes  \Gamma_{K_c}) \otimes (\Gamma^*_{L'_f} \otimes  \Gamma_{K_f}) = 
\Gamma_{c_B}\otimes\Gamma_{f_B} = {} & \Gamma_B  \label{Eq:blas_b}
\end{align}
}

and then multiply the sorted tensors $A'$ and $B'$ to produce the product tensor $C'$

\begin{align}
C \leftarrow C' : (\Gamma^*_{K'_f} \otimes  \Gamma_{L_f}) \otimes (\Gamma^*_{L'_f} \otimes  \Gamma_{K_f}) = 
\Gamma_{f_A}\otimes\Gamma_{f_B} = {} & \Gamma_C, \label{Eq:blas_c} 
\end{align}

Here we take into account that $\Gamma_{c_A} = \Gamma_{c_B}^*$. We then resort $C'$ tensors to obtain $C$.

We have included this extension to our previous contraction routines by communicating the irreps to which left and right input tensors belong to as two additional arguments. As before, the symmetry of the product tensor is obtained by taking direct product between the input irreps using the multiplication tables that were already available.

\section {EOMCC $\sigma$-vector equations} \label{work_eqs}
\begin{itemize}
\item [IP:] \begin{align}
^R \sigma_i = {} & - \sum_m \overline {F}^m_i  r_m + \sum_{me} \overline{F}^m_e  r_{im}^{e} + \sum_{m>n,e}  W^{mn}_{ie} r_{mn}^{e} \\
^R \sigma_{ij}^{a} = {} & - \sum_m W^{ma}_{ij}  r_m - P(ij) \sum_m \overline{F}^m_i r^a_{mj} \nonumber \\
                  {} & + \sum_{mn} W ^{mn}_{ij} r_{mn}^a - P(ij) \sum_{me} W^{ma}_{ie} r_{mj}^e \nonumber \\
                  {} &  + \sum_e \overline{F}^a_e r_{ij}^e + \sum_{m>n,ef} (W^{mn}_{ef} r_{mn}^{f}) t^{ea}_{ij}                 
\end{align}
\begin{align}
^L \sigma^i = {} & - \sum_m \overline {F}^i_m  l^m  + \sum_{m>n,e}  W^{ie}_{mn} l^{mn}_{e} \\
^L \sigma^{ij}_{a} = {} & - \sum_m W_{ma}^{ij}  l^m + \sum_e \overline{F}^e_a l_e^{ij} \nonumber \\
                  {} & - P(ij) \sum_m \overline{F}^j_m l_a^{im}  +  P(ij) \sum_{me} W_{am}^{je} l^{im}_e  \nonumber \\
                  {} & + \sum_{m>n} W ^{ij}_{mn} l^{mn}_a + \sum_{m>n,ef} V^{ij}_{ea} (l^{mn}_{f} t^{ef}_{mn})  + P(ij) l^i \overline{F}^j_a              
\end{align}

\item[EA:] \begin{align}
^R \sigma^a = {} & \sum_e \overline{F}^a_e r^e + \sum_{m,e} \overline{F}^m_e r^{ea}_m + \sum_{e>f, m} W^{am}_{fe} r^{ef}_m  \\
^R \sigma^{ab}_i = {} & \sum_e W^{ab}_{ei} r^e + P(ab) \sum_e \overline{F}^a_e r^{eb}_i  \nonumber  \\
         {} & + \sum_{e>f} W^{ab}_{ef} r^{ef}_i - P(ab) \sum_{em} W^{am}_{ei} r^{eb}_m - \sum_m \overline{F} ^m_i r ^{ab}_ m \nonumber \\
         {} & + \sum_{mn,e>f} (V^{mn}_{ef} r^{ef}_n) t^{ab}_{mi}
\end{align}
\begin{align}
^L \sigma_a = {} & \sum_e \overline{F}^e_a l_e - \sum_{m,e>f} W^{ef}_{am} l_{ef}^m  \\
^L \sigma_{ab}^i = {} & \sum_e W_{ba}^{ei} l_e + P(ab) \sum_e \overline{F}_a^e l_{eb}^i - \sum_m \overline{F} _m^i l_{ab}^m \nonumber \\
                   {} & + \sum_{e>f} W_{ab}^{ef} l_{ef}^i - P(ab) \sum_{em} W_{bm}^{ei} l_{ae}^m \nonumber \\ 
                   {} & - \sum_{mo,e>f} (l_{ef}^o t^{ef}_{mo}) V^{mi}_{ba} + P(ab) \overline{F}_a^i  l_b
\end{align}

\item[EE:] 
\begin{align}
^R \sigma_i^a = {} & \sum_e \overline{F}^a_e r^e_i - \sum_m \overline {F}^m_i  r_m^a + \sum_{me} \overline{F}^m_e r^{ea}_{mi} \nonumber  \\
                        {} &  + \sum_{me} W^{ma}_{ei} r^e_m + \sum_{m,e>f} W^{am}_{ef} r^{ef}_{im} - \sum_{m>n,e}  W^{mn}_{ie} r_{mn}^{ea} \\
^R \sigma_{ij}^{ab} = {} & - P(ab) \sum_m W^{mb}_{ij} r^a_m + P(ij) \sum_e W^{ab}_{ej} r^e_i \nonumber \\
                   {} & + P(ab) \sum_{efm} (W^{bm}_{fe} r^e_m) t^{af}_{ij} + P(ij) \sum_{mne} (W^{nm}_{je} r^e_m) t^{ab}_{in} \nonumber \\
                   {} & + P(ab) \sum_e \overline{F}^b_e r^{ae}_{ij} - P(ij) \sum_m \overline{F}^m_j r^{ab}_{im} \nonumber \\
                   {} & + \sum_{m>n} W^{mn}_{ij} r ^{ab}_{mn} + P(ab)P(ij) \sum_{me} W^{mb}_{ej} r^{ae}_{im} \nonumber\\
                   {} & - P(ab) \sum_{ef,m>n} (V^{nm}_{fe} r^{ea}_{mn}) t^{fb}_{ij}  \nonumber \\
                   {} & - P(ij) \sum_{e>f, mn} (V^{nm}_{fe} r ^{fe}_{im}) t^{ba}_{jn} + \sum_{e>f} W^{ab}_{ef} r^{ef}_{ij}                          
\end{align}

\begin{align}
^L \sigma^i_a = {} & \overline{F}^i_a + \sum_e l^i_e \overline{F}^e_a - \sum_m l^m_a \overline{F}^i_m  + \sum_{m,e<f} l^{im}_{ef} W ^{ef}_{am} \nonumber \\
                          {} & - \sum_{e,f} G^{f}_{e} W^{ei}_{fa} - \sum_{mn} G^n_m W^{mi}_{na} + \sum_{me} l^m_e W ^ {ie} _{am} \nonumber \\
                          {} & - \sum_{m>n,e} l^{mn}_{ae} W^{ie}_{mn} \\
^L \sigma^{ij}_{ab} = {} & V^{ij}_{ab} + P(ab) \sum_e l^{ij}_{ae} \overline{F}^e_b - P(ij) \sum_m l^{im}_{ab} \overline{F}^j_m \nonumber \\ 
                                 {} & + \sum_{m>n} l^{mn}_{ab} W^{ij}_{mn} + P(ij) P(ab) \sum_{me} l^ {im}_{ae} W^{je}_{bm} \nonumber \\
                                 {} & + P(ab) \sum_e V^{ij}_{ae} G^e_b -  \sum_{m} l^m_a W ^{ij}_{mb}  - P(ij) \sum_m V^{im}_{ab} G^j_m  \nonumber  \\ 
                                 {} & + P(ij) \sum_{m,e} V^{mj}_{ab} (l ^i_e t^e_m) + P(ij) \sum_e l_e^i V^{ej}_{ab} \nonumber \\
                                 {} & + P(ij)P(ab) l^i_a \overline{F}^j_b + \sum_{e>f} l^{ij}_{ef} W ^{ef}_{ab}
\end{align}

\end{itemize}

The $\overline{F}$, $W$ and $G$ intermediates are defined as follows:

\begin{align}\label{eq:fintm1}
\overline{F}^{i}_{m} =  {} & f^{i}_{m} + \sum_e f^i_e t^e_m + \sum_{en} V^{in}_{me} t ^e_n + \sum_{n,e>f} V^{in}_{ef} \tau^{ef}_{mn} \\
\overline{F}^{e}_{a} =  {} & f^{e}_{a} - \sum _m f^m_a t_m^e -  {\sum_{mf} V^{me}_{fa} t_m^f}  - \sum_{m>n,f} V^{mn}_{af} \tau^{ef}_{mn}\\
\overline{F}^{m}_{e} = {} & f^m_e + \sum_{nf} v^{mn}_{ef} t^f_n,
\end{align}

\begin{align}
   W^{ij}_{mn} = {} & V^ {ij}_{mn} + P (mn) \sum_e V^ {ij}_{en} t_m^e + \sum_{e>f} V^{ij}_{ef} \tau^ {ef}_{mn} \\
   W^{mb}_{ej} = {} & V^ {mb}_{ej} + {\sum_f V^{mb}_{ef} t_j^f} - \sum_n V^{mn}_{ej} t_n^b \nonumber \\
                 {} & + \sum_{nf} V ^ {mn}_{ef} (t^{fb}_{jn} - t^f_j t^b_n) \\
   W^{ie}_{mn} = {} & V^{ie}_{mn} + \sum_{f} \overline{F}^{i}_{f} t^{ef}_{mn} - \sum_o W^{io}_{mn} t^e_o +  {\sum_{f>g} V^{ie}_{fg} \tau ^{fg}_{mn}} \nonumber \\
                 {} & +  P(mn) \sum_f \overline{W}^{ie}_{fn} t^f_m  + {P(mn) \sum_{of} V^{io}_{mf} t^{ef}_{no}}\\
   W^{ef}_{am} = {} &  {V^{ef}_{am} + P (ef) \sum _{ng} V^{en}_{ag}  \tau^{gf}_{mn} + \sum_g W^{ef}_{ag} t^g_m} \nonumber \\
                 {} & { + \sum_n \overline{F}^n_a t^{ef}_{mn} + \sum_{n>o} V_{am}^{no} \tau^{ef}_{no} - P(ef) \sum_n \overline{W}^{nf}_{am}  t^e_n} \label{Eq:3-virt} \\
   W^{ef}_{ab} = {} &   {V^{ef}_{ab} - P(ef) \sum_m V^{mf}_{ab} t^e_m + \sum_{m>n} V^{mn}_{ab} \tau ^{ef}_{mn}} \label{Eq:4-virt}\\ 
   \overline {W}^{mb}_{ej} = {} & V^{mb}_{ej} - \sum_{nf} V^{mn}_{ef} t^{bf}_{nj} \\
   W^{mn}_{ie} = {} & V^{mn}_{ie} + \sum_f t^f_i V^{mn}_{fe} \\
   W^{am}_{ef} = {} & V^{am}_{ef} - \sum_n V^{nm}_{ef} t^a_n \label{eq:fintm2} \\
   G^e_a       = {} & - \sum_{m>n,f} \lambda^{mn}_{af} t^{ef}_{mn} \\
   G^i_m       = {} & \sum_{n,e>f} \lambda^{in}_{ef} t^{ef}_{mn} 
\end{align}


\begin{thebibliography}{163}%
\makeatletter
\providecommand \@ifxundefined [1]{%
 \@ifx{#1\undefined}
}%
\providecommand \@ifnum [1]{%
 \ifnum #1\expandafter \@firstoftwo
 \else \expandafter \@secondoftwo
 \fi
}%
\providecommand \@ifx [1]{%
 \ifx #1\expandafter \@firstoftwo
 \else \expandafter \@secondoftwo
 \fi
}%
\providecommand \natexlab [1]{#1}%
\providecommand \enquote  [1]{``#1''}%
\providecommand \bibnamefont  [1]{#1}%
\providecommand \bibfnamefont [1]{#1}%
\providecommand \citenamefont [1]{#1}%
\providecommand \href@noop [0]{\@secondoftwo}%
\providecommand \href [0]{\begingroup \@sanitize@url \@href}%
\providecommand \@href[1]{\@@startlink{#1}\@@href}%
\providecommand \@@href[1]{\endgroup#1\@@endlink}%
\providecommand \@sanitize@url [0]{\catcode `\\12\catcode `\$12\catcode
  `\&12\catcode `\#12\catcode `\^12\catcode `\_12\catcode `\%12\relax}%
\providecommand \@@startlink[1]{}%
\providecommand \@@endlink[0]{}%
\providecommand \url  [0]{\begingroup\@sanitize@url \@url }%
\providecommand \@url [1]{\endgroup\@href {#1}{\urlprefix }}%
\providecommand \urlprefix  [0]{URL }%
\providecommand \Eprint [0]{\href }%
\providecommand \doibase [0]{http://dx.doi.org/}%
\providecommand \selectlanguage [0]{\@gobble}%
\providecommand \bibinfo  [0]{\@secondoftwo}%
\providecommand \bibfield  [0]{\@secondoftwo}%
\providecommand \translation [1]{[#1]}%
\providecommand \BibitemOpen [0]{}%
\providecommand \bibitemStop [0]{}%
\providecommand \bibitemNoStop [0]{.\EOS\space}%
\providecommand \EOS [0]{\spacefactor3000\relax}%
\providecommand \BibitemShut  [1]{\csname bibitem#1\endcsname}%
\let\auto@bib@innerbib\@empty
\bibitem [{\citenamefont {Grimme}(2004)}]{grimme_calculation_2004}%
  \BibitemOpen
  \bibfield  {author} {\bibinfo {author} {\bibfnamefont {S.}~\bibnamefont
  {Grimme}},\ }in\ \href {\doibase 10.1002/0471678856.ch3} {\emph {\bibinfo
  {booktitle} {Reviews in Computational Chemistry}}},\ Vol.~\bibinfo {volume}
  {20},\ \bibinfo {editor} {edited by\ \bibinfo {editor} {\bibfnamefont
  {K.~B.}\ \bibnamefont {Lipkowitz}}, \bibinfo {editor} {\bibfnamefont
  {R.}~\bibnamefont {Larter}}, \ and\ \bibinfo {editor} {\bibfnamefont {T.~R.}\
  \bibnamefont {Cundari}}}\ (\bibinfo  {publisher} {Wiley},\ \bibinfo {year}
  {2004})\ pp.\ \bibinfo {pages} {153--218}\BibitemShut {NoStop}%
\bibitem [{\citenamefont {Helgaker}\ \emph {et~al.}(2012)\citenamefont
  {Helgaker}, \citenamefont {Coriani}, \citenamefont {J{\o}rgensen},
  \citenamefont {Kristensen}, \citenamefont {Olsen},\ and\ \citenamefont
  {Ruud}}]{Helgaker2012}%
  \BibitemOpen
  \bibfield  {author} {\bibinfo {author} {\bibfnamefont {T.}~\bibnamefont
  {Helgaker}}, \bibinfo {author} {\bibfnamefont {S.}~\bibnamefont {Coriani}},
  \bibinfo {author} {\bibfnamefont {P.}~\bibnamefont {J{\o}rgensen}}, \bibinfo
  {author} {\bibfnamefont {K.}~\bibnamefont {Kristensen}}, \bibinfo {author}
  {\bibfnamefont {J.}~\bibnamefont {Olsen}}, \ and\ \bibinfo {author}
  {\bibfnamefont {K.}~\bibnamefont {Ruud}},\ }\href {\doibase
  10.1021/cr2002239} {\bibfield  {journal} {\bibinfo  {journal} {Chem. Rev.}\
  }\textbf {\bibinfo {volume} {112}},\ \bibinfo {pages} {543} (\bibinfo {year}
  {2012})}\BibitemShut {NoStop}%
\bibitem [{\citenamefont {Szalay}\ \emph {et~al.}(2012)\citenamefont {Szalay},
  \citenamefont {M{\"u}ller}, \citenamefont {Gidofalvi}, \citenamefont
  {Lischka},\ and\ \citenamefont
  {Shepard}}]{correlation-szalay-cr2011-112-108}%
  \BibitemOpen
  \bibfield  {author} {\bibinfo {author} {\bibfnamefont {P.~G.}\ \bibnamefont
  {Szalay}}, \bibinfo {author} {\bibfnamefont {T.}~\bibnamefont {M{\"u}ller}},
  \bibinfo {author} {\bibfnamefont {G.}~\bibnamefont {Gidofalvi}}, \bibinfo
  {author} {\bibfnamefont {H.}~\bibnamefont {Lischka}}, \ and\ \bibinfo
  {author} {\bibfnamefont {R.}~\bibnamefont {Shepard}},\ }\href {\doibase
  10.1021/cr200137a} {\bibfield  {journal} {\bibinfo  {journal} {Chem. Rev.}\
  }\textbf {\bibinfo {volume} {112}},\ \bibinfo {pages} {108} (\bibinfo {year}
  {2012})}\BibitemShut {NoStop}%
\bibitem [{\citenamefont {Lyakh}\ \emph {et~al.}(2012)\citenamefont {Lyakh},
  \citenamefont {Musia{\l}}, \citenamefont {Lotrich},\ and\ \citenamefont
  {Bartlett}}]{Lyakh:2012cn}%
  \BibitemOpen
  \bibfield  {author} {\bibinfo {author} {\bibfnamefont {D.~I.}\ \bibnamefont
  {Lyakh}}, \bibinfo {author} {\bibfnamefont {M.}~\bibnamefont {Musia{\l}}},
  \bibinfo {author} {\bibfnamefont {V.~F.}\ \bibnamefont {Lotrich}}, \ and\
  \bibinfo {author} {\bibfnamefont {R.~J.}\ \bibnamefont {Bartlett}},\ }\href
  {\doibase 10.1021/cr2001417} {\bibfield  {journal} {\bibinfo  {journal}
  {Chem. Rev.}\ }\textbf {\bibinfo {volume} {112}},\ \bibinfo {pages} {182}
  (\bibinfo {year} {2012})}\BibitemShut {NoStop}%
\bibitem [{\citenamefont {Casida}\ and\ \citenamefont
  {Huix-Rotllant}(2012)}]{Casida:2012gy}%
  \BibitemOpen
  \bibfield  {author} {\bibinfo {author} {\bibfnamefont {M.~E.}\ \bibnamefont
  {Casida}}\ and\ \bibinfo {author} {\bibfnamefont {M.}~\bibnamefont
  {Huix-Rotllant}},\ }\href {\doibase 10.1146/annurev-physchem-032511-143803}
  {\bibfield  {journal} {\bibinfo  {journal} {Annu. Rev. Phys. Chem.}\ }\textbf
  {\bibinfo {volume} {63}},\ \bibinfo {pages} {287} (\bibinfo {year}
  {2012})}\BibitemShut {NoStop}%
\bibitem [{\citenamefont {Gibson}\ and\ \citenamefont
  {Marcalo}(2006)}]{GIBSON:2006hd}%
  \BibitemOpen
  \bibfield  {author} {\bibinfo {author} {\bibfnamefont {J.~K.}\ \bibnamefont
  {Gibson}}\ and\ \bibinfo {author} {\bibfnamefont {J.}~\bibnamefont
  {Marcalo}},\ }\href {\doibase 10.1016/j.ccr.2005.09.011} {\bibfield
  {journal} {\bibinfo  {journal} {Coord. Chem. Rev.}\ }\textbf {\bibinfo
  {volume} {250}},\ \bibinfo {pages} {776} (\bibinfo {year}
  {2006})}\BibitemShut {NoStop}%
\bibitem [{\citenamefont {Senn}\ and\ \citenamefont
  {Thiel}(2009)}]{senn_qm/mm_2009}%
  \BibitemOpen
  \bibfield  {author} {\bibinfo {author} {\bibfnamefont {H.~M.}\ \bibnamefont
  {Senn}}\ and\ \bibinfo {author} {\bibfnamefont {W.}~\bibnamefont {Thiel}},\
  }\href {\doibase 10.1002/anie.200802019} {\bibfield  {journal} {\bibinfo
  {journal} {Angew. Chem., Int. Ed.}\ }\textbf {\bibinfo {volume} {48}},\
  \bibinfo {pages} {1198} (\bibinfo {year} {2009})}\BibitemShut {NoStop}%
\bibitem [{\citenamefont {Gomes}\ and\ \citenamefont
  {Jacob}(2012)}]{env-gomes-arpcspc2012-108-222}%
  \BibitemOpen
  \bibfield  {author} {\bibinfo {author} {\bibfnamefont {A.~S.~P.}\
  \bibnamefont {Gomes}}\ and\ \bibinfo {author} {\bibfnamefont {C.~R.}\
  \bibnamefont {Jacob}},\ }\href {\doibase 10.1039/C2PC90007F} {\bibfield
  {journal} {\bibinfo  {journal} {Annu. Rep. Prog. Chem.{,} Sect. C: Phys.
  Chem.}\ }\textbf {\bibinfo {volume} {108}},\ \bibinfo {pages} {222} (\bibinfo
  {year} {2012})}\BibitemShut {NoStop}%
\bibitem [{\citenamefont {Gordon}\ \emph {et~al.}(2012)\citenamefont {Gordon},
  \citenamefont {Fedorov}, \citenamefont {Pruitt},\ and\ \citenamefont
  {Slipchenko}}]{Gordon:2012dk}%
  \BibitemOpen
  \bibfield  {author} {\bibinfo {author} {\bibfnamefont {M.~S.}\ \bibnamefont
  {Gordon}}, \bibinfo {author} {\bibfnamefont {D.~G.}\ \bibnamefont {Fedorov}},
  \bibinfo {author} {\bibfnamefont {S.~R.}\ \bibnamefont {Pruitt}}, \ and\
  \bibinfo {author} {\bibfnamefont {L.~V.}\ \bibnamefont {Slipchenko}},\ }\href
  {\doibase 10.1021/cr200093j} {\bibfield  {journal} {\bibinfo  {journal}
  {Chem. Rev.}\ }\textbf {\bibinfo {volume} {112}},\ \bibinfo {pages} {632}
  (\bibinfo {year} {2012})}\BibitemShut {NoStop}%
\bibitem [{\citenamefont {Bagus}, \citenamefont {Ilton},\ and\ \citenamefont
  {Nelin}(2013)}]{core-bagus-ssr2013-68-273}%
  \BibitemOpen
  \bibfield  {author} {\bibinfo {author} {\bibfnamefont {P.~S.}\ \bibnamefont
  {Bagus}}, \bibinfo {author} {\bibfnamefont {E.~S.}\ \bibnamefont {Ilton}}, \
  and\ \bibinfo {author} {\bibfnamefont {C.~J.}\ \bibnamefont {Nelin}},\ }\href
  {\doibase 10.1016/j.surfrep.2013.03.001} {\bibfield  {journal} {\bibinfo
  {journal} {Surf. Sci. Rep.}\ }\textbf {\bibinfo {volume} {68}},\ \bibinfo
  {pages} {273} (\bibinfo {year} {2013})}\BibitemShut {NoStop}%
\bibitem [{\citenamefont {Geckeis}\ \emph {et~al.}(2013)\citenamefont
  {Geckeis}, \citenamefont {L{\"u}tzenkirchen}, \citenamefont {Polly},
  \citenamefont {Rabung},\ and\ \citenamefont {Schmidt}}]{Geckeis:2013jb}%
  \BibitemOpen
  \bibfield  {author} {\bibinfo {author} {\bibfnamefont {H.}~\bibnamefont
  {Geckeis}}, \bibinfo {author} {\bibfnamefont {J.}~\bibnamefont
  {L{\"u}tzenkirchen}}, \bibinfo {author} {\bibfnamefont {R.}~\bibnamefont
  {Polly}}, \bibinfo {author} {\bibfnamefont {T.}~\bibnamefont {Rabung}}, \
  and\ \bibinfo {author} {\bibfnamefont {M.}~\bibnamefont {Schmidt}},\ }\href
  {\doibase 10.1021/cr300370h} {\bibfield  {journal} {\bibinfo  {journal}
  {Chem. Rev.}\ }\textbf {\bibinfo {volume} {113}},\ \bibinfo {pages} {1016}
  (\bibinfo {year} {2013})}\BibitemShut {NoStop}%
\bibitem [{\citenamefont {Wilson}, \citenamefont {De~Sio},\ and\ \citenamefont
  {Vallet}(2016)}]{Wilson:2016hb}%
  \BibitemOpen
  \bibfield  {author} {\bibinfo {author} {\bibfnamefont {R.~E.}\ \bibnamefont
  {Wilson}}, \bibinfo {author} {\bibfnamefont {S.}~\bibnamefont {De~Sio}}, \
  and\ \bibinfo {author} {\bibfnamefont {V.}~\bibnamefont {Vallet}},\ }\href
  {\doibase 10.1002/ejic.201600981} {\bibfield  {journal} {\bibinfo  {journal}
  {Eur. J. Inorg. Chem.}\ }\textbf {\bibinfo {volume} {2016}},\ \bibinfo
  {pages} {5467} (\bibinfo {year} {2016})}\BibitemShut {NoStop}%
\bibitem [{\citenamefont {Seidel}, \citenamefont {Winter},\ and\ \citenamefont
  {Bradforth}(2016)}]{Seidel:2016iga}%
  \BibitemOpen
  \bibfield  {author} {\bibinfo {author} {\bibfnamefont {R.}~\bibnamefont
  {Seidel}}, \bibinfo {author} {\bibfnamefont {B.}~\bibnamefont {Winter}}, \
  and\ \bibinfo {author} {\bibfnamefont {S.~E.}\ \bibnamefont {Bradforth}},\
  }\href {\doibase 10.1146/annurev-physchem-040513-103715} {\bibfield
  {journal} {\bibinfo  {journal} {Annu. Rev. Phys. Chem.}\ }\textbf {\bibinfo
  {volume} {67}},\ \bibinfo {pages} {283} (\bibinfo {year} {2016})}\BibitemShut
  {NoStop}%
\bibitem [{\citenamefont {Purgel}\ \emph {et~al.}(2011)\citenamefont {Purgel},
  \citenamefont {Maliarik}, \citenamefont {Glaser}, \citenamefont
  {Platas-Iglesias}, \citenamefont {Persson},\ and\ \citenamefont
  {T{\'o}th}}]{Purgel:2011ds}%
  \BibitemOpen
  \bibfield  {author} {\bibinfo {author} {\bibfnamefont {M.}~\bibnamefont
  {Purgel}}, \bibinfo {author} {\bibfnamefont {M.}~\bibnamefont {Maliarik}},
  \bibinfo {author} {\bibfnamefont {J.}~\bibnamefont {Glaser}}, \bibinfo
  {author} {\bibfnamefont {C.}~\bibnamefont {Platas-Iglesias}}, \bibinfo
  {author} {\bibfnamefont {I.}~\bibnamefont {Persson}}, \ and\ \bibinfo
  {author} {\bibfnamefont {I.}~\bibnamefont {T{\'o}th}},\ }\href {\doibase
  10.1021/ic200417q} {\bibfield  {journal} {\bibinfo  {journal} {Inorg. Chem.}\
  }\textbf {\bibinfo {volume} {50}},\ \bibinfo {pages} {6163} (\bibinfo {year}
  {2011})}\BibitemShut {NoStop}%
\bibitem [{\citenamefont {Sures}\ \emph {et~al.}(2017)\citenamefont {Sures},
  \citenamefont {Serapian}, \citenamefont {Kozma}, \citenamefont {Molina},
  \citenamefont {Bo},\ and\ \citenamefont {Nyman}}]{Sures:2017hd}%
  \BibitemOpen
  \bibfield  {author} {\bibinfo {author} {\bibfnamefont {D.~J.}\ \bibnamefont
  {Sures}}, \bibinfo {author} {\bibfnamefont {S.~A.}\ \bibnamefont {Serapian}},
  \bibinfo {author} {\bibfnamefont {K.}~\bibnamefont {Kozma}}, \bibinfo
  {author} {\bibfnamefont {P.~I.}\ \bibnamefont {Molina}}, \bibinfo {author}
  {\bibfnamefont {C.}~\bibnamefont {Bo}}, \ and\ \bibinfo {author}
  {\bibfnamefont {M.}~\bibnamefont {Nyman}},\ }\href {\doibase
  10.1039/C6CP08454K} {\bibfield  {journal} {\bibinfo  {journal} {Phys. Chem.
  Chem. Phys.}\ }\textbf {\bibinfo {volume} {19}},\ \bibinfo {pages} {8715}
  (\bibinfo {year} {2017})}\BibitemShut {NoStop}%
\bibitem [{\citenamefont {Liu}\ and\ \citenamefont {Fang}(2009)}]{Liu:2009hf}%
  \BibitemOpen
  \bibfield  {author} {\bibinfo {author} {\bibfnamefont {Y.-J.}\ \bibnamefont
  {Liu}}\ and\ \bibinfo {author} {\bibfnamefont {W.-H.}\ \bibnamefont {Fang}},\
  }\href {\doibase 10.1016/S0065-3276(08)00401-2} {\bibfield  {journal}
  {\bibinfo  {journal} {Adv. Quant. Chem.}\ }\textbf {\bibinfo {volume} {56}},\
  \bibinfo {pages} {1 } (\bibinfo {year} {2009})}\BibitemShut {NoStop}%
\bibitem [{\citenamefont {Saiz-Lopez}\ \emph {et~al.}(2012)\citenamefont
  {Saiz-Lopez}, \citenamefont {Plane}, \citenamefont {Baker}, \citenamefont
  {Carpenter}, \citenamefont {von Glasow}, \citenamefont
  {G{\'o}mez~Mart{\'\i}n}, \citenamefont {McFiggans},\ and\ \citenamefont
  {Saunders}}]{SaizLopez:2012is}%
  \BibitemOpen
  \bibfield  {author} {\bibinfo {author} {\bibfnamefont {A.}~\bibnamefont
  {Saiz-Lopez}}, \bibinfo {author} {\bibfnamefont {J.~M.~C.}\ \bibnamefont
  {Plane}}, \bibinfo {author} {\bibfnamefont {A.~R.}\ \bibnamefont {Baker}},
  \bibinfo {author} {\bibfnamefont {L.~J.}\ \bibnamefont {Carpenter}}, \bibinfo
  {author} {\bibfnamefont {R.}~\bibnamefont {von Glasow}}, \bibinfo {author}
  {\bibfnamefont {J.~C.}\ \bibnamefont {G{\'o}mez~Mart{\'\i}n}}, \bibinfo
  {author} {\bibfnamefont {G.}~\bibnamefont {McFiggans}}, \ and\ \bibinfo
  {author} {\bibfnamefont {R.~W.}\ \bibnamefont {Saunders}},\ }\href {\doibase
  10.1021/cr200029u} {\bibfield  {journal} {\bibinfo  {journal} {Chem. Rev.}\
  }\textbf {\bibinfo {volume} {112}},\ \bibinfo {pages} {1773} (\bibinfo {year}
  {2012})}\BibitemShut {NoStop}%
\bibitem [{\citenamefont {Xing}, \citenamefont {Rey-de Castro},\ and\
  \citenamefont {Rabitz}(2014)}]{Xing:2014gp}%
  \BibitemOpen
  \bibfield  {author} {\bibinfo {author} {\bibfnamefont {X.}~\bibnamefont
  {Xing}}, \bibinfo {author} {\bibfnamefont {R.}~\bibnamefont {Rey-de Castro}},
  \ and\ \bibinfo {author} {\bibfnamefont {H.}~\bibnamefont {Rabitz}},\ }\href
  {\doibase 10.1088/1367-2630/16/12/125004} {\bibfield  {journal} {\bibinfo
  {journal} {New J. Phys.}\ }\textbf {\bibinfo {volume} {16}},\ \bibinfo
  {pages} {125004} (\bibinfo {year} {2014})}\BibitemShut {NoStop}%
\bibitem [{\citenamefont {Starik}\ \emph {et~al.}(2015)\citenamefont {Starik},
  \citenamefont {Loukhovitski}, \citenamefont {Sharipov},\ and\ \citenamefont
  {Titova}}]{Starik:2015gz}%
  \BibitemOpen
  \bibfield  {author} {\bibinfo {author} {\bibfnamefont {A.~M.}\ \bibnamefont
  {Starik}}, \bibinfo {author} {\bibfnamefont {B.~I.}\ \bibnamefont
  {Loukhovitski}}, \bibinfo {author} {\bibfnamefont {A.~S.}\ \bibnamefont
  {Sharipov}}, \ and\ \bibinfo {author} {\bibfnamefont {N.~S.}\ \bibnamefont
  {Titova}},\ }\href {\doibase 10.1098/rsta.2014.0341} {\bibfield  {journal}
  {\bibinfo  {journal} {Phil. Trans. R. Soc. A}\ }\textbf {\bibinfo {volume}
  {373}},\ \bibinfo {pages} {20140341} (\bibinfo {year} {2015})}\BibitemShut
  {NoStop}%
\bibitem [{\citenamefont {Hochlaf}(2017)}]{Hochlaf:2017cw}%
  \BibitemOpen
  \bibfield  {author} {\bibinfo {author} {\bibfnamefont {M.}~\bibnamefont
  {Hochlaf}},\ }\href {\doibase 10.1039/C7CP01980G} {\bibfield  {journal}
  {\bibinfo  {journal} {Phys. Chem. Chem. Phys.}\ }\textbf {\bibinfo {volume}
  {145}},\ \bibinfo {pages} {120901} (\bibinfo {year} {2017})}\BibitemShut
  {NoStop}%
\bibitem [{\citenamefont {Shavitt}\ and\ \citenamefont
  {Bartlett}(2009)}]{Shavitt2009}%
  \BibitemOpen
  \bibfield  {author} {\bibinfo {author} {\bibfnamefont {I.}~\bibnamefont
  {Shavitt}}\ and\ \bibinfo {author} {\bibfnamefont {R.~J.}\ \bibnamefont
  {Bartlett}},\ }\href
  {https://books.google.com/books?hl=en{\&}lr={\&}id=SWw6ac1NHZYC{\&}pgis=1}
  {\emph {\bibinfo {title} {{Many-Body Methods in Chemistry and Physics: MBPT
  and Coupled-Cluster Theory}}}}\ (\bibinfo  {publisher} {Cambridge University
  Press},\ \bibinfo {year} {2009})\ p.\ \bibinfo {pages} {532}\BibitemShut
  {NoStop}%
\bibitem [{\citenamefont {Pyykk{\"o}}(1988)}]{Pyykko1988}%
  \BibitemOpen
  \bibfield  {author} {\bibinfo {author} {\bibfnamefont {P.}~\bibnamefont
  {Pyykk{\"o}}},\ }\href {\doibase 10.1021/cr00085a006} {\bibfield  {journal}
  {\bibinfo  {journal} {Chem. Rev.}\ }\textbf {\bibinfo {volume} {88}},\
  \bibinfo {pages} {563} (\bibinfo {year} {1988})}\BibitemShut {NoStop}%
\bibitem [{\citenamefont {Saue}(2011)}]{saue:hamprimer}%
  \BibitemOpen
  \bibfield  {author} {\bibinfo {author} {\bibfnamefont {T.}~\bibnamefont
  {Saue}},\ }\href {\doibase 10.1002/cphc.201100682} {\bibfield  {journal}
  {\bibinfo  {journal} {ChemPhysChem}\ }\textbf {\bibinfo {volume} {12}},\
  \bibinfo {pages} {3077} (\bibinfo {year} {2011})}\BibitemShut {NoStop}%
\bibitem [{\citenamefont {Pyykk{\"o}}(2012{\natexlab{a}})}]{pyykko:2012bt}%
  \BibitemOpen
  \bibfield  {author} {\bibinfo {author} {\bibfnamefont {P.}~\bibnamefont
  {Pyykk{\"o}}},\ }\href {\doibase 10.1146/annurev-physchem-032511-143755}
  {\bibfield  {journal} {\bibinfo  {journal} {Ann.~Rev.~Phys.~Chem}\ }\textbf
  {\bibinfo {volume} {63}},\ \bibinfo {pages} {45} (\bibinfo {year}
  {2012}{\natexlab{a}})}\BibitemShut {NoStop}%
\bibitem [{\citenamefont {Pyykk{\"o}}(2012{\natexlab{b}})}]{pyykko:2012kh}%
  \BibitemOpen
  \bibfield  {author} {\bibinfo {author} {\bibfnamefont {P.}~\bibnamefont
  {Pyykk{\"o}}},\ }\href {\doibase 10.1021/cr200042e} {\bibfield  {journal}
  {\bibinfo  {journal} {Chem. Rev.}\ }\textbf {\bibinfo {volume} {112}},\
  \bibinfo {pages} {371} (\bibinfo {year} {2012}{\natexlab{b}})}\BibitemShut
  {NoStop}%
\bibitem [{\citenamefont {Autschbach}(2012)}]{Autschbach2012}%
  \BibitemOpen
  \bibfield  {author} {\bibinfo {author} {\bibfnamefont {J.}~\bibnamefont
  {Autschbach}},\ }\href {\doibase 10.1063/1.3702628} {\bibfield  {journal}
  {\bibinfo  {journal} {J. Chem. Phys.}\ }\textbf {\bibinfo {volume} {136}},\
  \bibinfo {pages} {150902} (\bibinfo {year} {2012})}\BibitemShut {NoStop}%
\bibitem [{\citenamefont {Dyall}\ and\ \citenamefont {{Faegri
  Jr.}}(2007)}]{dyall-faegri-book-2007}%
  \BibitemOpen
  \bibfield  {author} {\bibinfo {author} {\bibfnamefont {K.~G.}\ \bibnamefont
  {Dyall}}\ and\ \bibinfo {author} {\bibfnamefont {K.}~\bibnamefont {{Faegri
  Jr.}}},\ }\href@noop {} {\emph {\bibinfo {title} {{Introduction to
  Relativistic Quantum Chemistry}}}}\ (\bibinfo  {publisher} {Oxford University
  Press},\ \bibinfo {year} {2007})\BibitemShut {NoStop}%
\bibitem [{\citenamefont {Reiher}\ and\ \citenamefont
  {Wolf}(2009)}]{reiher_book}%
  \BibitemOpen
  \bibfield  {author} {\bibinfo {author} {\bibfnamefont {M.}~\bibnamefont
  {Reiher}}\ and\ \bibinfo {author} {\bibfnamefont {A.}~\bibnamefont {Wolf}},\
  }\href@noop {} {\emph {\bibinfo {title} {Relativistic Quantum Chemistry.
  {T}he Fundamental Theory of Molecular Science}}}\ (\bibinfo  {publisher}
  {Wiley},\ \bibinfo {year} {2009})\BibitemShut {NoStop}%
\bibitem [{\citenamefont {Saue}(2002)}]{saue:molprop}%
  \BibitemOpen
  \bibfield  {author} {\bibinfo {author} {\bibfnamefont {T.}~\bibnamefont
  {Saue}},\ }in\ \href {\doibase 10.1016/S1380-7323(02)80033-4} {\emph
  {\bibinfo {booktitle} {Relativistic Electronic Structure Theory. Part 1.
  Fundamentals}}},\ \bibinfo {editor} {edited by\ \bibinfo {editor}
  {\bibfnamefont {P.}~\bibnamefont {Schwerdtfeger}}}\ (\bibinfo  {publisher}
  {Elsevier},\ \bibinfo {address} {Amsterdam},\ \bibinfo {year} {2002})\ p.\
  \bibinfo {pages} {332}\BibitemShut {NoStop}%
\bibitem [{\citenamefont {Saue}(2005)}]{Saue2005}%
  \BibitemOpen
  \bibfield  {author} {\bibinfo {author} {\bibfnamefont {T.}~\bibnamefont
  {Saue}},\ }\href {\doibase 10.1016/S0065-3276(05)48020-X} {\bibfield
  {journal} {\bibinfo  {journal} {Adv. Quant. Chem.}\ }\textbf {\bibinfo
  {volume} {48}},\ \bibinfo {pages} {383} (\bibinfo {year} {2005})}\BibitemShut
  {NoStop}%
\bibitem [{\citenamefont {Norman}, \citenamefont {Ruud},\ and\ \citenamefont
  {Saue}(2018)}]{nrsbook}%
  \BibitemOpen
  \bibfield  {author} {\bibinfo {author} {\bibfnamefont {P.}~\bibnamefont
  {Norman}}, \bibinfo {author} {\bibfnamefont {K.}~\bibnamefont {Ruud}}, \ and\
  \bibinfo {author} {\bibfnamefont {T.}~\bibnamefont {Saue}},\ }\href@noop {}
  {\emph {\bibinfo {title} {Principles and Practices of Molecular Properties:
  Theory, Modeling and Simulations}}}\ (\bibinfo  {publisher} {Wiley},\
  \bibinfo {address} {Hoboken, NJ},\ \bibinfo {year} {2018})\BibitemShut
  {NoStop}%
\bibitem [{\citenamefont {Fleig}(2012)}]{Timo_overview}%
  \BibitemOpen
  \bibfield  {author} {\bibinfo {author} {\bibfnamefont {T.}~\bibnamefont
  {Fleig}},\ }\href {\doibase 10.1016/j.chemphys.2011.06.032} {\bibfield
  {journal} {\bibinfo  {journal} {Chem.~Phys.}\ }\textbf {\bibinfo {volume}
  {395}},\ \bibinfo {pages} {2} (\bibinfo {year} {2012})}\BibitemShut {NoStop}%
\bibitem [{\citenamefont {Bartlett}\ and\ \citenamefont
  {Musial}(2007)}]{bartlett:reviewcc}%
  \BibitemOpen
  \bibfield  {author} {\bibinfo {author} {\bibfnamefont {R.~J.}\ \bibnamefont
  {Bartlett}}\ and\ \bibinfo {author} {\bibfnamefont {M.}~\bibnamefont
  {Musial}},\ }\href {\doibase 10.1103/RevModPhys.79.291} {\bibfield  {journal}
  {\bibinfo  {journal} {Rev. Mod. Phys.}\ }\textbf {\bibinfo {volume} {79}},\
  \bibinfo {pages} {291} (\bibinfo {year} {2007})}\BibitemShut {NoStop}%
\bibitem [{\citenamefont {Visscher}, \citenamefont {Eliav},\ and\ \citenamefont
  {Kaldor}(2001)}]{correlation-Visscher-JCP2001-115-9720}%
  \BibitemOpen
  \bibfield  {author} {\bibinfo {author} {\bibfnamefont {L.}~\bibnamefont
  {Visscher}}, \bibinfo {author} {\bibfnamefont {E.}~\bibnamefont {Eliav}}, \
  and\ \bibinfo {author} {\bibfnamefont {U.}~\bibnamefont {Kaldor}},\ }\href
  {\doibase 10.1063/1.1415746} {\bibfield  {journal} {\bibinfo  {journal} {J.
  Chem. Phys.}\ }\textbf {\bibinfo {volume} {115}},\ \bibinfo {pages} {9720}
  (\bibinfo {year} {2001})}\BibitemShut {NoStop}%
\bibitem [{\citenamefont {Landau}\ \emph {et~al.}(2000)\citenamefont {Landau},
  \citenamefont {Eliav}, \citenamefont {Ishikawa},\ and\ \citenamefont
  {Kaldor}}]{correlation-Landau-JCP2000-113-9905}%
  \BibitemOpen
  \bibfield  {author} {\bibinfo {author} {\bibfnamefont {A.}~\bibnamefont
  {Landau}}, \bibinfo {author} {\bibfnamefont {E.}~\bibnamefont {Eliav}},
  \bibinfo {author} {\bibfnamefont {Y.}~\bibnamefont {Ishikawa}}, \ and\
  \bibinfo {author} {\bibfnamefont {U.}~\bibnamefont {Kaldor}},\ }\href
  {\doibase 10.1063/1.1323258} {\bibfield  {journal} {\bibinfo  {journal} {J.
  Chem. Phys.}\ }\textbf {\bibinfo {volume} {113}},\ \bibinfo {pages} {9905}
  (\bibinfo {year} {2000})}\BibitemShut {NoStop}%
\bibitem [{\citenamefont {Landau}\ \emph {et~al.}(2001)\citenamefont {Landau},
  \citenamefont {Eliav}, \citenamefont {Ishikawa},\ and\ \citenamefont
  {Kaldor}}]{correlation-Landau-JCP2001-115-6862}%
  \BibitemOpen
  \bibfield  {author} {\bibinfo {author} {\bibfnamefont {A.}~\bibnamefont
  {Landau}}, \bibinfo {author} {\bibfnamefont {E.}~\bibnamefont {Eliav}},
  \bibinfo {author} {\bibfnamefont {Y.}~\bibnamefont {Ishikawa}}, \ and\
  \bibinfo {author} {\bibfnamefont {U.}~\bibnamefont {Kaldor}},\ }\href
  {\doibase 10.1063/1.1405005} {\bibfield  {journal} {\bibinfo  {journal} {J.
  Chem. Phys.}\ }\textbf {\bibinfo {volume} {115}},\ \bibinfo {pages} {6862}
  (\bibinfo {year} {2001})}\BibitemShut {NoStop}%
\bibitem [{\citenamefont {Landau}\ \emph {et~al.}(2004)\citenamefont {Landau},
  \citenamefont {Eliav}, \citenamefont {Ishikawa},\ and\ \citenamefont
  {Kaldor}}]{Landau:2004dd}%
  \BibitemOpen
  \bibfield  {author} {\bibinfo {author} {\bibfnamefont {A.}~\bibnamefont
  {Landau}}, \bibinfo {author} {\bibfnamefont {E.}~\bibnamefont {Eliav}},
  \bibinfo {author} {\bibfnamefont {Y.}~\bibnamefont {Ishikawa}}, \ and\
  \bibinfo {author} {\bibfnamefont {U.}~\bibnamefont {Kaldor}},\ }\href
  {\doibase 10.1063/1.1788652} {\bibfield  {journal} {\bibinfo  {journal} {J.
  Chem. Phys.}\ }\textbf {\bibinfo {volume} {121}},\ \bibinfo {pages} {6634}
  (\bibinfo {year} {2004})}\BibitemShut {NoStop}%
\bibitem [{\citenamefont {Eliav}\ and\ \citenamefont
  {Kaldor}(2012)}]{Eliav:2012gp}%
  \BibitemOpen
  \bibfield  {author} {\bibinfo {author} {\bibfnamefont {E.}~\bibnamefont
  {Eliav}}\ and\ \bibinfo {author} {\bibfnamefont {U.}~\bibnamefont {Kaldor}},\
  }\href {\doibase 10.1016/j.chemphys.2011.10.019} {\bibfield  {journal}
  {\bibinfo  {journal} {Chem. Phys.}\ }\textbf {\bibinfo {volume} {392}},\
  \bibinfo {pages} {78} (\bibinfo {year} {2012})}\BibitemShut {NoStop}%
\bibitem [{\citenamefont {Infante}, \citenamefont {Gomes},\ and\ \citenamefont
  {Visscher}(2006)}]{actinide-infante-jcp2006-125-074301}%
  \BibitemOpen
  \bibfield  {author} {\bibinfo {author} {\bibfnamefont {I.}~\bibnamefont
  {Infante}}, \bibinfo {author} {\bibfnamefont {A.~S.~P.}\ \bibnamefont
  {Gomes}}, \ and\ \bibinfo {author} {\bibfnamefont {L.}~\bibnamefont
  {Visscher}},\ }\href {\doibase 10.1063/1.2244564} {\bibfield  {journal}
  {\bibinfo  {journal} {J. Chem. Phys.}\ }\textbf {\bibinfo {volume} {125}},\
  \bibinfo {pages} {074301} (\bibinfo {year} {2006})}\BibitemShut {NoStop}%
\bibitem [{\citenamefont {Infante}\ \emph {et~al.}(2007)\citenamefont
  {Infante}, \citenamefont {Eliav}, \citenamefont {Vilkas}, \citenamefont
  {Ishikawa}, \citenamefont {Kaldor},\ and\ \citenamefont
  {Visscher}}]{actinide-Infante-JCP2007-127-124308}%
  \BibitemOpen
  \bibfield  {author} {\bibinfo {author} {\bibfnamefont {I.}~\bibnamefont
  {Infante}}, \bibinfo {author} {\bibfnamefont {E.}~\bibnamefont {Eliav}},
  \bibinfo {author} {\bibfnamefont {M.~J.}\ \bibnamefont {Vilkas}}, \bibinfo
  {author} {\bibfnamefont {Y.}~\bibnamefont {Ishikawa}}, \bibinfo {author}
  {\bibfnamefont {U.}~\bibnamefont {Kaldor}}, \ and\ \bibinfo {author}
  {\bibfnamefont {L.}~\bibnamefont {Visscher}},\ }\href {\doibase
  10.1063/1.2770699} {\bibfield  {journal} {\bibinfo  {journal} {J. Chem.
  Phys.}\ }\textbf {\bibinfo {volume} {127}},\ \bibinfo {pages} {124308}
  (\bibinfo {year} {2007})}\BibitemShut {NoStop}%
\bibitem [{\citenamefont {Gomes}, \citenamefont {Jacob},\ and\ \citenamefont
  {Visscher}(2008)}]{actinide-Gomes-PCCP2008-10-5353}%
  \BibitemOpen
  \bibfield  {author} {\bibinfo {author} {\bibfnamefont {A.~S.~P.}\
  \bibnamefont {Gomes}}, \bibinfo {author} {\bibfnamefont {C.~R.}\ \bibnamefont
  {Jacob}}, \ and\ \bibinfo {author} {\bibfnamefont {L.}~\bibnamefont
  {Visscher}},\ }\href {\doibase 10.1039/b805739g} {\bibfield  {journal}
  {\bibinfo  {journal} {Phys. Chem. Chem. Phys.}\ }\textbf {\bibinfo {volume}
  {10}},\ \bibinfo {pages} {5353} (\bibinfo {year} {2008})}\BibitemShut
  {NoStop}%
\bibitem [{\citenamefont {Ruip{\'e}rez}\ \emph {et~al.}(2009)\citenamefont
  {Ruip{\'e}rez}, \citenamefont {Danilo}, \citenamefont {R{\'e}al},
  \citenamefont {Flament}, \citenamefont {Vallet},\ and\ \citenamefont
  {Wahlgren}}]{ruiprez_ab_2009}%
  \BibitemOpen
  \bibfield  {author} {\bibinfo {author} {\bibfnamefont {F.}~\bibnamefont
  {Ruip{\'e}rez}}, \bibinfo {author} {\bibfnamefont {C.}~\bibnamefont
  {Danilo}}, \bibinfo {author} {\bibfnamefont {F.}~\bibnamefont {R{\'e}al}},
  \bibinfo {author} {\bibfnamefont {J.-P.}\ \bibnamefont {Flament}}, \bibinfo
  {author} {\bibfnamefont {V.}~\bibnamefont {Vallet}}, \ and\ \bibinfo {author}
  {\bibfnamefont {U.}~\bibnamefont {Wahlgren}},\ }\href {\doibase
  10.1021/jp809108h} {\bibfield  {journal} {\bibinfo  {journal} {J. Phys. Chem.
  A}\ }\textbf {\bibinfo {volume} {113}},\ \bibinfo {pages} {1420} (\bibinfo
  {year} {2009})}\BibitemShut {NoStop}%
\bibitem [{\citenamefont {Rota}\ \emph {et~al.}(2011)\citenamefont {Rota},
  \citenamefont {Knecht}, \citenamefont {Fleig}, \citenamefont {Ganyushin},
  \citenamefont {Saue}, \citenamefont {Neese},\ and\ \citenamefont
  {Bolvin}}]{ATO-Rota-JCP2011-135-114106}%
  \BibitemOpen
  \bibfield  {author} {\bibinfo {author} {\bibfnamefont {J.-B.}\ \bibnamefont
  {Rota}}, \bibinfo {author} {\bibfnamefont {S.}~\bibnamefont {Knecht}},
  \bibinfo {author} {\bibfnamefont {T.}~\bibnamefont {Fleig}}, \bibinfo
  {author} {\bibfnamefont {D.}~\bibnamefont {Ganyushin}}, \bibinfo {author}
  {\bibfnamefont {T.}~\bibnamefont {Saue}}, \bibinfo {author} {\bibfnamefont
  {F.}~\bibnamefont {Neese}}, \ and\ \bibinfo {author} {\bibfnamefont
  {H.}~\bibnamefont {Bolvin}},\ }\href {\doibase 10.1063/1.3636084} {\bibfield
  {journal} {\bibinfo  {journal} {J. Chem. Phys.}\ }\textbf {\bibinfo {volume}
  {135}},\ \bibinfo {eid} {114106} (\bibinfo {year} {2011})}\BibitemShut
  {NoStop}%
\bibitem [{\citenamefont {Gomes}\ \emph {et~al.}(2014)\citenamefont {Gomes},
  \citenamefont {R{\'e}al}, \citenamefont {Galland}, \citenamefont {Angeli},
  \citenamefont {Cimiraglia},\ and\ \citenamefont
  {Vallet}}]{PereiraGomes:2014iy}%
  \BibitemOpen
  \bibfield  {author} {\bibinfo {author} {\bibfnamefont {A.~S.~P.}\
  \bibnamefont {Gomes}}, \bibinfo {author} {\bibfnamefont {F.}~\bibnamefont
  {R{\'e}al}}, \bibinfo {author} {\bibfnamefont {N.}~\bibnamefont {Galland}},
  \bibinfo {author} {\bibfnamefont {C.}~\bibnamefont {Angeli}}, \bibinfo
  {author} {\bibfnamefont {R.}~\bibnamefont {Cimiraglia}}, \ and\ \bibinfo
  {author} {\bibfnamefont {V.}~\bibnamefont {Vallet}},\ }\href {\doibase
  10.1039/C3CP55294B} {\bibfield  {journal} {\bibinfo  {journal} {Phys. Chem.
  Chem. Phys.}\ }\textbf {\bibinfo {volume} {16}},\ \bibinfo {pages} {9238}
  (\bibinfo {year} {2014})}\BibitemShut {NoStop}%
\bibitem [{\citenamefont {Denis}\ \emph {et~al.}(2015)\citenamefont {Denis},
  \citenamefont {N{\o}rby}, \citenamefont {Jensen}, \citenamefont {Gomes},
  \citenamefont {Nayak}, \citenamefont {Knecht},\ and\ \citenamefont
  {Fleig}}]{Denis:2015ez}%
  \BibitemOpen
  \bibfield  {author} {\bibinfo {author} {\bibfnamefont {M.}~\bibnamefont
  {Denis}}, \bibinfo {author} {\bibfnamefont {M.~S.}\ \bibnamefont {N{\o}rby}},
  \bibinfo {author} {\bibfnamefont {H.~J.~A.}\ \bibnamefont {Jensen}}, \bibinfo
  {author} {\bibfnamefont {A.~S.~P.}\ \bibnamefont {Gomes}}, \bibinfo {author}
  {\bibfnamefont {M.~K.}\ \bibnamefont {Nayak}}, \bibinfo {author}
  {\bibfnamefont {S.}~\bibnamefont {Knecht}}, \ and\ \bibinfo {author}
  {\bibfnamefont {T.}~\bibnamefont {Fleig}},\ }\href {\doibase
  10.1088/1367-2630/17/4/043005} {\bibfield  {journal} {\bibinfo  {journal}
  {New J. Phys.}\ }\textbf {\bibinfo {volume} {17}},\ \bibinfo {pages} {043005}
  (\bibinfo {year} {2015})}\BibitemShut {NoStop}%
\bibitem [{\citenamefont {Parmar}, \citenamefont {Peterson},\ and\
  \citenamefont {Clark}(2014)}]{Parmar:2014kn}%
  \BibitemOpen
  \bibfield  {author} {\bibinfo {author} {\bibfnamefont {P.}~\bibnamefont
  {Parmar}}, \bibinfo {author} {\bibfnamefont {K.~A.}\ \bibnamefont
  {Peterson}}, \ and\ \bibinfo {author} {\bibfnamefont {A.~E.}\ \bibnamefont
  {Clark}},\ }\href {\doibase 10.1063/1.4903792} {\bibfield  {journal}
  {\bibinfo  {journal} {J. Chem. Phys.}\ }\textbf {\bibinfo {volume} {141}},\
  \bibinfo {pages} {234304} (\bibinfo {year} {2014})}\BibitemShut {NoStop}%
\bibitem [{\citenamefont {Figgen}\ \emph {et~al.}(2008)\citenamefont {Figgen},
  \citenamefont {Wedig}, \citenamefont {Stoll}, \citenamefont {Dolg},
  \citenamefont {Eliav},\ and\ \citenamefont {Kaldor}}]{Figgen:2008hd}%
  \BibitemOpen
  \bibfield  {author} {\bibinfo {author} {\bibfnamefont {D.}~\bibnamefont
  {Figgen}}, \bibinfo {author} {\bibfnamefont {A.}~\bibnamefont {Wedig}},
  \bibinfo {author} {\bibfnamefont {H.}~\bibnamefont {Stoll}}, \bibinfo
  {author} {\bibfnamefont {M.}~\bibnamefont {Dolg}}, \bibinfo {author}
  {\bibfnamefont {E.}~\bibnamefont {Eliav}}, \ and\ \bibinfo {author}
  {\bibfnamefont {U.}~\bibnamefont {Kaldor}},\ }\href {\doibase
  10.1063/1.2823053} {\bibfield  {journal} {\bibinfo  {journal} {J. Chem.
  Phys.}\ }\textbf {\bibinfo {volume} {128}},\ \bibinfo {pages} {024106}
  (\bibinfo {year} {2008})}\BibitemShut {NoStop}%
\bibitem [{\citenamefont {Weigand}\ \emph {et~al.}(2009)\citenamefont
  {Weigand}, \citenamefont {Cao}, \citenamefont {Vallet}, \citenamefont
  {Flament},\ and\ \citenamefont {Dolg}}]{Weigand:2009dq}%
  \BibitemOpen
  \bibfield  {author} {\bibinfo {author} {\bibfnamefont {A.}~\bibnamefont
  {Weigand}}, \bibinfo {author} {\bibfnamefont {X.}~\bibnamefont {Cao}},
  \bibinfo {author} {\bibfnamefont {V.}~\bibnamefont {Vallet}}, \bibinfo
  {author} {\bibfnamefont {J.-P.}\ \bibnamefont {Flament}}, \ and\ \bibinfo
  {author} {\bibfnamefont {M.}~\bibnamefont {Dolg}},\ }\href {\doibase
  10.1021/jp902693b} {\bibfield  {journal} {\bibinfo  {journal} {J. Phys. Chem.
  A}\ }\textbf {\bibinfo {volume} {113}},\ \bibinfo {pages} {11509} (\bibinfo
  {year} {2009})}\BibitemShut {NoStop}%
\bibitem [{\citenamefont {Nikoobakht}, \citenamefont {Siebert},\ and\
  \citenamefont {Pernpointner}(2015)}]{Nikoobakht:2015ko}%
  \BibitemOpen
  \bibfield  {author} {\bibinfo {author} {\bibfnamefont {B.}~\bibnamefont
  {Nikoobakht}}, \bibinfo {author} {\bibfnamefont {M.}~\bibnamefont {Siebert}},
  \ and\ \bibinfo {author} {\bibfnamefont {M.}~\bibnamefont {Pernpointner}},\
  }\href {\doibase 10.1080/00268976.2015.1031839} {\bibfield  {journal}
  {\bibinfo  {journal} {Mol. Phys.}\ }\textbf {\bibinfo {volume} {113}},\
  \bibinfo {pages} {3431} (\bibinfo {year} {2015})}\BibitemShut {NoStop}%
\bibitem [{\citenamefont {R\'{e}al}\ \emph {et~al.}(2009)\citenamefont
  {R\'{e}al}, \citenamefont {Gomes}, \citenamefont {Visscher}, \citenamefont
  {Vallet},\ and\ \citenamefont {Eliav}}]{real09}%
  \BibitemOpen
  \bibfield  {author} {\bibinfo {author} {\bibfnamefont {F.}~\bibnamefont
  {R\'{e}al}}, \bibinfo {author} {\bibfnamefont {A.~S.~P.}\ \bibnamefont
  {Gomes}}, \bibinfo {author} {\bibfnamefont {L.}~\bibnamefont {Visscher}},
  \bibinfo {author} {\bibfnamefont {V.}~\bibnamefont {Vallet}}, \ and\ \bibinfo
  {author} {\bibfnamefont {E.}~\bibnamefont {Eliav}},\ }\href {\doibase
  10.1021/jp903758c} {\bibfield  {journal} {\bibinfo  {journal} {J. Phys. Chem.
  A}\ }\textbf {\bibinfo {volume} {113}},\ \bibinfo {pages} {12504} (\bibinfo
  {year} {2009})}\BibitemShut {NoStop}%
\bibitem [{\citenamefont {Gomes}\ \emph {et~al.}(2010)\citenamefont {Gomes},
  \citenamefont {Visscher}, \citenamefont {Bolvin}, \citenamefont {Saue},
  \citenamefont {Knecht}, \citenamefont {Fleig},\ and\ \citenamefont
  {Eliav}}]{gomes10}%
  \BibitemOpen
  \bibfield  {author} {\bibinfo {author} {\bibfnamefont {A.~S.~P.}\
  \bibnamefont {Gomes}}, \bibinfo {author} {\bibfnamefont {L.}~\bibnamefont
  {Visscher}}, \bibinfo {author} {\bibfnamefont {H.}~\bibnamefont {Bolvin}},
  \bibinfo {author} {\bibfnamefont {T.}~\bibnamefont {Saue}}, \bibinfo {author}
  {\bibfnamefont {S.}~\bibnamefont {Knecht}}, \bibinfo {author} {\bibfnamefont
  {T.}~\bibnamefont {Fleig}}, \ and\ \bibinfo {author} {\bibfnamefont
  {E.}~\bibnamefont {Eliav}},\ }\href {\doibase 10.1063/1.3474571} {\bibfield
  {journal} {\bibinfo  {journal} {J. Chem. Phys.}\ }\textbf {\bibinfo {volume}
  {133}},\ \bibinfo {pages} {064305} (\bibinfo {year} {2010})}\BibitemShut
  {NoStop}%
\bibitem [{\citenamefont {Tecmer}\ \emph {et~al.}(2011)\citenamefont {Tecmer},
  \citenamefont {Gomes}, \citenamefont {Ekstr\"om},\ and\ \citenamefont
  {Visscher}}]{pawel1}%
  \BibitemOpen
  \bibfield  {author} {\bibinfo {author} {\bibfnamefont {P.}~\bibnamefont
  {Tecmer}}, \bibinfo {author} {\bibfnamefont {A.~S.~P.}\ \bibnamefont
  {Gomes}}, \bibinfo {author} {\bibfnamefont {U.}~\bibnamefont {Ekstr\"om}}, \
  and\ \bibinfo {author} {\bibfnamefont {L.}~\bibnamefont {Visscher}},\ }\href
  {\doibase 10.1039/C0CP02534H} {\bibfield  {journal} {\bibinfo  {journal}
  {Phys. Chem. Chem. Phys.}\ }\textbf {\bibinfo {volume} {13}},\ \bibinfo
  {pages} {6249} (\bibinfo {year} {2011})}\BibitemShut {NoStop}%
\bibitem [{\citenamefont {Tecmer}\ \emph {et~al.}(2012)\citenamefont {Tecmer},
  \citenamefont {van Lingen}, \citenamefont {Gomes},\ and\ \citenamefont
  {Visscher}}]{Tecmer:2012fc}%
  \BibitemOpen
  \bibfield  {author} {\bibinfo {author} {\bibfnamefont {P.}~\bibnamefont
  {Tecmer}}, \bibinfo {author} {\bibfnamefont {H.}~\bibnamefont {van Lingen}},
  \bibinfo {author} {\bibfnamefont {A.~S.~P.}\ \bibnamefont {Gomes}}, \ and\
  \bibinfo {author} {\bibfnamefont {L.}~\bibnamefont {Visscher}},\ }\href
  {\doibase 10.1063/1.4742765} {\bibfield  {journal} {\bibinfo  {journal} {J.
  Chem. Phys.}\ }\textbf {\bibinfo {volume} {137}},\ \bibinfo {pages} {084308}
  (\bibinfo {year} {2012})}\BibitemShut {NoStop}%
\bibitem [{\citenamefont {Gomes}\ \emph {et~al.}(2013)\citenamefont {Gomes},
  \citenamefont {Jacob}, \citenamefont {R{\'e}al}, \citenamefont {Vallet},\
  and\ \citenamefont {Visscher}}]{actinide-Gomes-PCCP2013-15-15153}%
  \BibitemOpen
  \bibfield  {author} {\bibinfo {author} {\bibfnamefont {A.~S.~P.}\
  \bibnamefont {Gomes}}, \bibinfo {author} {\bibfnamefont {C.~R.}\ \bibnamefont
  {Jacob}}, \bibinfo {author} {\bibfnamefont {F.}~\bibnamefont {R{\'e}al}},
  \bibinfo {author} {\bibfnamefont {V.}~\bibnamefont {Vallet}}, \ and\ \bibinfo
  {author} {\bibfnamefont {L.}~\bibnamefont {Visscher}},\ }\href {\doibase
  10.1039/C3CP52090K} {\bibfield  {journal} {\bibinfo  {journal} {Phys. Chem.
  Chem. Phys.}\ }\textbf {\bibinfo {volume} {15}},\ \bibinfo {pages} {15153}
  (\bibinfo {year} {2013})}\BibitemShut {NoStop}%
\bibitem [{\citenamefont {Tecmer}\ \emph {et~al.}(2014)\citenamefont {Tecmer},
  \citenamefont {Gomes}, \citenamefont {Knecht},\ and\ \citenamefont
  {Visscher}}]{Tecmer:2014fs}%
  \BibitemOpen
  \bibfield  {author} {\bibinfo {author} {\bibfnamefont {P.}~\bibnamefont
  {Tecmer}}, \bibinfo {author} {\bibfnamefont {A.~S.~P.}\ \bibnamefont
  {Gomes}}, \bibinfo {author} {\bibfnamefont {S.}~\bibnamefont {Knecht}}, \
  and\ \bibinfo {author} {\bibfnamefont {L.}~\bibnamefont {Visscher}},\ }\href
  {\doibase 10.1063/1.4891801} {\bibfield  {journal} {\bibinfo  {journal} {J.
  Chem. Phys.}\ }\textbf {\bibinfo {volume} {141}},\ \bibinfo {pages} {041107}
  (\bibinfo {year} {2014})}\BibitemShut {NoStop}%
\bibitem [{\citenamefont {Haque}\ and\ \citenamefont
  {Mukherjee}(1984)}]{Haque:1984dk}%
  \BibitemOpen
  \bibfield  {author} {\bibinfo {author} {\bibfnamefont {M.~A.}\ \bibnamefont
  {Haque}}\ and\ \bibinfo {author} {\bibfnamefont {D.}~\bibnamefont
  {Mukherjee}},\ }\href {\doibase 10.1063/1.446574} {\bibfield  {journal}
  {\bibinfo  {journal} {J. Chem. Phys.}\ }\textbf {\bibinfo {volume} {80}},\
  \bibinfo {pages} {5058} (\bibinfo {year} {1984})}\BibitemShut {NoStop}%
\bibitem [{\citenamefont {Lindgren}(1974)}]{Lindgren:1974kc}%
  \BibitemOpen
  \bibfield  {author} {\bibinfo {author} {\bibfnamefont {I.}~\bibnamefont
  {Lindgren}},\ }\href {\doibase 10.1088/0022-3700/7/18/010} {\bibfield
  {journal} {\bibinfo  {journal} {J. Phys. B: At. Mol. Phys.}\ }\textbf
  {\bibinfo {volume} {7}},\ \bibinfo {pages} {2441} (\bibinfo {year}
  {1974})}\BibitemShut {NoStop}%
\bibitem [{\citenamefont {Lindgren}\ and\ \citenamefont
  {Mukherjee}(1987)}]{Lindgren:1987in}%
  \BibitemOpen
  \bibfield  {author} {\bibinfo {author} {\bibfnamefont {I.}~\bibnamefont
  {Lindgren}}\ and\ \bibinfo {author} {\bibfnamefont {D.}~\bibnamefont
  {Mukherjee}},\ }\href {\doibase 10.1016/0370-1573(87)90073-1} {\bibfield
  {journal} {\bibinfo  {journal} {Phys. Rep.}\ }\textbf {\bibinfo {volume}
  {151}},\ \bibinfo {pages} {93} (\bibinfo {year} {1987})}\BibitemShut
  {NoStop}%
\bibitem [{\citenamefont {Kaldor}(1988)}]{Kaldor:1988wp}%
  \BibitemOpen
  \bibfield  {author} {\bibinfo {author} {\bibfnamefont {U.}~\bibnamefont
  {Kaldor}},\ }\href {\doibase 10.1103/PhysRevA.38.6013} {\bibfield  {journal}
  {\bibinfo  {journal} {Phys. Rev. A}\ }\textbf {\bibinfo {volume} {38}},\
  \bibinfo {pages} {6013} (\bibinfo {year} {1988})}\BibitemShut {NoStop}%
\bibitem [{\citenamefont {Kaldor}(1991)}]{Kaldor:1991cd}%
  \BibitemOpen
  \bibfield  {author} {\bibinfo {author} {\bibfnamefont {U.}~\bibnamefont
  {Kaldor}},\ }\href {\doibase 10.1007/BF01119664} {\bibfield  {journal}
  {\bibinfo  {journal} {Theor. Chim. Acta}\ }\textbf {\bibinfo {volume} {80}},\
  \bibinfo {pages} {427} (\bibinfo {year} {1991})}\BibitemShut {NoStop}%
\bibitem [{\citenamefont {Malrieu}, \citenamefont {Durand},\ and\ \citenamefont
  {Daudey}(1985)}]{Malrieu:1985ce}%
  \BibitemOpen
  \bibfield  {author} {\bibinfo {author} {\bibfnamefont {J.~P.}\ \bibnamefont
  {Malrieu}}, \bibinfo {author} {\bibfnamefont {P.}~\bibnamefont {Durand}}, \
  and\ \bibinfo {author} {\bibfnamefont {J.~P.}\ \bibnamefont {Daudey}},\
  }\href {\doibase 10.1088/0305-4470/18/5/014} {\bibfield  {journal} {\bibinfo
  {journal} {J. Phys. A: Math. Gen.}\ }\textbf {\bibinfo {volume} {18}},\
  \bibinfo {pages} {809} (\bibinfo {year} {1985})}\BibitemShut {NoStop}%
\bibitem [{\citenamefont {Bartlett}(2011)}]{Bartlett:2011ho}%
  \BibitemOpen
  \bibfield  {author} {\bibinfo {author} {\bibfnamefont {R.~J.}\ \bibnamefont
  {Bartlett}},\ }\href {\doibase 10.1002/wcms.76} {\bibfield  {journal}
  {\bibinfo  {journal} {Wiley Interdisciplinary Reviews: Computational
  Molecular Science}\ }\textbf {\bibinfo {volume} {2}},\ \bibinfo {pages} {126}
  (\bibinfo {year} {2011})}\BibitemShut {NoStop}%
\bibitem [{\citenamefont {Meissner}(2010)}]{Meissner:2010fq}%
  \BibitemOpen
  \bibfield  {author} {\bibinfo {author} {\bibfnamefont {L.}~\bibnamefont
  {Meissner}},\ }\href {\doibase 10.1080/00268976.2010.522605} {\bibfield
  {journal} {\bibinfo  {journal} {Mol. Phys.}\ }\textbf {\bibinfo {volume}
  {108}},\ \bibinfo {pages} {2961} (\bibinfo {year} {2010})}\BibitemShut
  {NoStop}%
\bibitem [{\citenamefont {Mukhopadhyay}\ \emph {et~al.}(1991)\citenamefont
  {Mukhopadhyay}, \citenamefont {Mukhopadhyay}, \citenamefont {Chaudhuri},\
  and\ \citenamefont {Mukherjee}}]{Mukhopadhyay:1991fp}%
  \BibitemOpen
  \bibfield  {author} {\bibinfo {author} {\bibfnamefont {D.}~\bibnamefont
  {Mukhopadhyay}}, \bibinfo {author} {\bibfnamefont {S.}~\bibnamefont
  {Mukhopadhyay}}, \bibinfo {author} {\bibfnamefont {R.}~\bibnamefont
  {Chaudhuri}}, \ and\ \bibinfo {author} {\bibfnamefont {D.}~\bibnamefont
  {Mukherjee}},\ }\href {\doibase 10.1007/BF01119665} {\bibfield  {journal}
  {\bibinfo  {journal} {Theor. Chim. Acta}\ }\textbf {\bibinfo {volume} {80}},\
  \bibinfo {pages} {441} (\bibinfo {year} {1991})}\BibitemShut {NoStop}%
\bibitem [{\citenamefont {Musial}\ and\ \citenamefont
  {Bartlett}(2008{\natexlab{a}})}]{Musial:2008da}%
  \BibitemOpen
  \bibfield  {author} {\bibinfo {author} {\bibfnamefont {M.}~\bibnamefont
  {Musial}}\ and\ \bibinfo {author} {\bibfnamefont {R.~J.}\ \bibnamefont
  {Bartlett}},\ }\href {\doibase 10.1063/1.3046453} {\bibfield  {journal}
  {\bibinfo  {journal} {The Journal of Chemical Physics}\ }\textbf {\bibinfo
  {volume} {129}},\ \bibinfo {pages} {244111} (\bibinfo {year}
  {2008}{\natexlab{a}})}\BibitemShut {NoStop}%
\bibitem [{\citenamefont {Musial}\ and\ \citenamefont
  {Bartlett}(2008{\natexlab{b}})}]{fscc-vs-lrcc:musial:1}%
  \BibitemOpen
  \bibfield  {author} {\bibinfo {author} {\bibfnamefont {M.}~\bibnamefont
  {Musial}}\ and\ \bibinfo {author} {\bibfnamefont {R.~J.}\ \bibnamefont
  {Bartlett}},\ }\href {\doibase 10.1063/1.2952521} {\bibfield  {journal}
  {\bibinfo  {journal} {J. Chem. Phys.}\ }\textbf {\bibinfo {volume} {129}},\
  \bibinfo {pages} {044101} (\bibinfo {year} {2008}{\natexlab{b}})}\BibitemShut
  {NoStop}%
\bibitem [{\citenamefont {Musial}\ and\ \citenamefont
  {Bartlett}(2008{\natexlab{c}})}]{fscc-vs-lrcc:musial:2}%
  \BibitemOpen
  \bibfield  {author} {\bibinfo {author} {\bibfnamefont {M.}~\bibnamefont
  {Musial}}\ and\ \bibinfo {author} {\bibfnamefont {R.~J.}\ \bibnamefont
  {Bartlett}},\ }\href {\doibase 10.1063/1.2982788} {\bibfield  {journal}
  {\bibinfo  {journal} {J. Chem. Phys.}\ }\textbf {\bibinfo {volume} {129}},\
  \bibinfo {pages} {134105} (\bibinfo {year} {2008}{\natexlab{c}})}\BibitemShut
  {NoStop}%
\bibitem [{\citenamefont {Musial}\ and\ \citenamefont
  {Bartlett}(2008{\natexlab{d}})}]{fscc-vs-lrcc:musial:3}%
  \BibitemOpen
  \bibfield  {author} {\bibinfo {author} {\bibfnamefont {M.}~\bibnamefont
  {Musial}}\ and\ \bibinfo {author} {\bibfnamefont {R.~J.}\ \bibnamefont
  {Bartlett}},\ }\href {\doibase 10.1016/j.cplett.2008.04.004} {\bibfield
  {journal} {\bibinfo  {journal} {Chem. Phys. Lett.}\ }\textbf {\bibinfo
  {volume} {457}},\ \bibinfo {pages} {267} (\bibinfo {year}
  {2008}{\natexlab{d}})}\BibitemShut {NoStop}%
\bibitem [{\citenamefont {Pathak}\ \emph
  {et~al.}(2016{\natexlab{a}})\citenamefont {Pathak}, \citenamefont {Sasmal},
  \citenamefont {Nayak}, \citenamefont {Vaval},\ and\ \citenamefont
  {Pal}}]{Pathak:2016ck}%
  \BibitemOpen
  \bibfield  {author} {\bibinfo {author} {\bibfnamefont {H.}~\bibnamefont
  {Pathak}}, \bibinfo {author} {\bibfnamefont {S.}~\bibnamefont {Sasmal}},
  \bibinfo {author} {\bibfnamefont {M.~K.}\ \bibnamefont {Nayak}}, \bibinfo
  {author} {\bibfnamefont {N.}~\bibnamefont {Vaval}}, \ and\ \bibinfo {author}
  {\bibfnamefont {S.}~\bibnamefont {Pal}},\ }\href {\doibase
  10.1016/j.comptc.2015.12.015} {\bibfield  {journal} {\bibinfo  {journal}
  {Comput. Theor. Chem.}\ }\textbf {\bibinfo {volume} {1076}},\ \bibinfo
  {pages} {94} (\bibinfo {year} {2016}{\natexlab{a}})}\BibitemShut {NoStop}%
\bibitem [{\citenamefont {Pathak}\ \emph {et~al.}(2015)\citenamefont {Pathak},
  \citenamefont {Sahoo}, \citenamefont {Sengupta}, \citenamefont {Das},
  \citenamefont {Vaval},\ and\ \citenamefont {Pal}}]{Pathak:2015he}%
  \BibitemOpen
  \bibfield  {author} {\bibinfo {author} {\bibfnamefont {H.}~\bibnamefont
  {Pathak}}, \bibinfo {author} {\bibfnamefont {B.~K.}\ \bibnamefont {Sahoo}},
  \bibinfo {author} {\bibfnamefont {T.}~\bibnamefont {Sengupta}}, \bibinfo
  {author} {\bibfnamefont {B.~P.}\ \bibnamefont {Das}}, \bibinfo {author}
  {\bibfnamefont {N.}~\bibnamefont {Vaval}}, \ and\ \bibinfo {author}
  {\bibfnamefont {S.}~\bibnamefont {Pal}},\ }\href {\doibase
  10.1088/0953-4075/48/11/115009} {\bibfield  {journal} {\bibinfo  {journal}
  {J. Phys. B: Atom. Mol. Opt. Phys.}\ }\textbf {\bibinfo {volume} {48}},\
  \bibinfo {pages} {115009} (\bibinfo {year} {2015})}\BibitemShut {NoStop}%
\bibitem [{\citenamefont {Pathak}\ \emph
  {et~al.}(2014{\natexlab{a}})\citenamefont {Pathak}, \citenamefont {Sasmal},
  \citenamefont {Nayak}, \citenamefont {Vaval},\ and\ \citenamefont
  {Pal}}]{Pathak:2014gv}%
  \BibitemOpen
  \bibfield  {author} {\bibinfo {author} {\bibfnamefont {H.}~\bibnamefont
  {Pathak}}, \bibinfo {author} {\bibfnamefont {S.}~\bibnamefont {Sasmal}},
  \bibinfo {author} {\bibfnamefont {M.~K.}\ \bibnamefont {Nayak}}, \bibinfo
  {author} {\bibfnamefont {N.}~\bibnamefont {Vaval}}, \ and\ \bibinfo {author}
  {\bibfnamefont {S.}~\bibnamefont {Pal}},\ }\href {\doibase
  10.1103/PhysRevA.90.062501} {\bibfield  {journal} {\bibinfo  {journal} {Phys.
  Rev. A}\ }\textbf {\bibinfo {volume} {90}},\ \bibinfo {pages} {062501}
  (\bibinfo {year} {2014}{\natexlab{a}})}\BibitemShut {NoStop}%
\bibitem [{\citenamefont {Pathak}\ \emph
  {et~al.}(2014{\natexlab{b}})\citenamefont {Pathak}, \citenamefont {Sahoo},
  \citenamefont {Das}, \citenamefont {Vaval},\ and\ \citenamefont
  {Pal}}]{Pathak:2014dm}%
  \BibitemOpen
  \bibfield  {author} {\bibinfo {author} {\bibfnamefont {H.}~\bibnamefont
  {Pathak}}, \bibinfo {author} {\bibfnamefont {B.~K.}\ \bibnamefont {Sahoo}},
  \bibinfo {author} {\bibfnamefont {B.~P.}\ \bibnamefont {Das}}, \bibinfo
  {author} {\bibfnamefont {N.}~\bibnamefont {Vaval}}, \ and\ \bibinfo {author}
  {\bibfnamefont {S.}~\bibnamefont {Pal}},\ }\href {\doibase
  10.1103/PhysRevA.89.042510} {\bibfield  {journal} {\bibinfo  {journal} {Phys.
  Rev. A}\ }\textbf {\bibinfo {volume} {89}},\ \bibinfo {pages} {042510}
  (\bibinfo {year} {2014}{\natexlab{b}})}\BibitemShut {NoStop}%
\bibitem [{\citenamefont {Pathak}\ \emph
  {et~al.}(2016{\natexlab{b}})\citenamefont {Pathak}, \citenamefont {Sasmal},
  \citenamefont {Nayak}, \citenamefont {Vaval},\ and\ \citenamefont
  {Pal}}]{Pathak:2016dk}%
  \BibitemOpen
  \bibfield  {author} {\bibinfo {author} {\bibfnamefont {H.}~\bibnamefont
  {Pathak}}, \bibinfo {author} {\bibfnamefont {S.}~\bibnamefont {Sasmal}},
  \bibinfo {author} {\bibfnamefont {M.~K.}\ \bibnamefont {Nayak}}, \bibinfo
  {author} {\bibfnamefont {N.}~\bibnamefont {Vaval}}, \ and\ \bibinfo {author}
  {\bibfnamefont {S.}~\bibnamefont {Pal}},\ }\href {\doibase 10.1063/1.4960954}
  {\bibfield  {journal} {\bibinfo  {journal} {J. Chem. Phys.}\ }\textbf
  {\bibinfo {volume} {145}},\ \bibinfo {pages} {074110} (\bibinfo {year}
  {2016}{\natexlab{b}})}\BibitemShut {NoStop}%
\bibitem [{\citenamefont {Klein}\ and\ \citenamefont
  {Gauss}(2008)}]{Klein:2008hx}%
  \BibitemOpen
  \bibfield  {author} {\bibinfo {author} {\bibfnamefont {K.}~\bibnamefont
  {Klein}}\ and\ \bibinfo {author} {\bibfnamefont {J.}~\bibnamefont {Gauss}},\
  }\href {\doibase 10.1063/1.3013199} {\bibfield  {journal} {\bibinfo
  {journal} {J. Chem. Phys.}\ }\textbf {\bibinfo {volume} {129}},\ \bibinfo
  {pages} {194106} (\bibinfo {year} {2008})}\BibitemShut {NoStop}%
\bibitem [{\citenamefont {Yang}, \citenamefont {Wang},\ and\ \citenamefont
  {Guo}(2012)}]{Yang:2012gd}%
  \BibitemOpen
  \bibfield  {author} {\bibinfo {author} {\bibfnamefont {D.-D.}\ \bibnamefont
  {Yang}}, \bibinfo {author} {\bibfnamefont {F.}~\bibnamefont {Wang}}, \ and\
  \bibinfo {author} {\bibfnamefont {J.}~\bibnamefont {Guo}},\ }\href {\doibase
  10.1016/j.cplett.2012.02.014} {\bibfield  {journal} {\bibinfo  {journal}
  {Chem. Phys. Lett.}\ }\textbf {\bibinfo {volume} {531}},\ \bibinfo {pages}
  {236} (\bibinfo {year} {2012})}\BibitemShut {NoStop}%
\bibitem [{\citenamefont {Wang}\ \emph {et~al.}(2015)\citenamefont {Wang},
  \citenamefont {Hu}, \citenamefont {Wang},\ and\ \citenamefont
  {Guo}}]{Wang:2015jw}%
  \BibitemOpen
  \bibfield  {author} {\bibinfo {author} {\bibfnamefont {Z.}~\bibnamefont
  {Wang}}, \bibinfo {author} {\bibfnamefont {S.}~\bibnamefont {Hu}}, \bibinfo
  {author} {\bibfnamefont {F.}~\bibnamefont {Wang}}, \ and\ \bibinfo {author}
  {\bibfnamefont {J.}~\bibnamefont {Guo}},\ }\href {\doibase 10.1063/1.4917041}
  {\bibfield  {journal} {\bibinfo  {journal} {J. Chem. Phys.}\ }\textbf
  {\bibinfo {volume} {142}},\ \bibinfo {pages} {144109} (\bibinfo {year}
  {2015})}\BibitemShut {NoStop}%
\bibitem [{\citenamefont {Epifanovsky}\ \emph {et~al.}(2015)\citenamefont
  {Epifanovsky}, \citenamefont {Klein}, \citenamefont {Stopkowicz},
  \citenamefont {Gauss},\ and\ \citenamefont {Krylov}}]{Epifanovsky:2015hsa}%
  \BibitemOpen
  \bibfield  {author} {\bibinfo {author} {\bibfnamefont {E.}~\bibnamefont
  {Epifanovsky}}, \bibinfo {author} {\bibfnamefont {K.}~\bibnamefont {Klein}},
  \bibinfo {author} {\bibfnamefont {S.}~\bibnamefont {Stopkowicz}}, \bibinfo
  {author} {\bibfnamefont {J.}~\bibnamefont {Gauss}}, \ and\ \bibinfo {author}
  {\bibfnamefont {A.~I.}\ \bibnamefont {Krylov}},\ }\href {\doibase
  10.1063/1.4927785} {\bibfield  {journal} {\bibinfo  {journal} {J. Chem.
  Phys.}\ }\textbf {\bibinfo {volume} {143}},\ \bibinfo {pages} {064102}
  (\bibinfo {year} {2015})}\BibitemShut {NoStop}%
\bibitem [{\citenamefont {Cao}, \citenamefont {Wang},\ and\ \citenamefont
  {Yang}(2016)}]{Cao:2016fx}%
  \BibitemOpen
  \bibfield  {author} {\bibinfo {author} {\bibfnamefont {Z.}~\bibnamefont
  {Cao}}, \bibinfo {author} {\bibfnamefont {F.}~\bibnamefont {Wang}}, \ and\
  \bibinfo {author} {\bibfnamefont {M.}~\bibnamefont {Yang}},\ }\href {\doibase
  10.1063/1.4964859} {\bibfield  {journal} {\bibinfo  {journal} {J. Chem.
  Phys.}\ }\textbf {\bibinfo {volume} {145}},\ \bibinfo {pages} {154110}
  (\bibinfo {year} {2016})}\BibitemShut {NoStop}%
\bibitem [{\citenamefont {Cao}\ \emph {et~al.}(2017)\citenamefont {Cao},
  \citenamefont {Li}, \citenamefont {Wang},\ and\ \citenamefont
  {Liu}}]{Cao:2017id}%
  \BibitemOpen
  \bibfield  {author} {\bibinfo {author} {\bibfnamefont {Z.}~\bibnamefont
  {Cao}}, \bibinfo {author} {\bibfnamefont {Z.}~\bibnamefont {Li}}, \bibinfo
  {author} {\bibfnamefont {F.}~\bibnamefont {Wang}}, \ and\ \bibinfo {author}
  {\bibfnamefont {W.}~\bibnamefont {Liu}},\ }\href {\doibase
  10.1039/C6CP07588F} {\bibfield  {journal} {\bibinfo  {journal} {Phys. Chem.
  Chem. Phys.}\ }\textbf {\bibinfo {volume} {19}},\ \bibinfo {pages} {3713}
  (\bibinfo {year} {2017})}\BibitemShut {NoStop}%
\bibitem [{\citenamefont {Zhang}\ and\ \citenamefont
  {Wang}(2017)}]{Zhang:2017jj}%
  \BibitemOpen
  \bibfield  {author} {\bibinfo {author} {\bibfnamefont {S.}~\bibnamefont
  {Zhang}}\ and\ \bibinfo {author} {\bibfnamefont {F.}~\bibnamefont {Wang}},\
  }\href {\doibase 10.1021/acs.jpca.7b02985} {\bibfield  {journal} {\bibinfo
  {journal} {J. Phys. Chem. A}\ }\textbf {\bibinfo {volume} {121}},\ \bibinfo
  {pages} {3966} (\bibinfo {year} {2017})}\BibitemShut {NoStop}%
\bibitem [{\citenamefont {Akinaga}\ and\ \citenamefont
  {Nakajima}(2017)}]{Akinaga:2017hv}%
  \BibitemOpen
  \bibfield  {author} {\bibinfo {author} {\bibfnamefont {Y.}~\bibnamefont
  {Akinaga}}\ and\ \bibinfo {author} {\bibfnamefont {T.}~\bibnamefont
  {Nakajima}},\ }\href {\doibase 10.1021/acs.jpca.6b10921} {\bibfield
  {journal} {\bibinfo  {journal} {J. Phys. Chem. A}\ }\textbf {\bibinfo
  {volume} {121}},\ \bibinfo {pages} {827} (\bibinfo {year}
  {2017})}\BibitemShut {NoStop}%
\bibitem [{\citenamefont {Wang}(2016)}]{Wang:2016hx}%
  \BibitemOpen
  \bibfield  {author} {\bibinfo {author} {\bibfnamefont {F.}~\bibnamefont
  {Wang}},\ }in\ \href {\doibase 10.1007/978-3-642-41611-8_33-1} {\emph
  {\bibinfo {booktitle} {Handbook of Relativistic Quantum Chemistry}}},\
  \bibinfo {editor} {edited by\ \bibinfo {editor} {\bibfnamefont
  {W.}~\bibnamefont {Liu}}}\ (\bibinfo  {publisher} {Springer Berlin
  Heidelberg},\ \bibinfo {address} {Berlin, Heidelberg},\ \bibinfo {year}
  {2016})\ pp.\ \bibinfo {pages} {1--27}\BibitemShut {NoStop}%
\bibitem [{\citenamefont {He{\ss}}\ \emph {et~al.}(1996)\citenamefont
  {He{\ss}}, \citenamefont {Marian}, \citenamefont {Wahlgren},\ and\
  \citenamefont {Gropen}}]{so-Hes-CPL1996-251-365-371}%
  \BibitemOpen
  \bibfield  {author} {\bibinfo {author} {\bibfnamefont {B.~A.}\ \bibnamefont
  {He{\ss}}}, \bibinfo {author} {\bibfnamefont {C.~M.}\ \bibnamefont {Marian}},
  \bibinfo {author} {\bibfnamefont {U.}~\bibnamefont {Wahlgren}}, \ and\
  \bibinfo {author} {\bibfnamefont {O.}~\bibnamefont {Gropen}},\ }\href
  {\doibase 10.1016/0009-2614(96)00119-4} {\bibfield  {journal} {\bibinfo
  {journal} {Chem. Phys. Lett.}\ }\textbf {\bibinfo {volume} {251}},\ \bibinfo
  {pages} {365} (\bibinfo {year} {1996})}\BibitemShut {NoStop}%
\bibitem [{\citenamefont {Sikkema}\ \emph {et~al.}(2009)\citenamefont
  {Sikkema}, \citenamefont {Visscher}, \citenamefont {Saue},\ and\
  \citenamefont {Ilias}}]{Sikkema2009}%
  \BibitemOpen
  \bibfield  {author} {\bibinfo {author} {\bibfnamefont {J.}~\bibnamefont
  {Sikkema}}, \bibinfo {author} {\bibfnamefont {L.}~\bibnamefont {Visscher}},
  \bibinfo {author} {\bibfnamefont {T.}~\bibnamefont {Saue}}, \ and\ \bibinfo
  {author} {\bibfnamefont {M.}~\bibnamefont {Ilias}},\ }\href {\doibase
  10.1063/1.3239505} {\bibfield  {journal} {\bibinfo  {journal} {J. Chem.
  Phys.}\ }\textbf {\bibinfo {volume} {131}},\ \bibinfo {pages} {124116}
  (\bibinfo {year} {2009})}\BibitemShut {NoStop}%
\bibitem [{DIR()}]{DIRAC17}%
  \BibitemOpen
  \href@noop {} {}\bibinfo {note} {{DIRAC}, a relativistic ab initio electronic
  structure program, Release {DIRAC17} (2017), written by L.~Visscher,
  H.~J.~{\relax Aa}.~Jensen, R.~Bast, and T.~Saue, with contributions from
  V.~Bakken, K.~G.~Dyall, S.~Dubillard, U.~Ekstr{\"o}m, E.~Eliav,
  T.~Enevoldsen, E.~Fa{\ss}hauer, T.~Fleig, O.~Fossgaard, A.~S.~P.~Gomes,
  E.~D.~Hedeg{\aa}rd, T.~Helgaker, J.~Henriksson, M.~Ilia{\v{s}}, Ch.~R.~Jacob,
  S.~Knecht, S.~Komorovsk{\'y}, O.~Kullie, J.~K.~L{\ae}rdahl, C.~V.~Larsen,
  Y.~S.~Lee, H.~S.~Nataraj, M.~K.~Nayak, P.~Norman, G.~Olejniczak, J.~Olsen,
  J.~M.~H.~Olsen, Y.~C.~Park, J.~K.~Pedersen, M.~Pernpointner, R.~di~Remigio,
  K.~Ruud, P.~Sa{\l}ek, B.~Schimmelpfennig, A.~Shee, J.~Sikkema,
  A.~J.~Thorvaldsen, J.~Thyssen, J.~van~Stralen, S.~Villaume, O.~Visser,
  T.~Winther, and S.~Yamamoto (see
  \url{http://www.diracprogram.org})}\BibitemShut {NoStop}%
\bibitem [{\citenamefont {Saiz-Lopez}\ \emph {et~al.}(2016)\citenamefont
  {Saiz-Lopez}, \citenamefont {Plane}, \citenamefont {Cuevas}, \citenamefont
  {Mahajan}, \citenamefont {Lamarque},\ and\ \citenamefont
  {Kinnison}}]{SaizLopez:2016fx}%
  \BibitemOpen
  \bibfield  {author} {\bibinfo {author} {\bibfnamefont {A.}~\bibnamefont
  {Saiz-Lopez}}, \bibinfo {author} {\bibfnamefont {J.~M.~C.}\ \bibnamefont
  {Plane}}, \bibinfo {author} {\bibfnamefont {C.~A.}\ \bibnamefont {Cuevas}},
  \bibinfo {author} {\bibfnamefont {A.~S.}\ \bibnamefont {Mahajan}}, \bibinfo
  {author} {\bibfnamefont {J.-F.}\ \bibnamefont {Lamarque}}, \ and\ \bibinfo
  {author} {\bibfnamefont {D.~E.}\ \bibnamefont {Kinnison}},\ }\href {\doibase
  10.5194/acp-16-15593-2016} {\bibfield  {journal} {\bibinfo  {journal} {Atmos.
  Chem. Phys.}\ }\textbf {\bibinfo {volume} {16}},\ \bibinfo {pages} {15593}
  (\bibinfo {year} {2016})}\BibitemShut {NoStop}%
\bibitem [{\citenamefont {Burkholder}, \citenamefont {Cox},\ and\ \citenamefont
  {Ravishankara}(2015)}]{Burkholder:2015js}%
  \BibitemOpen
  \bibfield  {author} {\bibinfo {author} {\bibfnamefont {J.~B.}\ \bibnamefont
  {Burkholder}}, \bibinfo {author} {\bibfnamefont {R.~A.}\ \bibnamefont {Cox}},
  \ and\ \bibinfo {author} {\bibfnamefont {A.~R.}\ \bibnamefont
  {Ravishankara}},\ }\href {\doibase 10.1021/cr5006759} {\bibfield  {journal}
  {\bibinfo  {journal} {Chem. Rev.}\ }\textbf {\bibinfo {volume} {115}},\
  \bibinfo {pages} {3704} (\bibinfo {year} {2015})}\BibitemShut {NoStop}%
\bibitem [{\citenamefont {Saiz-Lopez}\ \emph {et~al.}(2014)\citenamefont
  {Saiz-Lopez}, \citenamefont {Fernandez}, \citenamefont {Ord{\'o}{\~n}ez},
  \citenamefont {Kinnison}, \citenamefont {G{\'o}mez~Mart{\'\i}n},
  \citenamefont {Lamarque},\ and\ \citenamefont {Tilmes}}]{SaizLopez:2014fr}%
  \BibitemOpen
  \bibfield  {author} {\bibinfo {author} {\bibfnamefont {A.}~\bibnamefont
  {Saiz-Lopez}}, \bibinfo {author} {\bibfnamefont {R.~P.}\ \bibnamefont
  {Fernandez}}, \bibinfo {author} {\bibfnamefont {C.}~\bibnamefont
  {Ord{\'o}{\~n}ez}}, \bibinfo {author} {\bibfnamefont {D.~E.}\ \bibnamefont
  {Kinnison}}, \bibinfo {author} {\bibfnamefont {J.~C.}\ \bibnamefont
  {G{\'o}mez~Mart{\'\i}n}}, \bibinfo {author} {\bibfnamefont {J.~F.}\
  \bibnamefont {Lamarque}}, \ and\ \bibinfo {author} {\bibfnamefont
  {S.}~\bibnamefont {Tilmes}},\ }\href {\doibase 10.5194/acp-14-13119-2014}
  {\bibfield  {journal} {\bibinfo  {journal} {Atmos. Chem. Phys.}\ }\textbf
  {\bibinfo {volume} {14}},\ \bibinfo {pages} {13119} (\bibinfo {year}
  {2014})}\BibitemShut {NoStop}%
\bibitem [{\citenamefont {G{\'o}mez~Mart{\'\i}n}\ \emph
  {et~al.}(2013)\citenamefont {G{\'o}mez~Mart{\'\i}n}, \citenamefont
  {G{\'a}lvez}, \citenamefont {Baeza-Romero}, \citenamefont {Ingham},
  \citenamefont {Plane},\ and\ \citenamefont {Blitz}}]{GomezMartin:2013fa}%
  \BibitemOpen
  \bibfield  {author} {\bibinfo {author} {\bibfnamefont {J.~C.}\ \bibnamefont
  {G{\'o}mez~Mart{\'\i}n}}, \bibinfo {author} {\bibfnamefont {O.}~\bibnamefont
  {G{\'a}lvez}}, \bibinfo {author} {\bibfnamefont {M.~T.}\ \bibnamefont
  {Baeza-Romero}}, \bibinfo {author} {\bibfnamefont {T.}~\bibnamefont
  {Ingham}}, \bibinfo {author} {\bibfnamefont {J.~M.~C.}\ \bibnamefont
  {Plane}}, \ and\ \bibinfo {author} {\bibfnamefont {M.~A.}\ \bibnamefont
  {Blitz}},\ }\href {\doibase 10.1039/C3CP51217G} {\bibfield  {journal}
  {\bibinfo  {journal} {Phys. Chem. Chem. Phys.}\ }\textbf {\bibinfo {volume}
  {15}},\ \bibinfo {pages} {15612} (\bibinfo {year} {2013})}\BibitemShut
  {NoStop}%
\bibitem [{\citenamefont {Mehboob}\ and\ \citenamefont
  {Aljohani}(2016)}]{Mehboob:2016kc}%
  \BibitemOpen
  \bibfield  {author} {\bibinfo {author} {\bibfnamefont {K.}~\bibnamefont
  {Mehboob}}\ and\ \bibinfo {author} {\bibfnamefont {M.~S.}\ \bibnamefont
  {Aljohani}},\ }\href {\doibase 10.1016/j.pnucene.2015.11.013} {\bibfield
  {journal} {\bibinfo  {journal} {Progress in Nuclear Energy}\ }\textbf
  {\bibinfo {volume} {88}},\ \bibinfo {pages} {75} (\bibinfo {year}
  {2016})}\BibitemShut {NoStop}%
\bibitem [{\citenamefont {Funke}\ \emph {et~al.}(2012)\citenamefont {Funke},
  \citenamefont {Langrock}, \citenamefont {Kanzleiter}, \citenamefont {Poss},
  \citenamefont {Fischer}, \citenamefont {K{\"u}hnel}, \citenamefont {Weber},\
  and\ \citenamefont {Allelein}}]{Funke:2012cc}%
  \BibitemOpen
  \bibfield  {author} {\bibinfo {author} {\bibfnamefont {F.}~\bibnamefont
  {Funke}}, \bibinfo {author} {\bibfnamefont {G.}~\bibnamefont {Langrock}},
  \bibinfo {author} {\bibfnamefont {T.}~\bibnamefont {Kanzleiter}}, \bibinfo
  {author} {\bibfnamefont {G.}~\bibnamefont {Poss}}, \bibinfo {author}
  {\bibfnamefont {K.}~\bibnamefont {Fischer}}, \bibinfo {author} {\bibfnamefont
  {A.}~\bibnamefont {K{\"u}hnel}}, \bibinfo {author} {\bibfnamefont
  {G.}~\bibnamefont {Weber}}, \ and\ \bibinfo {author} {\bibfnamefont {H.~J.}\
  \bibnamefont {Allelein}},\ }\href {\doibase 10.1016/j.nucengdes.2012.01.005}
  {\bibfield  {journal} {\bibinfo  {journal} {Nuclear Engineering and Design}\
  }\textbf {\bibinfo {volume} {245}},\ \bibinfo {pages} {206} (\bibinfo {year}
  {2012})}\BibitemShut {NoStop}%
\bibitem [{\citenamefont {Ayed}\ \emph
  {et~al.}(2013{\natexlab{a}})\citenamefont {Ayed}, \citenamefont {R{\'e}al},
  \citenamefont {Montavon},\ and\ \citenamefont
  {Galland}}]{ato-ayed-jpc2013-117-10589}%
  \BibitemOpen
  \bibfield  {author} {\bibinfo {author} {\bibfnamefont {T.}~\bibnamefont
  {Ayed}}, \bibinfo {author} {\bibfnamefont {F.}~\bibnamefont {R{\'e}al}},
  \bibinfo {author} {\bibfnamefont {G.}~\bibnamefont {Montavon}}, \ and\
  \bibinfo {author} {\bibfnamefont {N.}~\bibnamefont {Galland}},\ }\href
  {\doibase 10.1021/jp406803e} {\bibfield  {journal} {\bibinfo  {journal} {J.
  Phys. Chem. B}\ }\textbf {\bibinfo {volume} {117}},\ \bibinfo {pages} {10589}
  (\bibinfo {year} {2013}{\natexlab{a}})}\BibitemShut {NoStop}%
\bibitem [{\citenamefont {Ayed}\ \emph
  {et~al.}(2013{\natexlab{b}})\citenamefont {Ayed}, \citenamefont {Seydou},
  \citenamefont {R{\'e}al}, \citenamefont {Montavon},\ and\ \citenamefont
  {Galland}}]{ato-ayed-jpc2013-117-5206}%
  \BibitemOpen
  \bibfield  {author} {\bibinfo {author} {\bibfnamefont {T.}~\bibnamefont
  {Ayed}}, \bibinfo {author} {\bibfnamefont {M.}~\bibnamefont {Seydou}},
  \bibinfo {author} {\bibfnamefont {F.}~\bibnamefont {R{\'e}al}}, \bibinfo
  {author} {\bibfnamefont {G.}~\bibnamefont {Montavon}}, \ and\ \bibinfo
  {author} {\bibfnamefont {N.}~\bibnamefont {Galland}},\ }\href {\doibase
  10.1021/jp401759p} {\bibfield  {journal} {\bibinfo  {journal} {J. Phys. Chem.
  B}\ }\textbf {\bibinfo {volume} {117}},\ \bibinfo {pages} {5206} (\bibinfo
  {year} {2013}{\natexlab{b}})}\BibitemShut {NoStop}%
\bibitem [{\citenamefont {Champion}\ \emph {et~al.}(2009)\citenamefont
  {Champion}, \citenamefont {Alliot}, \citenamefont {Huclier}, \citenamefont
  {Deniaud}, \citenamefont {Asfari},\ and\ \citenamefont
  {Montavon}}]{ato-champion-ica2009-362-2654}%
  \BibitemOpen
  \bibfield  {author} {\bibinfo {author} {\bibfnamefont {J.}~\bibnamefont
  {Champion}}, \bibinfo {author} {\bibfnamefont {C.}~\bibnamefont {Alliot}},
  \bibinfo {author} {\bibfnamefont {S.}~\bibnamefont {Huclier}}, \bibinfo
  {author} {\bibfnamefont {D.}~\bibnamefont {Deniaud}}, \bibinfo {author}
  {\bibfnamefont {W.}~\bibnamefont {Asfari}}, \ and\ \bibinfo {author}
  {\bibfnamefont {G.}~\bibnamefont {Montavon}},\ }\href {\doibase
  10.1016/j.ica.2008.12.005} {\bibfield  {journal} {\bibinfo  {journal} {Inorg.
  Chim. Acta}\ }\textbf {\bibinfo {volume} {362}},\ \bibinfo {pages} {2654}
  (\bibinfo {year} {2009})}\BibitemShut {NoStop}%
\bibitem [{\citenamefont {Champion}\ \emph {et~al.}(2010)\citenamefont
  {Champion}, \citenamefont {Alliot}, \citenamefont {Renault}, \citenamefont
  {Mokili}, \citenamefont {Ch{\'e}rel}, \citenamefont {Galland},\ and\
  \citenamefont {Montavon}}]{ato-champion-jpc2010-114-576}%
  \BibitemOpen
  \bibfield  {author} {\bibinfo {author} {\bibfnamefont {J.}~\bibnamefont
  {Champion}}, \bibinfo {author} {\bibfnamefont {C.}~\bibnamefont {Alliot}},
  \bibinfo {author} {\bibfnamefont {E.}~\bibnamefont {Renault}}, \bibinfo
  {author} {\bibfnamefont {B.~M.}\ \bibnamefont {Mokili}}, \bibinfo {author}
  {\bibfnamefont {M.}~\bibnamefont {Ch{\'e}rel}}, \bibinfo {author}
  {\bibfnamefont {N.}~\bibnamefont {Galland}}, \ and\ \bibinfo {author}
  {\bibfnamefont {G.}~\bibnamefont {Montavon}},\ }\href {\doibase
  10.1021/jp9077008} {\bibfield  {journal} {\bibinfo  {journal} {J. Phys. Chem.
  A}\ }\textbf {\bibinfo {volume} {114}},\ \bibinfo {pages} {576} (\bibinfo
  {year} {2010})}\BibitemShut {NoStop}%
\bibitem [{\citenamefont {Champion}\ \emph {et~al.}(2013)\citenamefont
  {Champion}, \citenamefont {Sabati{\'e}-Gogova}, \citenamefont {Bassal},
  \citenamefont {Ayed}, \citenamefont {Alliot}, \citenamefont {Galland},\ and\
  \citenamefont {Montavon}}]{ato-champion-jpc2013-117-1983}%
  \BibitemOpen
  \bibfield  {author} {\bibinfo {author} {\bibfnamefont {J.}~\bibnamefont
  {Champion}}, \bibinfo {author} {\bibfnamefont {A.}~\bibnamefont
  {Sabati{\'e}-Gogova}}, \bibinfo {author} {\bibfnamefont {F.}~\bibnamefont
  {Bassal}}, \bibinfo {author} {\bibfnamefont {T.}~\bibnamefont {Ayed}},
  \bibinfo {author} {\bibfnamefont {C.}~\bibnamefont {Alliot}}, \bibinfo
  {author} {\bibfnamefont {N.}~\bibnamefont {Galland}}, \ and\ \bibinfo
  {author} {\bibfnamefont {G.}~\bibnamefont {Montavon}},\ }\href {\doibase
  10.1021/jp3099413} {\bibfield  {journal} {\bibinfo  {journal} {J. Phys. Chem.
  A}\ }\textbf {\bibinfo {volume} {117}},\ \bibinfo {pages} {1983} (\bibinfo
  {year} {2013})}\BibitemShut {NoStop}%
\bibitem [{\citenamefont {Sergentu}\ \emph {et~al.}(2016)\citenamefont
  {Sergentu}, \citenamefont {David}, \citenamefont {Montavon}, \citenamefont
  {Maurice},\ and\ \citenamefont {Galland}}]{Sergentu:2016hn}%
  \BibitemOpen
  \bibfield  {author} {\bibinfo {author} {\bibfnamefont {D.-C.}\ \bibnamefont
  {Sergentu}}, \bibinfo {author} {\bibfnamefont {G.}~\bibnamefont {David}},
  \bibinfo {author} {\bibfnamefont {G.}~\bibnamefont {Montavon}}, \bibinfo
  {author} {\bibfnamefont {R.}~\bibnamefont {Maurice}}, \ and\ \bibinfo
  {author} {\bibfnamefont {N.}~\bibnamefont {Galland}},\ }\href {\doibase
  10.1002/jcc.24326} {\bibfield  {journal} {\bibinfo  {journal} {J. Comput.
  Chem.}\ }\textbf {\bibinfo {volume} {37}},\ \bibinfo {pages} {1345} (\bibinfo
  {year} {2016})}\BibitemShut {NoStop}%
\bibitem [{\citenamefont {Shee}\ \emph
  {et~al.}(2018{\natexlab{a}})\citenamefont {Shee}, \citenamefont {Saue},
  \citenamefont {Visscher},\ and\ \citenamefont {Gomes}}]{paper:dataset}%
  \BibitemOpen
  \bibfield  {author} {\bibinfo {author} {\bibfnamefont {A.}~\bibnamefont
  {Shee}}, \bibinfo {author} {\bibfnamefont {T.}~\bibnamefont {Saue}}, \bibinfo
  {author} {\bibfnamefont {L.}~\bibnamefont {Visscher}}, \ and\ \bibinfo
  {author} {\bibfnamefont {A.~S.~P.}\ \bibnamefont {Gomes}},\ }\href {\doibase
  10.5281/zenodo.1320320} {\enquote {\bibinfo {title} {Equation--of--motion
  coupled--cluster theory based on the 4--component dirac--coulomb(--gaunt)
  hamiltonian. energies for single electron detachment, attachment and
  electronically excited states: Dataset (version 1.0)},}\ }\bibinfo
  {howpublished} {Zenodo \url{http://dx.doi.org/10.5281/zenodo.1320320}}
  (\bibinfo {year} {2018}{\natexlab{a}})\BibitemShut {NoStop}%
\bibitem [{\citenamefont {Visscher}, \citenamefont {Lee},\ and\ \citenamefont
  {Dyall}(1996)}]{CCSD-Visscher-JCP1996-105-8769-8776}%
  \BibitemOpen
  \bibfield  {author} {\bibinfo {author} {\bibfnamefont {L.}~\bibnamefont
  {Visscher}}, \bibinfo {author} {\bibfnamefont {T.~J.}\ \bibnamefont {Lee}}, \
  and\ \bibinfo {author} {\bibfnamefont {K.~G.}\ \bibnamefont {Dyall}},\ }\href
  {\doibase 10.1063/1.472655} {\bibfield  {journal} {\bibinfo  {journal} {J.
  Chem. Phys.}\ }\textbf {\bibinfo {volume} {105}},\ \bibinfo {pages} {8769}
  (\bibinfo {year} {1996})}\BibitemShut {NoStop}%
\bibitem [{\citenamefont {Pernpointner}\ and\ \citenamefont
  {Visscher}(2003)}]{Pernpointner:2003fm}%
  \BibitemOpen
  \bibfield  {author} {\bibinfo {author} {\bibfnamefont {M.}~\bibnamefont
  {Pernpointner}}\ and\ \bibinfo {author} {\bibfnamefont {L.}~\bibnamefont
  {Visscher}},\ }\href {\doibase 10.1002/jcc.10215} {\bibfield  {journal}
  {\bibinfo  {journal} {J. Comput. Chem.}\ }\textbf {\bibinfo {volume} {24}},\
  \bibinfo {pages} {754} (\bibinfo {year} {2003})}\BibitemShut {NoStop}%
\bibitem [{\citenamefont {Gauss}\ \emph {et~al.}(1991)\citenamefont {Gauss},
  \citenamefont {Lauderdale}, \citenamefont {Stanton}, \citenamefont {Watts},\
  and\ \citenamefont {Bartlett}}]{Gauss1991}%
  \BibitemOpen
  \bibfield  {author} {\bibinfo {author} {\bibfnamefont {J.}~\bibnamefont
  {Gauss}}, \bibinfo {author} {\bibfnamefont {W.~J.}\ \bibnamefont
  {Lauderdale}}, \bibinfo {author} {\bibfnamefont {J.~F.}\ \bibnamefont
  {Stanton}}, \bibinfo {author} {\bibfnamefont {J.~D.}\ \bibnamefont {Watts}},
  \ and\ \bibinfo {author} {\bibfnamefont {R.~J.}\ \bibnamefont {Bartlett}},\
  }\href {\doibase 10.1016/0009-2614(91)80203-A} {\bibfield  {journal}
  {\bibinfo  {journal} {Chem. Phys. Lett.}\ }\textbf {\bibinfo {volume}
  {182}},\ \bibinfo {pages} {207} (\bibinfo {year} {1991})}\BibitemShut
  {NoStop}%
\bibitem [{\citenamefont {Gauss}, \citenamefont {Stanton},\ and\ \citenamefont
  {Bartlett}(1991)}]{Gauss1991b}%
  \BibitemOpen
  \bibfield  {author} {\bibinfo {author} {\bibfnamefont {J.}~\bibnamefont
  {Gauss}}, \bibinfo {author} {\bibfnamefont {J.~F.}\ \bibnamefont {Stanton}},
  \ and\ \bibinfo {author} {\bibfnamefont {R.~J.}\ \bibnamefont {Bartlett}},\
  }\href {\doibase 10.1063/1.460915} {\bibfield  {journal} {\bibinfo  {journal}
  {J. Chem. Phys.}\ }\textbf {\bibinfo {volume} {95}},\ \bibinfo {pages} {2623}
  (\bibinfo {year} {1991})}\BibitemShut {NoStop}%
\bibitem [{\citenamefont {Stanton}\ and\ \citenamefont
  {Bartlett}(1993)}]{Stanton:1993be}%
  \BibitemOpen
  \bibfield  {author} {\bibinfo {author} {\bibfnamefont {J.~F.}\ \bibnamefont
  {Stanton}}\ and\ \bibinfo {author} {\bibfnamefont {R.~J.}\ \bibnamefont
  {Bartlett}},\ }\href {\doibase 10.1063/1.464746} {\bibfield  {journal}
  {\bibinfo  {journal} {J. Chem. Phys.}\ }\textbf {\bibinfo {volume} {98}},\
  \bibinfo {pages} {7029} (\bibinfo {year} {1993})}\BibitemShut {NoStop}%
\bibitem [{\citenamefont {Gauss}\ and\ \citenamefont
  {Stanton}(1995)}]{Gauss1995}%
  \BibitemOpen
  \bibfield  {author} {\bibinfo {author} {\bibfnamefont {J.}~\bibnamefont
  {Gauss}}\ and\ \bibinfo {author} {\bibfnamefont {J.~F.}\ \bibnamefont
  {Stanton}},\ }\href {\doibase 10.1063/1.470240} {\bibfield  {journal}
  {\bibinfo  {journal} {J. Chem. Phys.}\ }\textbf {\bibinfo {volume} {103}},\
  \bibinfo {pages} {3561} (\bibinfo {year} {1995})}\BibitemShut {NoStop}%
\bibitem [{\citenamefont {Gwaltney}, \citenamefont {Nooijen},\ and\
  \citenamefont {Bartlett}(1996)}]{Gwaltney:1996jw}%
  \BibitemOpen
  \bibfield  {author} {\bibinfo {author} {\bibfnamefont {S.~R.}\ \bibnamefont
  {Gwaltney}}, \bibinfo {author} {\bibfnamefont {M.}~\bibnamefont {Nooijen}}, \
  and\ \bibinfo {author} {\bibfnamefont {R.~J.}\ \bibnamefont {Bartlett}},\
  }\href {\doibase 10.1016/0009-2614(95)01329-6} {\bibfield  {journal}
  {\bibinfo  {journal} {Chem. Phys. Lett.}\ }\textbf {\bibinfo {volume}
  {248}},\ \bibinfo {pages} {189} (\bibinfo {year} {1996})}\BibitemShut
  {NoStop}%
\bibitem [{\citenamefont {Shee}, \citenamefont {Visscher},\ and\ \citenamefont
  {Saue}(2016)}]{Shee_JCP2016}%
  \BibitemOpen
  \bibfield  {author} {\bibinfo {author} {\bibfnamefont {A.}~\bibnamefont
  {Shee}}, \bibinfo {author} {\bibfnamefont {L.}~\bibnamefont {Visscher}}, \
  and\ \bibinfo {author} {\bibfnamefont {T.}~\bibnamefont {Saue}},\ }\href
  {\doibase 10.1063/1.4966643} {\bibfield  {journal} {\bibinfo  {journal} {J.
  Chem. Phys.}\ }\textbf {\bibinfo {volume} {145}},\ \bibinfo {pages} {184107}
  (\bibinfo {year} {2016})}\BibitemShut {NoStop}%
\bibitem [{\citenamefont {Stanton}\ and\ \citenamefont
  {Gauss}(1999)}]{stanton:1999continuumorbital}%
  \BibitemOpen
  \bibfield  {author} {\bibinfo {author} {\bibfnamefont {J.~F.}\ \bibnamefont
  {Stanton}}\ and\ \bibinfo {author} {\bibfnamefont {J.}~\bibnamefont
  {Gauss}},\ }\href {\doibase 10.1063/1.479673} {\bibfield  {journal} {\bibinfo
   {journal} {J. Chem. Phys.}\ }\textbf {\bibinfo {volume} {111}},\ \bibinfo
  {pages} {8785} (\bibinfo {year} {1999})}\BibitemShut {NoStop}%
\bibitem [{\citenamefont {Nooijen}\ and\ \citenamefont
  {Bartlett}(1995)}]{Nooijen_JCP95}%
  \BibitemOpen
  \bibfield  {author} {\bibinfo {author} {\bibfnamefont {M.}~\bibnamefont
  {Nooijen}}\ and\ \bibinfo {author} {\bibfnamefont {R.~J.}\ \bibnamefont
  {Bartlett}},\ }\href {\doibase 10.1063/1.468592} {\bibfield  {journal}
  {\bibinfo  {journal} {J. Chem. Phys.}\ }\textbf {\bibinfo {volume} {102}},\
  \bibinfo {pages} {3629} (\bibinfo {year} {1995})}\BibitemShut {NoStop}%
\bibitem [{\citenamefont {Davidson}(1975)}]{Davidson:1975db}%
  \BibitemOpen
  \bibfield  {author} {\bibinfo {author} {\bibfnamefont {E.~R.}\ \bibnamefont
  {Davidson}},\ }\href {\doibase 10.1016/0021-9991(75)90065-0} {\bibfield
  {journal} {\bibinfo  {journal} {J. Comput. Phys.}\ }\textbf {\bibinfo
  {volume} {17}},\ \bibinfo {pages} {87} (\bibinfo {year} {1975})}\BibitemShut
  {NoStop}%
\bibitem [{\citenamefont {Hirao}\ and\ \citenamefont
  {Nakatsuji}(1982)}]{Hirao:1982fc}%
  \BibitemOpen
  \bibfield  {author} {\bibinfo {author} {\bibfnamefont {K.}~\bibnamefont
  {Hirao}}\ and\ \bibinfo {author} {\bibfnamefont {H.}~\bibnamefont
  {Nakatsuji}},\ }\href {\doibase 10.1016/0021-9991(82)90119-X} {\bibfield
  {journal} {\bibinfo  {journal} {J. Comput. Phys.}\ }\textbf {\bibinfo
  {volume} {45}},\ \bibinfo {pages} {246} (\bibinfo {year} {1982})}\BibitemShut
  {NoStop}%
\bibitem [{\citenamefont {Morgan}(1992)}]{Morgan:1992iu}%
  \BibitemOpen
  \bibfield  {author} {\bibinfo {author} {\bibfnamefont {R.~B.}\ \bibnamefont
  {Morgan}},\ }\href {\doibase 10.1016/0021-9991(92)90006-K} {\bibfield
  {journal} {\bibinfo  {journal} {J. Comput. Phys.}\ }\textbf {\bibinfo
  {volume} {101}},\ \bibinfo {pages} {287} (\bibinfo {year}
  {1992})}\BibitemShut {NoStop}%
\bibitem [{\citenamefont {{Golub}}\ and\ \citenamefont {{van
  Loan}}(2013)}]{matrix-computations:book}%
  \BibitemOpen
  \bibfield  {author} {\bibinfo {author} {\bibfnamefont {G.~H.}\ \bibnamefont
  {{Golub}}}\ and\ \bibinfo {author} {\bibfnamefont {C.~F.}\ \bibnamefont {{van
  Loan}}},\ }\href@noop {} {\emph {\bibinfo {title} {{Matrix computations}}}}\
  (\bibinfo  {publisher} {Johns Hopkins University Press},\ \bibinfo {address}
  {Baltimore},\ \bibinfo {year} {2013})\BibitemShut {NoStop}%
\bibitem [{\citenamefont {Sherrill}\ and\ \citenamefont
  {Schaefer}(1999)}]{DAVIDSHERRILL1999143}%
  \BibitemOpen
  \bibfield  {author} {\bibinfo {author} {\bibfnamefont {C.~D.}\ \bibnamefont
  {Sherrill}}\ and\ \bibinfo {author} {\bibfnamefont {H.~F.}\ \bibnamefont
  {Schaefer}},\ }\href {\doibase 10.1016/S0065-3276(08)60532-8} {\bibfield
  {journal} {\bibinfo  {journal} {Adv. Quant. Chem.}\ }\textbf {\bibinfo
  {volume} {34}},\ \bibinfo {pages} {143 } (\bibinfo {year}
  {1999})}\BibitemShut {NoStop}%
\bibitem [{\citenamefont {Vallet}\ \emph {et~al.}(2000)\citenamefont {Vallet},
  \citenamefont {Maron}, \citenamefont {Teichteil},\ and\ \citenamefont
  {Flament}}]{vallet:2000epciso}%
  \BibitemOpen
  \bibfield  {author} {\bibinfo {author} {\bibfnamefont {V.}~\bibnamefont
  {Vallet}}, \bibinfo {author} {\bibfnamefont {L.}~\bibnamefont {Maron}},
  \bibinfo {author} {\bibfnamefont {C.}~\bibnamefont {Teichteil}}, \ and\
  \bibinfo {author} {\bibfnamefont {J.-P.}\ \bibnamefont {Flament}},\ }\href
  {\doibase 10.1063/1.481929} {\bibfield  {journal} {\bibinfo  {journal} {J.
  Chem. Phys.}\ }\textbf {\bibinfo {volume} {113}},\ \bibinfo {pages} {1391}
  (\bibinfo {year} {2000})}\BibitemShut {NoStop}%
\bibitem [{\citenamefont {Butscher}\ and\ \citenamefont
  {Kammer}(1976)}]{Butscher:1976ku}%
  \BibitemOpen
  \bibfield  {author} {\bibinfo {author} {\bibfnamefont {W.}~\bibnamefont
  {Butscher}}\ and\ \bibinfo {author} {\bibfnamefont {W.~E.}\ \bibnamefont
  {Kammer}},\ }\href {\doibase 10.1016/0021-9991(76)90084-X} {\bibfield
  {journal} {\bibinfo  {journal} {J. Comput. Phys.}\ }\textbf {\bibinfo
  {volume} {20}},\ \bibinfo {pages} {313} (\bibinfo {year} {1976})}\BibitemShut
  {NoStop}%
\bibitem [{\citenamefont {Zuev}\ \emph {et~al.}(2015)\citenamefont {Zuev},
  \citenamefont {Vecharynski}, \citenamefont {Yang}, \citenamefont {Orms},\
  and\ \citenamefont {Krylov}}]{Zuev:2014fc}%
  \BibitemOpen
  \bibfield  {author} {\bibinfo {author} {\bibfnamefont {D.}~\bibnamefont
  {Zuev}}, \bibinfo {author} {\bibfnamefont {E.}~\bibnamefont {Vecharynski}},
  \bibinfo {author} {\bibfnamefont {C.}~\bibnamefont {Yang}}, \bibinfo {author}
  {\bibfnamefont {N.}~\bibnamefont {Orms}}, \ and\ \bibinfo {author}
  {\bibfnamefont {A.~I.}\ \bibnamefont {Krylov}},\ }\href {\doibase
  10.1002/jcc.23800} {\bibfield  {journal} {\bibinfo  {journal} {J. Comput.
  Chem.}\ }\textbf {\bibinfo {volume} {36}},\ \bibinfo {pages} {273} (\bibinfo
  {year} {2015})}\BibitemShut {NoStop}%
\bibitem [{\citenamefont {Dyall}(2002)}]{basis-Dyall-TCA2002-108-335}%
  \BibitemOpen
  \bibfield  {author} {\bibinfo {author} {\bibfnamefont {K.~G.}\ \bibnamefont
  {Dyall}},\ }\href {\doibase 10.1007/s00214-002-0388-0} {\bibfield  {journal}
  {\bibinfo  {journal} {Theor. Chem. Acc.}\ }\textbf {\bibinfo {volume}
  {108}},\ \bibinfo {pages} {335} (\bibinfo {year} {2002})}\BibitemShut
  {NoStop}%
\bibitem [{\citenamefont {Dyall}(2003)}]{basis-Dyall-TCA2003-109-284}%
  \BibitemOpen
  \bibfield  {author} {\bibinfo {author} {\bibfnamefont {K.~G.}\ \bibnamefont
  {Dyall}},\ }\href {\doibase 10.1007/s00214-003-0433-7} {\bibfield  {journal}
  {\bibinfo  {journal} {Theor. Chem. Acc.}\ }\textbf {\bibinfo {volume}
  {109}},\ \bibinfo {pages} {284} (\bibinfo {year} {2003})}\BibitemShut
  {NoStop}%
\bibitem [{\citenamefont {Dyall}(2012)}]{basis-Dyall-TCA2012-131-3962}%
  \BibitemOpen
  \bibfield  {author} {\bibinfo {author} {\bibfnamefont {K.~G.}\ \bibnamefont
  {Dyall}},\ }\href {\doibase 10.1007/s00214-012-1172-4} {\bibfield  {journal}
  {\bibinfo  {journal} {Theor. Chem. Acc.}\ }\textbf {\bibinfo {volume}
  {131}},\ \bibinfo {pages} {1172} (\bibinfo {year} {2012})}\BibitemShut
  {NoStop}%
\bibitem [{\citenamefont {Dyall}(2016)}]{dyallbasis}%
  \BibitemOpen
  \bibfield  {author} {\bibinfo {author} {\bibfnamefont {K.~G.}\ \bibnamefont
  {Dyall}},\ }\href {\doibase 10.1007/s00214-016-1884-y} {\bibfield  {journal}
  {\bibinfo  {journal} {Theor. Chem. Acc.}\ }\textbf {\bibinfo {volume}
  {135}},\ \bibinfo {pages} {128} (\bibinfo {year} {2016})}\BibitemShut
  {NoStop}%
\bibitem [{\citenamefont {Kendall}, \citenamefont {Dunning},\ and\
  \citenamefont {Harrison}(1992)}]{basis-Kendall-JCP1992-96-6796-6806}%
  \BibitemOpen
  \bibfield  {author} {\bibinfo {author} {\bibfnamefont {R.~A.}\ \bibnamefont
  {Kendall}}, \bibinfo {author} {\bibfnamefont {T.~H.}\ \bibnamefont {Dunning},
  \bibfnamefont {Jr.}}, \ and\ \bibinfo {author} {\bibfnamefont {R.~J.}\
  \bibnamefont {Harrison}},\ }\href {\doibase 10.1063/1.462569} {\bibfield
  {journal} {\bibinfo  {journal} {J. Chem. Phys.}\ }\textbf {\bibinfo {volume}
  {96}},\ \bibinfo {pages} {6796} (\bibinfo {year} {1992})}\BibitemShut
  {NoStop}%
\bibitem [{\citenamefont {Visscher}(1997)}]{relat-Visscher-TCA1997-98-68}%
  \BibitemOpen
  \bibfield  {author} {\bibinfo {author} {\bibfnamefont {L.}~\bibnamefont
  {Visscher}},\ }\href {\doibase 10.1007/s002140050280} {\bibfield  {journal}
  {\bibinfo  {journal} {Theor. Chem. Acc.}\ }\textbf {\bibinfo {volume} {98}},\
  \bibinfo {pages} {68} (\bibinfo {year} {1997})}\BibitemShut {NoStop}%
\bibitem [{\citenamefont {Fonseca~Guerra}\ \emph {et~al.}(1998)\citenamefont
  {Fonseca~Guerra}, \citenamefont {Snijders}, \citenamefont {te~Velde},\ and\
  \citenamefont {Baerends}}]{FonsecaGuerra1998}%
  \BibitemOpen
  \bibfield  {author} {\bibinfo {author} {\bibfnamefont {C.}~\bibnamefont
  {Fonseca~Guerra}}, \bibinfo {author} {\bibfnamefont {J.~G.}\ \bibnamefont
  {Snijders}}, \bibinfo {author} {\bibfnamefont {G.}~\bibnamefont {te~Velde}},
  \ and\ \bibinfo {author} {\bibfnamefont {E.~J.}\ \bibnamefont {Baerends}},\
  }\href {\doibase 10.1007/s002140050353} {\bibfield  {journal} {\bibinfo
  {journal} {Theor. Chem. Acc.}\ }\textbf {\bibinfo {volume} {99}},\ \bibinfo
  {pages} {391} (\bibinfo {year} {1998})}\BibitemShut {NoStop}%
\bibitem [{\citenamefont {te~Velde}\ \emph {et~al.}(2001)\citenamefont
  {te~Velde}, \citenamefont {Bickelhaupt}, \citenamefont {Baerends},
  \citenamefont {Fonseca~Guerra}, \citenamefont {van Gisbergen}, \citenamefont
  {Snijders},\ and\ \citenamefont {Ziegler}}]{ADF2001}%
  \BibitemOpen
  \bibfield  {author} {\bibinfo {author} {\bibfnamefont {G.}~\bibnamefont
  {te~Velde}}, \bibinfo {author} {\bibfnamefont {F.~M.}\ \bibnamefont
  {Bickelhaupt}}, \bibinfo {author} {\bibfnamefont {E.~J.}\ \bibnamefont
  {Baerends}}, \bibinfo {author} {\bibfnamefont {C.}~\bibnamefont
  {Fonseca~Guerra}}, \bibinfo {author} {\bibfnamefont {S.~J.~A.}\ \bibnamefont
  {van Gisbergen}}, \bibinfo {author} {\bibfnamefont {J.~G.}\ \bibnamefont
  {Snijders}}, \ and\ \bibinfo {author} {\bibfnamefont {T.}~\bibnamefont
  {Ziegler}},\ }\href {\doibase 10.1002/jcc.1056} {\bibfield  {journal}
  {\bibinfo  {journal} {J. Comput. Chem.}\ }\textbf {\bibinfo {volume} {22}},\
  \bibinfo {pages} {931} (\bibinfo {year} {2001})}\BibitemShut {NoStop}%
\bibitem [{\citenamefont {Baerends}\ \emph {et~al.}()\citenamefont {Baerends},
  \citenamefont {Ziegler}, \citenamefont {Atkins}, \citenamefont {Autschbach},
  \citenamefont {Bashford}, \citenamefont {Baseggio}, \citenamefont
  {B{\'{e}}rces}, \citenamefont {Bickelhaupt}, \citenamefont {Bo},
  \citenamefont {Boerritger}, \citenamefont {Cavallo}, \citenamefont {Daul},
  \citenamefont {Chong}, \citenamefont {Chulhai}, \citenamefont {Deng},
  \citenamefont {Dickson}, \citenamefont {Dieterich}, \citenamefont {Ellis},
  \citenamefont {van Faassen}, \citenamefont {Ghysels}, \citenamefont
  {Giammona}, \citenamefont {van Gisbergen}, \citenamefont {Goez},
  \citenamefont {G{\"{o}}tz}, \citenamefont {Gusarov}, \citenamefont {Harris},
  \citenamefont {van~den Hoek}, \citenamefont {Hu}, \citenamefont {Jacob},
  \citenamefont {Jacobsen}, \citenamefont {Jensen}, \citenamefont {Joubert},
  \citenamefont {Kaminski}, \citenamefont {van Kessel}, \citenamefont
  {K{\"{o}}nig}, \citenamefont {Kootstra}, \citenamefont {Kovalenko},
  \citenamefont {Krykunov}, \citenamefont {van Lenthe}, \citenamefont
  {McCormack}, \citenamefont {Michalak}, \citenamefont {Mitoraj}, \citenamefont
  {Morton}, \citenamefont {Neugebauer}, \citenamefont {Nicu}, \citenamefont
  {Noodleman}, \citenamefont {Osinga}, \citenamefont {Patchkovskii},
  \citenamefont {Pavanello}, \citenamefont {Peeples}, \citenamefont
  {Philipsen}, \citenamefont {Post}, \citenamefont {Pye}, \citenamefont
  {Ramanantoanina}, \citenamefont {Ramos}, \citenamefont {Ravenek},
  \citenamefont {Rodr{\'{i}}guez}, \citenamefont {Ros}, \citenamefont
  {R{\"{u}}ger}, \citenamefont {Schipper}, \citenamefont {Schl{\"{u}}ns},
  \citenamefont {van Schoot}, \citenamefont {Schreckenbach}, \citenamefont
  {Seldenthuis}, \citenamefont {Seth}, \citenamefont {Snijders}, \citenamefont
  {Sol{\`{a}}}, \citenamefont {M.}, \citenamefont {Swart}, \citenamefont
  {Swerhone}, \citenamefont {te~Velde}, \citenamefont {Tognetti}, \citenamefont
  {Vernooijs}, \citenamefont {Versluis}, \citenamefont {Visscher},
  \citenamefont {Visser}, \citenamefont {Wang}, \citenamefont {Wesolowski},
  \citenamefont {van Wezenbeek}, \citenamefont {Wiesenekker}, \citenamefont
  {Wolff}, \citenamefont {Woo},\ and\ \citenamefont
  {Yakovlev}}]{ADF2017authors}%
  \BibitemOpen
  \bibfield  {author} {\bibinfo {author} {\bibfnamefont {E.~J.}\ \bibnamefont
  {Baerends}}, \bibinfo {author} {\bibfnamefont {T.}~\bibnamefont {Ziegler}},
  \bibinfo {author} {\bibfnamefont {A.~J.}\ \bibnamefont {Atkins}}, \bibinfo
  {author} {\bibfnamefont {J.}~\bibnamefont {Autschbach}}, \bibinfo {author}
  {\bibfnamefont {D.}~\bibnamefont {Bashford}}, \bibinfo {author}
  {\bibfnamefont {O.}~\bibnamefont {Baseggio}}, \bibinfo {author}
  {\bibfnamefont {A.}~\bibnamefont {B{\'{e}}rces}}, \bibinfo {author}
  {\bibfnamefont {F.~M.}\ \bibnamefont {Bickelhaupt}}, \bibinfo {author}
  {\bibfnamefont {C.}~\bibnamefont {Bo}}, \bibinfo {author} {\bibfnamefont
  {P.~M.}\ \bibnamefont {Boerritger}}, \bibinfo {author} {\bibfnamefont
  {L.}~\bibnamefont {Cavallo}}, \bibinfo {author} {\bibfnamefont
  {C.}~\bibnamefont {Daul}}, \bibinfo {author} {\bibfnamefont {D.~P.}\
  \bibnamefont {Chong}}, \bibinfo {author} {\bibfnamefont {D.~V.}\ \bibnamefont
  {Chulhai}}, \bibinfo {author} {\bibfnamefont {L.}~\bibnamefont {Deng}},
  \bibinfo {author} {\bibfnamefont {R.~M.}\ \bibnamefont {Dickson}}, \bibinfo
  {author} {\bibfnamefont {J.~M.}\ \bibnamefont {Dieterich}}, \bibinfo {author}
  {\bibfnamefont {D.~E.}\ \bibnamefont {Ellis}}, \bibinfo {author}
  {\bibfnamefont {M.}~\bibnamefont {van Faassen}}, \bibinfo {author}
  {\bibfnamefont {A.}~\bibnamefont {Ghysels}}, \bibinfo {author} {\bibfnamefont
  {A.}~\bibnamefont {Giammona}}, \bibinfo {author} {\bibfnamefont {S.~J.~A.}\
  \bibnamefont {van Gisbergen}}, \bibinfo {author} {\bibfnamefont
  {A.}~\bibnamefont {Goez}}, \bibinfo {author} {\bibfnamefont {A.~W.}\
  \bibnamefont {G{\"{o}}tz}}, \bibinfo {author} {\bibfnamefont
  {S.}~\bibnamefont {Gusarov}}, \bibinfo {author} {\bibfnamefont {F.~E.}\
  \bibnamefont {Harris}}, \bibinfo {author} {\bibfnamefont {P.}~\bibnamefont
  {van~den Hoek}}, \bibinfo {author} {\bibfnamefont {Z.}~\bibnamefont {Hu}},
  \bibinfo {author} {\bibfnamefont {C.~R.}\ \bibnamefont {Jacob}}, \bibinfo
  {author} {\bibfnamefont {H.}~\bibnamefont {Jacobsen}}, \bibinfo {author}
  {\bibfnamefont {L.}~\bibnamefont {Jensen}}, \bibinfo {author} {\bibfnamefont
  {L.}~\bibnamefont {Joubert}}, \bibinfo {author} {\bibfnamefont {J.~W.}\
  \bibnamefont {Kaminski}}, \bibinfo {author} {\bibfnamefont {G.}~\bibnamefont
  {van Kessel}}, \bibinfo {author} {\bibfnamefont {C.}~\bibnamefont
  {K{\"{o}}nig}}, \bibinfo {author} {\bibfnamefont {F.}~\bibnamefont
  {Kootstra}}, \bibinfo {author} {\bibfnamefont {A.}~\bibnamefont {Kovalenko}},
  \bibinfo {author} {\bibfnamefont {M.}~\bibnamefont {Krykunov}}, \bibinfo
  {author} {\bibfnamefont {E.}~\bibnamefont {van Lenthe}}, \bibinfo {author}
  {\bibfnamefont {D.~A.}\ \bibnamefont {McCormack}}, \bibinfo {author}
  {\bibfnamefont {A.}~\bibnamefont {Michalak}}, \bibinfo {author}
  {\bibfnamefont {M.}~\bibnamefont {Mitoraj}}, \bibinfo {author} {\bibfnamefont
  {S.~M.}\ \bibnamefont {Morton}}, \bibinfo {author} {\bibfnamefont
  {J.}~\bibnamefont {Neugebauer}}, \bibinfo {author} {\bibfnamefont {V.~P.}\
  \bibnamefont {Nicu}}, \bibinfo {author} {\bibfnamefont {L.}~\bibnamefont
  {Noodleman}}, \bibinfo {author} {\bibfnamefont {V.~P.}\ \bibnamefont
  {Osinga}}, \bibinfo {author} {\bibfnamefont {S.}~\bibnamefont
  {Patchkovskii}}, \bibinfo {author} {\bibfnamefont {M.}~\bibnamefont
  {Pavanello}}, \bibinfo {author} {\bibfnamefont {C.~A.}\ \bibnamefont
  {Peeples}}, \bibinfo {author} {\bibfnamefont {P.~H.~T.}\ \bibnamefont
  {Philipsen}}, \bibinfo {author} {\bibfnamefont {D.}~\bibnamefont {Post}},
  \bibinfo {author} {\bibfnamefont {C.~C.}\ \bibnamefont {Pye}}, \bibinfo
  {author} {\bibfnamefont {H.}~\bibnamefont {Ramanantoanina}}, \bibinfo
  {author} {\bibfnamefont {P.}~\bibnamefont {Ramos}}, \bibinfo {author}
  {\bibfnamefont {W.}~\bibnamefont {Ravenek}}, \bibinfo {author} {\bibfnamefont
  {J.~I.}\ \bibnamefont {Rodr{\'{i}}guez}}, \bibinfo {author} {\bibfnamefont
  {P.}~\bibnamefont {Ros}}, \bibinfo {author} {\bibfnamefont {R.}~\bibnamefont
  {R{\"{u}}ger}}, \bibinfo {author} {\bibfnamefont {P.~R.~T.}\ \bibnamefont
  {Schipper}}, \bibinfo {author} {\bibfnamefont {D.}~\bibnamefont
  {Schl{\"{u}}ns}}, \bibinfo {author} {\bibfnamefont {H.}~\bibnamefont {van
  Schoot}}, \bibinfo {author} {\bibfnamefont {G.}~\bibnamefont
  {Schreckenbach}}, \bibinfo {author} {\bibfnamefont {J.~S.}\ \bibnamefont
  {Seldenthuis}}, \bibinfo {author} {\bibfnamefont {M.}~\bibnamefont {Seth}},
  \bibinfo {author} {\bibfnamefont {J.~G.}\ \bibnamefont {Snijders}}, \bibinfo
  {author} {\bibfnamefont {M.}~\bibnamefont {Sol{\`{a}}}}, \bibinfo {author}
  {\bibfnamefont {S.}~\bibnamefont {M.}}, \bibinfo {author} {\bibfnamefont
  {M.}~\bibnamefont {Swart}}, \bibinfo {author} {\bibfnamefont
  {D.}~\bibnamefont {Swerhone}}, \bibinfo {author} {\bibfnamefont
  {G.}~\bibnamefont {te~Velde}}, \bibinfo {author} {\bibfnamefont
  {V.}~\bibnamefont {Tognetti}}, \bibinfo {author} {\bibfnamefont
  {P.}~\bibnamefont {Vernooijs}}, \bibinfo {author} {\bibfnamefont
  {L.}~\bibnamefont {Versluis}}, \bibinfo {author} {\bibfnamefont
  {L.}~\bibnamefont {Visscher}}, \bibinfo {author} {\bibfnamefont
  {O.}~\bibnamefont {Visser}}, \bibinfo {author} {\bibfnamefont
  {F.}~\bibnamefont {Wang}}, \bibinfo {author} {\bibfnamefont {T.~A.}\
  \bibnamefont {Wesolowski}}, \bibinfo {author} {\bibfnamefont {E.~M.}\
  \bibnamefont {van Wezenbeek}}, \bibinfo {author} {\bibfnamefont
  {G.}~\bibnamefont {Wiesenekker}}, \bibinfo {author} {\bibfnamefont {S.~K.}\
  \bibnamefont {Wolff}}, \bibinfo {author} {\bibfnamefont {T.~K.}\ \bibnamefont
  {Woo}}, \ and\ \bibinfo {author} {\bibfnamefont {A.~L.}\ \bibnamefont
  {Yakovlev}},\ }\href@noop {} {\enquote {\bibinfo {title} {{ADF2017, SCM,
  Theoretical Chemistry, Vrije Universiteit, Amsterdam, The Netherlands,
  https://www.scm.com}},}\ }\BibitemShut {NoStop}%
\bibitem [{\citenamefont {Visscher}(1996)}]{luuk:sym}%
  \BibitemOpen
  \bibfield  {author} {\bibinfo {author} {\bibfnamefont {L.}~\bibnamefont
  {Visscher}},\ }\href {\doibase 10.1016/0009-2614(96)00234-5} {\bibfield
  {journal} {\bibinfo  {journal} {Chem. Phys. Lett.}\ }\textbf {\bibinfo
  {volume} {253}},\ \bibinfo {pages} {20} (\bibinfo {year} {1996})}\BibitemShut
  {NoStop}%
\bibitem [{\citenamefont {F{\ae}gri}\ and\ \citenamefont
  {Saue}(2001)}]{Faegri_JCP2001}%
  \BibitemOpen
  \bibfield  {author} {\bibinfo {author} {\bibfnamefont {K.}~\bibnamefont
  {F{\ae}gri}}\ and\ \bibinfo {author} {\bibfnamefont {T.}~\bibnamefont
  {Saue}},\ }\href {\doibase 10.1063/1.1385366} {\bibfield  {journal} {\bibinfo
   {journal} {J. Chem. Phys.}\ }\textbf {\bibinfo {volume} {115}},\ \bibinfo
  {pages} {2456} (\bibinfo {year} {2001})}\BibitemShut {NoStop}%
\bibitem [{\citenamefont {Gilles}, \citenamefont {Polak},\ and\ \citenamefont
  {Lineberger}(1992)}]{IO-Gilles-JCP1992-96-8012}%
  \BibitemOpen
  \bibfield  {author} {\bibinfo {author} {\bibfnamefont {M.~K.}\ \bibnamefont
  {Gilles}}, \bibinfo {author} {\bibfnamefont {M.~L.}\ \bibnamefont {Polak}}, \
  and\ \bibinfo {author} {\bibfnamefont {W.~C.}\ \bibnamefont {Lineberger}},\
  }\href {\doibase 10.1063/1.462352} {\bibfield  {journal} {\bibinfo  {journal}
  {J. Chem. Phys.}\ }\textbf {\bibinfo {volume} {96}},\ \bibinfo {pages} {8012}
  (\bibinfo {year} {1992})}\BibitemShut {NoStop}%
\bibitem [{\citenamefont {Peterson}\ \emph {et~al.}(2006)\citenamefont
  {Peterson}, \citenamefont {Shepler}, \citenamefont {Figgen},\ and\
  \citenamefont {Stoll}}]{IO-Peterson-JPC2006-110-13877-13883}%
  \BibitemOpen
  \bibfield  {author} {\bibinfo {author} {\bibfnamefont {K.~A.}\ \bibnamefont
  {Peterson}}, \bibinfo {author} {\bibfnamefont {B.~C.}\ \bibnamefont
  {Shepler}}, \bibinfo {author} {\bibfnamefont {D.}~\bibnamefont {Figgen}}, \
  and\ \bibinfo {author} {\bibfnamefont {H.}~\bibnamefont {Stoll}},\ }\href
  {\doibase 10.1021/jp065887l} {\bibfield  {journal} {\bibinfo  {journal} {J.
  Phys. Chem. A}\ }\textbf {\bibinfo {volume} {110}},\ \bibinfo {pages} {13877}
  (\bibinfo {year} {2006})}\BibitemShut {NoStop}%
\bibitem [{\citenamefont {Zhou}\ \emph {et~al.}(2017)\citenamefont {Zhou},
  \citenamefont {Shi}, \citenamefont {Sun},\ and\ \citenamefont
  {Zhu}}]{Zhou:2017ky}%
  \BibitemOpen
  \bibfield  {author} {\bibinfo {author} {\bibfnamefont {D.}~\bibnamefont
  {Zhou}}, \bibinfo {author} {\bibfnamefont {D.}~\bibnamefont {Shi}}, \bibinfo
  {author} {\bibfnamefont {J.}~\bibnamefont {Sun}}, \ and\ \bibinfo {author}
  {\bibfnamefont {Z.}~\bibnamefont {Zhu}},\ }\href {\doibase
  10.1016/j.comptc.2017.03.037} {\bibfield  {journal} {\bibinfo  {journal}
  {Comput. Theor. Chem.}\ }\textbf {\bibinfo {volume} {1112}},\ \bibinfo
  {pages} {94} (\bibinfo {year} {2017})}\BibitemShut {NoStop}%
\bibitem [{\citenamefont {Cheng}\ \emph {et~al.}(2018)\citenamefont {Cheng},
  \citenamefont {Wang}, \citenamefont {Stanton},\ and\ \citenamefont
  {Gauss}}]{Cheng:2018ji}%
  \BibitemOpen
  \bibfield  {author} {\bibinfo {author} {\bibfnamefont {L.}~\bibnamefont
  {Cheng}}, \bibinfo {author} {\bibfnamefont {F.}~\bibnamefont {Wang}},
  \bibinfo {author} {\bibfnamefont {J.~F.}\ \bibnamefont {Stanton}}, \ and\
  \bibinfo {author} {\bibfnamefont {J.}~\bibnamefont {Gauss}},\ }\href
  {\doibase 10.1063/1.5012041} {\bibfield  {journal} {\bibinfo  {journal} {J.
  Chem. Phys.}\ }\textbf {\bibinfo {volume} {148}},\ \bibinfo {pages} {044108}
  (\bibinfo {year} {2018})}\BibitemShut {NoStop}%
\bibitem [{\citenamefont {Dyall}(1994)}]{Dyall1994}%
  \BibitemOpen
  \bibfield  {author} {\bibinfo {author} {\bibfnamefont {K.~G.}\ \bibnamefont
  {Dyall}},\ }\href {\doibase 10.1063/1.466508} {\bibfield  {journal} {\bibinfo
   {journal} {J. Chem. Phys.}\ }\textbf {\bibinfo {volume} {100}},\ \bibinfo
  {pages} {2118} (\bibinfo {year} {1994})}\BibitemShut {NoStop}%
\bibitem [{\citenamefont {Shee}\ \emph
  {et~al.}(2018{\natexlab{b}})\citenamefont {Shee}, \citenamefont {Saue},
  \citenamefont {Visscher},\ and\ \citenamefont {Gomes}}]{paper:figures}%
  \BibitemOpen
  \bibfield  {author} {\bibinfo {author} {\bibfnamefont {A.}~\bibnamefont
  {Shee}}, \bibinfo {author} {\bibfnamefont {T.}~\bibnamefont {Saue}}, \bibinfo
  {author} {\bibfnamefont {L.}~\bibnamefont {Visscher}}, \ and\ \bibinfo
  {author} {\bibfnamefont {A.~S.~P.}\ \bibnamefont {Gomes}},\ }\href {\doibase
  10.5281/zenodo.1320786} {\enquote {\bibinfo {title} {{Equation--of--Motion
  Coupled--Cluster Theory based on the 4--component Dirac--Coulomb(--Gaunt)
  Hamiltonian. Energies for single electron detachment, attachment and
  electronically excited states: Figures (Version 1.0)}},}\ }\bibinfo
  {howpublished} {Zenodo \url{http://dx.doi.org/10.5281/zenodo.1320786}}
  (\bibinfo {year} {2018}{\natexlab{b}})\BibitemShut {NoStop}%
\bibitem [{\citenamefont {Wang}, \citenamefont {Tu},\ and\ \citenamefont
  {Wang}(2014)}]{Wang:2014es}%
  \BibitemOpen
  \bibfield  {author} {\bibinfo {author} {\bibfnamefont {Z.}~\bibnamefont
  {Wang}}, \bibinfo {author} {\bibfnamefont {Z.}~\bibnamefont {Tu}}, \ and\
  \bibinfo {author} {\bibfnamefont {F.}~\bibnamefont {Wang}},\ }\href {\doibase
  10.1021/ct500854m} {\bibfield  {journal} {\bibinfo  {journal} {J. Chem.
  Theory Comput.}\ }\textbf {\bibinfo {volume} {10}},\ \bibinfo {pages} {5567}
  (\bibinfo {year} {2014})}\BibitemShut {NoStop}%
\bibitem [{\citenamefont {Choi}\ \emph
  {et~al.}(2000{\natexlab{a}})\citenamefont {Choi}, \citenamefont {Bise},
  \citenamefont {Hoops},\ and\ \citenamefont {Neumark}}]{Choi:2000jv}%
  \BibitemOpen
  \bibfield  {author} {\bibinfo {author} {\bibfnamefont {H.}~\bibnamefont
  {Choi}}, \bibinfo {author} {\bibfnamefont {R.~T.}\ \bibnamefont {Bise}},
  \bibinfo {author} {\bibfnamefont {A.~A.}\ \bibnamefont {Hoops}}, \ and\
  \bibinfo {author} {\bibfnamefont {D.~M.}\ \bibnamefont {Neumark}},\ }\href
  {\doibase 10.1063/1.482040} {\bibfield  {journal} {\bibinfo  {journal} {J.
  Chem. Phys.}\ }\textbf {\bibinfo {volume} {113}},\ \bibinfo {pages} {2255}
  (\bibinfo {year} {2000}{\natexlab{a}})}\BibitemShut {NoStop}%
\bibitem [{\citenamefont {Zhu}\ \emph {et~al.}(2001)\citenamefont {Zhu},
  \citenamefont {Takahashi}, \citenamefont {Saeki}, \citenamefont {Tsukuda},\
  and\ \citenamefont {Nagata}}]{Zhu:2001fy}%
  \BibitemOpen
  \bibfield  {author} {\bibinfo {author} {\bibfnamefont {L.}~\bibnamefont
  {Zhu}}, \bibinfo {author} {\bibfnamefont {K.}~\bibnamefont {Takahashi}},
  \bibinfo {author} {\bibfnamefont {M.}~\bibnamefont {Saeki}}, \bibinfo
  {author} {\bibfnamefont {T.}~\bibnamefont {Tsukuda}}, \ and\ \bibinfo
  {author} {\bibfnamefont {T.}~\bibnamefont {Nagata}},\ }\href {\doibase
  10.1016/S0009-2614(01)01288-X} {\bibfield  {journal} {\bibinfo  {journal}
  {Chem. Phys. Lett.}\ }\textbf {\bibinfo {volume} {350}},\ \bibinfo {pages}
  {233} (\bibinfo {year} {2001})}\BibitemShut {NoStop}%
\bibitem [{\citenamefont {Choi}\ \emph
  {et~al.}(2000{\natexlab{b}})\citenamefont {Choi}, \citenamefont {Taylor},
  \citenamefont {Bise}, \citenamefont {Hoops},\ and\ \citenamefont
  {Neumark}}]{Choi:2000eo}%
  \BibitemOpen
  \bibfield  {author} {\bibinfo {author} {\bibfnamefont {H.}~\bibnamefont
  {Choi}}, \bibinfo {author} {\bibfnamefont {T.~R.}\ \bibnamefont {Taylor}},
  \bibinfo {author} {\bibfnamefont {R.~T.}\ \bibnamefont {Bise}}, \bibinfo
  {author} {\bibfnamefont {A.~A.}\ \bibnamefont {Hoops}}, \ and\ \bibinfo
  {author} {\bibfnamefont {D.~M.}\ \bibnamefont {Neumark}},\ }\href {\doibase
  10.1063/1.1318755} {\bibfield  {journal} {\bibinfo  {journal} {J. Chem.
  Phys.}\ }\textbf {\bibinfo {volume} {113}},\ \bibinfo {pages} {8608}
  (\bibinfo {year} {2000}{\natexlab{b}})}\BibitemShut {NoStop}%
\bibitem [{\citenamefont {Potts}\ \emph {et~al.}(1970)\citenamefont {Potts},
  \citenamefont {Lempka}, \citenamefont {Streets},\ and\ \citenamefont
  {Price}}]{Potts:79nRiJUm}%
  \BibitemOpen
  \bibfield  {author} {\bibinfo {author} {\bibfnamefont {A.~W.}\ \bibnamefont
  {Potts}}, \bibinfo {author} {\bibfnamefont {H.~J.}\ \bibnamefont {Lempka}},
  \bibinfo {author} {\bibfnamefont {D.~G.}\ \bibnamefont {Streets}}, \ and\
  \bibinfo {author} {\bibfnamefont {W.~C.}\ \bibnamefont {Price}},\ }\href
  {\doibase 10.1098/rsta.1970.0061} {\bibfield  {journal} {\bibinfo  {journal}
  {Phil. Trans. R. Soc. A}\ }\textbf {\bibinfo {volume} {268}},\ \bibinfo
  {pages} {59} (\bibinfo {year} {1970})}\BibitemShut {NoStop}%
\bibitem [{\citenamefont {Cartoni}\ \emph {et~al.}(2015)\citenamefont
  {Cartoni}, \citenamefont {Casavola}, \citenamefont {Bolognesi}, \citenamefont
  {Borocci},\ and\ \citenamefont {Avaldi}}]{Cartoni:2015ex}%
  \BibitemOpen
  \bibfield  {author} {\bibinfo {author} {\bibfnamefont {A.}~\bibnamefont
  {Cartoni}}, \bibinfo {author} {\bibfnamefont {A.~R.}\ \bibnamefont
  {Casavola}}, \bibinfo {author} {\bibfnamefont {P.}~\bibnamefont {Bolognesi}},
  \bibinfo {author} {\bibfnamefont {S.}~\bibnamefont {Borocci}}, \ and\
  \bibinfo {author} {\bibfnamefont {L.}~\bibnamefont {Avaldi}},\ }\href
  {\doibase 10.1021/jp511067d} {\bibfield  {journal} {\bibinfo  {journal} {J.
  Phys. Chem. A}\ }\textbf {\bibinfo {volume} {119}},\ \bibinfo {pages} {3704}
  (\bibinfo {year} {2015})}\BibitemShut {NoStop}%
\bibitem [{\citenamefont {Satta}\ \emph {et~al.}(2016)\citenamefont {Satta},
  \citenamefont {Bolognesi}, \citenamefont {Cartoni}, \citenamefont {Casavola},
  \citenamefont {Catone}, \citenamefont {Markus},\ and\ \citenamefont
  {Avaldi}}]{Satta:2015gp}%
  \BibitemOpen
  \bibfield  {author} {\bibinfo {author} {\bibfnamefont {M.}~\bibnamefont
  {Satta}}, \bibinfo {author} {\bibfnamefont {P.}~\bibnamefont {Bolognesi}},
  \bibinfo {author} {\bibfnamefont {A.}~\bibnamefont {Cartoni}}, \bibinfo
  {author} {\bibfnamefont {A.~R.}\ \bibnamefont {Casavola}}, \bibinfo {author}
  {\bibfnamefont {D.}~\bibnamefont {Catone}}, \bibinfo {author} {\bibfnamefont
  {P.}~\bibnamefont {Markus}}, \ and\ \bibinfo {author} {\bibfnamefont
  {L.}~\bibnamefont {Avaldi}},\ }\href {\doibase 10.1063/1.4937425} {\bibfield
  {journal} {\bibinfo  {journal} {J. Chem. Phys.}\ }\textbf {\bibinfo {volume}
  {143}},\ \bibinfo {pages} {244312} (\bibinfo {year} {2016})}\BibitemShut
  {NoStop}%
\bibitem [{\citenamefont {Tsal}\ \emph {et~al.}(1975)\citenamefont {Tsal},
  \citenamefont {Baer}, \citenamefont {Werner},\ and\ \citenamefont
  {Lin}}]{Tsal:1975er}%
  \BibitemOpen
  \bibfield  {author} {\bibinfo {author} {\bibfnamefont {B.~P.}\ \bibnamefont
  {Tsal}}, \bibinfo {author} {\bibfnamefont {T.}~\bibnamefont {Baer}}, \bibinfo
  {author} {\bibfnamefont {A.~S.}\ \bibnamefont {Werner}}, \ and\ \bibinfo
  {author} {\bibfnamefont {S.~F.}\ \bibnamefont {Lin}},\ }\href {\doibase
  10.1063/1.440184} {\bibfield  {journal} {\bibinfo  {journal} {J. Phys.
  Chem.}\ }\textbf {\bibinfo {volume} {79}},\ \bibinfo {pages} {570} (\bibinfo
  {year} {1975})}\BibitemShut {NoStop}%
\bibitem [{\citenamefont {Lago}\ \emph {et~al.}(2005)\citenamefont {Lago},
  \citenamefont {Kercher}, \citenamefont {B{\"o}di}, \citenamefont
  {Szt{\'a}ray}, \citenamefont {Miller}, \citenamefont {Wurzelmann},\ and\
  \citenamefont {Baer}}]{Lago:2005kr}%
  \BibitemOpen
  \bibfield  {author} {\bibinfo {author} {\bibfnamefont {A.~F.}\ \bibnamefont
  {Lago}}, \bibinfo {author} {\bibfnamefont {J.~P.}\ \bibnamefont {Kercher}},
  \bibinfo {author} {\bibfnamefont {A.}~\bibnamefont {B{\"o}di}}, \bibinfo
  {author} {\bibfnamefont {B.}~\bibnamefont {Szt{\'a}ray}}, \bibinfo {author}
  {\bibfnamefont {B.}~\bibnamefont {Miller}}, \bibinfo {author} {\bibfnamefont
  {D.}~\bibnamefont {Wurzelmann}}, \ and\ \bibinfo {author} {\bibfnamefont
  {T.}~\bibnamefont {Baer}},\ }\href {\doibase 10.1021/jp045337s} {\bibfield
  {journal} {\bibinfo  {journal} {J. Phys. Chem. A}\ }\textbf {\bibinfo
  {volume} {109}},\ \bibinfo {pages} {1802} (\bibinfo {year}
  {2005})}\BibitemShut {NoStop}%
\bibitem [{\citenamefont {Zhao}\ \emph {et~al.}(2014)\citenamefont {Zhao},
  \citenamefont {S{\'a}ndor}, \citenamefont {Rozgonyi},\ and\ \citenamefont
  {Weinacht}}]{Zhao:2014cu}%
  \BibitemOpen
  \bibfield  {author} {\bibinfo {author} {\bibfnamefont {A.}~\bibnamefont
  {Zhao}}, \bibinfo {author} {\bibfnamefont {P.}~\bibnamefont {S{\'a}ndor}},
  \bibinfo {author} {\bibfnamefont {T.}~\bibnamefont {Rozgonyi}}, \ and\
  \bibinfo {author} {\bibfnamefont {T.}~\bibnamefont {Weinacht}},\ }\href
  {\doibase 10.1088/0953-4075/47/20/204023} {\bibfield  {journal} {\bibinfo
  {journal} {J. Phys. B: Atom. Mol. Opt. Phys.}\ }\textbf {\bibinfo {volume}
  {47}},\ \bibinfo {pages} {204023} (\bibinfo {year} {2014})}\BibitemShut
  {NoStop}%
\bibitem [{\citenamefont {S{\'a}ndor}\ \emph {et~al.}(2016)\citenamefont
  {S{\'a}ndor}, \citenamefont {Tagliamonti}, \citenamefont {Zhao},
  \citenamefont {Rozgonyi}, \citenamefont {Ruckenbauer}, \citenamefont
  {Marquetand},\ and\ \citenamefont {Weinacht}}]{Sandor:2016eq}%
  \BibitemOpen
  \bibfield  {author} {\bibinfo {author} {\bibfnamefont {P.}~\bibnamefont
  {S{\'a}ndor}}, \bibinfo {author} {\bibfnamefont {V.}~\bibnamefont
  {Tagliamonti}}, \bibinfo {author} {\bibfnamefont {A.}~\bibnamefont {Zhao}},
  \bibinfo {author} {\bibfnamefont {T.}~\bibnamefont {Rozgonyi}}, \bibinfo
  {author} {\bibfnamefont {M.}~\bibnamefont {Ruckenbauer}}, \bibinfo {author}
  {\bibfnamefont {P.}~\bibnamefont {Marquetand}}, \ and\ \bibinfo {author}
  {\bibfnamefont {T.}~\bibnamefont {Weinacht}},\ }\href {\doibase
  10.1103/PhysRevLett.116.063002} {\bibfield  {journal} {\bibinfo  {journal}
  {Phys. Rev. Lett.}\ }\textbf {\bibinfo {volume} {116}},\ \bibinfo {pages}
  {063002} (\bibinfo {year} {2016})}\BibitemShut {NoStop}%
\bibitem [{\citenamefont {Lee}\ \emph {et~al.}(2005)\citenamefont {Lee},
  \citenamefont {Kim}, \citenamefont {Lee},\ and\ \citenamefont
  {Kim}}]{Lee:2005jt}%
  \BibitemOpen
  \bibfield  {author} {\bibinfo {author} {\bibfnamefont {M.}~\bibnamefont
  {Lee}}, \bibinfo {author} {\bibfnamefont {H.}~\bibnamefont {Kim}}, \bibinfo
  {author} {\bibfnamefont {Y.~S.}\ \bibnamefont {Lee}}, \ and\ \bibinfo
  {author} {\bibfnamefont {M.~S.}\ \bibnamefont {Kim}},\ }\href {\doibase
  10.1063/1.1954770} {\bibfield  {journal} {\bibinfo  {journal} {J. Chem.
  Phys.}\ }\textbf {\bibinfo {volume} {123}},\ \bibinfo {pages} {024310}
  (\bibinfo {year} {2005})}\BibitemShut {NoStop}%
\bibitem [{\citenamefont {Kudchadker}\ and\ \citenamefont
  {Kudchadker}(1975)}]{Kudchadker:1975cp}%
  \BibitemOpen
  \bibfield  {author} {\bibinfo {author} {\bibfnamefont {S.~A.}\ \bibnamefont
  {Kudchadker}}\ and\ \bibinfo {author} {\bibfnamefont {A.~P.}\ \bibnamefont
  {Kudchadker}},\ }\href {\doibase 10.1063/1.555522} {\bibfield  {journal}
  {\bibinfo  {journal} {J. Phys. Chem. Ref. Data}\ }\textbf {\bibinfo {volume}
  {4}},\ \bibinfo {pages} {457} (\bibinfo {year} {1975})}\BibitemShut {NoStop}%
\bibitem [{\citenamefont {von Niessen}, \citenamefont {{\AA}sbrink},\ and\
  \citenamefont {Bieri}(1982)}]{vonNiessen:1982fz}%
  \BibitemOpen
  \bibfield  {author} {\bibinfo {author} {\bibfnamefont {W.}~\bibnamefont {von
  Niessen}}, \bibinfo {author} {\bibfnamefont {L.}~\bibnamefont {{\AA}sbrink}},
  \ and\ \bibinfo {author} {\bibfnamefont {G.}~\bibnamefont {Bieri}},\ }\href
  {\doibase 10.1016/0368-2048(82)85065-2} {\bibfield  {journal} {\bibinfo
  {journal} {Journal of Electron Spectroscopy and Related Phenomena}\ }\textbf
  {\bibinfo {volume} {26}},\ \bibinfo {pages} {173} (\bibinfo {year}
  {1982})}\BibitemShut {NoStop}%
\bibitem [{\citenamefont {S{\'a}ndor}\ \emph {et~al.}(2014)\citenamefont
  {S{\'a}ndor}, \citenamefont {Zhao}, \citenamefont {Rozgonyi},\ and\
  \citenamefont {Weinacht}}]{Sandor:2014cu}%
  \BibitemOpen
  \bibfield  {author} {\bibinfo {author} {\bibfnamefont {P.}~\bibnamefont
  {S{\'a}ndor}}, \bibinfo {author} {\bibfnamefont {A.}~\bibnamefont {Zhao}},
  \bibinfo {author} {\bibfnamefont {T.}~\bibnamefont {Rozgonyi}}, \ and\
  \bibinfo {author} {\bibfnamefont {T.}~\bibnamefont {Weinacht}},\ }\href
  {\doibase 10.1088/0953-4075/47/12/124021} {\bibfield  {journal} {\bibinfo
  {journal} {J. Phys. B: Atom. Mol. Opt. Phys.}\ }\textbf {\bibinfo {volume}
  {47}},\ \bibinfo {pages} {124021} (\bibinfo {year} {2014})}\BibitemShut
  {NoStop}%
\bibitem [{\citenamefont {Gonz{\'a}lez-V{\'a}zquez}\ \emph
  {et~al.}(2010)\citenamefont {Gonz{\'a}lez-V{\'a}zquez}, \citenamefont
  {Gonz{\'a}lez}, \citenamefont {Nichols}, \citenamefont {Weinacht},\ and\
  \citenamefont {Rozgonyi}}]{GonzalezVazquez:2010ki}%
  \BibitemOpen
  \bibfield  {author} {\bibinfo {author} {\bibfnamefont {J.}~\bibnamefont
  {Gonz{\'a}lez-V{\'a}zquez}}, \bibinfo {author} {\bibfnamefont
  {L.}~\bibnamefont {Gonz{\'a}lez}}, \bibinfo {author} {\bibfnamefont {S.~R.}\
  \bibnamefont {Nichols}}, \bibinfo {author} {\bibfnamefont {T.~C.}\
  \bibnamefont {Weinacht}}, \ and\ \bibinfo {author} {\bibfnamefont
  {T.}~\bibnamefont {Rozgonyi}},\ }\href {\doibase 10.1039/C0CP00303D}
  {\bibfield  {journal} {\bibinfo  {journal} {Phys. Chem. Chem. Phys.}\
  }\textbf {\bibinfo {volume} {12}},\ \bibinfo {pages} {14203} (\bibinfo {year}
  {2010})}\BibitemShut {NoStop}%
\bibitem [{\citenamefont {Gei{\ss}ler}\ \emph {et~al.}(2011)\citenamefont
  {Gei{\ss}ler}, \citenamefont {Rozgonyi}, \citenamefont
  {Gonz{\'a}lez-V{\'a}zquez}, \citenamefont {Gonz{\'a}lez}, \citenamefont
  {Marquetand},\ and\ \citenamefont {Weinacht}}]{Geissler:2011dy}%
  \BibitemOpen
  \bibfield  {author} {\bibinfo {author} {\bibfnamefont {D.}~\bibnamefont
  {Gei{\ss}ler}}, \bibinfo {author} {\bibfnamefont {T.}~\bibnamefont
  {Rozgonyi}}, \bibinfo {author} {\bibfnamefont {J.}~\bibnamefont
  {Gonz{\'a}lez-V{\'a}zquez}}, \bibinfo {author} {\bibfnamefont
  {L.}~\bibnamefont {Gonz{\'a}lez}}, \bibinfo {author} {\bibfnamefont
  {P.}~\bibnamefont {Marquetand}}, \ and\ \bibinfo {author} {\bibfnamefont
  {T.~C.}\ \bibnamefont {Weinacht}},\ }\href {\doibase
  10.1103/PhysRevA.84.053422} {\bibfield  {journal} {\bibinfo  {journal} {Phys.
  Rev. A}\ }\textbf {\bibinfo {volume} {84}},\ \bibinfo {pages} {053422}
  (\bibinfo {year} {2011})}\BibitemShut {NoStop}%
\bibitem [{\citenamefont {Modelli}\ \emph {et~al.}(1992)\citenamefont
  {Modelli}, \citenamefont {Scagnolari}, \citenamefont {Distefano},
  \citenamefont {Jones},\ and\ \citenamefont {Guerra}}]{Modelli:1992jp}%
  \BibitemOpen
  \bibfield  {author} {\bibinfo {author} {\bibfnamefont {A.}~\bibnamefont
  {Modelli}}, \bibinfo {author} {\bibfnamefont {F.}~\bibnamefont {Scagnolari}},
  \bibinfo {author} {\bibfnamefont {G.}~\bibnamefont {Distefano}}, \bibinfo
  {author} {\bibfnamefont {D.}~\bibnamefont {Jones}}, \ and\ \bibinfo {author}
  {\bibfnamefont {M.}~\bibnamefont {Guerra}},\ }\href {\doibase
  10.1063/1.462058} {\bibfield  {journal} {\bibinfo  {journal} {J. Chem.
  Phys.}\ }\textbf {\bibinfo {volume} {96}},\ \bibinfo {pages} {2061} (\bibinfo
  {year} {1992})}\BibitemShut {NoStop}%
\bibitem [{\citenamefont {Guerra}\ \emph {et~al.}(1991)\citenamefont {Guerra},
  \citenamefont {Jones}, \citenamefont {Distefano}, \citenamefont
  {Scagnolari},\ and\ \citenamefont {Modelli}}]{Guerra:1991jp}%
  \BibitemOpen
  \bibfield  {author} {\bibinfo {author} {\bibfnamefont {M.}~\bibnamefont
  {Guerra}}, \bibinfo {author} {\bibfnamefont {D.}~\bibnamefont {Jones}},
  \bibinfo {author} {\bibfnamefont {G.}~\bibnamefont {Distefano}}, \bibinfo
  {author} {\bibfnamefont {F.}~\bibnamefont {Scagnolari}}, \ and\ \bibinfo
  {author} {\bibfnamefont {A.}~\bibnamefont {Modelli}},\ }\href {\doibase
  10.1063/1.460364} {\bibfield  {journal} {\bibinfo  {journal} {J. Chem.
  Phys.}\ }\textbf {\bibinfo {volume} {94}},\ \bibinfo {pages} {484} (\bibinfo
  {year} {1991})}\BibitemShut {NoStop}%
\bibitem [{\citenamefont {Liu}\ \emph {et~al.}(2007)\citenamefont {Liu},
  \citenamefont {De~Vico}, \citenamefont {Lindh},\ and\ \citenamefont
  {Fang}}]{Liu:2007cg}%
  \BibitemOpen
  \bibfield  {author} {\bibinfo {author} {\bibfnamefont {Y.-J.}\ \bibnamefont
  {Liu}}, \bibinfo {author} {\bibfnamefont {L.}~\bibnamefont {De~Vico}},
  \bibinfo {author} {\bibfnamefont {R.}~\bibnamefont {Lindh}}, \ and\ \bibinfo
  {author} {\bibfnamefont {W.-H.}\ \bibnamefont {Fang}},\ }\href {\doibase
  10.1002/cphc.200600737} {\bibfield  {journal} {\bibinfo  {journal}
  {ChemPhysChem}\ }\textbf {\bibinfo {volume} {8}},\ \bibinfo {pages} {890}
  (\bibinfo {year} {2007})}\BibitemShut {NoStop}%
\bibitem [{\citenamefont {Liu}\ \emph {et~al.}(2006)\citenamefont {Liu},
  \citenamefont {Ajitha}, \citenamefont {Krogh}, \citenamefont {Tarnovsky},\
  and\ \citenamefont {Lindh}}]{Liu:2006cp}%
  \BibitemOpen
  \bibfield  {author} {\bibinfo {author} {\bibfnamefont {Y.-J.}\ \bibnamefont
  {Liu}}, \bibinfo {author} {\bibfnamefont {D.}~\bibnamefont {Ajitha}},
  \bibinfo {author} {\bibfnamefont {J.~W.}\ \bibnamefont {Krogh}}, \bibinfo
  {author} {\bibfnamefont {A.~N.}\ \bibnamefont {Tarnovsky}}, \ and\ \bibinfo
  {author} {\bibfnamefont {R.}~\bibnamefont {Lindh}},\ }\href {\doibase
  10.1002/cphc.200500654} {\bibfield  {journal} {\bibinfo  {journal}
  {ChemPhysChem}\ }\textbf {\bibinfo {volume} {7}},\ \bibinfo {pages} {955}
  (\bibinfo {year} {2006})}\BibitemShut {NoStop}%
\bibitem [{\citenamefont {Ito}, \citenamefont {Huang},\ and\ \citenamefont
  {Kosower}(1961)}]{Ito:1961eh}%
  \BibitemOpen
  \bibfield  {author} {\bibinfo {author} {\bibfnamefont {M.}~\bibnamefont
  {Ito}}, \bibinfo {author} {\bibfnamefont {P.-k.~C.}\ \bibnamefont {Huang}}, \
  and\ \bibinfo {author} {\bibfnamefont {E.~M.}\ \bibnamefont {Kosower}},\
  }\href {\doibase 10.1039/TF9615701662} {\bibfield  {journal} {\bibinfo
  {journal} {Trans. Faraday Soc.}\ }\textbf {\bibinfo {volume} {57}},\ \bibinfo
  {pages} {1662} (\bibinfo {year} {1961})}\BibitemShut {NoStop}%
\bibitem [{\citenamefont {Kawasaki}, \citenamefont {Lee},\ and\ \citenamefont
  {Bersohn}(1975)}]{Kawasaki:1975gb}%
  \BibitemOpen
  \bibfield  {author} {\bibinfo {author} {\bibfnamefont {M.}~\bibnamefont
  {Kawasaki}}, \bibinfo {author} {\bibfnamefont {S.~J.}\ \bibnamefont {Lee}}, \
  and\ \bibinfo {author} {\bibfnamefont {R.}~\bibnamefont {Bersohn}},\ }\href
  {\doibase 10.1063/1.431361} {\bibfield  {journal} {\bibinfo  {journal} {J.
  Chem. Phys.}\ }\textbf {\bibinfo {volume} {63}},\ \bibinfo {pages} {809}
  (\bibinfo {year} {1975})}\BibitemShut {NoStop}%
\bibitem [{\citenamefont {Gedanken}\ and\ \citenamefont
  {Rowe}(1979)}]{Gedanken:1979fa}%
  \BibitemOpen
  \bibfield  {author} {\bibinfo {author} {\bibfnamefont {A.}~\bibnamefont
  {Gedanken}}\ and\ \bibinfo {author} {\bibfnamefont {M.~D.}\ \bibnamefont
  {Rowe}},\ }\href {\doibase 10.1016/0301-0104(79)85004-1} {\bibfield
  {journal} {\bibinfo  {journal} {Chem. Phys.}\ }\textbf {\bibinfo {volume}
  {36}},\ \bibinfo {pages} {181} (\bibinfo {year} {1979})}\BibitemShut
  {NoStop}%
\bibitem [{\citenamefont {Butler}\ \emph {et~al.}(1987)\citenamefont {Butler},
  \citenamefont {Hintsa}, \citenamefont {Shane},\ and\ \citenamefont
  {Lee}}]{Butler:1987il}%
  \BibitemOpen
  \bibfield  {author} {\bibinfo {author} {\bibfnamefont {L.~J.}\ \bibnamefont
  {Butler}}, \bibinfo {author} {\bibfnamefont {E.~J.}\ \bibnamefont {Hintsa}},
  \bibinfo {author} {\bibfnamefont {S.~F.}\ \bibnamefont {Shane}}, \ and\
  \bibinfo {author} {\bibfnamefont {Y.~T.}\ \bibnamefont {Lee}},\ }\href
  {\doibase 10.1063/1.452155} {\bibfield  {journal} {\bibinfo  {journal} {J.
  Chem. Phys.}\ }\textbf {\bibinfo {volume} {86}},\ \bibinfo {pages} {2051}
  (\bibinfo {year} {1987})}\BibitemShut {NoStop}%
\bibitem [{\citenamefont {Lee}\ and\ \citenamefont
  {Bersohn}(1982)}]{Lee:1982ko}%
  \BibitemOpen
  \bibfield  {author} {\bibinfo {author} {\bibfnamefont {S.~J.}\ \bibnamefont
  {Lee}}\ and\ \bibinfo {author} {\bibfnamefont {R.}~\bibnamefont {Bersohn}},\
  }\href {\doibase 10.1021/j100394a028} {\bibfield  {journal} {\bibinfo
  {journal} {J. Phys. Chem.}\ }\textbf {\bibinfo {volume} {86}},\ \bibinfo
  {pages} {728} (\bibinfo {year} {1982})}\BibitemShut {NoStop}%
\bibitem [{\citenamefont {Watts}\ and\ \citenamefont
  {Bartlett}(1996)}]{Watts:1996bi}%
  \BibitemOpen
  \bibfield  {author} {\bibinfo {author} {\bibfnamefont {J.~D.}\ \bibnamefont
  {Watts}}\ and\ \bibinfo {author} {\bibfnamefont {R.~J.}\ \bibnamefont
  {Bartlett}},\ }\href {\doibase 10.1016/0009-2614(96)00708-7} {\bibfield
  {journal} {\bibinfo  {journal} {Chemical Physics Letters}\ }\textbf {\bibinfo
  {volume} {258}},\ \bibinfo {pages} {581} (\bibinfo {year}
  {1996})}\BibitemShut {NoStop}%
\bibitem [{\citenamefont {Stanton}\ and\ \citenamefont
  {Gauss}(1996)}]{Stanton:1996hw}%
  \BibitemOpen
  \bibfield  {author} {\bibinfo {author} {\bibfnamefont {J.~F.}\ \bibnamefont
  {Stanton}}\ and\ \bibinfo {author} {\bibfnamefont {J.}~\bibnamefont
  {Gauss}},\ }\href {\doibase 10.1007/BF01127508} {\bibfield  {journal}
  {\bibinfo  {journal} {Theoretica Chimica Acta}\ }\textbf {\bibinfo {volume}
  {93}},\ \bibinfo {pages} {303} (\bibinfo {year} {1996})}\BibitemShut
  {NoStop}%
\bibitem [{\citenamefont {Manohar}, \citenamefont {Stanton},\ and\
  \citenamefont {Krylov}(2009)}]{Manohar:2009kza}%
  \BibitemOpen
  \bibfield  {author} {\bibinfo {author} {\bibfnamefont {P.~U.}\ \bibnamefont
  {Manohar}}, \bibinfo {author} {\bibfnamefont {J.~F.}\ \bibnamefont
  {Stanton}}, \ and\ \bibinfo {author} {\bibfnamefont {A.~I.}\ \bibnamefont
  {Krylov}},\ }\href {\doibase 10.1063/1.3231133} {\bibfield  {journal}
  {\bibinfo  {journal} {The Journal of Chemical Physics}\ }\textbf {\bibinfo
  {volume} {131}},\ \bibinfo {pages} {114112} (\bibinfo {year}
  {2009})}\BibitemShut {NoStop}%
\bibitem [{\citenamefont {Matthews}\ and\ \citenamefont
  {Stanton}(2016)}]{Matthews:2016bi}%
  \BibitemOpen
  \bibfield  {author} {\bibinfo {author} {\bibfnamefont {D.~A.}\ \bibnamefont
  {Matthews}}\ and\ \bibinfo {author} {\bibfnamefont {J.~F.}\ \bibnamefont
  {Stanton}},\ }\href {\doibase 10.1063/1.4962910} {\bibfield  {journal}
  {\bibinfo  {journal} {The Journal of Chemical Physics}\ }\textbf {\bibinfo
  {volume} {145}},\ \bibinfo {pages} {124102} (\bibinfo {year}
  {2016})}\BibitemShut {NoStop}%
\end{thebibliography}
\end{document}


\title{Supplementary information for ``Equation-of-Motion Coupled-Cluster Theory based on the 4-component Dirac--Coulomb(--Gaunt) Hamiltonian. Energies for single electron detachment, attachment and electronically excited states''}

\author{Avijit Shee}
\email{ashee@umich.edu}
\affiliation{Department of Chemistry, University of Michigan, 930 N.\ University, Ann Arbor, MI 48109-1055, USA}
\affiliation{Universit\'e de Lille, CNRS, UMR 8523 -- PhLAM -- Physique des Lasers, Atomes et Mol\'ecules, F-59000 Lille, France Fax: +33-3-2033-7020; Tel: +33-3-2043-4163}

\author {Trond Saue}
\email{trond.saue@irsamc.ups-tlse.fr}
\affiliation{Laboratoire de Chimie et Physique Quantiques, UMR 5626 CNRS --- Universit\'e Toulouse III--Paul Sabatier, 118 route de Narbonne, F-31062 Toulouse, France Fax: +33-5-6155-6065; Tel: +33-5-6155-6031}

\author {Lucas Visscher}
\email{l.visscher@vu.nl}
\affiliation{Division of Theoretical Chemistry, Faculty of Sciences, Vrije Universiteit Amsterdam, De Boelelaan 1083, 1081 HV Amsterdam, The Netherlands, Tel: +31-20-598-7624}

\author{Andr\'e Severo Pereira Gomes}
\email{andre.gomes@univ-lille.fr (corresponding author)}
\affiliation{Universit\'e de Lille, CNRS, UMR 8523 -- PhLAM -- Physique des Lasers, Atomes et Mol\'ecules, F-59000 Lille, France Fax: +33-3-2033-7020; Tel: +33-3-2043-4163}

\date{\today}
\revised{\today}
\maketitle

\clearpage

\section{Halogen monoxides}

\begin{figure}[h]
    \centering
\begin{minipage}[b]{0.47\linewidth}
\centering
\includegraphics[width=\textwidth]{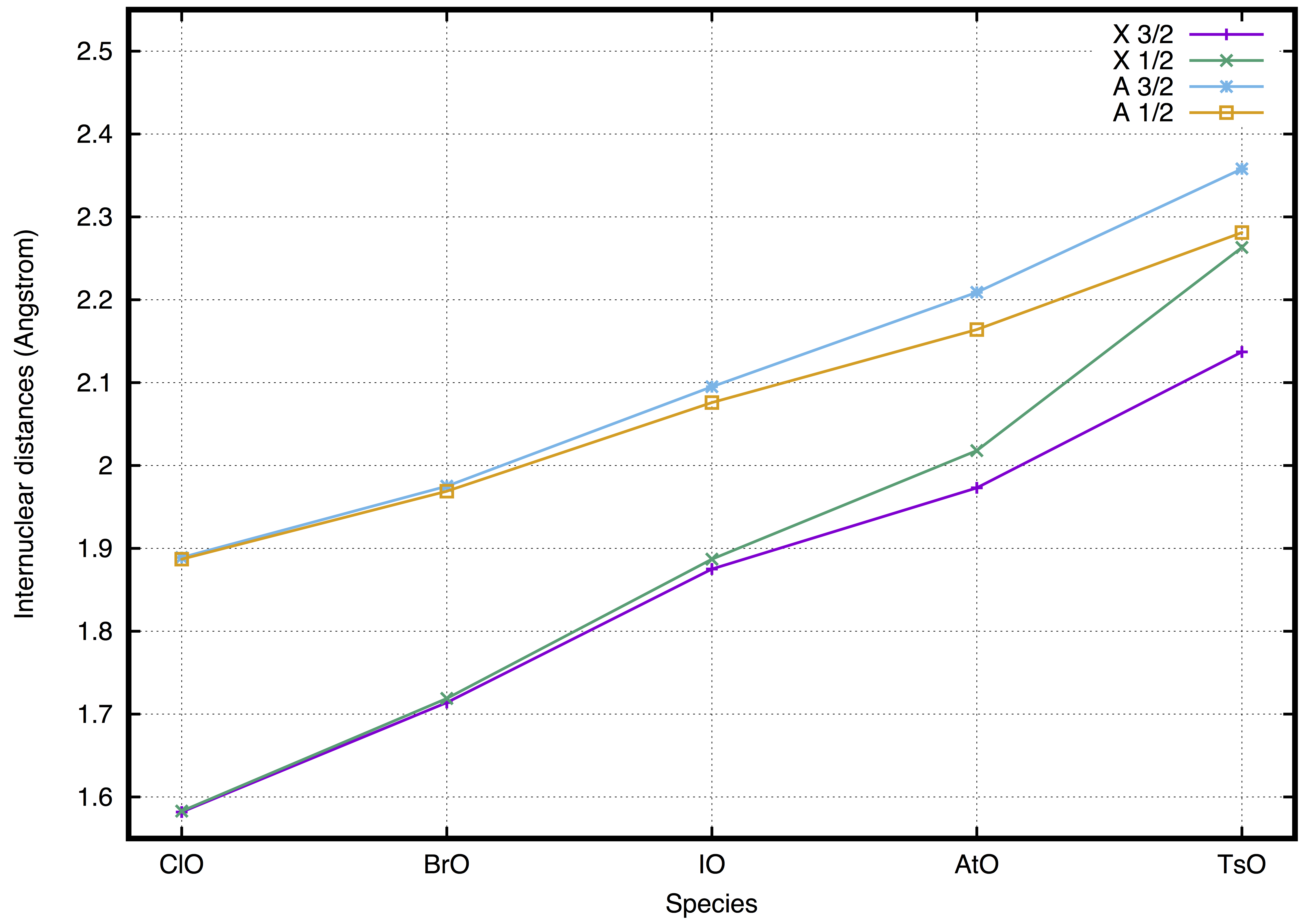}
\label{fig:distances-xo}
\end{minipage}
\begin{minipage}[b]{0.47\linewidth}
\centering
\includegraphics[width=\textwidth]{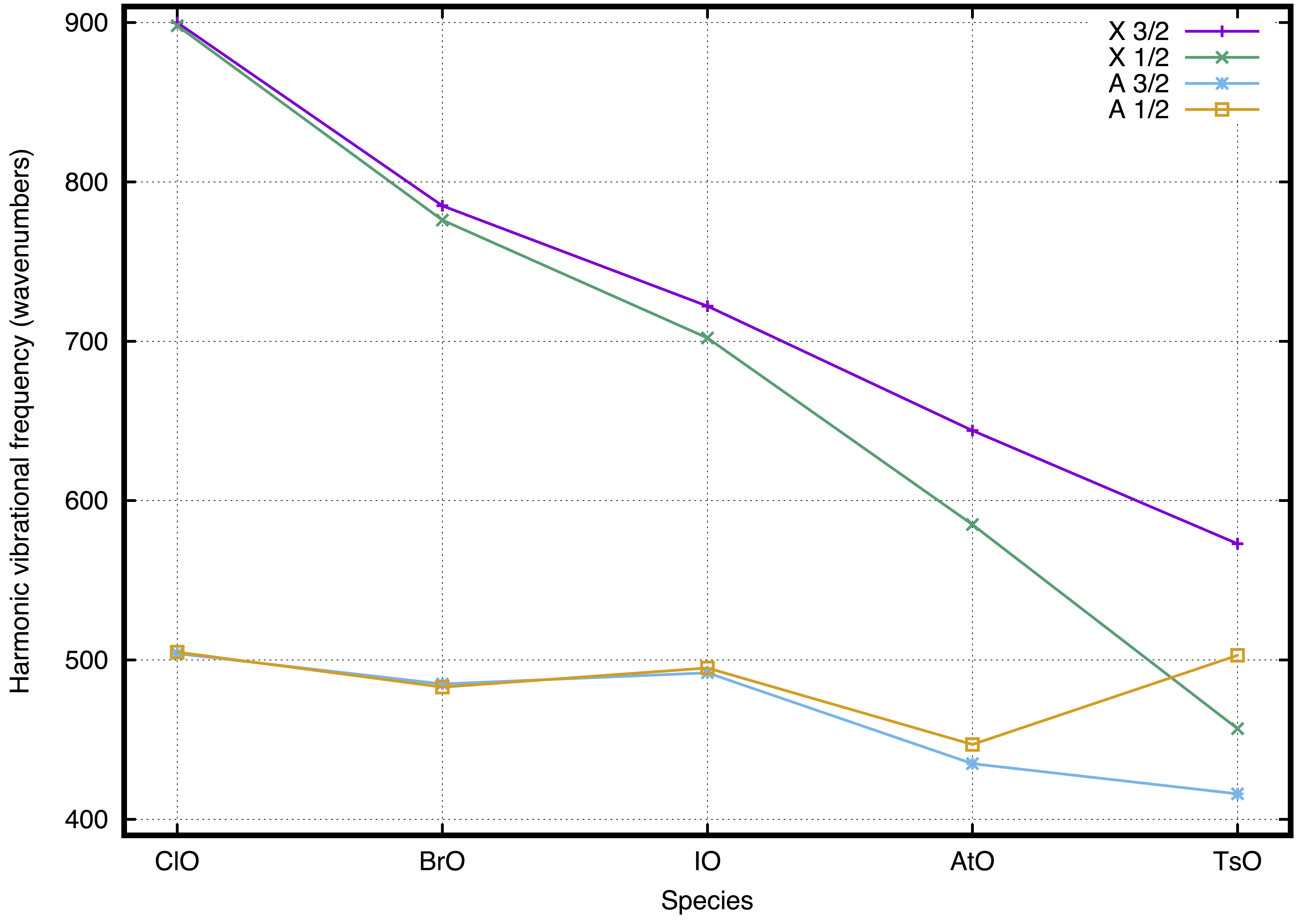}
\label{fig:frequencies-xo}
\end{minipage}
\begin{minipage}[b]{0.47\linewidth}
\centering
\includegraphics[width=\textwidth]{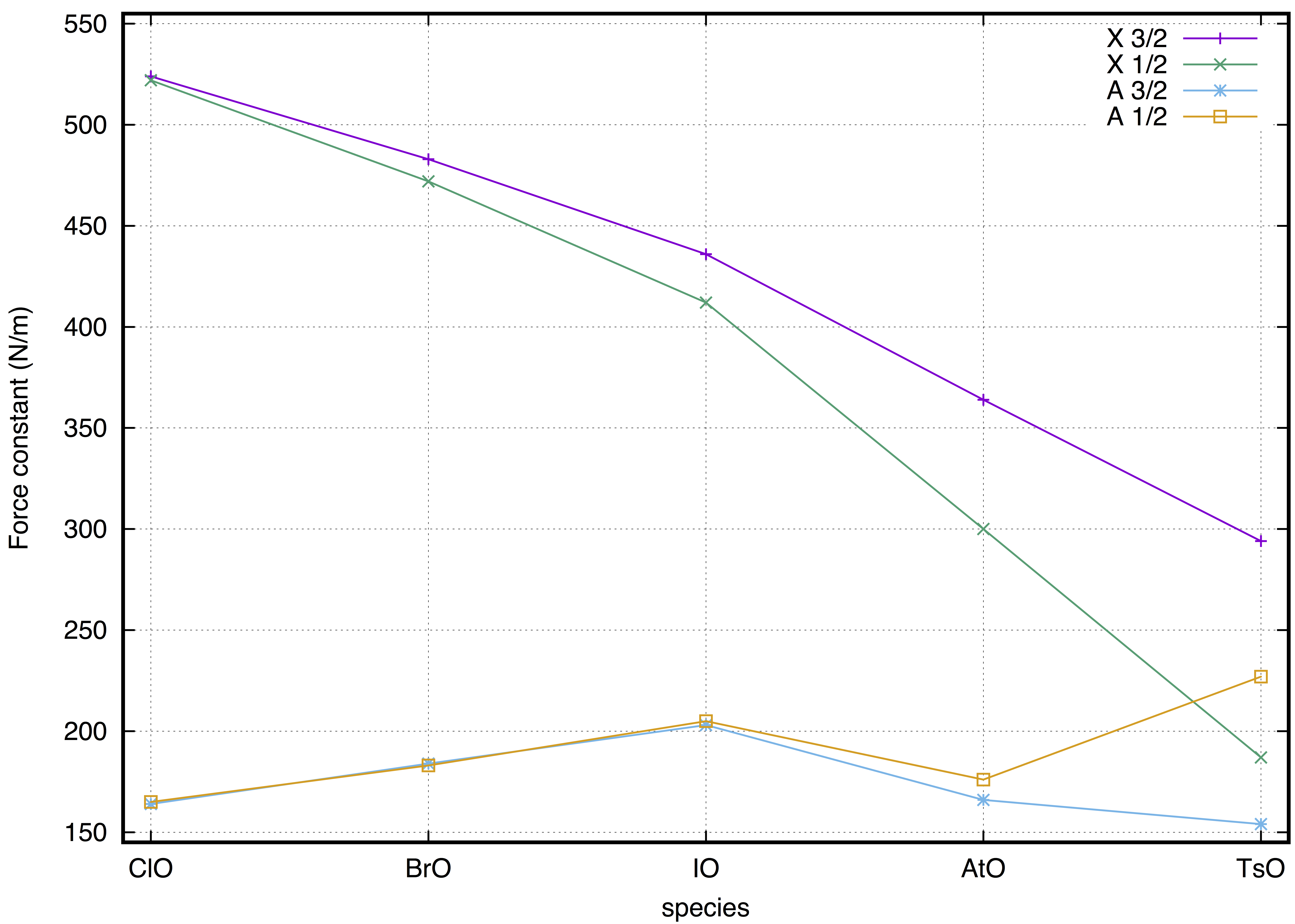}
\label{fig:so-force-constant-xo}
\end{minipage}
\begin{minipage}[b]{0.47\linewidth}
\centering
\includegraphics[width=\textwidth]{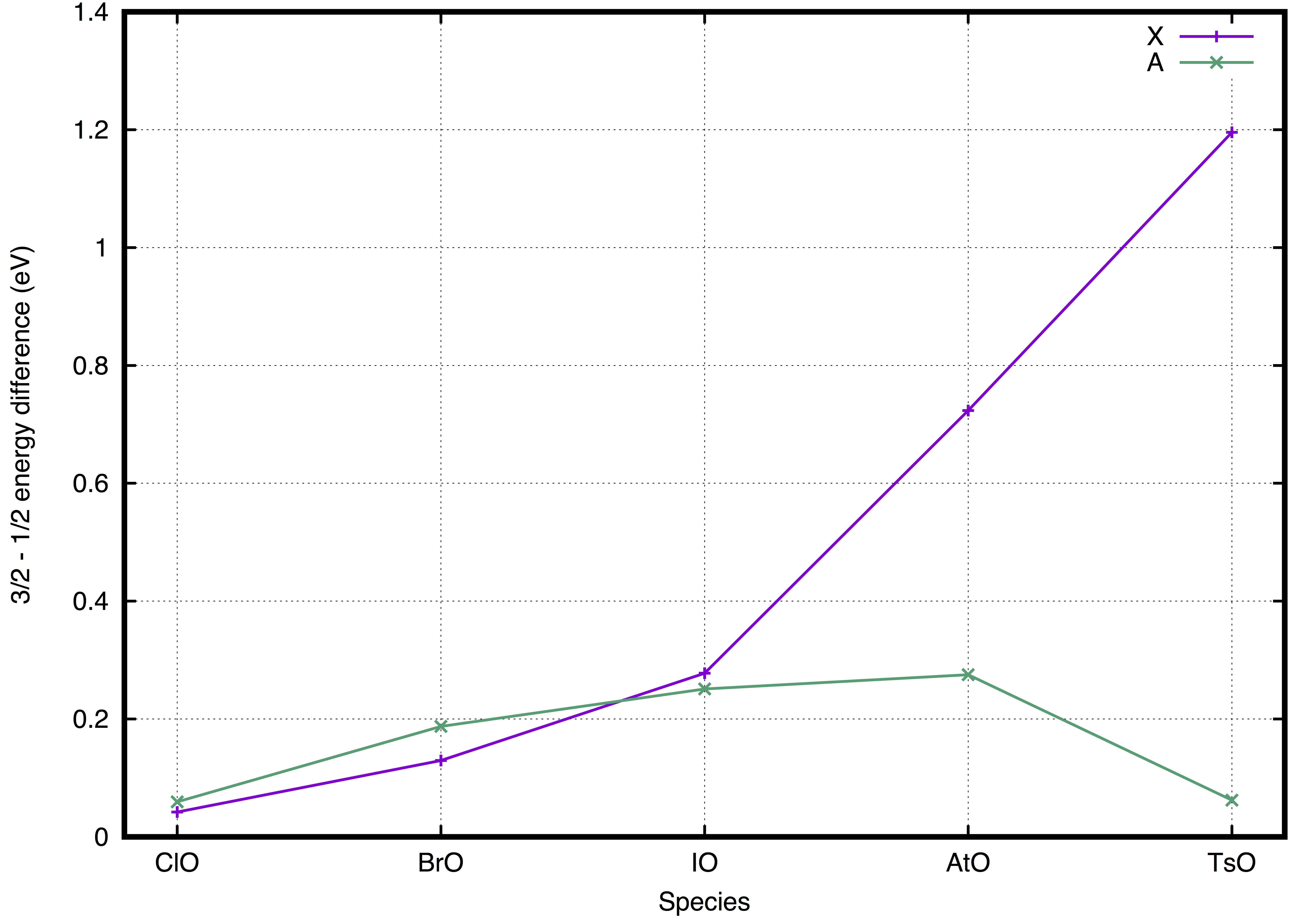}
\label{fig:so-splitting-xo}
\end{minipage}

    \caption{Internuclear distances (in \AA), harmonic vibrational frequencies (in cm$^{-1}$) and force constants (in N/m), and  the vertical $\Omega = 3/2 - 1/2$ energy difference  (in eV) for the $X ^2\Pi$ and $A ^2\Pi$ states of the XO molecules, obtained with EOM-IP and the $^2$DCG$^M$ Hamiltonian.~\cite{paper:figures}}
    \label{relativistic-effects-xo}
\end{figure}

In Table~\ref{tab:spectro-monoxide-radicals} we present the vertical and adiabatic electronic spectra of the halogen monoxide radicals, as well as equilibrium geometries and vibrational frequencies of the individual states obtained with EOM-IP and IHFS(1h0p) with the $^2$DC$^M$ and $^2$DCG$^M$ Hamiltonians, whereas in Table~\ref{tab:spectro-monoxide-radical-dc-vs-2dcm} we compare the results for the Dirac-Coulomb and $^2$DC$^M$
Hamiltonians. In Figure~\ref{relativistic-effects-xo} the trends along the series for the internuclear distances, vibrational frequencies and vertical $\Omega = 3/2 - 1/2$ energy difference for the $X ^2\Pi$ and $A ^2\Pi$ states, based on the data in Table~\ref{tab:spectro-monoxide-radicals} for the  $^2$DCG$^M$ Hamiltonian, are also presented.

In complement to that, in Tables~\ref{tab:clo-state-wfs} to~\ref{tab:tso-state-wfs} we present the states' 
wavefunctions for the different states of each species, in terms of the dominant configurations and their expansion coefficients in the respective wavefunction parametrizations ($r_i$ 
and eventually $r_{ij}^a$ for EOM-IP, as well as $c_i$ IHFS(1h0p), which has the same role as $r_i$). We note we have chosen, for the wavefunctions, to present only the $^2$DC$^M$ results as there are no qualitative differences 
between the Hamiltonians. The $^2$DCG$^M$ values can be retrieved from the output files provided as part of the supplementary information.

In Tables~\ref{tab:clo-state-prjana}, \ref{tab:bro-state-prjana}, \ref{tab:io-state-prjana}, \ref{tab:ato-state-prjana} and \ref{tab:tso-state-prjana} we present a projection analysis \cite{Faegri_JCP2001} as well as the $\sigma$- and $\pi$-character of the valence orbitals of the anions (XO$^-$) from 4-component Hartree--Fock calculations based on the Dirac--Coulomb--Gaunt Hamiltonian. The dominant configuration of the corresponding XO radicals are given in Tables~\ref{tab:clo-state-wfs}, \ref{tab:bro-state-wfs}, \ref{tab:io-state-wfs}, \ref{tab:ato-state-wfs} and \ref{tab:tso-state-wfs}. In all cases the 
$\pi^{*}_{3/2}$, $\pi^{*}_{1/2}$, $\sigma_{1/2}$, $\pi_{3/2}$ and $\pi_{1/2}$ spinors are occupied in the SCF reference wavefunctions for the XO$^-$ species. 

\begin{figure}[h]
\centering
\begin{minipage}[b]{0.19\linewidth}
\centering
\includegraphics[width=\textwidth]{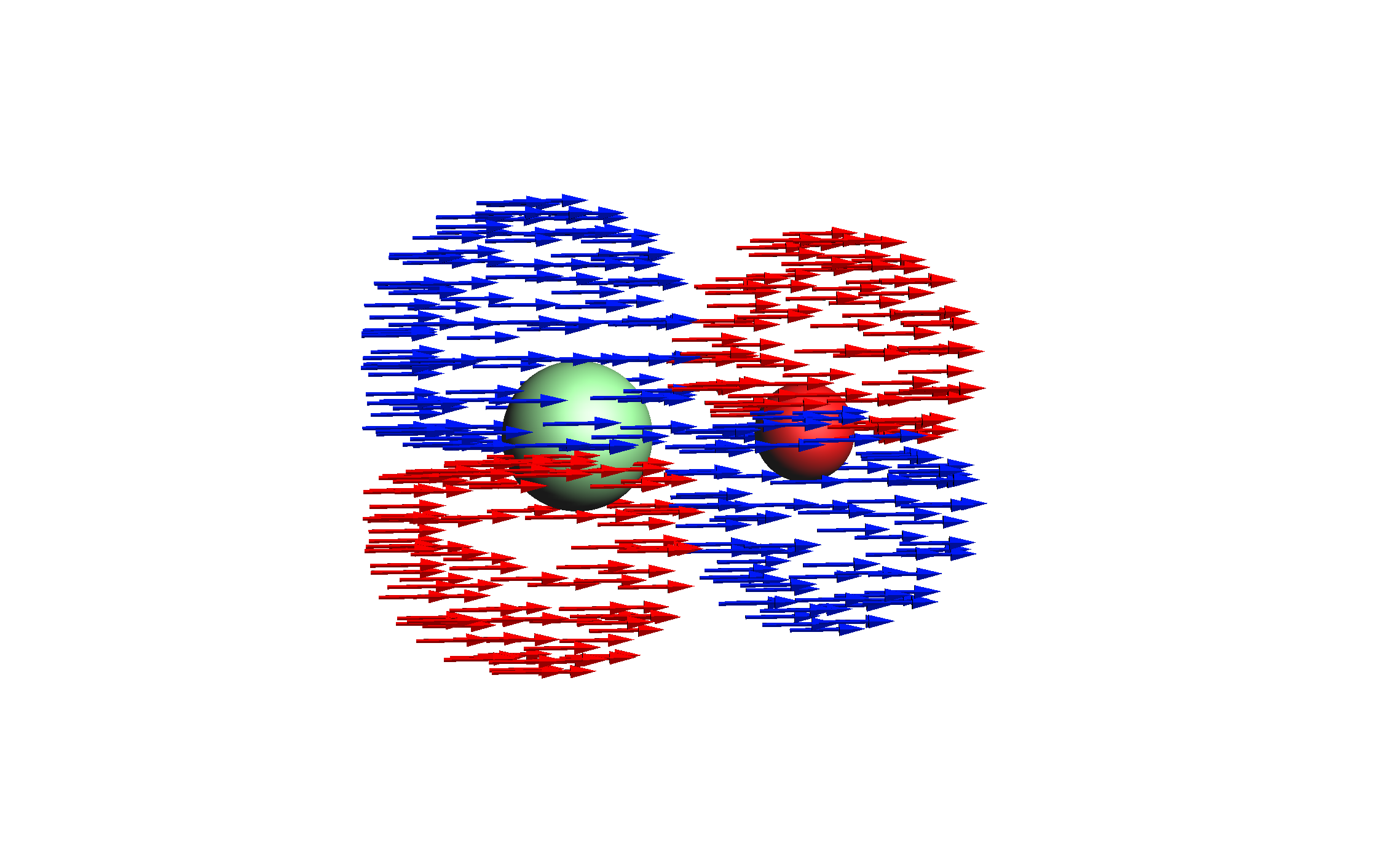}
\caption*{\scriptsize $\pi^{*}_{3/2}$: -0.116}
\end{minipage}
\begin{minipage}[b]{0.19\linewidth}
\centering
\includegraphics[width=\textwidth]{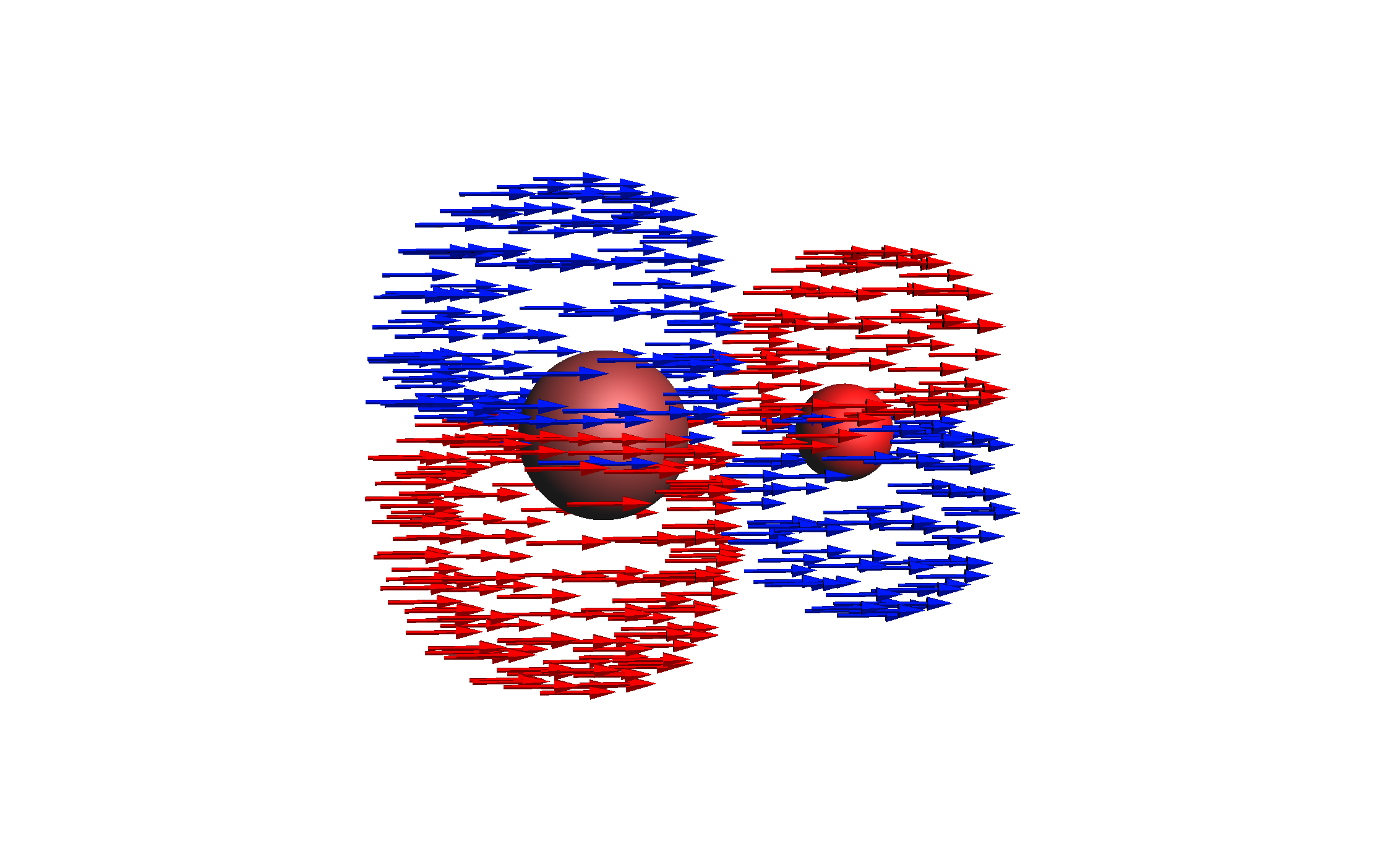}
\caption*{\scriptsize $\pi^{*}_{3/2}$: -0.116}
\end{minipage}
\begin{minipage}[b]{0.19\linewidth}
\centering
\includegraphics[width=\textwidth]{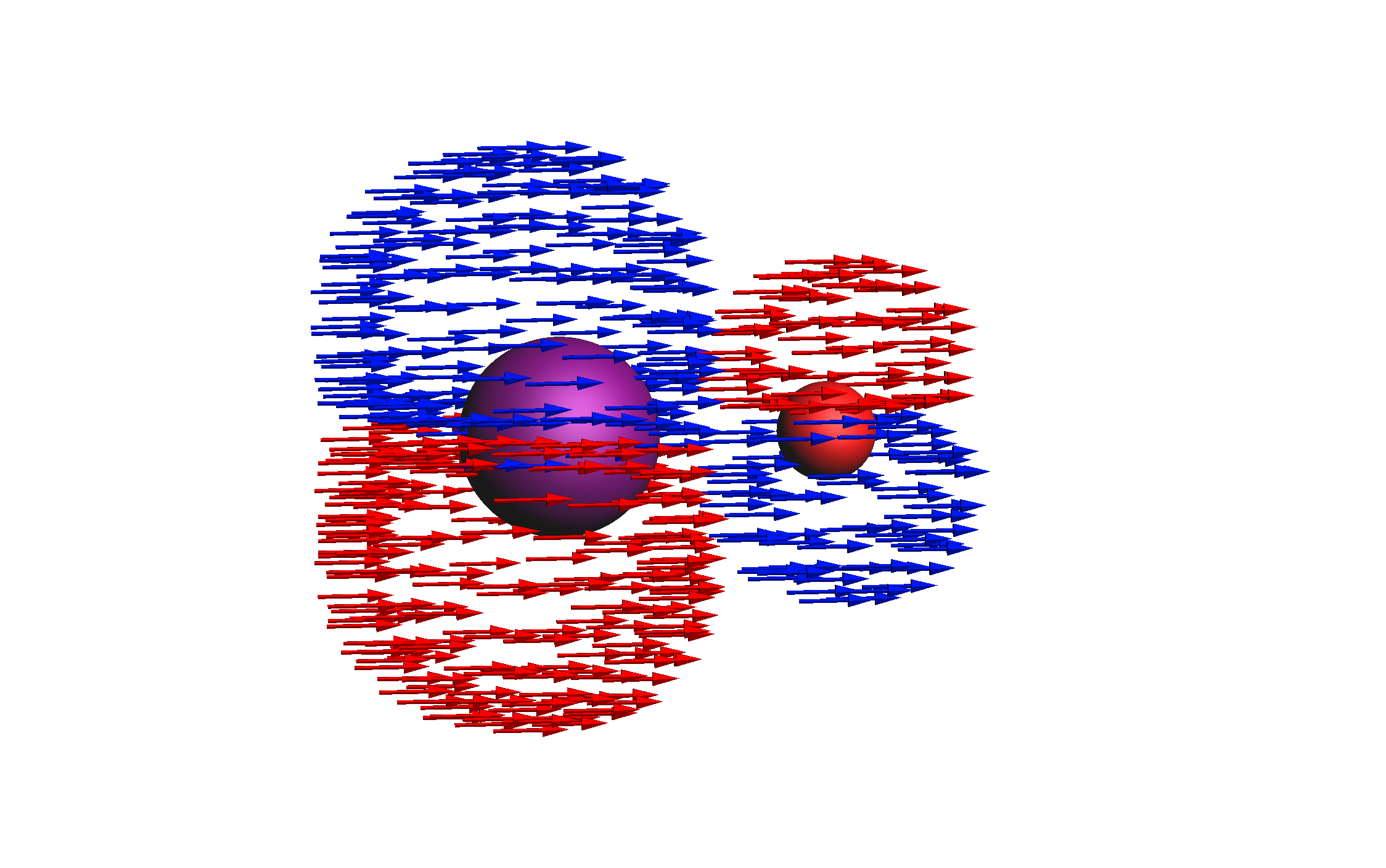}
\caption*{\scriptsize $\pi^{*}_{3/2}$: -0.109}
\end{minipage}
\begin{minipage}[b]{0.19\linewidth}
\centering
\includegraphics[width=\textwidth]{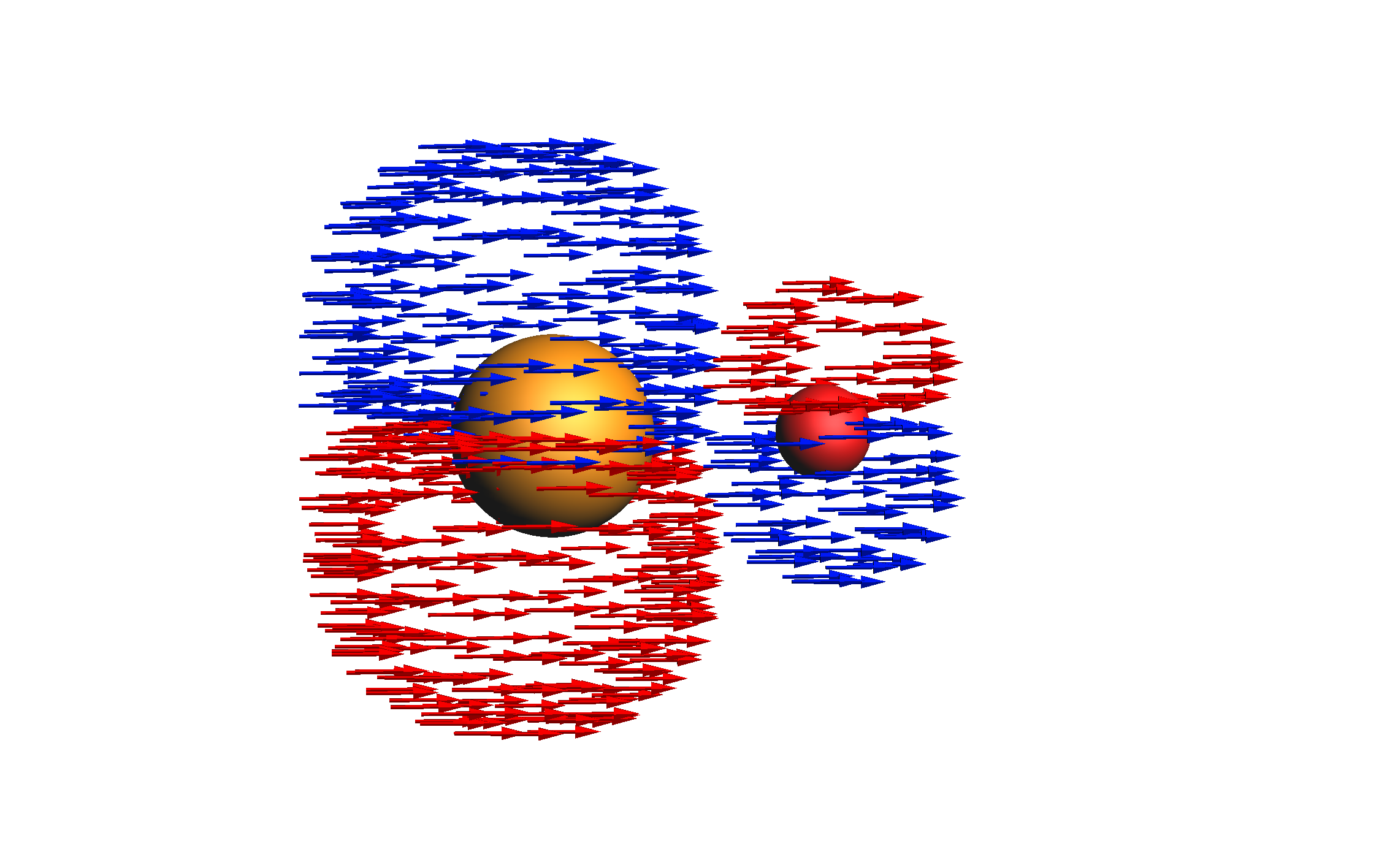}
\caption*{\scriptsize $\pi^{*}_{3/2}$: -0.087}
\end{minipage}
\centering
\begin{minipage}[b]{0.19\linewidth}
\includegraphics[width=\textwidth]{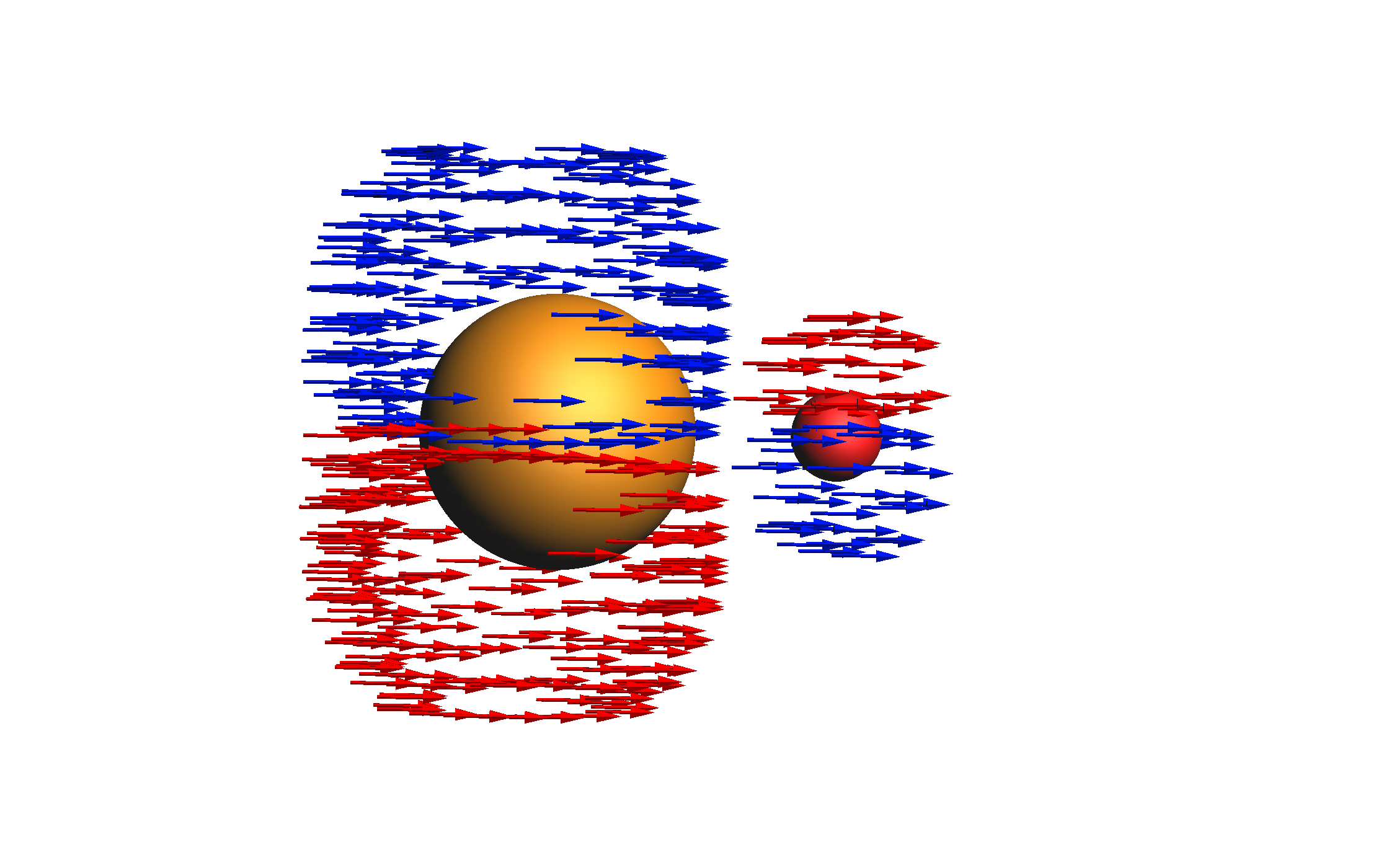}
\caption*{\scriptsize $\pi^{*}_{3/2}$: -0.050}
\end{minipage}
    \centering
\begin{minipage}[b]{0.19\linewidth}
\centering
\includegraphics[width=\textwidth]{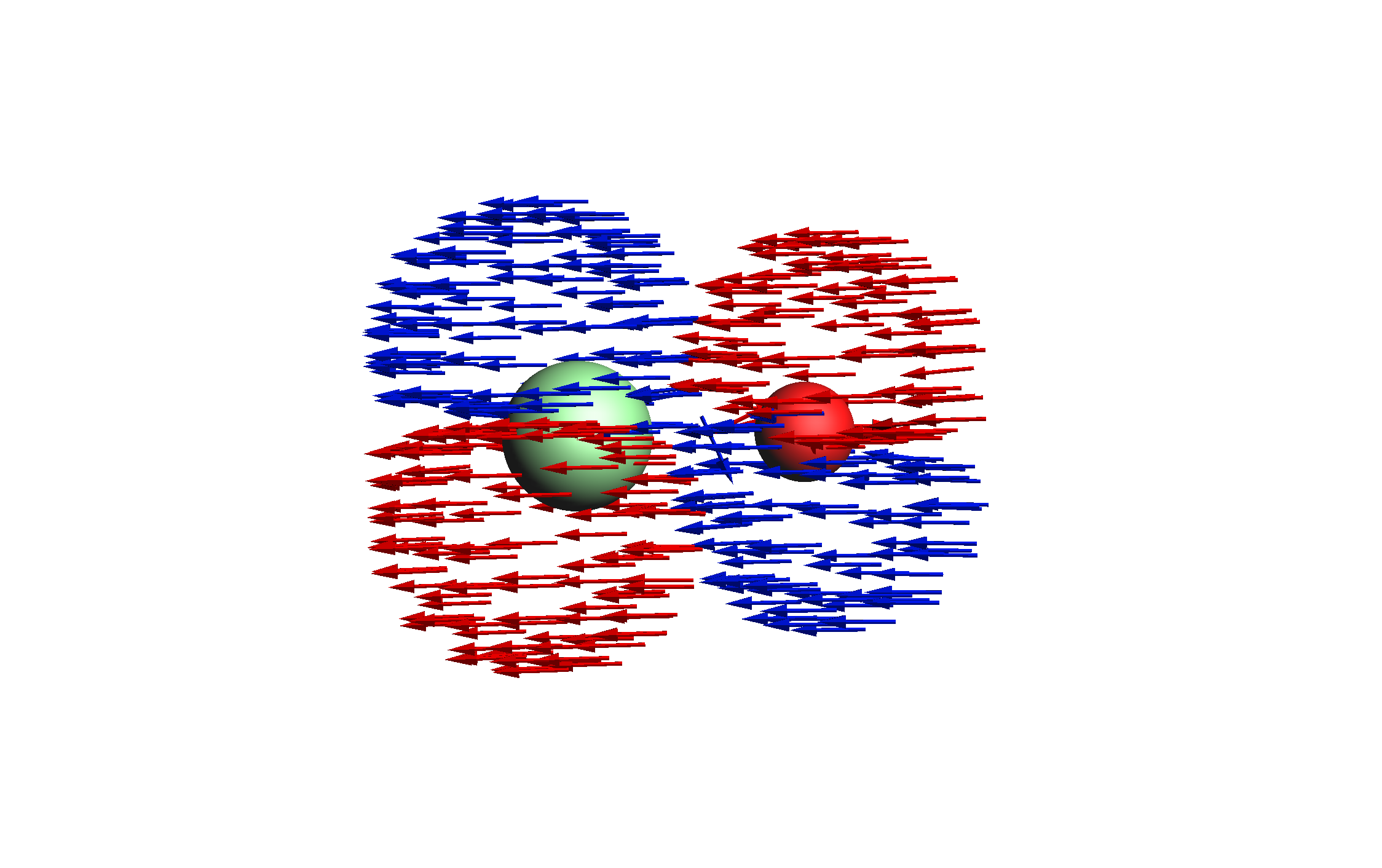}
\caption*{\scriptsize $\pi^{*}_{1/2}$: -0.119}
\end{minipage}
\begin{minipage}[b]{0.19\linewidth}
\centering
\includegraphics[width=\textwidth]{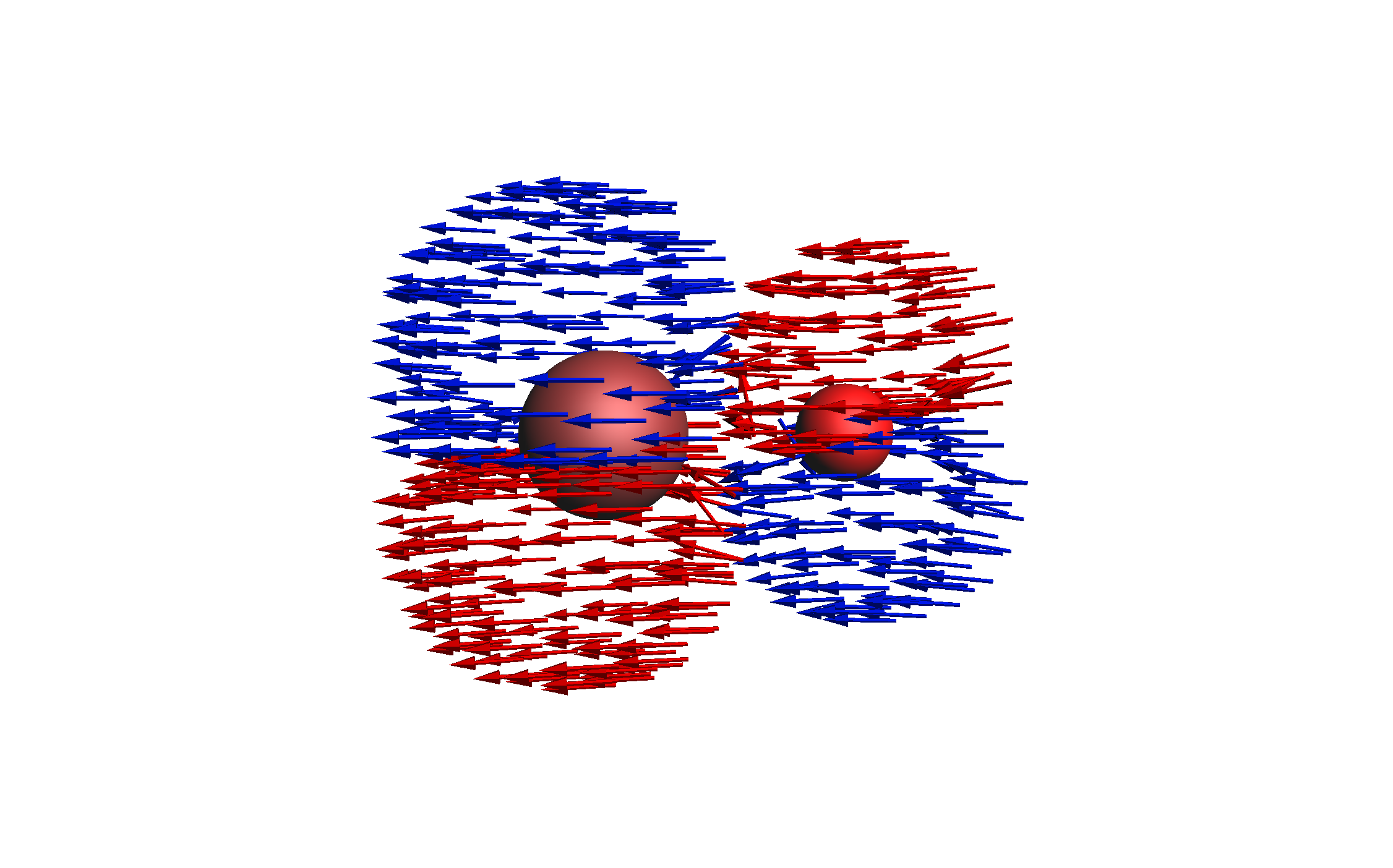}
\caption*{\scriptsize $\pi^{*}_{1/2}$: -0.125}
\end{minipage}
\begin{minipage}[b]{0.19\linewidth}
\centering
\includegraphics[width=\textwidth]{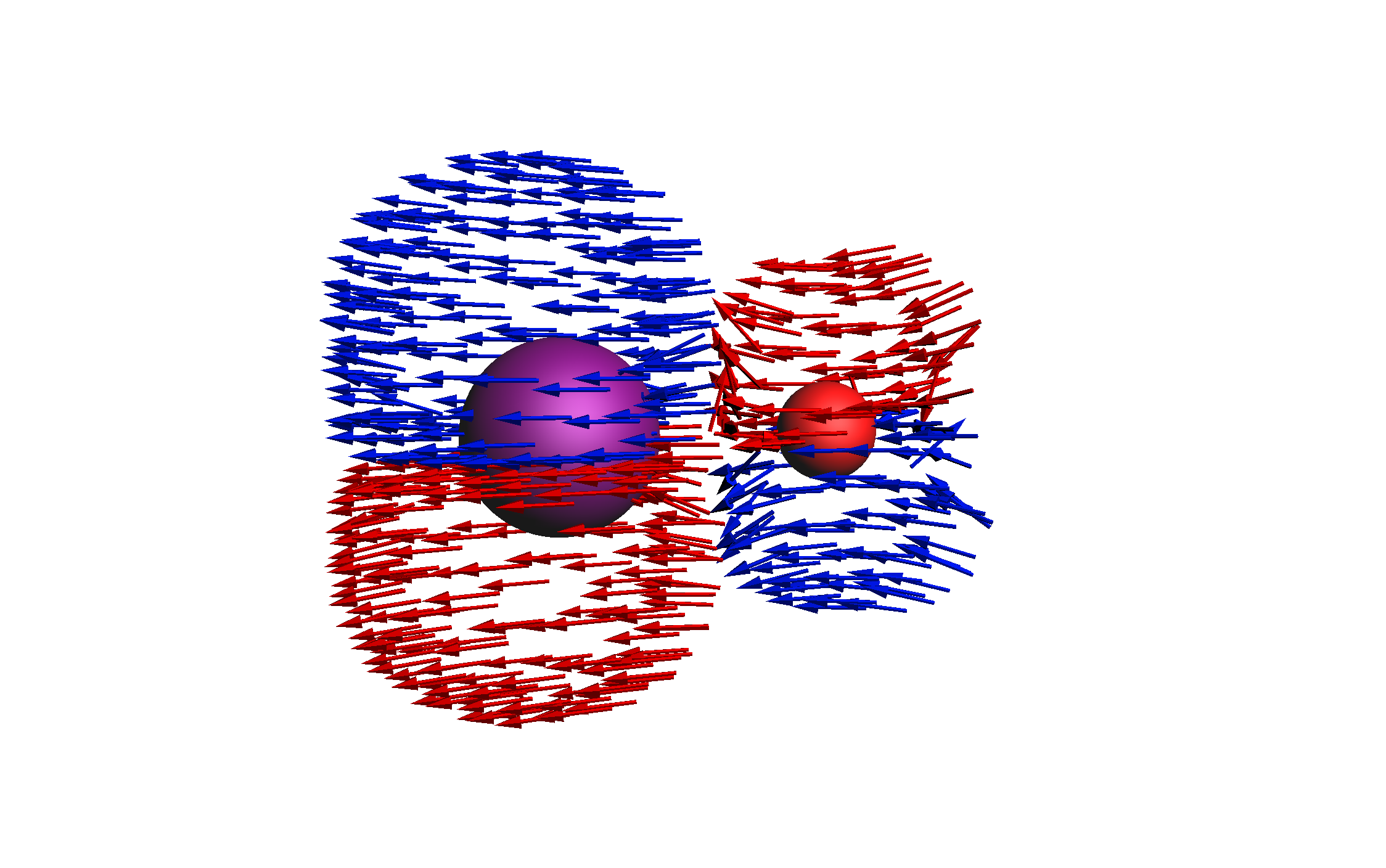}
\caption*{\scriptsize $\pi^{*}_{1/2}$: -0.129}
\end{minipage}
\begin{minipage}[b]{0.19\linewidth}
\centering
\includegraphics[width=\textwidth]{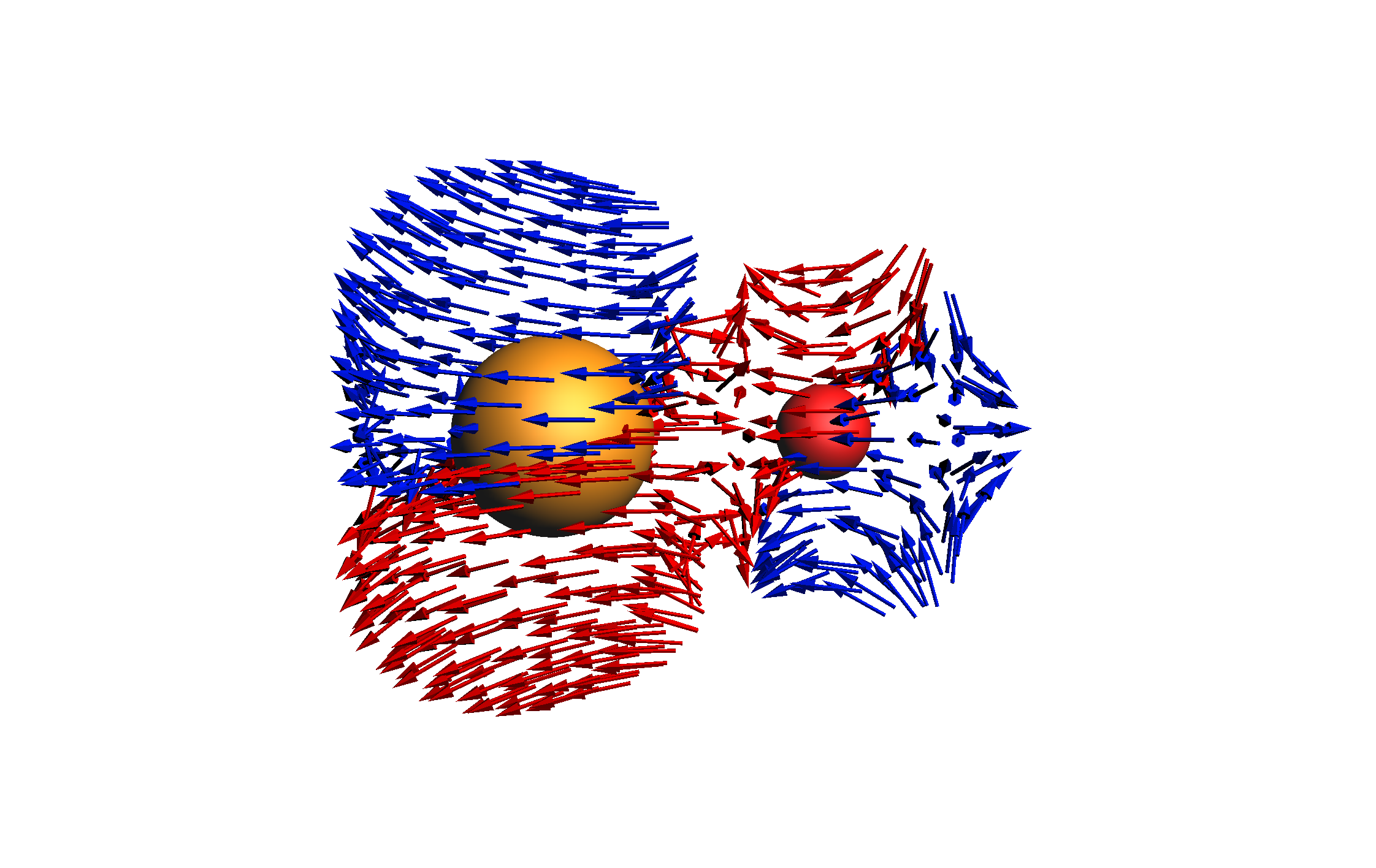}
\caption*{\scriptsize $\pi^{*}_{1/2}$: -0.141}
\end{minipage}
\centering
\begin{minipage}[b]{0.19\linewidth}
\includegraphics[width=\textwidth]{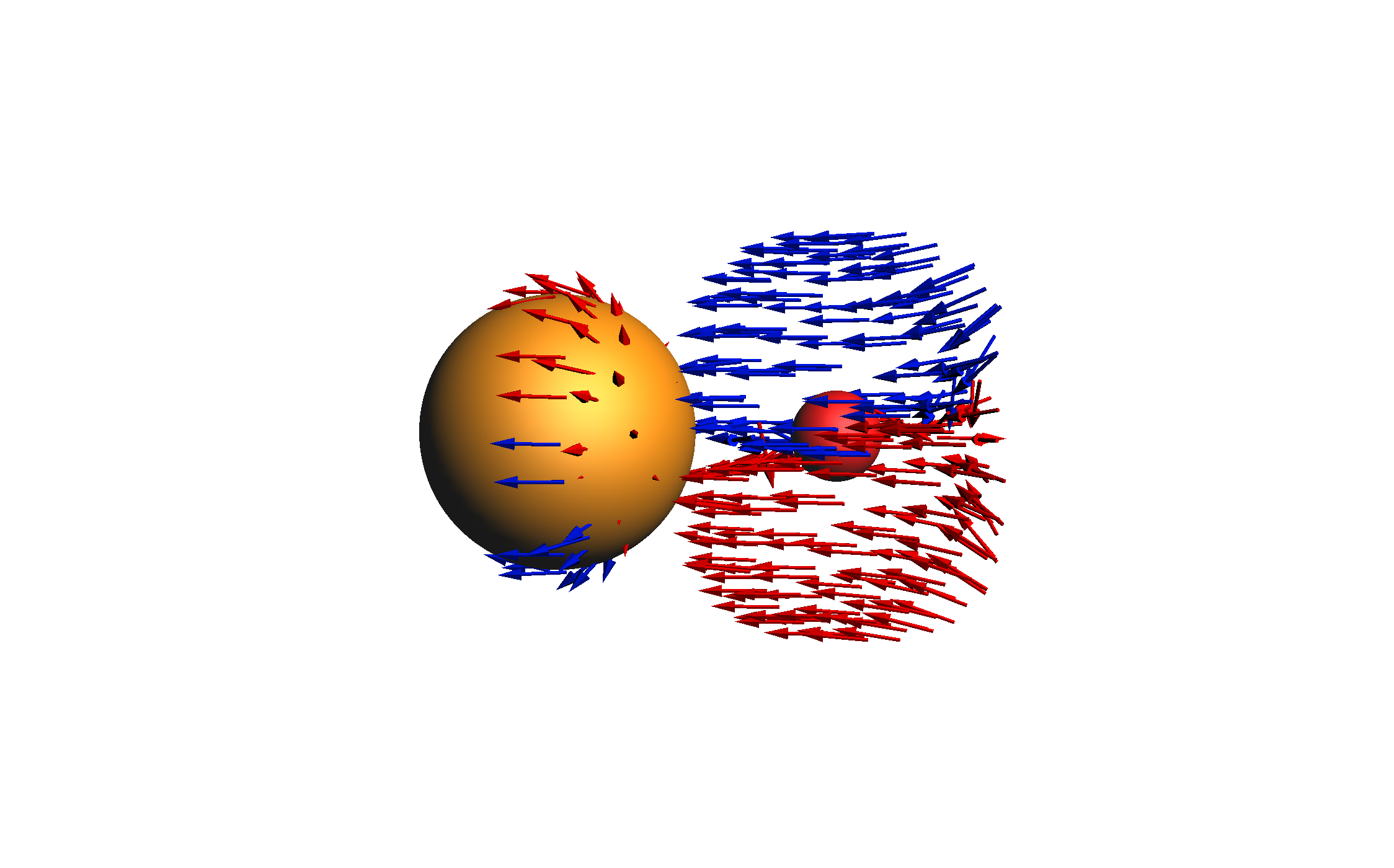}
\caption*{\scriptsize $\pi^{*}_{1/2}$: -0.190}
\end{minipage}
\centering
\begin{minipage}[b]{0.19\linewidth}
\centering
\includegraphics[width=\textwidth]{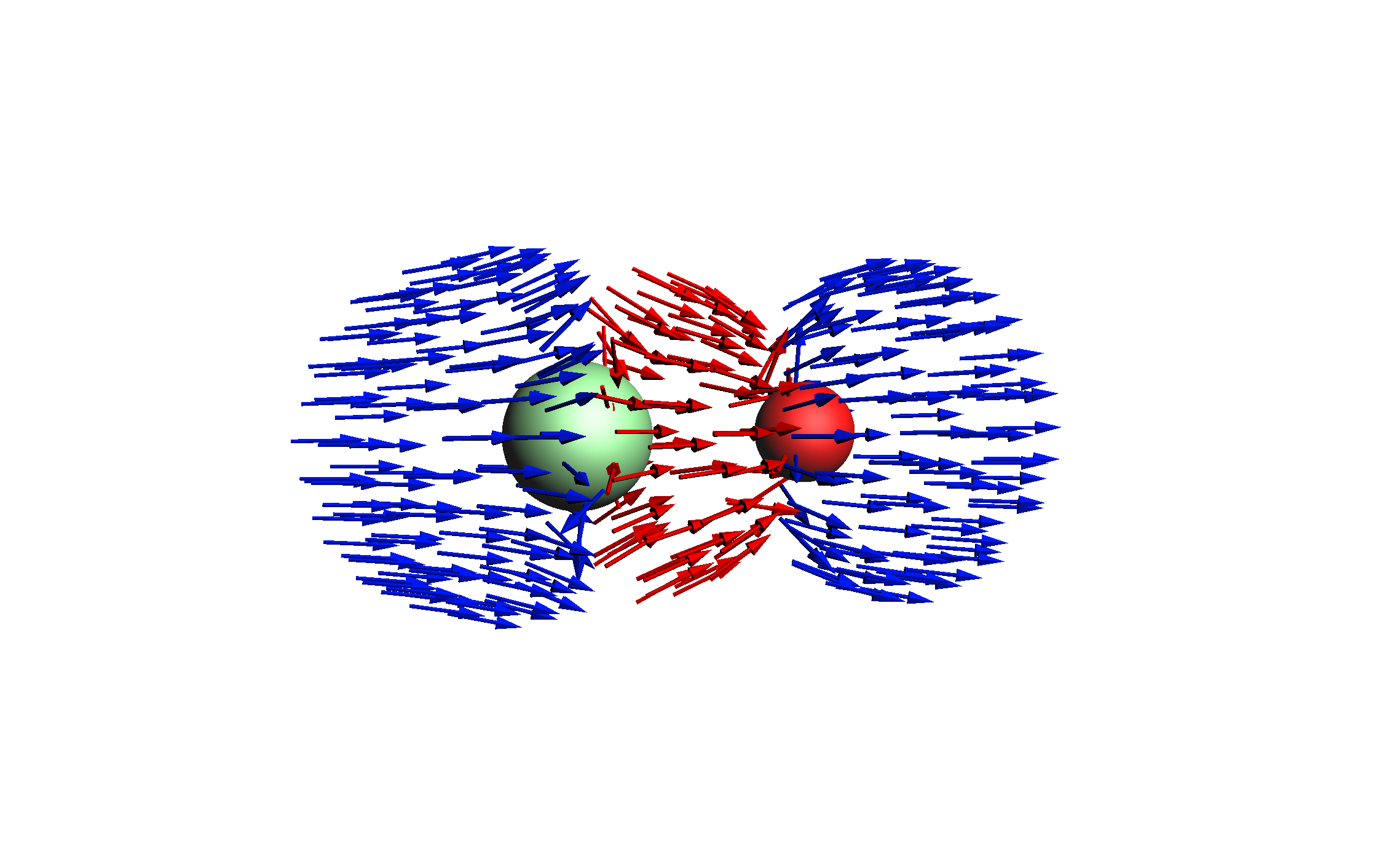}
\caption*{\scriptsize $\sigma_{1/2}$: -0.296}
\end{minipage}
\begin{minipage}[b]{0.19\linewidth}
\centering
\includegraphics[width=\textwidth]{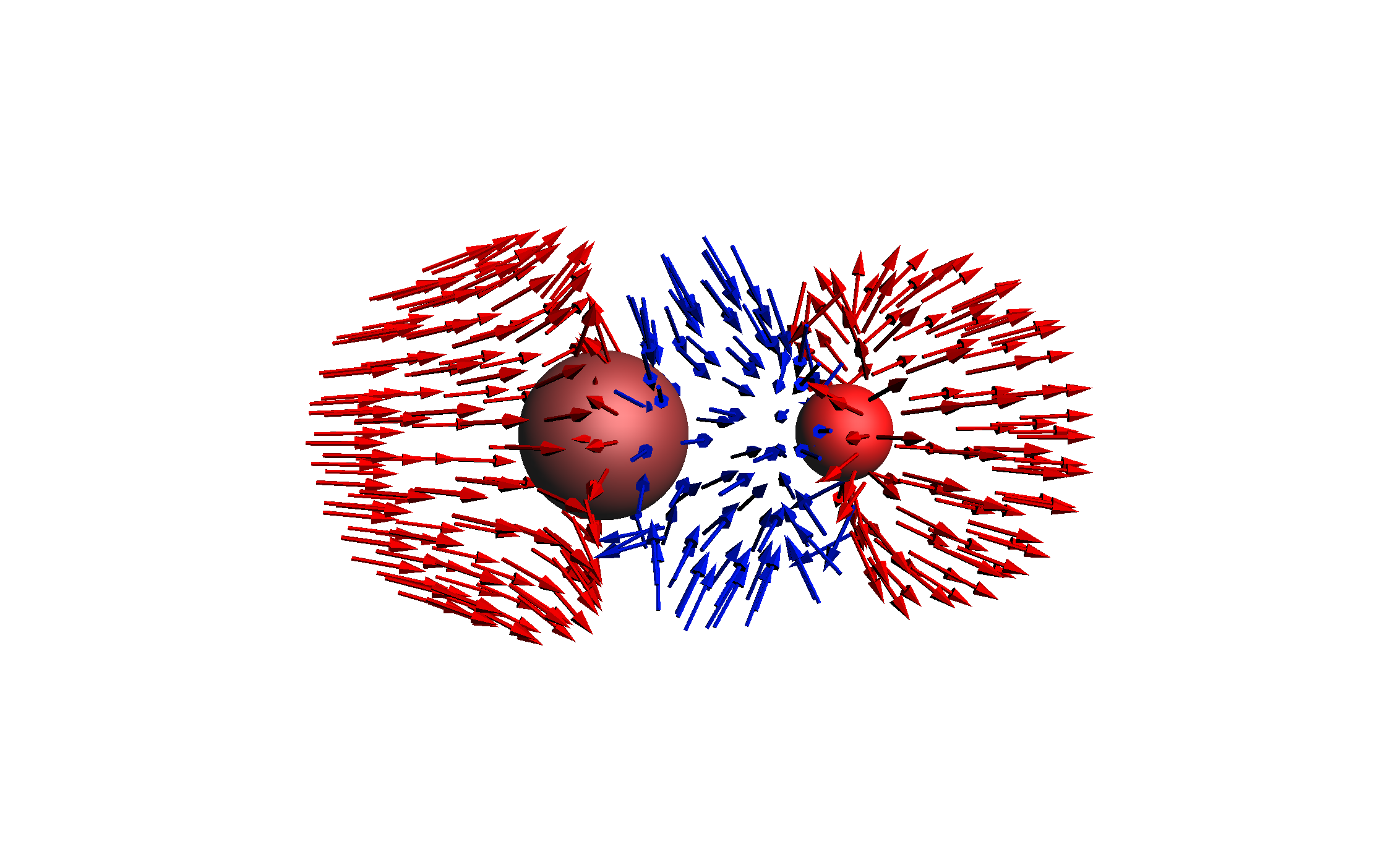}
\caption*{\scriptsize $\sigma_{1/2}$: -0.266}
\end{minipage}
\begin{minipage}[b]{0.19\linewidth}
\centering
\includegraphics[width=\textwidth]{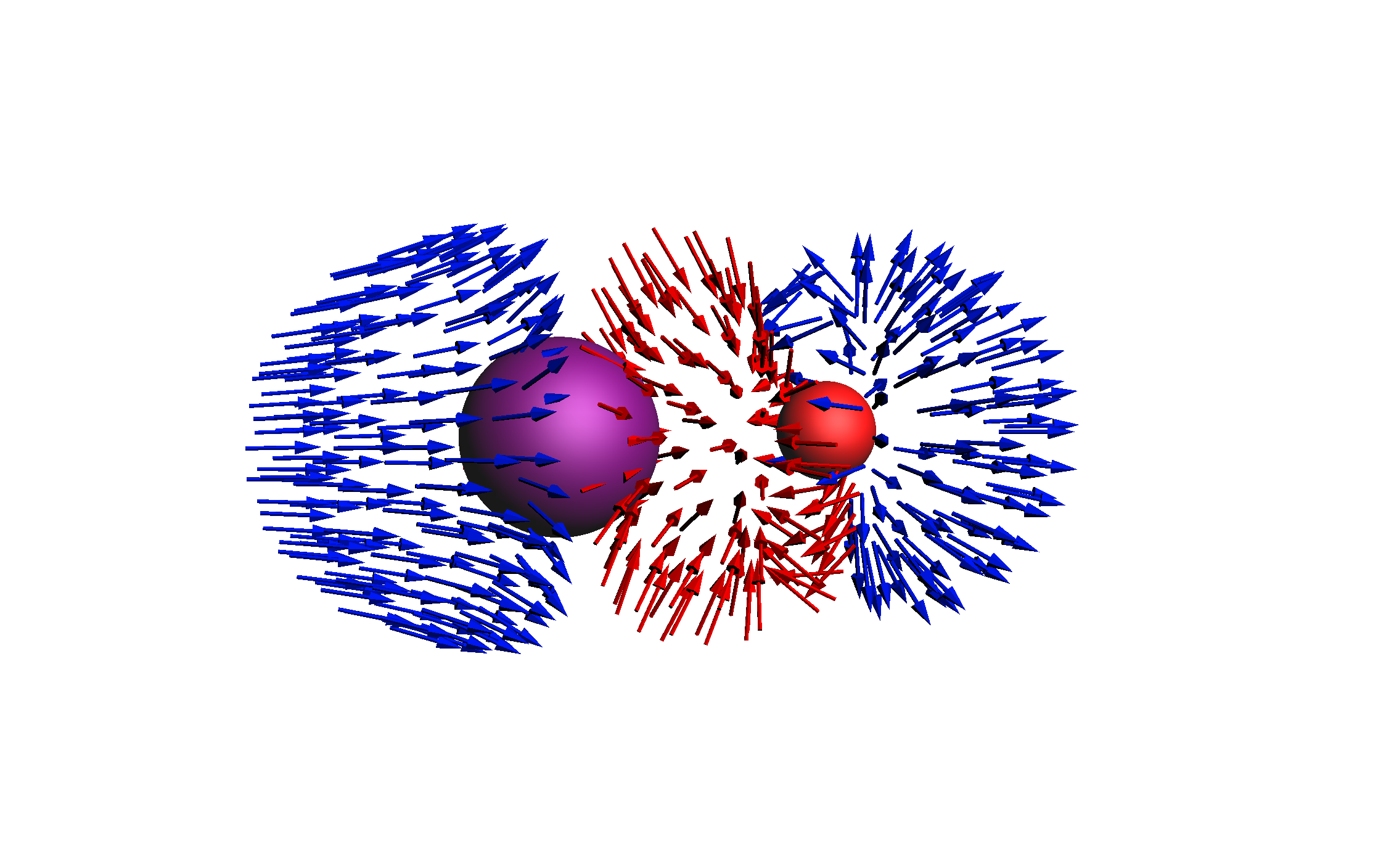}
\caption*{\scriptsize $\sigma_{1/2}$: -0.231}
\end{minipage}
\begin{minipage}[b]{0.19\linewidth}
\centering
\includegraphics[width=\textwidth]{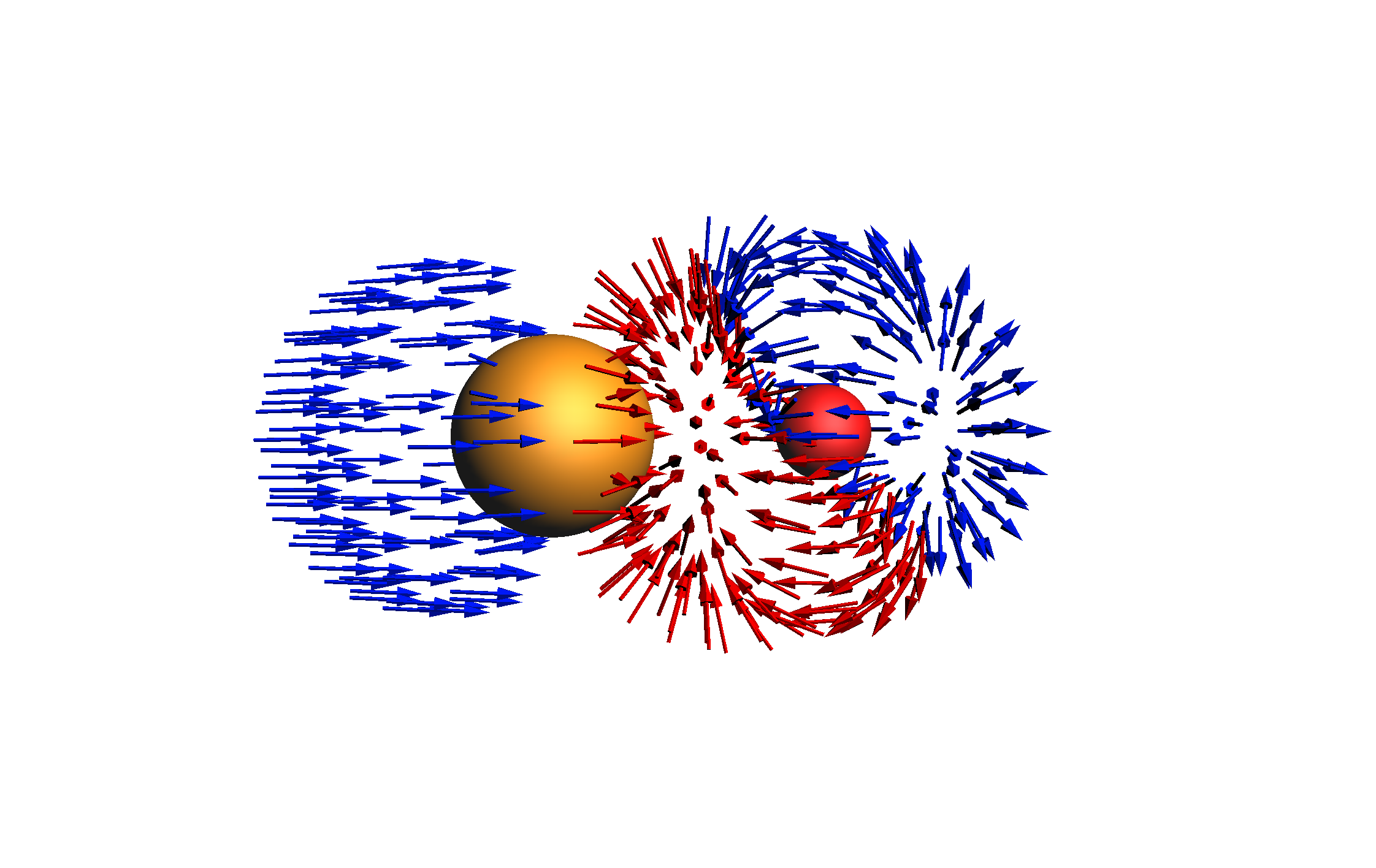}
\caption*{\scriptsize $\sigma_{1/2}$: -0.208}
\end{minipage}
\centering
\begin{minipage}[b]{0.19\linewidth}
\includegraphics[width=\textwidth]{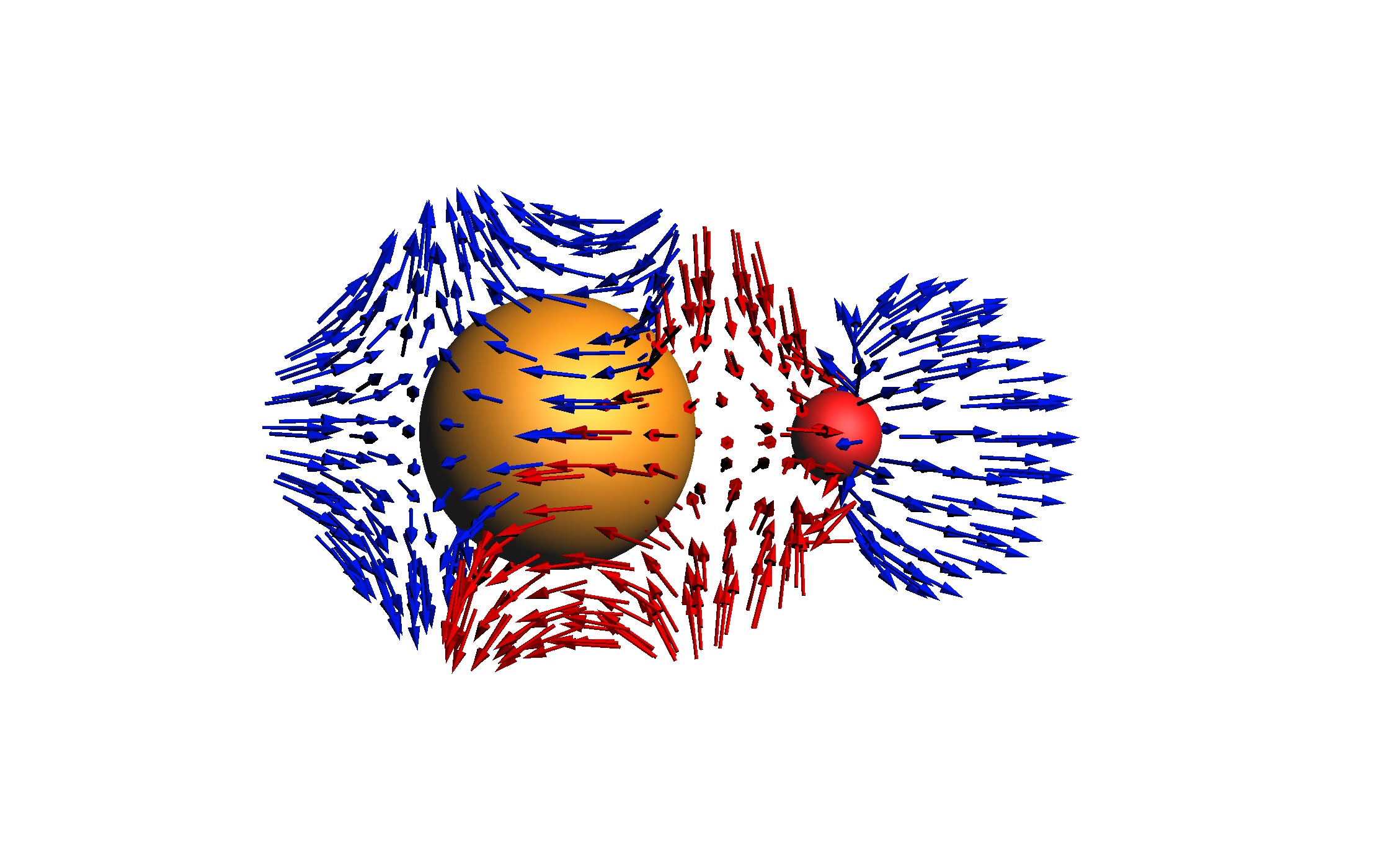}
\caption*{\scriptsize $\sigma_{1/2}$: -0.124}
\end{minipage}
\centering
\begin{minipage}[b]{0.19\linewidth}
\centering
\includegraphics[width=\textwidth]{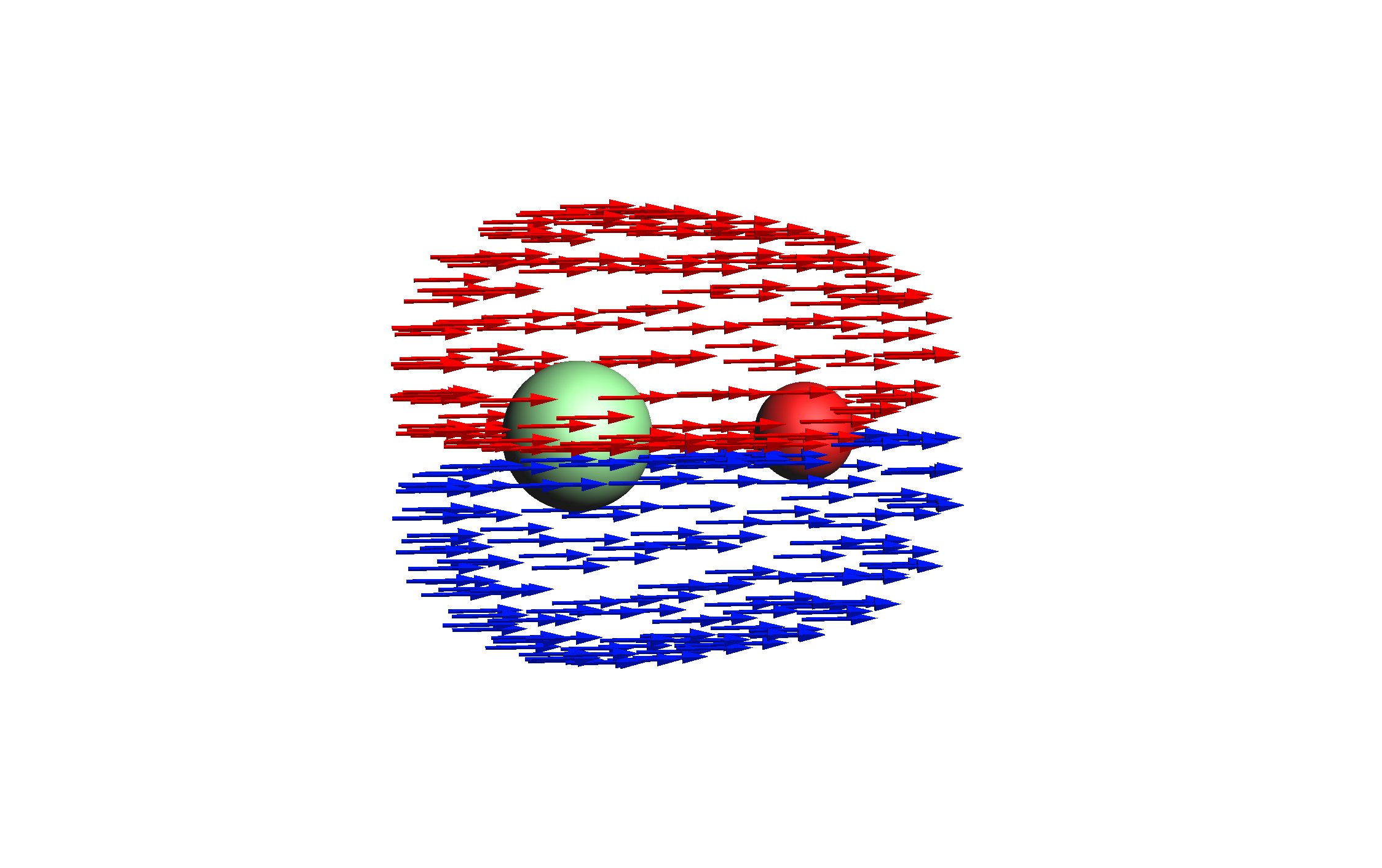}
\caption*{\scriptsize $\pi_{3/2}$: -0.304}
\end{minipage}
\begin{minipage}[b]{0.19\linewidth}
\centering
\includegraphics[width=\textwidth]{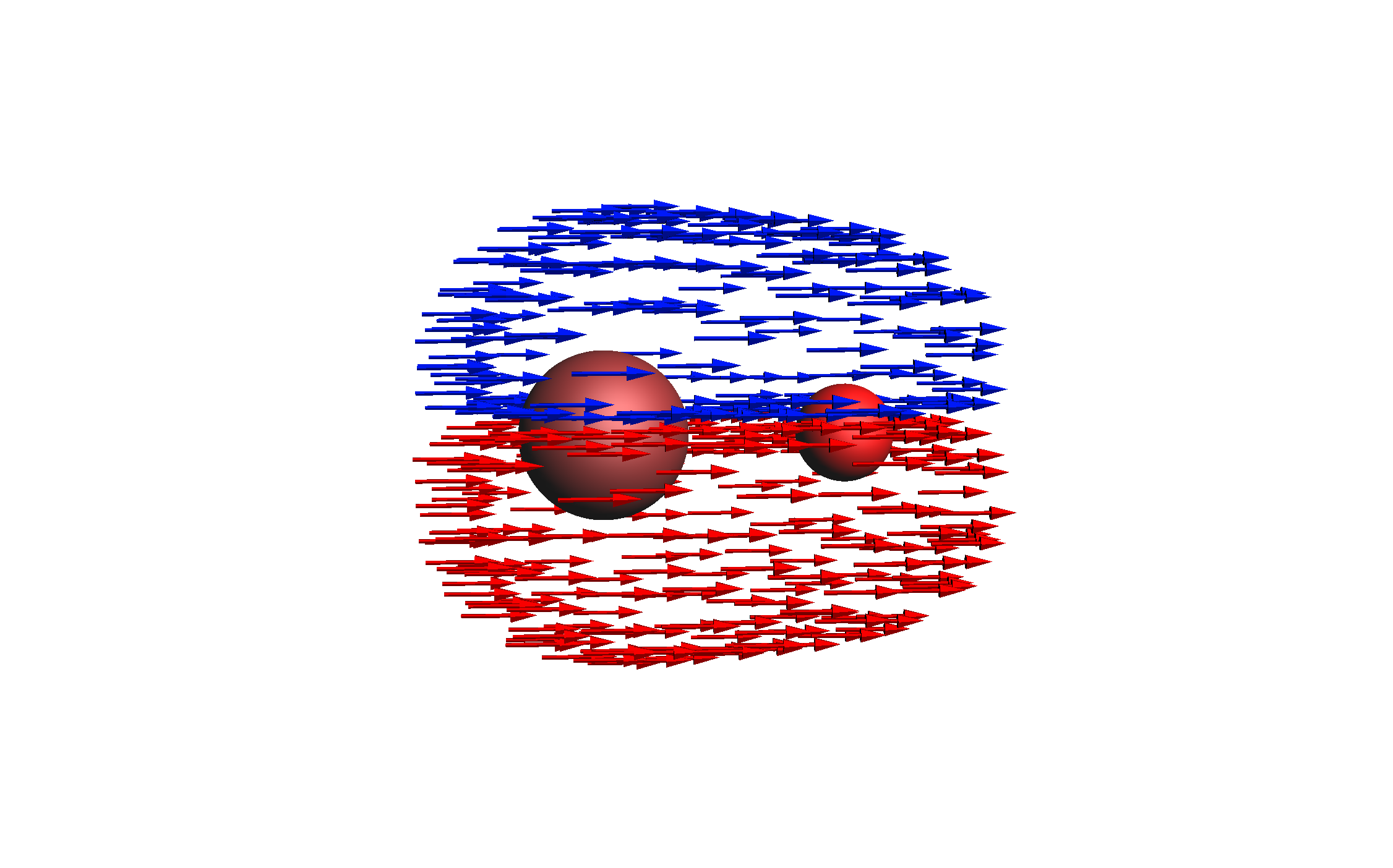}
\caption*{\scriptsize $\pi_{3/2}$: -0.279}
\end{minipage}
\begin{minipage}[b]{0.19\linewidth}
\centering
\includegraphics[width=\textwidth]{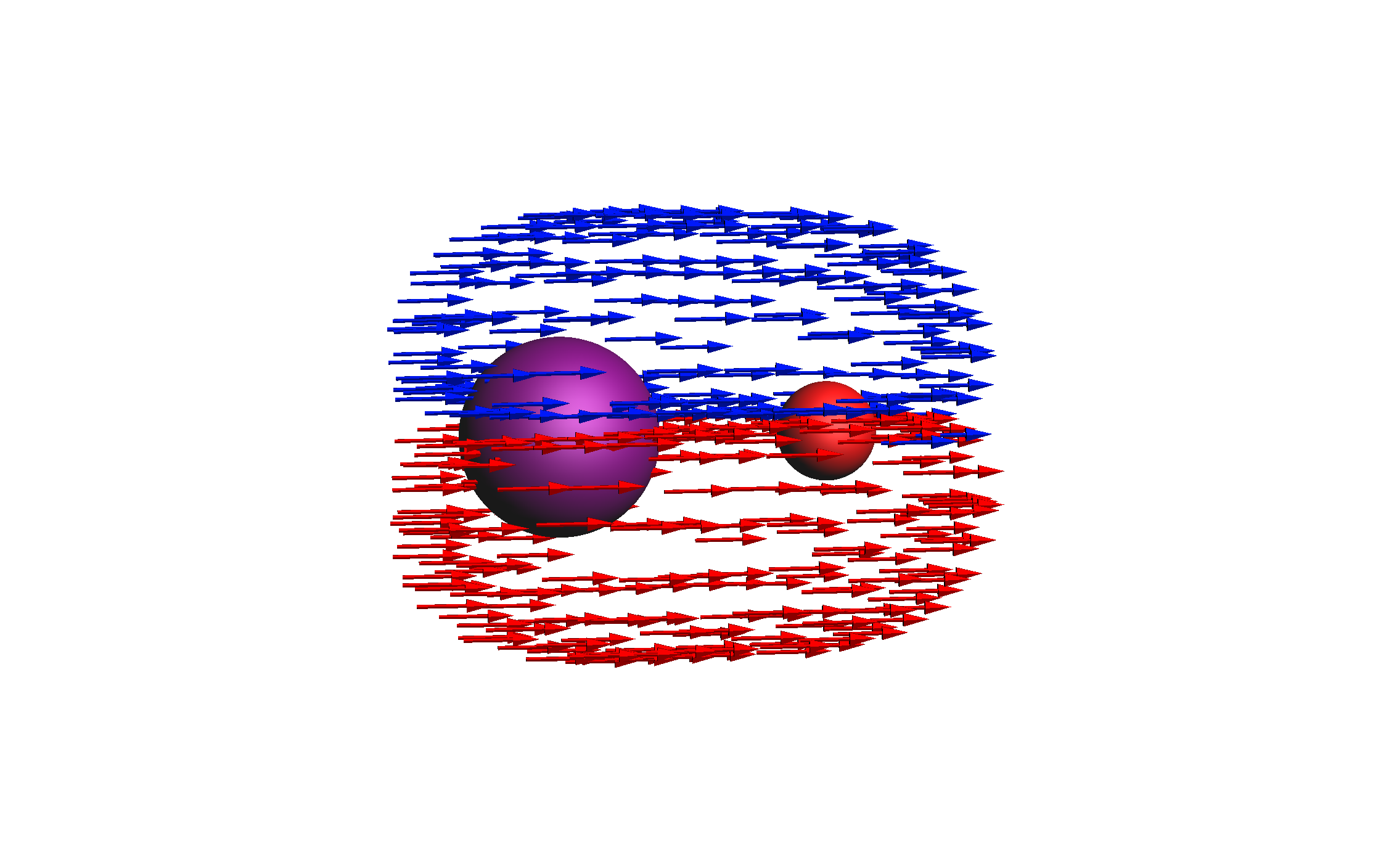}
\caption*{\scriptsize $\pi_{3/2}$: -0.260}
\end{minipage}
\begin{minipage}[b]{0.19\linewidth}
\centering
\includegraphics[width=\textwidth]{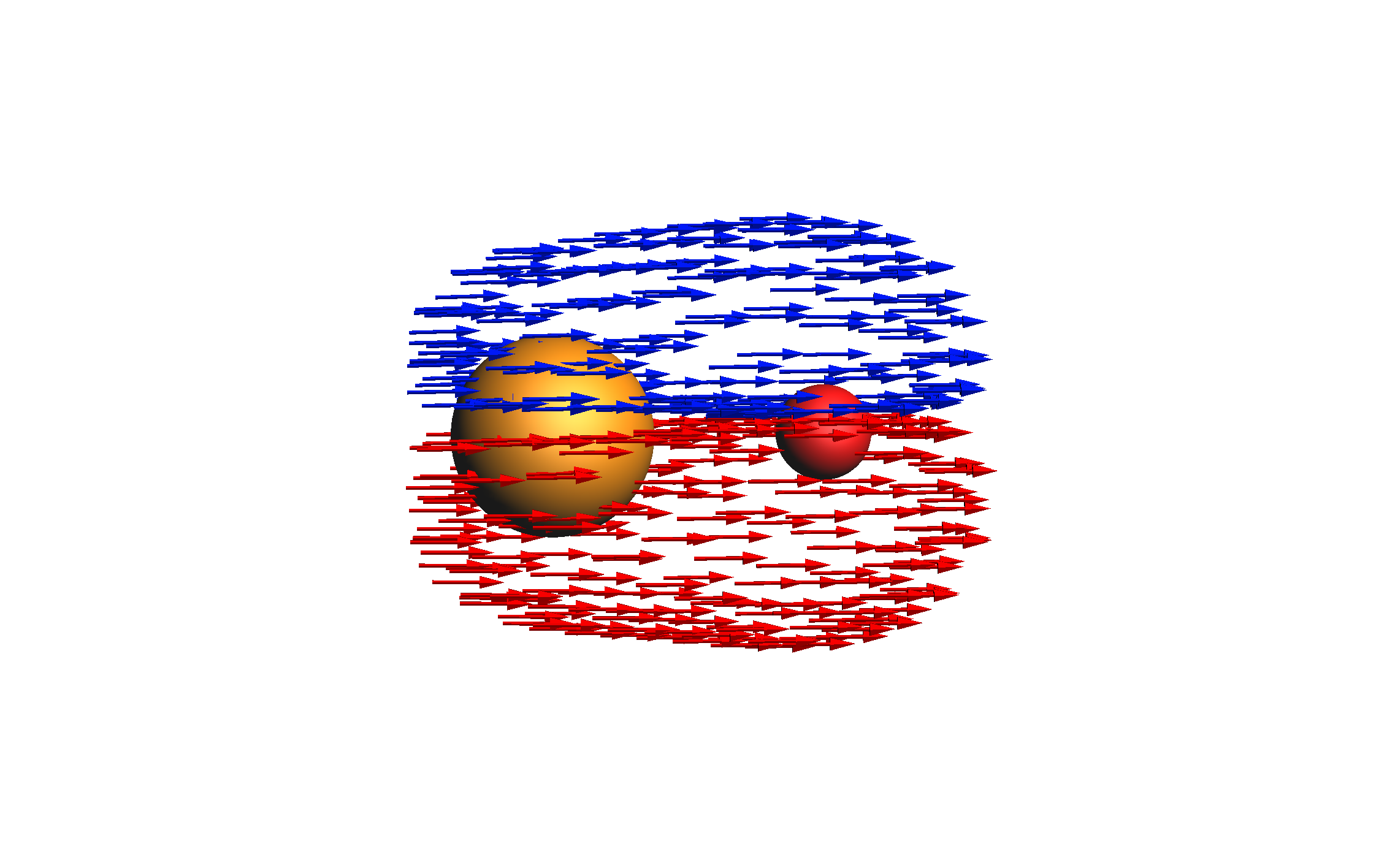}
\caption*{\scriptsize $\pi_{3/2}$: -0.241}
\end{minipage}
\centering
\begin{minipage}[b]{0.19\linewidth}
\includegraphics[width=\textwidth]{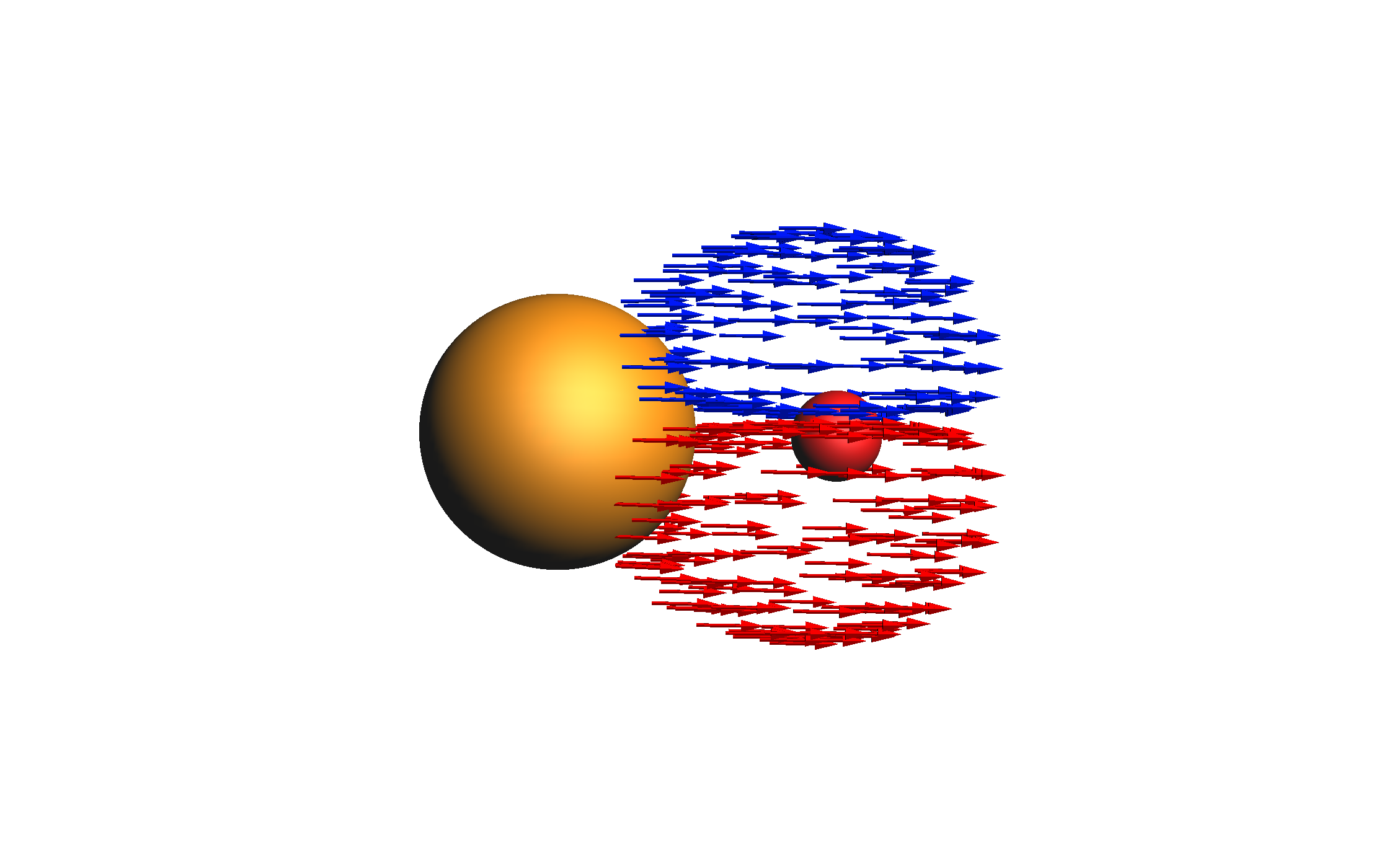}
\caption*{\scriptsize $\pi_{3/2}$: -0.213}
\end{minipage}
\centering
\begin{minipage}[b]{0.19\linewidth}
\centering
\includegraphics[width=\textwidth]{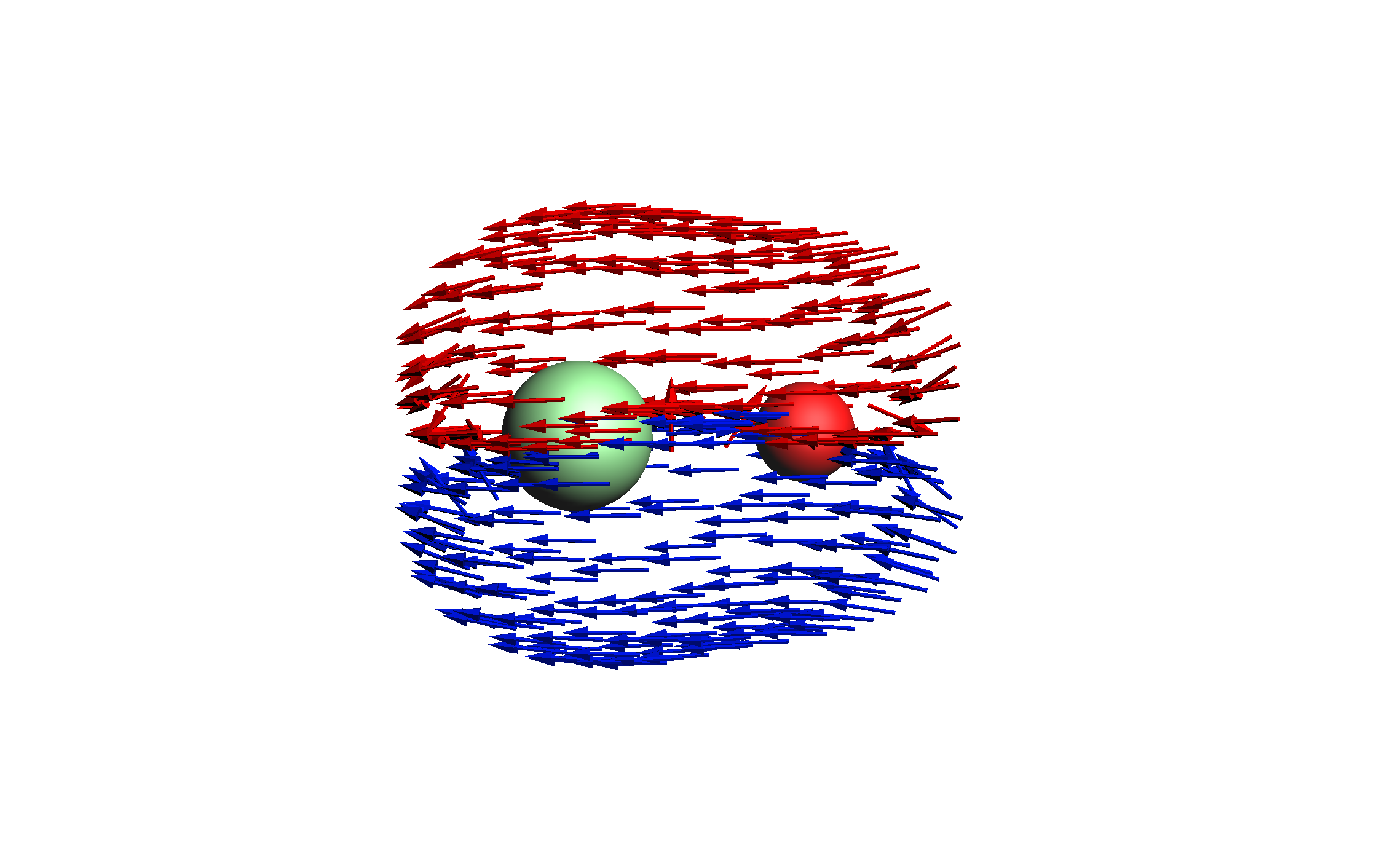}
\caption*{\scriptsize $\pi_{1/2}$: -0.307}
\end{minipage}
\begin{minipage}[b]{0.19\linewidth}
\centering
\includegraphics[width=\textwidth]{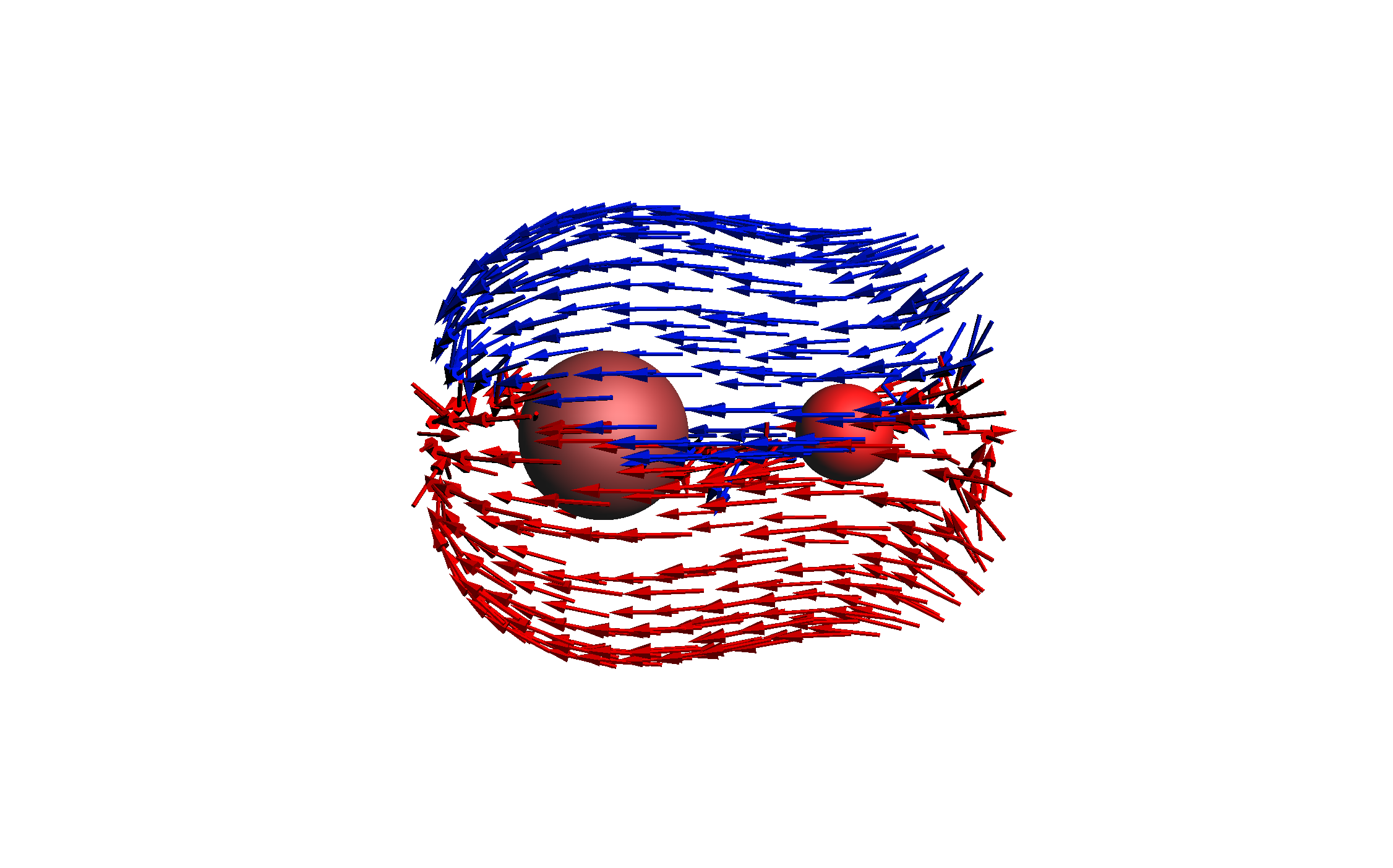}
\caption*{\scriptsize $\pi_{1/2}$: -0.287}
\end{minipage}
\begin{minipage}[b]{0.19\linewidth}
\centering
\includegraphics[width=\textwidth]{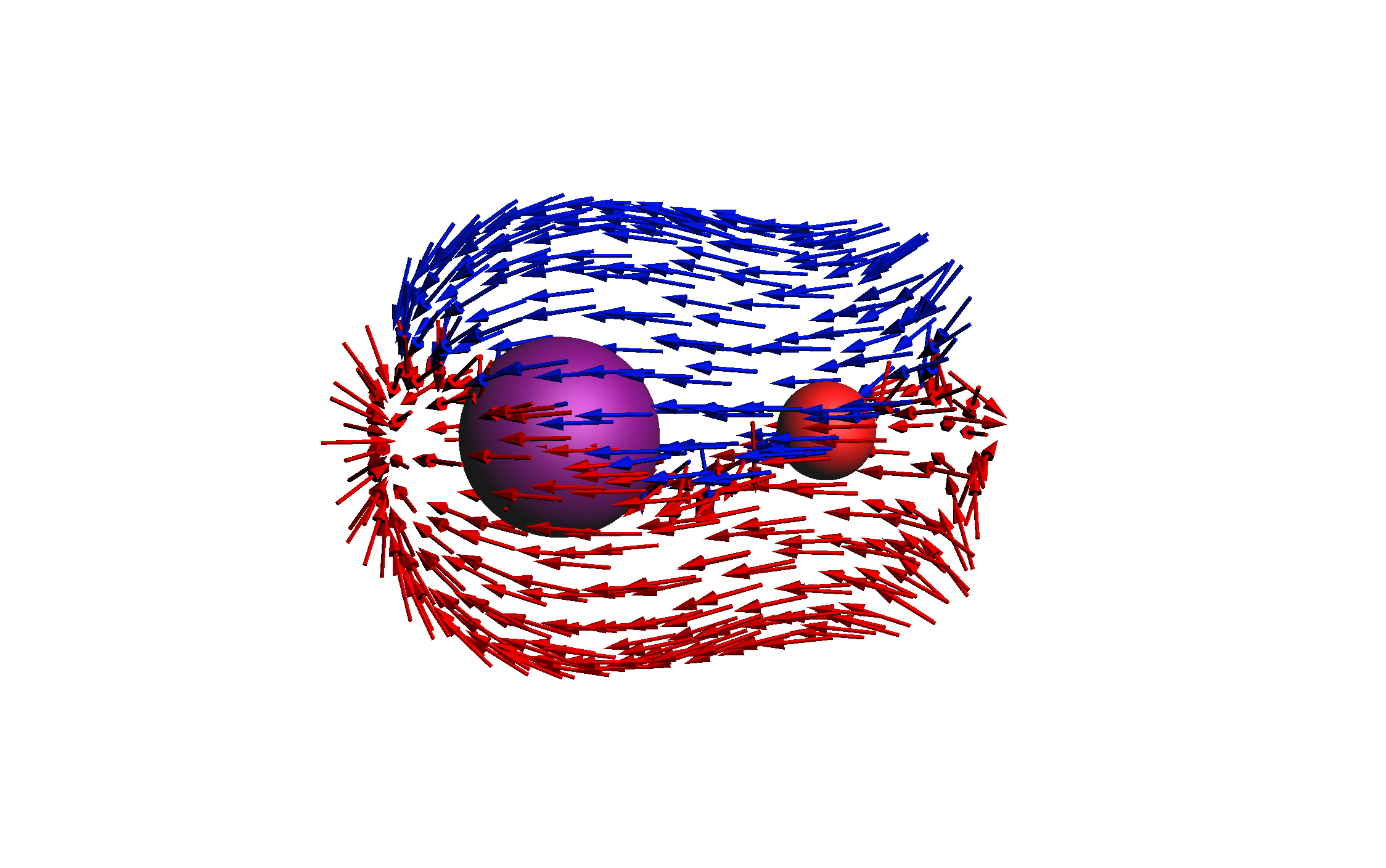}
\caption*{\scriptsize $\pi_{1/2}$: -0.275}
\end{minipage}
\begin{minipage}[b]{0.19\linewidth}
\centering
\includegraphics[width=\textwidth]{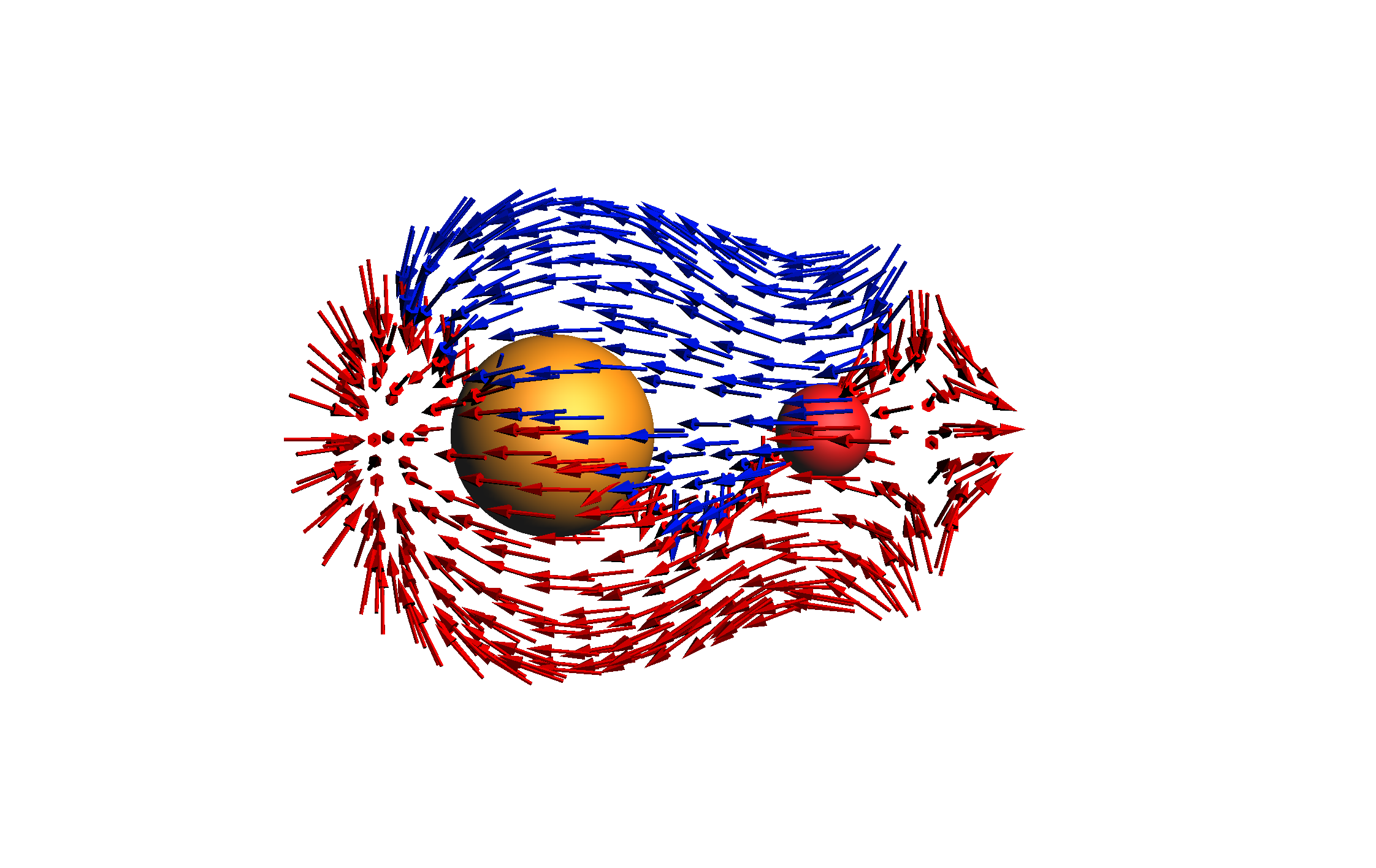}
\caption*{\scriptsize $\pi_{1/2}$: -0.295}
\end{minipage}
\centering
\begin{minipage}[b]{0.19\linewidth}
\includegraphics[width=\textwidth]{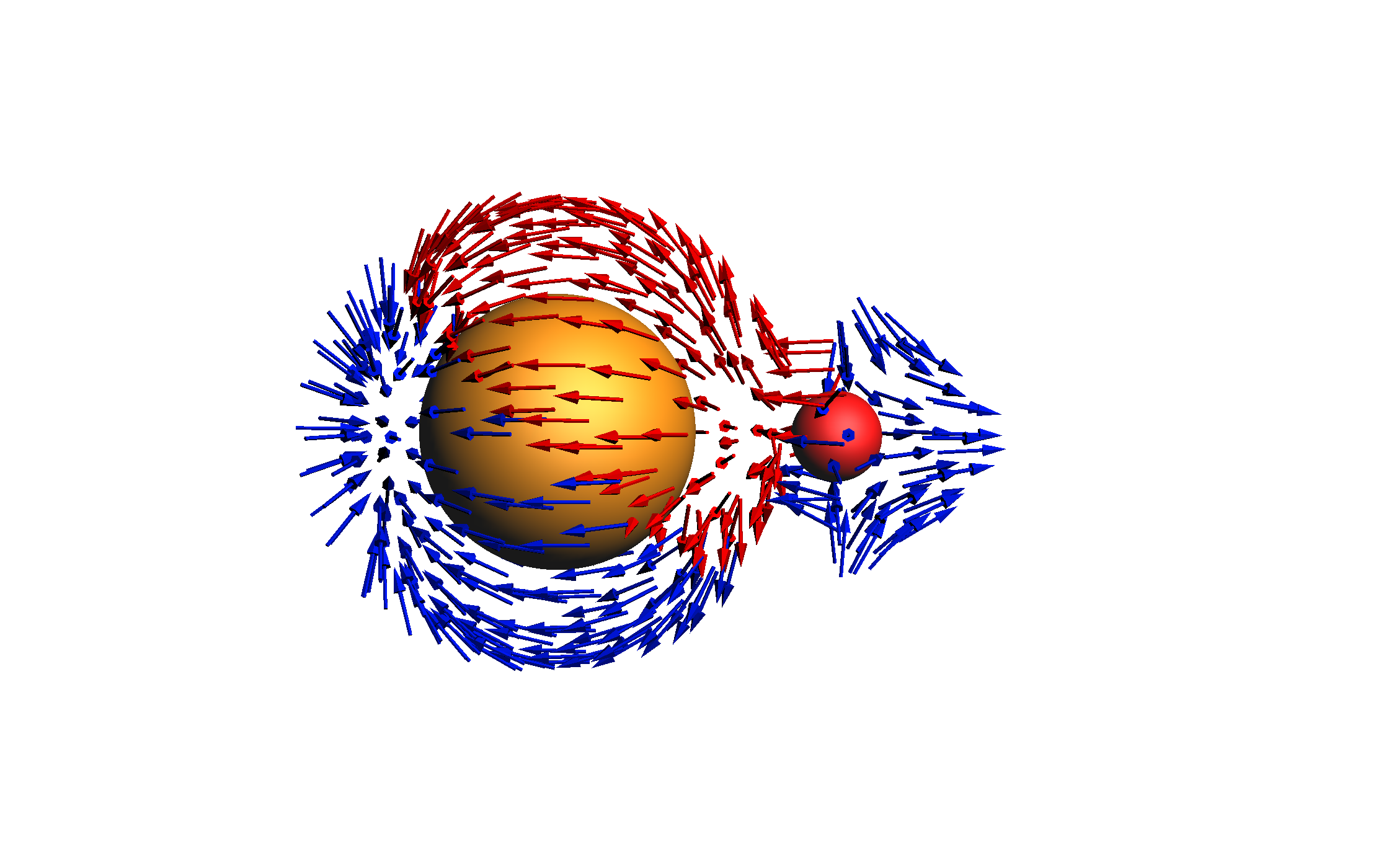}
\caption*{\scriptsize $\pi_{1/2}$: -0.440}
\end{minipage}

    \caption{SO-ZORA/QZ4P/Hartree-Fock (ADF) spinor magnetization plots (isosurfaces at 0.03 a.u.) and energies (in E$_h$) for the valence spinors of the XO$^-$ species (from left to right: X = Cl, Br, I, At, Ts).~\cite{paper:figures}}
    \label{spinor-magnetization-xo-all}
\end{figure}

In Figure~\ref{spinor-magnetization-xo-all}, through the plots of the spinor spin magnetization, we provide a graphical representation not of the aforementioned four-component spinors but the nearly equivalent SO-ZORA/Hartree-Fock/QZ4P two-component ones obtained with ADF.  As indicated in the ADF documentation (\url{http://www.scm.com}), these two-component spinors
\begin{equation}
\psi(\mathbf{r}) = 
\begin{bmatrix}
    \phi^{Re}_\alpha(\mathbf{r}) + i\phi^{Im}_\alpha(\mathbf{r}) \\
     \phi^{Re}_\beta(\mathbf{r}) + i\phi^{Im}_\beta(\mathbf{r}) 
\end{bmatrix}
\end{equation}
with $\{\phi^{Re}_a, \phi^{Im}_a\}$ the real and imaginary parts of the $a$($=\alpha/\beta$)-spin functions,
are represented by arrows showing the direction of the spin magnetization
\begin{equation}
\mathbf{m}(\mathbf{r}) = \psi^\dagger(\mathbf{r}) \boldsymbol{\Sigma} \psi(\mathbf{r}), \qquad \boldsymbol{\Sigma} = (\sigma_x, \sigma_y, \sigma_z) 
\end{equation}
($\sigma_x, \sigma_y, \sigma_z$ are the Pauli matrices) over space, drawn starting from points in space where the square root of the density
$\rho(\mathbf{r}) = \psi^\dagger(\mathbf{r}) \psi(\mathbf{r})$ is equal to a given isosurface value. The arrow colors represent
the sign (red: minus; blue: plus) of the phase factor $e^{i\theta}$ which, along with $\mathbf{m}$, fully determine the spinor.
In this representation, the direction of the arrows indicate, for instance, to which extent the spinor is an eigenfunction of spin: when this is verified the arrows
all point to the same direction.

\section{Triiodide}

In Table~\ref{tab:spectro-i3-dc-vs-2dcm} we compare the results for the Dirac-Coulomb and $^2$DC$^M$ Hamiltonians for the I$_3^-$, I$_3$ and I$_3^{2-}$ species.

\section{Halomethanes}

In Table~\ref{tab:xyz-halomethanes} we provide the Cartesian coordinates of the CH$_2$I$_2$ and CH$_2$IBr obtained by geometry optimization at the SO-ZORA/PBE/TZP level of theory. In Table~\ref{tab:bonds-angles-halomethanes} we provide the interatomic distances and angles for the same species, alongside the experimental numbers for CH$_2$I$_2$.


%

\onecolumngrid

\begin{sidewaystable}[htb]
\caption{Equilibrium structures ($R_e$, in \AA), harmonic frequencies ($\omega_e$, in cm$^{-1}$), vertical and adiabatic excitation energies ($T_v$ and $T_e$ respectively, in eV)
for the first four electronic states of the halogen monoxide radicals for EOM-IP and IHFS(1h0p) approaches, employing the molecular mean-field Hamiltonians with ($^2$DCG$^M$)
and without ($^2$DC$^M$) the Gaunt interaction.}
\begin{center}
\begin{tabular}{l p{5mm} c p{5mm} rrrr p{5mm} rrrr p{5mm} rrrr p{5mm} rrrr}
\hline
\hline
     &&          && \multicolumn{9}{c}{EOM-IP}      && \multicolumn{9}{c}{IHFS(1h0p)} \\ 
\cline{5-13}
\cline{15-23}
     &&          && \multicolumn{4}{c}{$^2$DC$^M$}  &&
                  \multicolumn{4}{c}{$^2$DCG$^M$} &&
                  \multicolumn{4}{c}{$^2$DC$^M$}  &&
                  \multicolumn{4}{c}{$^2$DCG$^M$} \\
\cline{5-8}
\cline{10-13}
\cline{15-18}
\cline{20-23}
Species	&& State $_\Omega$ && $R_e$  &$\omega_e$&$T_e$&$T_v$&&  $R_e$ &$\omega_e$&$T_e$&$T_v$  &&
                  $R_e$	 &$\omega_e$&$T_e$&$T_v$&&  $R_e$ &$\omega_e$&$T_e$&$T_v$  \\
\hline
\hline
ClO     && X$_{ 3/2}$    && 1.582  &	900	&	     0	&	     0	&&	1.582	&	900	&	     0	&	     0	&&	1.582	&	900      &           0	&	     0	&&	1.582	&	900	&	     0	&	     0	\\
        && X$_{ 1/2}$    && 1.583  &	898	&	0.0423	&	0.0423	&&	1.583	&	898	&	0.0396	&	0.0396	&&	1.583	&	898      &      0.0423	&	0.0423	&&	1.583	&	898	&	0.0396	&	0.0396	\\
        && A$_{ 3/2}$    && 1.889  &	504	&	3.9530	&	4.8206	&&	1.889	&	504	&	3.9520	&	4.8210	&&	1.889	&	504      &      3.9535	&	4.8209	&&	1.889	&	504	&	3.9525	&	4.8213	\\
        && A$_{ 1/2}$    && 1.887  &	505	&	4.0192	&	4.8800	&&	1.887	&	505	&	4.0153	&	4.8776	&&	1.887	&	505      &      4.0197	&	4.8804	&&	1.888	&	505	&	4.0157	&	4.8780	\\
\hline
BrO     && X$_{ 3/2}$    && 1.714  &	785	&	     0	&	     0	&&	1.714	&	785	&	     0	&	     0	&&	1.714	&	785      &           0	&	     0	&&	1.714	&	785	&	     0	&	     0	\\
        && X$_{ 1/2}$    && 1.719  &	776	&	0.1291	&	0.1295	&&	1.719	&	776	&	0.1256	&	0.1260	&&	1.719	&	776      &      0.1291	&	0.1294	&&	1.719	&	776	&	0.1256	&	0.1260	\\
        && A$_{ 3/2}$    && 1.975  &	485	&	3.3429	&	3.9831	&&	1.975	&	485	&	3.3421	&	3.9837	&&	1.975	&	484      &      3.3430	&	3.9831	&&	1.975	&	485	&	3.3422	&	3.9837	\\
        && A$_{ 1/2}$    && 1.969  &	483	&	3.5636	&	4.1707	&&	1.969	&	483	&	3.5586	&	4.1676	&&	1.969	&	483      &      3.5643	&	4.1723	&&	1.969	&	483	&	3.5593	&	4.1692	\\
\hline
IO      && X$_{ 3/2}$    && 1.875  &	722	&	     0	&	     0	&&	1.876	&	722	&	     0	&	     0	&&	1.875	&	722      &           0	&	     0	&&	1.876	&	722	&	     0	&	     0	\\
        && X$_{ 1/2}$    && 1.887  &	702	&	0.2757	&	0.2777	&&	1.887	&	702	&	0.2716	&	0.2734	&&	1.887	&	702      &      0.2758	&	0.2777	&&	1.887	&	702	&	0.2717	&	0.2734	\\
        && A$_{ 3/2}$    && 2.095  &	492	&	2.7626	&	3.2206	&&	2.095	&	492	&	2.7620	&	3.2163	&&	2.095	&	492      &      2.7626	&	3.2206	&&	2.095	&	492	&	2.7620	&	3.2163	\\
        && A$_{ 1/2}$    && 2.076  &	495	&	3.0915	&	3.4716	&&	2.077	&	495	&	3.0870	&	3.4645	&&	2.077	&	495      &      3.0948	&	3.4773	&&	2.077	&	495	&	3.0903	&	3.4701	\\
\hline
AtO     && X$_{ 3/2}$    && 1.973  &	644	&	     0	&	     0	&&	1.973	&	644	&	     0	&	     0	&&	1.973	&	644      &           0	&	     0	&&	1.973	&	644	&	     0	&	     0	\\
        && X$_{ 1/2}$    && 2.018  &	585	&	0.7029	&	0.7233	&&	2.018	&	585	&	0.6985	&	0.7189	&&	2.018	&	585      &      0.7034	&	0.7238	&&	2.018	&	585	&	0.6990	&	0.7194	\\
        && A$_{ 3/2}$    && 2.209  &	435	&	2.3436	&	2.7841	&&	2.209	&	435	&	2.3442	&	2.7857	&&	2.209	&	435      &      2.3437	&	2.7841	&&	2.209	&	435	&	2.3442	&	2.7857	\\
        && A$_{ 1/2}$    && 2.164  &	447	&	2.7670	&	3.0594	&&	2.165	&	448	&	2.7647	&	3.0580	&&	2.166	&	445      &      2.7789	&	3.0730	&&	2.166	&	446	&	2.7765	&	3.0715	\\
\hline
TsO     && X$_{ 3/2}$    && 2.137  &	573	&	     0	&	     0	&&	2.138	&	573	&	     0	&	     0	&&	2.137	&	573      &           0	&	     0	&&	2.138	&	573	&	     0	&	     0	\\
        && X$_{ 1/2}$    && 2.263  &	457	&	1.0825	&	1.1954	&&	2.263	&	457	&	1.0808	&	1.1915	&&	2.259	&	467      &      1.1106	&	1.2194	&&	2.258	&	468	&	1.1103	&	1.2165	\\
        && A$_{ 3/2}$    && 2.358  &	416	&	1.8899	&	2.2279	&&	2.358	&	415	&	1.8905	&	2.2265	&&	2.358	&	413      &      1.8897  &	2.2277  &&	2.358	&	413 	&	1.8904  &	2.2262 \\
        && B$_{ 1/2}$    && 2.281  &	503	&	2.1102	&	2.2901	&&	2.281	&	504	&	2.1085	&	2.2870	&&	2.279	&	508      &      2.1229	&	2.3014	&&	2.280	&	508	&	2.1213	&	2.2983	\\
\hline
\hline
\end{tabular}
\end{center}
\label{tab:spectro-monoxide-radicals}
\end{sidewaystable}%

\begin{table}[htb]
\caption{Comparison of total energies (in $E_h$) obtained with the $ {^2}$DC$^M$ and DC Hamiltonians for the XO and XO$^-$ species. The $\Delta E$ values correspond to the difference (in eV)
         between the tho Hamiltonians, whereas $\Delta(\Delta E_a)$ and $\Delta(\Delta E_b)$ correspond to the differences between Hamiltonians to 
         relative energies with respect to the anion and neutral ground state, respectively (in eV).}
\centering
\begin{tabular}{l c c cc p{0.2cm} ccc}
\hline
\hline
species	&	R(\AA)	& $\Omega$	&$ {^2}$DC$^M$ ($E_h$)	&	DC ($E_h$)	&&   $\Delta E$ (eV)	&$\Delta(\Delta E_a)$ (eV)   & $\Delta(\Delta E_b)$ (eV)\\
\hline
ClO$^-$	&	1.582	&	0	&	-536.405099	&	-536.405224	&&	3.41E-03	&		        &		\\
ClO	&		&	3/2     &	-536.330849	&	-536.330975	&&	3.43E-03	&	1.60E-05	&		\\
	&		&	1/2     &	-536.329299	&	-536.329421	&&	3.33E-03	&	-8.61E-05	&	-1.02E-04	\\
	&		&	3/2     &	-536.153696	&	-536.153822	&&	3.42E-03	&	9.85E-06	&	-6.11E-06	\\
	&		&	1/2     &	-536.151516	&	-536.151639	&&	3.35E-03	&	-5.96E-05	&	-7.56E-05	\\
\cline{2-9}
BrO$^-$	&	1.714	&	0       &	-2680.807249	&	-2680.807526	&&	7.56E-03	&		        &		\\
BrO	&		&	3/2     &	-2680.728854	&	-2680.729132	&&	7.58E-03	&	1.74E-05	&		\\
	&		&	1/2     &	-2680.724099	&	-2680.724374	&&	7.48E-03	&	-8.09E-05	&	-9.83E-05	\\
	&		&	3/2     &	-2680.582479	&	-2680.582757	&&	7.57E-03	&	9.26E-06	&	-8.18E-06	\\
	&		&	1/2     &	-2680.575584	&	-2680.575860	&&	7.51E-03	&	-4.95E-05	&	-6.69E-05	\\
\cline{2-9}
IO$^-$	&	1.875	&	0       &	-7191.595926	&	-7191.596051	&&	3.38E-03	&		        &		\\
IO	&		&	3/2     &	-7191.514120	&	-7191.514245	&&	3.39E-03	&	1.07E-05	&		\\
	&		&	1/2     &	-7191.503920	&	-7191.504041	&&	3.31E-03	&	-7.05E-05	&	-8.11E-05	\\
	&		&	3/2     &	-7191.395764	&	-7191.395889	&&	3.39E-03	&	8.12E-06	&	-2.55E-06	\\
	&		&	1/2     &	-7191.386543	&	-7191.386666	&&	3.33E-03	&	-4.62E-05	&	-5.68E-05	\\
\cline{2-9}
AtO$^-$	&	1.973	&	0       &	-22989.611030	&	-22989.611463	&&	1.18E-02	&		        &		\\
AtO	&		&	3/2     &	-22989.541893	&	-22989.542326	&&	1.18E-02	&	-7.28E-07	&		\\
	&		&	1/2     &	-22989.515315	&	-22989.515745	&&	1.17E-02	&	-8.41E-05	&	-8.33E-05	\\
	&		&	3/2     &	-22989.439577	&	-22989.440011	&&	1.18E-02	&	4.25E-07	&	1.15E-06	\\
	&		&	1/2     &	-22989.429464	&	-22989.429896	&&	1.18E-02	&	-3.43E-05	&	-3.36E-05	\\
\cline{2-9}
TsO$^-$	&	2.137	&	0       &	-53601.220032	&	-53601.220265	&&	6.33E-03	&		        &		\\
TsO	&		&	3/2     &	-53601.172734	&	-53601.172966	&&	6.31E-03	&	-2.10E-05	&		\\
	&		&	1/2     &	-53601.128806	&	-53601.129035	&&	6.24E-03	&	-9.05E-05	&	-6.95E-05	\\
	&		&	3/2     &	-53601.090860	&	-53601.091091	&&	6.31E-03	&	-2.26E-05	&	-1.63E-06	\\
	&		&	1/2     &	-53601.088574	&	-53601.088807	&&	6.32E-03	&	-5.69E-06	&	1.53E-05	\\
\hline
\hline
\end{tabular}
\label{tab:spectro-monoxide-radical-dc-vs-2dcm}
\end{table}

\begin{table}[htb]
\caption{Composition of the ground and excited-state $^2$DC$^M$ EOM-IP and IHFS(1h0p) wavefunctions for ClO at $R_e$ = 1.582 \AA. Only main contributions ($(r_i, c_i) > 0.1$) are shown.} 
\centering
\begin{tabular}{l ccc c ccc}
\hline
            & \multicolumn{3}{c}{EOM-IP}
           && \multicolumn{3}{c}{IHFS(1h0p)}\\
\cline{2-4}
\cline{6-8}
 $\Omega$   & E (eV) & configuration & $r_i$
           && E (eV) & configuration & $c_i$ \\
\hline
3/2 & 0.0000 & $\pi^{(2)}_{1/2}\;\pi^{(1)}_{3/2}\;\sigma^{(2)}_{1/2}\;\pi^{*(2)}_{1/2}\;\pi^{*(2)}_{3/2}$ & -0.141  && 0.0000 & $\pi^{(2)}_{1/2}\;\pi^{(1)}_{3/2}\;\sigma^{(2)}_{1/2}\;\pi^{*(2)}_{1/2}\;\pi^{*(2)}_{3/2}$ & -0.164  \\
    &        & $\pi^{(2)}_{1/2}\;\pi^{(2)}_{3/2}\;\sigma^{(2)}_{1/2}\;\pi^{*(2)}_{1/2}\;\pi^{*(1)}_{3/2}$ & -0.953  &&        & $\pi^{(2)}_{1/2}\;\pi^{(2)}_{3/2}\;\sigma^{(2)}_{1/2}\;\pi^{*(2)}_{1/2}\;\pi^{*(1)}_{3/2}$ & -0.986  \\
\cline{2-8}
1/2 & 0.0423 & $\pi^{(1)}_{1/2}\;\pi^{(2)}_{3/2}\;\sigma^{(2)}_{1/2}\;\pi^{*(2)}_{1/2}\;\pi^{*(2)}_{3/2}$ &  0.140  && 0.0423 & $\pi^{(1)}_{1/2}\;\pi^{(2)}_{3/2}\;\sigma^{(2)}_{1/2}\;\pi^{*(2)}_{1/2}\;\pi^{*(2)}_{3/2}$ &  0.162  \\
    &        & $\pi^{(2)}_{1/2}\;\pi^{(2)}_{3/2}\;\sigma^{(2)}_{1/2}\;\pi^{*(1)}_{1/2}\;\pi^{*(2)}_{3/2}$ & -0.953  &&        & $\pi^{(2)}_{1/2}\;\pi^{(2)}_{3/2}\;\sigma^{(2)}_{1/2}\;\pi^{*(1)}_{1/2}\;\pi^{*(2)}_{3/2}$ & -0.987  \\
\cline{2-8}
3/2 & 4.8206 & $\pi^{(2)}_{1/2}\;\pi^{(1)}_{3/2}\;\sigma^{(2)}_{1/2}\;\pi^{*(2)}_{1/2}\;\pi^{*(2)}_{3/2}$ &  0.945  && 4.8209 & $\pi^{(2)}_{1/2}\;\pi^{(1)}_{3/2}\;\sigma^{(2)}_{1/2}\;\pi^{*(2)}_{1/2}\;\pi^{*(2)}_{3/2}$ & -0.989  \\
    &        & $\pi^{(2)}_{1/2}\;\pi^{(2)}_{3/2}\;\sigma^{(2)}_{1/2}\;\pi^{*(2)}_{1/2}\;\pi^{*(1)}_{3/2}$ & -0.157  &&        & $\pi^{(2)}_{1/2}\;\pi^{(2)}_{3/2}\;\sigma^{(2)}_{1/2}\;\pi^{*(2)}_{1/2}\;\pi^{*(1)}_{3/2}$ &  0.147  \\
\cline{2-8}
1/2 & 4.8800 & $\pi^{(1)}_{1/2}\;\pi^{(2)}_{3/2}\;\sigma^{(2)}_{1/2}\;\pi^{*(2)}_{1/2}\;\pi^{*(2)}_{3/2}$ & -0.934  && 4.8804 & $\pi^{(1)}_{1/2}\;\pi^{(2)}_{3/2}\;\sigma^{(2)}_{1/2}\;\pi^{*(2)}_{1/2}\;\pi^{*(2)}_{3/2}$ &  0.978  \\
    &        & $\pi^{(2)}_{1/2}\;\pi^{(2)}_{3/2}\;\sigma^{(1)}_{1/2}\;\pi^{*(2)}_{1/2}\;\pi^{*(2)}_{3/2}$ & -0.148  &&        & $\pi^{(2)}_{1/2}\;\pi^{(2)}_{3/2}\;\sigma^{(1)}_{1/2}\;\pi^{*(2)}_{1/2}\;\pi^{*(2)}_{3/2}$ &  0.151  \\
    &        & $\pi^{(2)}_{1/2}\;\pi^{(2)}_{3/2}\;\sigma^{(2)}_{1/2}\;\pi^{*(1)}_{1/2}\;\pi^{*(2)}_{3/2}$ & -0.156  &&        & $\pi^{(2)}_{1/2}\;\pi^{(2)}_{3/2}\;\sigma^{(2)}_{1/2}\;\pi^{*(1)}_{1/2}\;\pi^{*(2)}_{3/2}$ &  0.146  \\
\cline{2-8}
1/2 & 5.4701 & $\pi^{(1)}_{1/2}\;\pi^{(2)}_{3/2}\;\sigma^{(2)}_{1/2}\;\pi^{*(2)}_{1/2}\;\pi^{*(2)}_{3/2}$ &  0.149  && 5.4975 & $\pi^{(1)}_{1/2}\;\pi^{(2)}_{3/2}\;\sigma^{(2)}_{1/2}\;\pi^{*(2)}_{1/2}\;\pi^{*(2)}_{3/2}$ &  0.154  \\
    &        & $\pi^{(2)}_{1/2}\;\pi^{(2)}_{3/2}\;\sigma^{(1)}_{1/2}\;\pi^{*(2)}_{1/2}\;\pi^{*(2)}_{3/2}$ & -0.952  &&        & $\pi^{(2)}_{1/2}\;\pi^{(2)}_{3/2}\;\sigma^{(1)}_{1/2}\;\pi^{*(2)}_{1/2}\;\pi^{*(2)}_{3/2}$ & -0.988  \\
\hline\hline
\end{tabular}
\label{tab:clo-state-wfs}
\end{table}

\begin{table}[htb]
\caption{Projection analysis\cite{Faegri_JCP2001} and $\sigma$- and $\pi$-character of valence orbitals of ClO$^-$  from 4-component Hartree--Fock calculations based on the Dirac--Coulomb--Gaunt Hamiltonian. Only the most significant contributions are listed. The number before the colons is the gross population of the atom.}
\centering
\begin{tabular}{lcccll}
  \hline
  \hline
 Spinor& $\epsilon$(E$_h$) & $\sigma$ & $\pi$ & Cl & O\tabularnewline\hline
$\pi^{*}_{3/2}$  &-0.117 & 0.00 & 1.00 & 0.46: +0.78 $3p_{1/2}$ & 0.52: -0.81 $2p_{3/2}$\tabularnewline
$\pi^{*}_{1/2}$  &-0.120 & 0.00 & 1.00 & 0.45: +0.63 $3p_{1/2}$ +0.45 $3p_{1/2}$ & 0.52: -0.66 $2p_{1/2}$ -0.48 $2p_{13/2}$\tabularnewline
$\sigma_{1/2}$   &-0.296 & 1.00 & 0.00 & 0.45: -0.56 $3p_{1/2}$ + 0.32 $3p_{1/2}$ -0.24 $3s_{1/2}$ & 0.52: +0.47 $2p_{3/2}$ -0.41 $2p_{1/2}$ +0.38 $2s_{1/2}$\tabularnewline
$\pi_{3/2}$      & -0.306 & 0.00 & 1.00 & 0.53: +0.66 $3p_{3/2}$ & 0.46: +0.61 $2p_{3/2}$\tabularnewline
$\pi_{1/2}$      & -0.308 & 0.00 & 1.00 & 0.54: +0.58 $3p_{1/2}$ +0.33 $3p_{3/2}$ & 0.45: +0.45 $2p_{1/2}$ +0.39 $2s_{1/2}$\tabularnewline
\hline\hline
\end{tabular}
\label{tab:clo-state-prjana}
\end{table}

\begin{table}[htb]
\caption{Composition of the ground and excited-state $^2$DC$^M$ EOM-IP and IHFS(1h0p) wavefunctions for BrO at $R_e$ = 1.714 \AA. Only main contributions ($(r_i, c_i) > 0.1$) are shown.}
\centering
\begin{tabular}{l ccc c ccc}
\hline
            & \multicolumn{3}{c}{EOM-IP}
           && \multicolumn{3}{c}{IHFS(1h0p)}\\
\cline{2-4}
\cline{6-8}
 $\Omega$   & E (eV) & configuration & $r_i$
           && E (eV) & configuration & $c_i$ \\
\hline
3/2 & 0.0000 & $\pi^{(2)}_{1/2}\;\pi^{(1)}_{3/2}\;\sigma^{(2)}_{1/2}\;\pi^{*(2)}_{1/2}\;\pi^{*(2)}_{3/2}$ &  0.209  && 0.0000 & $\pi^{(2)}_{1/2}\;\pi^{(1)}_{3/2}\;\sigma^{(2)}_{1/2}\;\pi^{*(2)}_{1/2}\;\pi^{*(2)}_{3/2}$ & -0.238  \\
    &        & $\pi^{(2)}_{1/2}\;\pi^{(2)}_{3/2}\;\sigma^{(2)}_{1/2}\;\pi^{*(2)}_{1/2}\;\pi^{*(1)}_{3/2}$ &  0.940  &&        & $\pi^{(2)}_{1/2}\;\pi^{(2)}_{3/2}\;\sigma^{(2)}_{1/2}\;\pi^{*(2)}_{1/2}\;\pi^{*(1)}_{3/2}$ & -0.971  \\
\cline{2-8}
1/2 & 0.1295 & $\pi^{(1)}_{1/2}\;\pi^{(2)}_{3/2}\;\sigma^{(2)}_{1/2}\;\pi^{*(2)}_{1/2}\;\pi^{*(2)}_{3/2}$ & -0.193  && 0.1294 & $\pi^{(1)}_{1/2}\;\pi^{(2)}_{3/2}\;\sigma^{(2)}_{1/2}\;\pi^{*(2)}_{1/2}\;\pi^{*(2)}_{3/2}$ &  0.220  \\
    &        & $\pi^{(2)}_{1/2}\;\pi^{(2)}_{3/2}\;\sigma^{(2)}_{1/2}\;\pi^{*(1)}_{1/2}\;\pi^{*(2)}_{3/2}$ &  0.940  &&        & $\pi^{(2)}_{1/2}\;\pi^{(2)}_{3/2}\;\sigma^{(2)}_{1/2}\;\pi^{*(1)}_{1/2}\;\pi^{*(2)}_{3/2}$ & -0.972  \\
\cline{2-8}
3/2 & 3.9831 & $\pi^{(2)}_{1/2}\;\pi^{(1)}_{3/2}\;\sigma^{(2)}_{1/2}\;\pi^{*(2)}_{1/2}\;\pi^{*(2)}_{3/2}$ &  0.931  && 3.9831 & $\pi^{(2)}_{1/2}\;\pi^{(1)}_{3/2}\;\sigma^{(2)}_{1/2}\;\pi^{*(2)}_{1/2}\;\pi^{*(2)}_{3/2}$ & -0.976  \\
    &        & $\pi^{(2)}_{1/2}\;\pi^{(2)}_{3/2}\;\sigma^{(2)}_{1/2}\;\pi^{*(2)}_{1/2}\;\pi^{*(1)}_{3/2}$ & -0.228  &&        & $\pi^{(2)}_{1/2}\;\pi^{(2)}_{3/2}\;\sigma^{(2)}_{1/2}\;\pi^{*(2)}_{1/2}\;\pi^{*(1)}_{3/2}$ &  0.217  \\
\cline{2-8}
1/2 & 4.1707 & $\pi^{(1)}_{1/2}\;\pi^{(2)}_{3/2}\;\sigma^{(2)}_{1/2}\;\pi^{*(2)}_{1/2}\;\pi^{*(2)}_{3/2}$ & -0.801  && 4.1723 & $\pi^{(1)}_{1/2}\;\pi^{(2)}_{3/2}\;\sigma^{(2)}_{1/2}\;\pi^{*(2)}_{1/2}\;\pi^{*(2)}_{3/2}$ &  0.846  \\
    &        & $\pi^{(2)}_{1/2}\;\pi^{(2)}_{3/2}\;\sigma^{(1)}_{1/2}\;\pi^{*(2)}_{1/2}\;\pi^{*(2)}_{3/2}$ & -0.479  &&        & $\pi^{(2)}_{1/2}\;\pi^{(2)}_{3/2}\;\sigma^{(1)}_{1/2}\;\pi^{*(2)}_{1/2}\;\pi^{*(2)}_{3/2}$ &  0.489 \\
    &        & $\pi^{(2)}_{1/2}\;\pi^{(2)}_{3/2}\;\sigma^{(2)}_{1/2}\;\pi^{*(1)}_{1/2}\;\pi^{*(2)}_{3/2}$ & -0.223  &&        & $\pi^{(2)}_{1/2}\;\pi^{(2)}_{3/2}\;\sigma^{(2)}_{1/2}\;\pi^{*(1)}_{1/2}\;\pi^{*(2)}_{3/2}$ &  0.213 \\
\cline{2-8}
1/2 & 4.8198 & $\pi^{(1)}_{1/2}\;\pi^{(2)}_{3/2}\;\sigma^{(2)}_{1/2}\;\pi^{*(2)}_{1/2}\;\pi^{*(2)}_{3/2}$ & -0.483  && 4.8467 & $\pi^{(1)}_{1/2}\;\pi^{(2)}_{3/2}\;\sigma^{(2)}_{1/2}\;\pi^{*(2)}_{1/2}\;\pi^{*(2)}_{3/2}$ &  0.498 \\
    &        & $\pi^{(2)}_{1/2}\;\pi^{(2)}_{3/2}\;\sigma^{(1)}_{1/2}\;\pi^{*(2)}_{1/2}\;\pi^{*(2)}_{3/2}$ &  0.832  &&        & $\pi^{(2)}_{1/2}\;\pi^{(2)}_{3/2}\;\sigma^{(1)}_{1/2}\;\pi^{*(2)}_{1/2}\;\pi^{*(2)}_{3/2}$ & -0.866  \\
\hline\hline
\end{tabular}
\label{tab:bro-state-wfs}
\end{table}

\begin{table}[htb]
\caption{Projection analysis\cite{Faegri_JCP2001} and $\sigma$- and $\pi$-character of valence orbitals of BrO$^-$  from 4-component Hartree--Fock calculations based on the Dirac--Coulomb--Gaunt Hamiltonian. Only the most significant contributions are listed. The number before the colons is the gross population of the atom.}
\centering
\begin{tabular}{lcccll}
\hline\hline
Spinor   &$\epsilon$(E$_h$) & $\sigma$ & $\pi$ & Br & O\tabularnewline\hline
$\pi^{*}_{3/2}$ & -0.117 & 0.00 & 1.00 & 0.58: -0.85 $4p_{3/2}$ & 0.40: +0.73 $2p_{3/2}$\tabularnewline
$\pi^{*}_{1/2}$ & -0.126 & 0.00 & 1.00 & 0.54: -0.66 $4p_{1/2}$ +0.48 $4p_{3/2}$ & 0.44: +0.60 $2p_{1/2}$ +0.47 $2p_{3/2}$\tabularnewline
$\sigma_{1/2}$ & -0.267 & 0.91 & 0.09 & 0.44: - 0.58 $4p_{3/2}$ -0.23 $4p_{1/2}$ +0.23 $4s_{1/2}$  & 0.53: +0.54 $2p_{1/2}$ -0.37 $2p_{3/2}$ -0.33 $2s_{1/2}$\tabularnewline
$\pi_{3/2}$ & -0.280 & 0.00 & 1.00 & 0.41: -0.57 $4p_{3/2}$  & 0.58: -0.70 $2p_{3/2}$\tabularnewline
$\pi_{1/2}$ & -0.289 & 0.08 & 0.92 & 0.47: +0.59 $4p_{1/2}$ -0.18 $4p_{3/2}$ & 0.53: +0.519 $2p_{3/2}$ +0.42 $2p_{1/2}$\tabularnewline
\hline\hline
\end{tabular}
\label{tab:bro-state-prjana}
\end{table}

\begin{table}[htb]
\caption{Composition of the ground and excited-state $^2$DC$^M$ EOM-IP and IHFS(1h0p) wavefunctions for IO at $R_e$ = 1.875 \AA. Only main contributions ($(r_i, c_i) > 0.1$) are shown.}
\centering
\begin{tabular}{l ccc c ccc}
\hline
            & \multicolumn{3}{c}{EOM-IP}
           && \multicolumn{3}{c}{IHFS(1h0p)}\\
\cline{2-4}
\cline{6-8}
 $\Omega$   & E (eV) & configuration & $r_i$
           && E (eV) & configuration & $c_i$ \\
\hline
3/2 & 0.0000 & $\pi^{(2)}_{1/2}\;\pi^{(1)}_{3/2}\;\sigma^{(2)}_{1/2}\;\pi^{*(2)}_{1/2}\;\pi^{*(2)}_{3/2}$ &  0.261  && 0.0000 & $\pi^{(2)}_{1/2}\;\pi^{(1)}_{3/2}\;\sigma^{(2)}_{1/2}\;\pi^{*(2)}_{1/2}\;\pi^{*(2)}_{3/2}$ & -0.293  \\
    &        & $\pi^{(2)}_{1/2}\;\pi^{(2)}_{3/2}\;\sigma^{(2)}_{1/2}\;\pi^{*(2)}_{1/2}\;\pi^{*(1)}_{3/2}$ &  0.928  &&        & $\pi^{(2)}_{1/2}\;\pi^{(2)}_{3/2}\;\sigma^{(2)}_{1/2}\;\pi^{*(2)}_{1/2}\;\pi^{*(1)}_{3/2}$ & -0.956  \\
\cline{2-8}
1/2 & 0.2777 & $\pi^{(1)}_{1/2}\;\pi^{(2)}_{3/2}\;\sigma^{(2)}_{1/2}\;\pi^{*(2)}_{1/2}\;\pi^{*(2)}_{3/2}$ & -0.240  && 0.2777 & $\pi^{(1)}_{1/2}\;\pi^{(2)}_{3/2}\;\sigma^{(2)}_{1/2}\;\pi^{*(2)}_{1/2}\;\pi^{*(2)}_{3/2}$ & -0.270  \\
    &        & $\pi^{(2)}_{1/2}\;\pi^{(2)}_{3/2}\;\sigma^{(1)}_{1/2}\;\pi^{*(2)}_{1/2}\;\pi^{*(2)}_{3/2}$ & -0.136  &&        & $\pi^{(2)}_{1/2}\;\pi^{(2)}_{3/2}\;\sigma^{(1)}_{1/2}\;\pi^{*(2)}_{1/2}\;\pi^{*(2)}_{3/2}$ & -0.151  \\
    &        & $\pi^{(2)}_{1/2}\;\pi^{(2)}_{3/2}\;\sigma^{(2)}_{1/2}\;\pi^{*(1)}_{1/2}\;\pi^{*(2)}_{3/2}$ &  0.921  &&        & $\pi^{(2)}_{1/2}\;\pi^{(2)}_{3/2}\;\sigma^{(2)}_{1/2}\;\pi^{*(1)}_{1/2}\;\pi^{*(2)}_{3/2}$ &  0.951  \\
\cline{2-8}
3/2 & 3.2206 & $\pi^{(2)}_{1/2}\;\pi^{(1)}_{3/2}\;\sigma^{(2)}_{1/2}\;\pi^{*(2)}_{1/2}\;\pi^{*(2)}_{3/2}$ & -0.914  && 3.2206 & $\pi^{(2)}_{1/2}\;\pi^{(1)}_{3/2}\;\sigma^{(2)}_{1/2}\;\pi^{*(2)}_{1/2}\;\pi^{*(2)}_{3/2}$ & -0.963  \\
    &        & $\pi^{(2)}_{1/2}\;\pi^{(2)}_{3/2}\;\sigma^{(2)}_{1/2}\;\pi^{*(2)}_{1/2}\;\pi^{*(1)}_{3/2}$ &  0.280  &&        & $\pi^{(2)}_{1/2}\;\pi^{(2)}_{3/2}\;\sigma^{(2)}_{1/2}\;\pi^{*(2)}_{1/2}\;\pi^{*(1)}_{3/2}$ &  0.271  \\
\cline{2-8}
1/2 & 3.4716 & $\pi^{(1)}_{1/2}\;\pi^{(2)}_{3/2}\;\sigma^{(2)}_{1/2}\;\pi^{*(2)}_{1/2}\;\pi^{*(2)}_{3/2}$ & -0.595  && 3.4773 & $\pi^{(1)}_{1/2}\;\pi^{(2)}_{3/2}\;\sigma^{(2)}_{1/2}\;\pi^{*(2)}_{1/2}\;\pi^{*(2)}_{3/2}$ & -0.642  \\
    &        & $\pi^{(2)}_{1/2}\;\pi^{(2)}_{3/2}\;\sigma^{(1)}_{1/2}\;\pi^{*(2)}_{1/2}\;\pi^{*(2)}_{3/2}$ & -0.697  &&        & $\pi^{(2)}_{1/2}\;\pi^{(2)}_{3/2}\;\sigma^{(1)}_{1/2}\;\pi^{*(2)}_{1/2}\;\pi^{*(2)}_{3/2}$ & -0.716  \\
    &        & $\pi^{(2)}_{1/2}\;\pi^{(2)}_{3/2}\;\sigma^{(2)}_{1/2}\;\pi^{*(1)}_{1/2}\;\pi^{*(2)}_{3/2}$ & -0.279  &&        & $\pi^{(2)}_{1/2}\;\pi^{(2)}_{3/2}\;\sigma^{(2)}_{1/2}\;\pi^{*(1)}_{1/2}\;\pi^{*(2)}_{3/2}$ & -0.273  \\
\cline{2-8}
1/2 & 4.2030 & $\pi^{(1)}_{1/2}\;\pi^{(2)}_{3/2}\;\sigma^{(2)}_{1/2}\;\pi^{*(2)}_{1/2}\;\pi^{*(2)}_{3/2}$ &  0.701  && 4.2232 & $\pi^{(1)}_{1/2}\;\pi^{(2)}_{3/2}\;\sigma^{(2)}_{1/2}\;\pi^{*(2)}_{1/2}\;\pi^{*(2)}_{3/2}$ &  0.730  \\
    &        & $\pi^{(2)}_{1/2}\;\pi^{(2)}_{3/2}\;\sigma^{(1)}_{1/2}\;\pi^{*(2)}_{1/2}\;\pi^{*(2)}_{3/2}$ & -0.646  &&        & $\pi^{(2)}_{1/2}\;\pi^{(2)}_{3/2}\;\sigma^{(1)}_{1/2}\;\pi^{*(2)}_{1/2}\;\pi^{*(2)}_{3/2}$ & -0.678  \\
\hline\hline
\end{tabular}
\label{tab:io-state-wfs}
\end{table}

\begin{table}[htb]
\caption{Projection analysis\cite{Faegri_JCP2001} and $\sigma$- and $\pi$-character of valence orbitals of IO$^-$  from 4-component Hartree--Fock calculations based on the Dirac--Coulomb--Gaunt Hamiltonian. Only the most significant contributions are listed.  The number before the colons is the gross population of the atom.}
\centering
\begin{tabular}{lcccll}\hline\hline
Spinor&$\epsilon$(E$_h$) & $\sigma$ & $\pi$ & I & O\tabularnewline\hline
$\pi^{*}_{3/2}$&-0.110 & 0.00 & 1.00 & 0.72: -0.92 $5p_{3/2}$ & 0.26: +0.62 $2p_{3/2}$\tabularnewline
$\pi^{*}_{1/2}$&-0.129 & 0.01 & 0.99 & 0.64: +0.70 $5p_{1/2}$ -0.52 $5p_{3/2}$ & 0.34: +0.49 $2p_{1/2}$ +0.48 $2s_{1/2}$\tabularnewline
$\sigma_{1/2}$&-0.241 & 0.86 & 0.14 & 0.39: -0.52 $5p_{3/2}$ -0.24 $5p_{1/2}$ +0.24 $5s_{1/2}$ & 0.59: -0.64 $2p_{1/2}$ +0.32 $2p_{3/2}$ +0.28 $2s_{1/2}$ \tabularnewline
$\pi_{3/2}$&-0.262 & 0.00 & 1.00 & 0.27: +0.44 $5p_{3/2}$ & 0.72: +0.81 $2p_{3/2}$\tabularnewline
$\pi_{1/2}$&-0.276 & 0.13 & 0.87 & 0.39: +0.55 $5p_{1/2}$ -0.11 $5p_{3/2}$ & 0.60: -0.58 $2p_{3/2}$ -0.42 $2p_{1/2}$ -0.12 $2s_{1/2}$ \tabularnewline\hline\hline
\end{tabular}
\label{tab:io-state-prjana}
\end{table}

\begin{table}[htb]
\caption{Composition of the ground and excited-state $^2$DC$^M$ EOM-IP and IHFS(1h0p) wavefunctions for AtO at $R_e$ = 1.973 \AA. Only main contributions ($(r_i, c_i) > 0.1$) are shown.}
\centering
\begin{tabular}{l ccc c ccc}
\hline
            & \multicolumn{3}{c}{EOM-IP}
           && \multicolumn{3}{c}{IHFS(1h0p)}\\
\cline{2-4}
\cline{6-8}
 $\Omega$   & E (eV) & configuration & $r_i$
           && E (eV) & configuration & $c_i$ \\
\hline
3/2 & 0.0000 & $\pi^{(2)}_{1/2}\;\pi^{(1)}_{3/2}\;\sigma^{(2)}_{1/2}\;\pi^{*(2)}_{1/2}\;\pi^{*(2)}_{3/2}$ &  0.305  && 0.0000 & $\pi^{(2)}_{1/2}\;\pi^{(1)}_{3/2}\;\sigma^{(2)}_{1/2}\;\pi^{*(2)}_{1/2}\;\pi^{*(2)}_{3/2}$ &  0.342  \\
    &        & $\pi^{(2)}_{1/2}\;\pi^{(2)}_{3/2}\;\sigma^{(2)}_{1/2}\;\pi^{*(2)}_{1/2}\;\pi^{*(1)}_{3/2}$ &  0.917  &&        & $\pi^{(2)}_{1/2}\;\pi^{(2)}_{3/2}\;\sigma^{(2)}_{1/2}\;\pi^{*(2)}_{1/2}\;\pi^{*(1)}_{3/2}$ & -0.940  \\
\cline{2-8}
1/2 & 0.7233 & $\pi^{(1)}_{1/2}\;\pi^{(2)}_{3/2}\;\sigma^{(2)}_{1/2}\;\pi^{*(2)}_{1/2}\;\pi^{*(2)}_{3/2}$ & -0.211  && 0.7238 & $\pi^{(1)}_{1/2}\;\pi^{(2)}_{3/2}\;\sigma^{(2)}_{1/2}\;\pi^{*(2)}_{1/2}\;\pi^{*(2)}_{3/2}$ & -0.245  \\
    &        & $\pi^{(2)}_{1/2}\;\pi^{(2)}_{3/2}\;\sigma^{(1)}_{1/2}\;\pi^{*(2)}_{1/2}\;\pi^{*(2)}_{3/2}$ & -0.402  &&        & $\pi^{(2)}_{1/2}\;\pi^{(2)}_{3/2}\;\sigma^{(1)}_{1/2}\;\pi^{*(2)}_{1/2}\;\pi^{*(2)}_{3/2}$ & -0.435  \\
    &        & $\pi^{(2)}_{1/2}\;\pi^{(2)}_{3/2}\;\sigma^{(2)}_{1/2}\;\pi^{*(1)}_{1/2}\;\pi^{*(2)}_{3/2}$ & -0.839  &&        & $\pi^{(2)}_{1/2}\;\pi^{(2)}_{3/2}\;\sigma^{(2)}_{1/2}\;\pi^{*(1)}_{1/2}\;\pi^{*(2)}_{3/2}$ &  0.866  \\
\cline{2-8}
3/2 & 2.7841 & $\pi^{(2)}_{1/2}\;\pi^{(1)}_{3/2}\;\sigma^{(2)}_{1/2}\;\pi^{*(2)}_{1/2}\;\pi^{*(2)}_{3/2}$ &  0.897  && 2.7841 & $\pi^{(2)}_{1/2}\;\pi^{(1)}_{3/2}\;\sigma^{(2)}_{1/2}\;\pi^{*(2)}_{1/2}\;\pi^{*(2)}_{3/2}$ & -0.949  \\
    &        & $\pi^{(2)}_{1/2}\;\pi^{(2)}_{3/2}\;\sigma^{(2)}_{1/2}\;\pi^{*(2)}_{1/2}\;\pi^{*(1)}_{3/2}$ & -0.327  &&        & $\pi^{(2)}_{1/2}\;\pi^{(2)}_{3/2}\;\sigma^{(2)}_{1/2}\;\pi^{*(2)}_{1/2}\;\pi^{*(1)}_{3/2}$ & -0.316  \\
\cline{2-8}
1/2 & 3.0594 & $\pi^{(1)}_{1/2}\;\pi^{(2)}_{3/2}\;\sigma^{(2)}_{1/2}\;\pi^{*(2)}_{1/2}\;\pi^{*(2)}_{3/2}$ &  0.169  && 3.0730 & $\pi^{(1)}_{1/2}\;\pi^{(2)}_{3/2}\;\sigma^{(2)}_{1/2}\;\pi^{*(2)}_{1/2}\;\pi^{*(2)}_{3/2}$ & -0.192  \\
    &        & $\pi^{(2)}_{1/2}\;\pi^{(2)}_{3/2}\;\sigma^{(1)}_{1/2}\;\pi^{*(2)}_{1/2}\;\pi^{*(2)}_{3/2}$ &  0.825  &&        & $\pi^{(2)}_{1/2}\;\pi^{(2)}_{3/2}\;\sigma^{(1)}_{1/2}\;\pi^{*(2)}_{1/2}\;\pi^{*(2)}_{3/2}$ & -0.865  \\
    &        & $\pi^{(2)}_{1/2}\;\pi^{(2)}_{3/2}\;\sigma^{(2)}_{1/2}\;\pi^{*(1)}_{1/2}\;\pi^{*(2)}_{3/2}$ & -0.461  &&        & $\pi^{(2)}_{1/2}\;\pi^{(2)}_{3/2}\;\sigma^{(2)}_{1/2}\;\pi^{*(1)}_{1/2}\;\pi^{*(2)}_{3/2}$ & -0.464  \\
\cline{2-8}
1/2 & 5.1108 & $\pi^{(1)}_{1/2}\;\pi^{(2)}_{3/2}\;\sigma^{(2)}_{1/2}\;\pi^{*(2)}_{1/2}\;\pi^{*(2)}_{3/2}$ &  0.910  && 5.1230 & $\pi^{(1)}_{1/2}\;\pi^{(2)}_{3/2}\;\sigma^{(2)}_{1/2}\;\pi^{*(2)}_{1/2}\;\pi^{*(2)}_{3/2}$ &  0.958  \\
    &        & $\pi^{(2)}_{1/2}\;\pi^{(2)}_{3/2}\;\sigma^{(1)}_{1/2}\;\pi^{*(2)}_{1/2}\;\pi^{*(2)}_{3/2}$ & -0.262  &&        & $\pi^{(2)}_{1/2}\;\pi^{(2)}_{3/2}\;\sigma^{(1)}_{1/2}\;\pi^{*(2)}_{1/2}\;\pi^{*(2)}_{3/2}$ & -0.265  \\
    &        & $\pi^{(2)}_{1/2}\;\pi^{(2)}_{3/2}\;\sigma^{(2)}_{1/2}\;\pi^{*(1)}_{1/2}\;\pi^{*(2)}_{3/2}$ & -0.127  &&        & $\pi^{(2)}_{1/2}\;\pi^{(2)}_{3/2}\;\sigma^{(2)}_{1/2}\;\pi^{*(1)}_{1/2}\;\pi^{*(2)}_{3/2}$ &  0.113  \\
\hline\hline
\end{tabular}
\label{tab:ato-state-wfs}
\end{table}

\begin{table}[htb]
\caption{Projection analysis\cite{Faegri_JCP2001} and $\sigma$- and $\pi$-character of valence orbitals of AtO$^-$  from 4-component Hartree--Fock calculations based on the Dirac--Coulomb--Gaunt Hamiltonian. Only the most significant contributions are listed. The number before the colons is the gross population of the atom.}
\centering
\begin{tabular}{lcccll}\hline\hline
Spinor&$\epsilon$(E$_h$) & $\sigma$ & $\pi$ & At & O\tabularnewline\hline
$\pi^{*}_{3/2}$&-0.088 & 0.00 & 1.00 & 0.81: +0.96 $6p_{3/2}$ & 0.16: +0.52 $2p_{3/2}$ \tabularnewline
$\pi^{*}_{1/2}$&-0.142 & 0.17 & 0.83 & 0.58: -0.59 $6p_{1/2}$ +0.56 $6p_{3/2}$ +0.14 $6s_{1/2}$ & 0.40: -0.65 $2p_{3/2}$ -0.27 $2p_{1/2}$\tabularnewline
$\sigma_{1/2}$&-0.209 & 0.55 & 0.45 & 0.26: -0.40 $6p_{3/2}$ +0.27 $6p_{1/2}$ +0.15 $6s_{1/2}$ & 0.72: 0.81 $2p_{1/2}$ -0.22 $2s_{1/2}$\tabularnewline
$\pi_{3/2}$&-0.242 & 0.00 & 1.00 & 0.17: +0.33 $6p_{1/2}$  & 0.82: -0.87 $2p_{3/2}$\tabularnewline
$\pi_{1/2}$&-0.295 & 0.30 & 0.70 & 0.57: -0.71 $6p_{1/2}$  & 0.42: +0.54 $2p_{3/2}$ +0.22 $2p_{1/2}$ +0.18 $2s_{1/2}$\tabularnewline\hline\hline
\end{tabular}
\label{tab:ato-state-prjana}
\end{table}

\begin{table}[htb]
\caption{Composition of the ground and excited-state $^2$DC$^M$ EOM-IP and IHFS(1h0p) wavefunctions for TsO at $R_e$ = 2.137 \AA. Only main contributions ($(r_i, c_i r_{ij}^a) > 0.1$) are shown.}
\centering
\begin{tabular}{l clrr c ccr}
\hline
            & \multicolumn{4}{c}{EOM-IP}
           && \multicolumn{3}{c}{IHFS(1h0p)}\\
\cline{2-5}
\cline{7-9}
 $\Omega$   & E (eV) & \multicolumn{1}{c}{configuration} & \multicolumn{1}{c}{$r_i$} & \multicolumn{1}{c}{$r_{ij}^a$} 
           && E (eV) & configuration & \multicolumn{1}{c}{$c_i$} \\
\hline
3/2 & 0.0000 & $\pi^{(2)}_{1/2}\;\pi^{(1)}_{3/2}\;\sigma^{(2)}_{1/2}\;\pi^{*(2)}_{1/2}\;\pi^{*(2)}_{3/2}$ & -0.336  &  && 0.0000 & $\pi^{(2)}_{1/2}\;\pi^{(1)}_{3/2}\;\sigma^{(2)}_{1/2}\;\pi^{*(2)}_{1/2}\;\pi^{*(2)}_{3/2}$ &  0.373  \\
    &        & $\pi^{(2)}_{1/2}\;\pi^{(2)}_{3/2}\;\sigma^{(2)}_{1/2}\;\pi^{*(2)}_{1/2}\;\pi^{*(1)}_{3/2}$ &  0.907  &  &&        & $\pi^{(2)}_{1/2}\;\pi^{(2)}_{3/2}\;\sigma^{(2)}_{1/2}\;\pi^{*(2)}_{1/2}\;\pi^{*(1)}_{3/2}$ & -0.928  \\
\cline{2-9}
1/2 & 1.1954 & $\pi^{(2)}_{1/2}\;\pi^{(2)}_{3/2}\;\sigma^{(2)}_{1/2}\;\pi^{*(1)}_{1/2}\;\pi^{*(2)}_{3/2}$ &  -0.937  & 
   && 1.2194 & $\pi^{(2)}_{1/2}\;\pi^{(2)}_{3/2}\;\sigma^{(2)}_{1/2}\;\pi^{*(1)}_{1/2}\;\pi^{*(2)}_{3/2}$ &  -0.995  \\
    &        & $\pi^{(2)}_{1/2}\;\pi^{(2)}_{3/2}\;\sigma^{(1)}_{1/2}\;\pi^{*(1)}_{1/2}\;\pi^{*(2)}_{3/2};{_4}\sigma^{*(1)}_{1/2}$ & & -0.120 
   &&        &                                                                                                                    & \\
\cline{2-9}
3/2 & 2.2279 & $\pi^{(2)}_{1/2}\;\pi^{(1)}_{3/2}\;\sigma^{(2)}_{1/2}\;\pi^{*(2)}_{1/2}\;\pi^{*(2)}_{3/2}$ & -0.882  & && 2.2277 & $\pi^{(2)}_{1/2}\;\pi^{(1)}_{3/2}\;\sigma^{(2)}_{1/2}\;\pi^{*(2)}_{1/2}\;\pi^{*(2)}_{3/2}$ & -0.938  \\
    &        & $\pi^{(2)}_{1/2}\;\pi^{(2)}_{3/2}\;\sigma^{(2)}_{1/2}\;\pi^{*(2)}_{1/2}\;\pi^{*(1)}_{3/2}$ & -0.354  & &&        & $\pi^{(2)}_{1/2}\;\pi^{(2)}_{3/2}\;\sigma^{(2)}_{1/2}\;\pi^{*(2)}_{1/2}\;\pi^{*(1)}_{3/2}$ & -0.348 \\
\cline{2-9}
1/2 & 2.2901 & $\pi^{(2)}_{1/2}\;\pi^{(2)}_{3/2}\;\sigma^{(1)}_{1/2}\;\pi^{*(2)}_{1/2}\;\pi^{*(2)}_{3/2}$ & -0.946 & 
   && 2.3014 & $\pi^{(2)}_{1/2}\;\pi^{(2)}_{3/2}\;\sigma^{(1)}_{1/2}\;\pi^{*(2)}_{1/2}\;\pi^{*(2)}_{3/2}$ & -0.997 \\
    &        & $\pi^{(2)}_{1/2}\;\pi^{(2)}_{3/2}\;\sigma^{(0)}_{1/2}\;\pi^{*(2)}_{1/2}\;\pi^{*(2)}_{3/2};{_4}\sigma^{*(1)}_{1/2}$ & & -0.131 
   &&        &                                                                                                                    & \\
\cline{2-9}
1/2 & 9.1319 & $\pi^{(1)}_{1/2}\;\pi^{(2)}_{3/2}\;\sigma^{(2)}_{1/2}\;\pi^{*(2)}_{1/2}\;\pi^{*(2)}_{3/2}$ &  0.929 & 
   &&10.7609 & $\pi^{(1)}_{1/2}\;\pi^{(2)}_{3/2}\;\sigma^{(2)}_{1/2}\;\pi^{*(2)}_{1/2}\;\pi^{*(2)}_{3/2}$ &  0.999 \\
    &        & $\pi^{(2)}_{1/2}\;\pi^{(2)}_{3/2}\;\sigma^{(0)}_{1/2}\;\pi^{*(2)}_{1/2}\;\pi^{*(2)}_{3/2};{_4}\sigma^{*(1)}_{1/2}$ & & -0.117
   &&        &                                                                                                                    & \\
\hline\hline
\end{tabular}
\label{tab:tso-state-wfs}
\end{table}

\begin{table}[htb]
\caption{Projection analysis\cite{Faegri_JCP2001} and $\sigma$- and $\pi$-character of valence orbitals of TsO$^-$  from 4-component Hartree--Fock calculations based on the Dirac--Coulomb--Gaunt Hamiltonian. Only the most significant contributions are listed. The number before the colons is the gross population of the atom.}
\centering
\begin{tabular}{lcccll}\hline\hline
$\epsilon$(E$_h$) & $\sigma$ & $\pi$ & Ts & O\tabularnewline\hline
$\pi^{*}_{3/2}$&-0.052 & 0.00 & 1.00 & 0.87: -0.97 $7p_{3/2}$ & 0.08: +0.40 $2p_{3/2}$\tabularnewline
$\sigma_{1/2}$&-0.126 & 0.72 & 0.28 & 0.44: -0.61 $7p_{3/2}$ -0.17 $7s_{1/2}$ +0.14 $7p_{1/2}$ & 0.54: +0.54 $2p_{1/2}$ -0.44 $2p_{3/2}$ -0.20 $2s_{1/2}$\tabularnewline
$\pi^{*}_{1/2}$&-0.189 & 0.07 & 0.93 & 0.05: +0.32 $7p_{1/2}$ & 0.94: -0.77 $2p_{3/2}$ -0.632 $2p_{1/2}$\tabularnewline
$\pi_{3/2}$&-0.212 & 0.00 & 1.00 & 0.09: +0.21 $7p_{3/2}$ & 0.91: +0.93 $2p_{3/2}$\tabularnewline
$\pi_{1/2}$&-0.439 & 0.34 & 0.66 & 0.90: +0.94 $7p_{1/2}$ & 0.10: +0.21 $2p_{3/2}$ +0.21 $2s_{1/2}$\tabularnewline\hline\hline
\end{tabular}
\label{tab:tso-state-prjana}
\end{table}

\begin{table}[htb]
\caption{Comparison of  total energies (in $E_h$) obtained with the $ {^2}$DC$^M$ and DC Hamiltonians for the I$_3$,  I$_3^-$ and I$_3^{2-}$ species. The $\Delta E$ values correspond to the difference (in eV)
         between the tho Hamiltonians, whereas $\Delta(\Delta E_a)$ and $\Delta(\Delta E_b)$ (in eV) correspond to the differences between Hamiltonians to 
         relative energies with respect to the I$_3^-$ ground state and the and I$_3^-$/I$_3$/I$_3^{2-}$ excited/electron attachment/detachment states, and 
         between the ground-states of I$_3$/I$_3^{2-}$ and their excited states.}
\centering
{\small\begin{tabular}{l c c cc p{0.2cm} ccc}
\hline
\hline
species	&	R(\AA)	& $\Omega$	&$ {^2}$DC$^M$ ($E_h$)	&	DC ($E_h$)	&&   $\Delta E$ (eV)	&$\Delta(\Delta E_a)$ (eV)   & $\Delta(\Delta E_b)$ (eV)\\
\hline
I$_3^-$	&	2.93	&	0g	&	-21349.033485	&	-21349.033561	&&	2.06E-03	&		&		\\
	&		&	2g	&	-21348.951086	&	-21348.951162	&&	2.07E-03	&	4.06E-06	&		\\
	&		&	0u	&	-21348.946379	&	-21348.946454	&&	2.06E-03	&	2.96E-06	&		\\
	&		&	1g	&	-21348.946210	&	-21348.946286	&&	2.06E-03	&	2.88E-06	&		\\
	&		&	1u	&	-21348.946131	&	-21348.946207	&&	2.06E-03	&	2.95E-06	&		\\
	&		&	0g	&	-21348.929175	&	-21348.929251	&&	2.06E-03	&	1.69E-06	&		\\
	&		&	0g	&	-21348.927148	&	-21348.927224	&&	2.06E-03	&	2.50E-06	&		\\
	&		&	1g	&	-21348.920795	&	-21348.920870	&&	2.05E-03	&	-8.11E-06	&		\\
	&		&	2u	&	-21348.911323	&	-21348.911398	&&	2.06E-03	&	-6.72E-07	&		\\
	&		&	1u	&	-21348.908207	&	-21348.908283	&&	2.06E-03	&	-1.95E-06	&		\\
	&		&	0u	&	-21348.898874	&	-21348.898949	&&	2.06E-03	&	-6.00E-06	&		\\
	&		&	0u	&	-21348.883557	&	-21348.883633	&&	2.07E-03	&	1.08E-05	&		\\
	&		&	2g	&	-21348.883223	&	-21348.883299	&&	2.07E-03	&	6.00E-06	&		\\
	&		&	1u	&	-21348.879740	&	-21348.879816	&&	2.07E-03	&	5.95E-06	&		\\
	&		&	1g	&	-21348.878818	&	-21348.878894	&&	2.07E-03	&	5.40E-06	&		\\
	&		&	0u	&	-21348.868504	&	-21348.868580	&&	2.07E-03	&	6.79E-06	&		\\
	&		&	0g	&	-21348.861167	&	-21348.861243	&&	2.07E-03	&	1.01E-05	&		\\
	&		&	0g	&	-21348.860747	&	-21348.860824	&&	2.07E-03	&	1.04E-05	&		\\
	&		&	1g	&	-21348.853322	&	-21348.853398	&&	2.06E-03	&	-9.03E-07	&		\\
\cline{2-9}															
I$_3$	&	2.93	&	3/2u	&	-21348.876116	&	-21348.876191	&&	2.04E-03	&	-1.78E-05	&		\\
	&		&	1/2g	&	-21348.869336	&	-21348.869411	&&	2.05E-03	&	-1.48E-05	&	2.99E-06	\\
	&		&	1/2u	&	-21348.852851	&	-21348.852926	&&	2.05E-03	&	-1.54E-05	&	2.45E-06	\\
	&		&	3/2g	&	-21348.849933	&	-21348.850008	&&	2.04E-03	&	-1.76E-05	&	2.72E-07	\\
\cline{2-9}															
I$_3^{2-}$	&	2.93	&	1/2u	&	-21348.941357	&	-21348.941433	&&	2.07E-03	&	7.48E-06	&		\\
	&		&	1/2u	&	-21348.899551	&	-21348.899627	&&	2.06E-03	&	2.04E-06	&	-5.44E-06	\\
	&		&	1/2g	&	-21348.890973	&	-21348.891049	&&	2.06E-03	&	-1.50E-06	&	-8.98E-06	\\
	&		&	1/2u	&	-21348.872185	&	-21348.872261	&&	2.06E-03	&	-6.86E-07	&	-8.16E-06	\\
\hline
\hline
\end{tabular}}
\label{tab:spectro-i3-dc-vs-2dcm}
\end{table}

\begin{table}[htb]
\caption{Atomic coordinates (in \AA) obtained from the geometry optimization 
at SO-ZORA/PBE/TZ2P level with the ADF code, for the CH$_2$I$_2$ and CH$_2$IBr
molecules.}
\begin{center}
\begin{tabular}{l l ccc}
\hline
\hline
        &      &  \multicolumn{3}{c}{Coordinates (\AA)}\\
\cline{3-5}
Species & Atom & X & Y & Z \\
\hline
CH$_2$I$_2$& C    &  0.000000   &    0.000000   &    0.216500 \\
           & H    &  0.000000   &   -0.905400   &    0.823100 \\
           & H    &  0.000000   &    0.905400   &    0.823100 \\
           & I    &  1.824300   &    0.000000   &   -0.931000 \\
           & I    & -1.824300   &    0.000000   &   -0.931000 \\
           &      &             &               &             \\ 
CH$_2$IBr  & C    &  0.094675   &    0.000000   &    0.175558 \\
           & H    &  0.103446   &   -0.906929   &    0.779625 \\
           & H    &  0.103446   &    0.906929   &    0.779625 \\
           & Br   &  1.698389   &    0.000000   &   -0.921358 \\
           & I    & -1.759397   &    0.000000   &   -0.932497 \\ 
\hline
\hline
\end{tabular}
\end{center}
\label{tab:xyz-halomethanes}
\end{table}%

\begin{table}[htb]
\caption{Interatomic distances (d, in \AA) and bond angles (a, in degrees) for CH$_2$I$_2$ and CH$_2$IBr,
obtained at SO-ZORA/PBE/TZ2P level, compared to the experimental results~\cite{Kudchadker:1975cp} where
available.}
\begin{center}
\begin{tabular}{lcc c lc}
\hline
\hline
\multicolumn{3}{c}{CH$_2$I$_2$} && \multicolumn{2}{c}{CH$_2$IBr} \\
\cline{1-3}\cline{5-6}
        & this work & Exp.~\cite{Kudchadker:1975cp} &&   &  this work \\
\hline
d(C-H)  & 1.09   & 1.09   &&  d(H1-C)    & 1.09   \\
d(C-I)  & 2.16   & 2.12   &&  d(H2-C)    & 1.09   \\
d(I-H)  & 2.69   &        &&  d(Br-C)    & 1.94   \\
d(I-I)  & 3.65   &        &&  d(I-C)     & 2.16   \\
        &        &        &&             &        \\
a(H-C-H)&  112.4 &  111.3 &&  d(H2-C-H1) & 112.7  \\
a(I-C-H)&  107.2 &        &&  d(Br-C-H1) & 107.8  \\
a(I-C-I)&  115.7 &  114.7 &&  d(Br-C-H2) & 107.8  \\
        &        &        &&  d(I-C-H1)  & 106.9  \\
        &        &        &&  d(I-C-H2)  & 106.9  \\
        &        &        &&  d(I-C-Br)  & 114.8  \\
\hline
\hline
\end{tabular}
\end{center}
\label{tab:bonds-angles-halomethanes}
\end{table}%